# TRANSACTION-ORIENTED SIMULATION IN

# AD HOC GRIDS

Gerald Krafft

This report is submitted in partial fulfilment of the requirements of the M.Sc. degree in Advanced Computer Science at the Westminster University.

Supervisor:   Vladimir Getov

Submitted on:   24[th] January 2007

# Abstract

Computer Simulation is an important area of Computer Science that is used in many other research areas like for instance engineering, military, biology and climate research. But the growing demand for more and more complex simulations can lead to long runtimes even on modern computer systems. Performing complex Computer Simulations in parallel, distributed across several processors or computing nodes within a network has proven to reduce the runtime of such complex simulations.

Large-scale parallel computer systems are usually very expensive. Grid Computing is a cost-effective way to perform resource intensive computing tasks because it allows several organisations to share their computing resources. Besides more traditional Computing Grids the concept of Ad Hoc Grids has emerged that offers a dynamic and transient resource-sharing infrastructure, suitable for short-term collaborations and with a very small administrative overhead to allow even small organisations or individual users to form Computing Grids. A Grid framework that fulfils the requirements of Ad Hoc Grids is ProActive.

This paper analyses the possibilities of performing parallel transaction-oriented simulations with a special focus on the space-parallel approach and discrete event simulation synchronisation algorithms that are suitable for transaction-oriented simulation and the target environment of Ad Hoc Grids. To demonstrate the findings a Java-based parallel transaction-oriented simulator is implemented on the basis of the promising Shock Resistant Time Warp synchronisation algorithm and using the Grid framework ProActive. The validation of this parallel simulator shows that the Shock Resistant Time Warp algorithm can successfully reduce the number of rolled back Transaction moves but it also reveals circumstances in which the Shock Resistant Time Warp algorithm can be outperformed by the normal Time Warp algorithm. The conclusion of this paper suggests possible improvements to the Shock Resistant Time Warp algorithm to avoid such problems.

# Table of Content

# **Abbreviations**

GFT        Global Furthest Time

GPSS      General Purpose Simulation System

GPW       Global Progress Window

GVT        Global Virtual Time

J2SE        Java 2 Platform, Standard Edition

JRE         Java Runtime Environment

JVM        Java Virtual Machine

LP          Logical Process

LPCC       Logical Process Control Component

LVT        Local Virtual Time

NAT        Network Address Translation

# Figures

# Tables

# 1    Introduction

Computer Simulation is one of the oldest areas in Computer Science. It provides answers about the behaviour of real or imaginary systems that otherwise could only be gained under great expenditure of time, with high costs or that could not be gained at all. Computer Simulation uses simulation models that are usually simpler than the systems they represent but that are expected to behave as analogue as possible or as required. The growing demand of complex Computer Simulations for instance in engineering, military, biology and climate research has also lead to a growing demand in computing power. One possibility to reduce the runtime of large, complex Computer Simulations is to perform such simulations distributed on several CPUs or computing nodes. This has induced the availability of high-performance parallel computer systems. Even so the performance of such systems has constantly increased, the ever-growing demand to simulate more and more complex systems means that suitable high-performance systems are still very expensive.

Grid computing promises to provide large-scale computing resources at lower costs by allowing several organisations to share their resources. But traditional Computing Grids are relatively static environments that require a dedicated administrative authority and are therefore less well suited for transient short-term collaborations and small organisations with fewer resources. Ad Hoc Grids provide such a dynamic and transient resource-sharing infrastructure that allows even small organisations or individual users to form Computing Grids. They will make Grid computing and Grid resources widely available to small organisations and mainstream users allowing them to perform resource intensive computing tasks like Computer Simulations.

There are several approaches to performing Computer Simulations distributed across a parallel computer system. The space-parallel approach [12] is one of these approaches that is robust, applicable to many different simulation types and that can be used to speed up single simulation runs. It requires the simulation model to be partitioned into relatively independent sub-systems that are then performed in parallel on several nodes. Synchronisation between these nodes is still required because the model sub-systems are not usually fully independent. A lot of past research has concentrated on different synchronisation algorithms for parallel simulation. Some of these are only suitable for certain types of parallel systems, like for instance shared memory systems.

This work investigates the possibility of performing parallel transaction-oriented simulation in an Ad Hoc Grid environment with the main focus on the aspects of parallel simulation. Potential synchronisation algorithms and other simulation aspects are analysed in respect of their suitability for transaction-oriented simulation and Ad Hoc Grids as the target environment and the chosen solutions are described and reasons for their choice given. A past attempt to investigate the parallelisation of transaction-oriented simulation was presented in [19] with the result that the synchronisation algorithm employed was not well suited for transaction-oriented simulation. Lessons from this past attempt have been learned and included in the considerations of this work. Furthermore this work outlines certain requirements that a Grid environment needs to fulfil in order to be appropriate for Ad Hoc Grids. The proposed solutions are demonstrated by implementing a Java-based parallel transaction-oriented simulator using the Grid middleware ProActive [15], which fulfils the requirements described before.

The specific simulation type transaction-oriented simulation was chosen because it is still taught at many universities and is therefore well known. It uses a relatively simple language for the modelling that does not require extensive programming skills and it is a special type of discrete event simulation so that most findings can also be applied to this wider simulation classification.

The remainder of this report is organised as follows. Section 2 introduces the fundamental concepts and terminology essential for the understanding of this work. In section 3 the specific requirements of Ad Hoc Grids are outlined and the Grid middleware ProActive is briefly described as an environment that fulfils these requirements. Section 4 focuses on the aspects of parallel simulation and their application to transaction-oriented simulation. Past research results are discussed, requirements for a suitable synchronisation algorithm outlined and the most promising algorithm selected. This section also addresses other points related to parallel transaction-oriented simulation like GVT calculation, handling of the simulation end, suitable cancellation techniques and the influence of the model partitioning. Section 5, which is the largest section of this report, describes the implementation of the parallel transaction-oriented simulator, starting from the initial implementation considerations and the implementation phases to specific details of the implementation and how the simulator is used. The functionality of the implemented parallel simulator is then validated in section 6 and the final conclusions presented in section 7.

# 2    Fundamental Concepts

This main section introduces the fundamental concepts and terminology essential for the understanding of this work. It covers areas like Grid Computing and the relation between the granularity of parallel algorithms and their expected target hardware architecture. It also describes the classification of simulation models as well as different approaches to parallel discrete event simulation and the main groups of synchronisation algorithms.

## 2.1  Grid Computing

The term "the Grid" first appeared in the mid-1990s in connection with a proposed distributed computing infrastructure for advanced science and engineering [9]. Today Grid computing is commonly used for a „distributed computing infrastructure that supports the creation and operation of virtual organizations by providing mechanisms for controlled, cross-organizational resource sharing" [9]. Similar to electric power grids Grid computing provides computational resources to clients using a network of multi organisational resource providers establishing a collaboration. In the context of Grid computing resource sharing means the "access to computers, software, data, and other resources" [9]. Control is needed for the sharing of resources that describes who is providing and who is consuming resources, what is shared and what are the conditions for the resource sharing to occur. These sharing rules and the group of organisations or individuals that are defined by it form a so-called Virtual Organisation (VO).

Grid computing technology has evolved and gone through several phases since it's beginning [9]. The first phase was characteristic for custom solutions to Grid computing problems. These were usually built directly on top of Internet protocols with limited functionality for security, scalability and robustness and interoperability was not considered to be important. From 1997 the emerging open source Globus Toolkit version 2 (GT2) became the de facto standard for Grid computing. It provided usability and interoperability via a set of protocols, APIs and services and was used in many Grid deployments worldwide. With the Open Grid Service Architecture (OGSA), which is a true community standard, came the shift of Grid computing towards a service-oriented architecture. In addition to a set of standard interfaces and services OGSA provides the framework in which a wide range of interoperable and portable services can be defined.

### 2.1.1 Ad Hoc Grids

Traditional computing Grids share certain characteristics [1]. They usually use a dedicated administrative authority, which often consists of a group of trained professionals to regulate and control membership and sharing rules of the Virtual Organisations. This includes administration of policy enforcement, monitoring and maintenance of the Grid resources. Well-defined policies are used for access privileges and the deployment of Grid applications and services.

It can be seen that these common characteristics are not ideal for a transient short-term collaboration with a dynamically changing structure because the administrative overhead for establishing and maintaining such a Virtual Organisation could outweigh its benefits [2]. Ad Hoc Grids provide this kind of dynamic and transient resource sharing infrastructure. According to [27] "An Ad Hoc Grid is a spontaneous formation of cooperating heterogeneous computing nodes into a logical community without a preconfigured fixed infrastructure and with only minimal administrative requirements". The transient dynamic structure of an Ad Hoc Grid means that new nodes can join or leave the collaboration at almost any time but Ad Hoc Grids can also contain permanent nodes.

Figure 1 [27] at the next page shows two example Ad Hoc Grids structures. Ad hoc Grid A is a collaboration of nodes from two organisations. It contains permanent nodes in form of dedicated high-performance computers but also transient nodes in form of non-dedicated workstations. Compared to this Ad hoc Grids B is an example for a more personal Grid system. It consists entirely of transient individual nodes. A practical example for the application of an Ad Hoc Grid is a group of scientists that for a specific scientific experiment want to collaborate and share computing resources. Using Ad Hoc Grid technology they can establish a short-term collaboration lasting only for the time of the experiment. These scientists might be part of research organisations but as the example of Ad hoc Grid B from Figure 1 shows Ad Hoc Grids allow even individuals to form Grid collaborations without the resources of large organisations. This way Ad Hoc Grids offer a way to more mainstream and personal Grid computing.

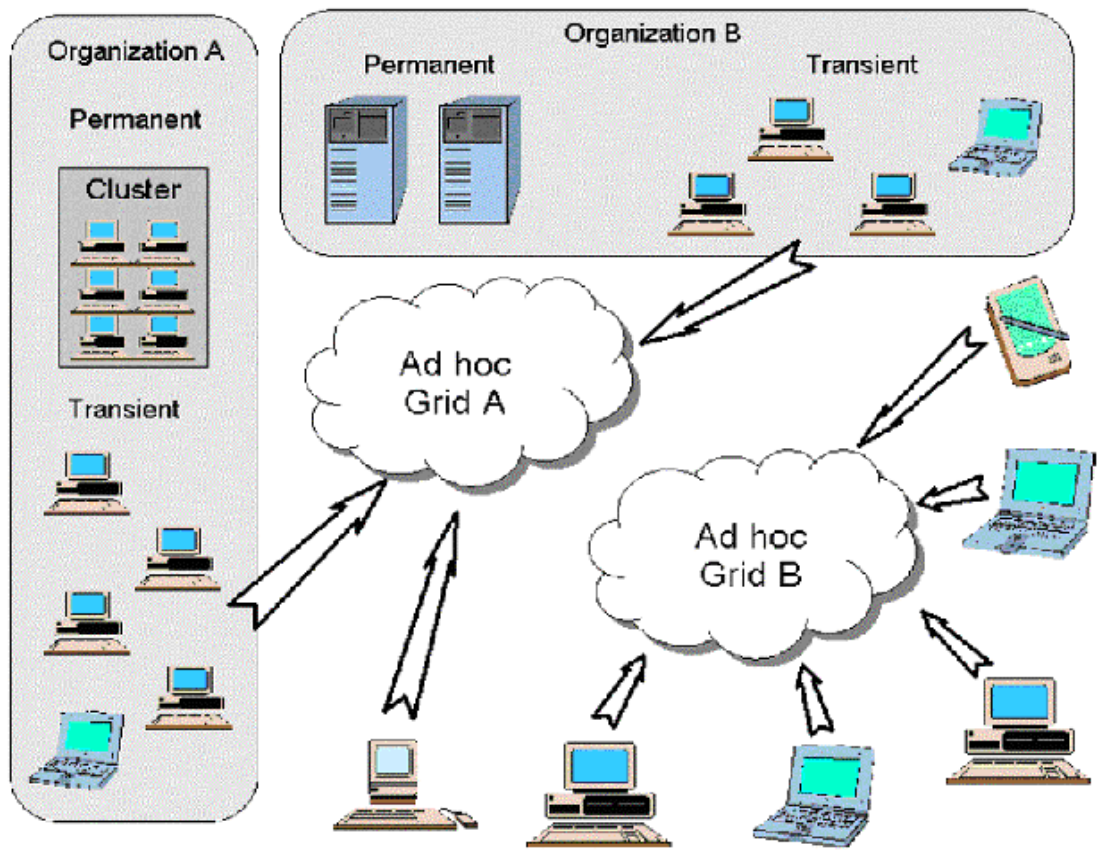

Figure 1: Ad Hoc Grid architecture overview [27]

## 2.2 Granularity and Hardware Architecture

When evaluating the suitability of different parallel algorithms for a specific parallel hardware architecture it is important to consider the granularity of the parallel algorithms and to compare the granularity to the processing and communication performance provided by the hardware architecture.

---

**Definition: granularity**

The granularity of a parallel algorithm can be defined as the ratio of the amount of computation to the amount of communication performed [18].

---

According to this definition parallel algorithms with a fine grained granularity perform a large amount of communication compared to the actual computation as apposed to parallel algorithms with a coarse grained granularity which only perform a small amount of communication compared to the computation. The following diagram illustrates the difference between fine grained and coarse grained granularity.

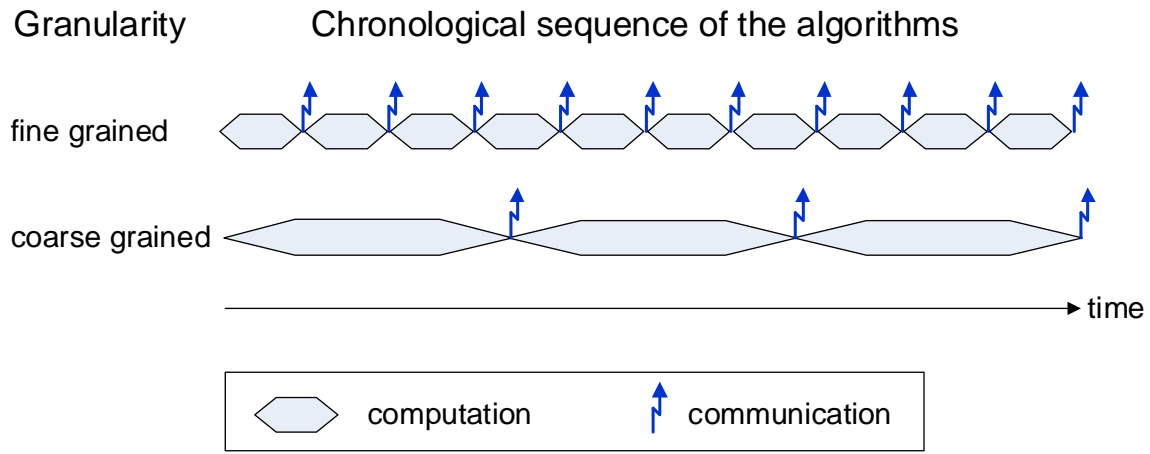

Figure 2: Comparison of fine grained and coarse grained granularity

Independent of the exact performance figures of a parallel hardware architecture it can be seen that for a hardware architecture with a high communication performance a fine grained parallel algorithm is well suited and that a hardware architecture with a low communication performance will require a coarse grained parallel algorithms [23].

## 2.3  Simulation Types

Simulation models are classified into continuous and discrete simulation according to when state transitions can occur [14]. Figure 3 below illustrates the classification of simulation types.

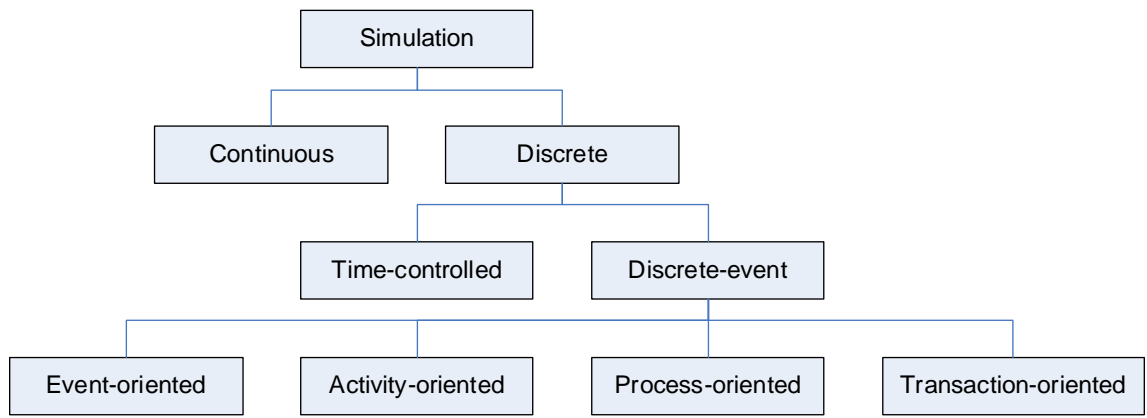

Figure 3: Classification of simulation types

In continuous simulation the state can change continuously with the time. This type of simulation uses differential equations, which are solved numerically to calculate the state. Continuous models are for instance used to simulate the streaming of liquids or gas.

Discrete simulation models allow the changing of the state only at discrete time intervals. They can be divided further according to whether the discrete time intervals are of fixed or variable length. In a time-controlled simulation model the time advances in fixed steps changing the state after each step as required.

But for many systems the state only changes in variable intervals, which are determined during the simulation. For these systems the discrete event simulation model is used. In discrete event simulation the state of the entities in the system is changed by events. Each event is linked to a specific simulated time and the simulation system keeps a list of events sorted by their time [35]. It then selects the next event according to its time stamp and executes it resulting in a change of the system state. The simulated time then jumps to the time of the next event that will be executed. The execution of an event can create new events with a time greater than the current simulated time that will be sorted into the event list according to their time stamp. Discrete event simulation is very flexible and can be applied to many groups of systems, which is why many general-purpose simulation systems use the discrete event model and a lot of research has gone into this model.

## 2.3.1 Transaction-Oriented Simulation and GPSS

A special case of the discrete event simulation is the transaction-oriented simulation. Transaction-oriented simulation uses two types of objects. There are stationary objects that make up the model of the system and then there are mobile objects called Transactions that move through the system and that can change the state of the stationary objects. The movement of a Transaction happens at a certain time (i.e. the time does not progress while the Transaction is moved), which is equivalent to an event in the discrete event model. But stationary objects can delay the movement of a Transaction by a random or fixed time. They can also spawn one Transaction into several or assemble several sub-Transactions back to one. The fact that transaction-oriented simulation systems are usually a bit simpler than full discrete event simulation systems makes them very useful for teaching purpose and academic use, especially because most discrete event simulation aspects can be applied to transaction-oriented simulation and vice versa.

The best-known transaction-oriented simulation language is GPSS, which stand for General Purpose Simulation System. GPSS was developed by Geoffrey Gordon at IBM around 1960 and has contributed important concepts to discrete event simulation. Later improved versions of the GPSS language were implemented in many systems, two of which are GPSS/H [36] and GPSS/PC. A detailed description of transaction-oriented simulation and the improved GPSS/H language can be found in [26].

## 2.4 Parallelisation of Discrete Event Simulation

Parallelisation of computer simulation is important because the growing performance of modern computer systems leads to a demand for the simulation of more and more complex systems that still result in excessive simulation time. Parallelisation reduces the time required for simulating such complex systems by performing different parts of the simulation in parallel on multiple CPUs or multiple computers within a network.

There are different approaches for the parallelisation of discrete event simulation that also cover different levels of parallelisation. One approach is to perform independent simulation runs in parallel [21]. There is only little communication needed for this approach, as it is limited to sending the model and a set of parameters to each node and collecting the simulation results after the simulation runs have finished. But this approach is relatively trivial and does not reduce the simulation time of a single simulation run. It can be used for simulations that consist of many shorter simulation runs. But these simulation runs have to be independent from each other (i.e. parameters for the simulation runs do not depend on results from each other).

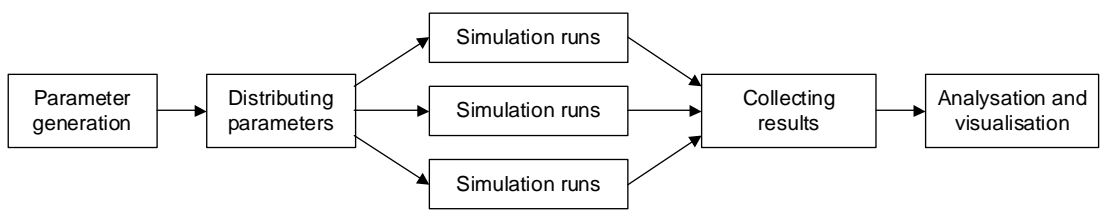

Figure 4: Parallelisation of independent simulation runs

Two other approaches for the parallelisation of discrete event simulation are the time-parallel approach and the space-parallel approach [12]. Both can be used to reduce the simulation time of single simulation runs. The time-parallel approach partitions the simulated time into intervals $[T_1, T_2]$, $[T_2, T_3]$, …, $[T_i, T_{i+1}]$. Each of these time intervals is then run on separate processors or nodes. This approach relies on being able to determine the starting state of each time interval before the simulation of the earlier time interval has been completed, e.g. it has to be possible to determine the state $T_2$ before the simulation of the time interval $[T_1, T_2]$ has been completed which is only possible for certain systems to be simulated, e.g. systems with state recurrences.

For the space-parallel approach the system model is partitioned into relatively independent sub-systems. Each of these sub-systems is then assigned and performed by a logical process (LP) with different LPs running on separate processors or nodes. In most cases these sub-systems will not be completely independent from each other, which is why the LPs will have to communicate with each other in order to exchange events. The space-parallel approach offers greater robustness and is applicable to most discrete event systems but the resulting speedup will depend on how the system is partitioned and how relative independent the resulting sub-systems are. A high dependency between the sub-systems will result in an increased synchronisation and communication overhead between the LPs. It will further depend on the synchronisation algorithm used.

## 2.5  Synchronisation Algorithms

The central problem for the space-parallel simulation approach is the synchronisation of the event execution. This synchronisation is also called time management. In discrete event simulation each event has a time stamp, which is the simulated time at which the event occurs. If two events are causal dependent on each other then they have to be performed in the correct order. Because causal dependent events could originate in different LPs synchronisation between the LPs becomes very important.

There are two main classes of algorithms for the event synchronisation between LPs, which are the classes of conservative and optimistic algorithms.

### 2.5.1 Conservative Algorithms

Conservative algorithms prevent that causal dependent events are executed out of order by executing only "safe" events [12]. An LP will consider an event to be "safe" if it is guaranteed that the LP cannot later receive an event with an earlier time stamp. The main task of conservative algorithms is to provide such guarantees so that LPs can determine which of the events are guaranteed and can be executed.

---

**Definition: guaranteed event**

An event e with the timestamp t which is to be executed in $LP_i$ is called guaranteed event if $LP_i$ knows all events with a timestamp t' < t that it will need to execute during the whole simulation.

---

One drawback of conservative algorithms is that LPs will have to wait or block if they don't have any "safe" events. This can even lead to deadlocks where all LPs are waiting for guarantees so that they can execute their events. Many of the conservative algorithms also require additional information about the simulation model like the communication topology[1] or lookahead[2] information. Further details about conservative algorithms can be found in [12], [34].

### 2.5.2 Optimistic Algorithms

Optimistic algorithms allow causal dependent events to be executed out of order first but they provide mechanisms to later recover from possible violations of the causal order. The best-known and most analysed optimistic algorithm is Time Warp [16] on which many other optimistic algorithms are based. In Time Warp an LP will first execute all events in its local event list but if it receives an event from another LP with a time stamp smaller than the ones already executed then it will rollback all its events that should have been executed after the event just received. State checkpointing[3] (also

---

[1] describes which LP can send events to which other LP

[2] is the models ability to predict the future course of events [5]

[3] the state of the simulation is saved into a state list together with the current simulation time after the execution of each event or in other defined intervals

known as state saving) is used in order to be able to rollback the state of the LP if required.

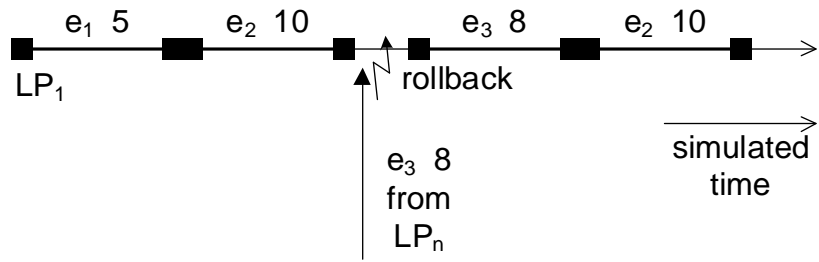

Figure 5: Event execution with rollback [21]

Figure 5 shows an example LP that performs the local events $e_1$ and $e_2$ with the time stamps of 5 and 10 but then receives another event $e_3$ with a time stamp of 8 from a different LP. At this point $LP_1$ will rollback the execution of event $e_2$ then execute the newly received event e3 and afterwards execute the event $e_2$ again in order to retain the causal order of the events. The rollback of already executed events can result in having to rollback events that have already been sent to other LPs. To archive this anti-events are sent to the same LPs like the original events, which will result in the original event being deleted if it has not been executed yet or in a rollback of the received event and all later events. These rollbacks and anti-events can lead to large cascaded rollbacks and many events having to be executed again. It is also possible that after the rollback the same events that have been rolled back are executed again sending out the same events to other LPs for which anti-events were sent during the rollback. In order to avoid this a different mechanism for cancelling events exists which is called *lazy cancellation* [13]. Compared to the original cancellation mechanism that is also called *aggressive cancellation* and was suggested by Jefferson [16], the *lazy cancellation* mechanism does not send out anti-events immediately during the rollback but instead keeps a history of the events sent that have been rolled back and only sends out anti-events when the event that was sent and rolled back is not re-executed. If for instance the LP is rolled back from the simulation time t' to the new simulation time t'' ≤ t' then the *lazy cancellation* mechanism will re-execute the events in the interval [t'',t'] and will only sent anti-events for events that had been sent during the first execution of that time interval but that were not generated during the re-execution. The difference between *aggressive* *cancellation* and *lazy cancellation* can be seen in the following diagram. In this diagram the event index is describing the scheduled time of the event.

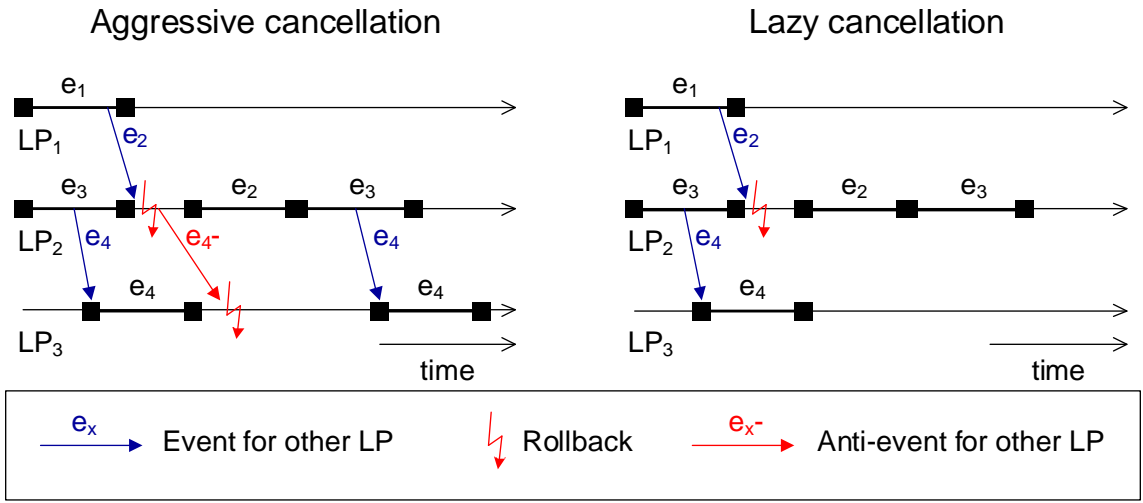

Figure 6: Comparison of aggressive and lazy cancellation

As shown in Figure 6 *lazy cancellation* can reduce cascaded rollbacks but it can also allow false events to propagate further and therefore lead to longer cascaded rollbacks when such false events are cancelled.

The concept of Global Virtual Time (GVT) is used to regain memory and to control the overall global progress of the simulation. The GVT is defined as the minimum of the local simulation time, also called Local Virtual Time (LVT), of all LPs and of the time stamps of all events that have been send but not yet processed by the receiving LP [16]. The GVT describes the minimum simulation time any unexecuted event can have at a particular point in real time. It therefore acts as a guarantee for all executed events with a time stamp smaller than the GVT, which can now be deleted. Further memory is freed by removing all state checkpoints with a virtual time less than the GVT except the one closest to the GVT.

Both conservative and optimistic algorithms have their advantages and disadvantages. The speedup of conservative algorithms can be limited because only guarantied events are executed. Compared to conservative algorithm optimistic algorithms can offer grater exploitation of parallelism [5] and they are less reliant on application specific information [11] or information about the communication topology. But optimistic algorithms have the overhead of maintaining rollback information and over-optimistic event execution in some LPs can lead to frequent and cascaded rollbacks and result in a degradation of the effective processing rate of events. Therefore research has focused on combining the advantages of conservative and optimistic algorithms creating so-called hybrid algorithms and on controlling the optimism in Time Warp. Such attempts to limit the optimism in Time Warp can be grouped into non-adaptive algorithms, adaptive algorithms with local state and adaptive algorithms with global state. Carl Tropper [34] and Samir R. Das [5] both give a good overview on algorithms in these categories.

The group of non-adaptive algorithms for instance contains algorithms that use time windows in order to limit how far ahead of the current GVT single LPs can process their events, which limits the frequency and length of rollbacks. Other algorithms in this group add conservative ideas to Time Warp. One example for this is the Breathing Time Buckets algorithm (also known as SPEEDES algorithm) [30]. Like Time Warp this algorithm executes all local events immediately and performs local rollbacks if required but it only sends events to other LPs that have been guaranteed by the current GVT and by doing so avoids cascaded rollbacks. The problem of all these algorithms is that either the effectiveness depends on finding the optimum value for static parameters like the window size or conservative aspects of the algorithm limit its effectiveness for models with certain characteristics. Finding the optimum value for such control parameters can be difficult for simulation modellers and many simulation models show a very dynamic behaviour of their characteristics, which would require different parameters at different times of the simulation.

Adaptive algorithms solve this problem by dynamically adapting the control parameters of the synchronisation algorithm according to "selected aspects of the state of the simulation" [24]. Some of these algorithms use mainly global state information like the Adaptive Memory Management algorithm [6], which uses the total amount of memory used by all LPs or the Near Perfect State Information algorithms [28] that are based on the availability of a reduced information set that almost perfectly describes the current global state of the simulation. Adaptive algorithms based on local state use only local information available to each LP in order to change the control parameters. They collect historic local state information and from these try to predict future local states and the required control parameter. Some examples for adaptive algorithms using local state information are Adaptive Time Warp [4], Probabilistic Direct Optimism Control [7] and the Shock Resistant Time Warp algorithm [8].

# 3    Ad Hoc Grid Aspects

There are certain requirements that a Grid environment needs to fulfil in order to be suitable for Ad Hoc Grids. These are outlined in this main section of the report and the Grid middleware ProActive [15] is chosen for the planned implementation of a Grid-based parallel simulator because it fulfils the requirements mentioned.

## 3.1  Considerations

In section 2.1.1 Ad Hoc Grids where described as dynamic, spontaneous and transient resource sharing infrastructures. The dynamic and transient structure of Ad Hoc Grids and the fact that Ad Hoc Grids should only have a minimal administrative overhead compared to traditional Grids creates special requirements that a Grid environment needs to fulfil in order to be suitable for Ad Hoc Grids. These requirements include automatic service deployment, service migration, fault tolerance and the discovery of resources.

### 3.1.1  Service Deployment

In a traditional Grid environment the deployment of Grid services is performed by an administrative authority that is also responsible for the usage policy and the monitoring of the Grid services. Grid services are usually deployed by installing a service factory onto the nodes. A service is then instantiated by calling the service factory for that service which will return a handle to the newly created service instance. In traditional Grid environments the deployment of service factories requires special access permissions and is performed by administrators.

Because of their dynamically changing structure Ad Hoc Grids need different ways of deploying Grid services that impose less administrative overhead. Automatic or hot service deployment has been suggested as a possible solution [10]. A Grid environment suitable for Ad Hoc Grids will have to provide means of installing services onto nodes either automatically or with very little administrative overhead.

### 3.1.2 Service Migration

Because Ad Hoc Grids allow a transient collaboration of nodes and the fact that nodes can join or leave the collaboration at different times a Grid application cannot rely on the discovered resources to be available for the whole runtime of the application. One solution to reach some degree of certainty about the availability of resources within an Ad Hoc Grid is the introduction of a scheme where individual nodes of the Grid guarantee the availability of the resources provided by them for a certain time as suggested in [1]. But such guaranties might not be possible for all transient nodes, especially for personal individual nodes as shown in the example Ad hoc Grid B in section 2.1.1.

Whether or not guarantees are used for the availability of resources an application running within an Ad Hoc Grid will have to migrate services or other resources from a node that wishes to leave the collaboration to another node that is available.

The migration of services or processes within distributed systems is a known problem and a detailed description can be found in [32]. A Grid environment for Ad Hoc Grids will have to support service migration in order to adapt to the dynamically changing structure of the Grid.

### 3.1.3 Fault Tolerance

Ad Hoc Grids can contain transient nodes like personal computer and there might be no guarantee for how long such nodes are available to the Grid application. In addition Ad Hoc Grids might be based on off-the-shelf computing and networking hardware that is more susceptible to hardware faults than special purpose build hardware.

A Grid environment suitable for Ad Hoc Grids will therefore have to provide mechanisms that offer fault tolerance and that can handle the loss of the connection to a node or the unexpected disappearing of a node in a manner that is transparent to Grid applications using the Ad Hoc Grid.

### 3.1.4 Resource Discovery

Resource discovery is one of the main tasks of Grid environments. It is often implemented by a special resource discovery service that keeps a directory of available resources and their specifications. But in an Ad Hoc Grid this becomes more of a challenge because of its dynamically changing structure. The task of the resource discovery can be divided further into the sub tasks of node discovery and node property assessment [27]. The node discovery task deals with the detection of new nodes that are joining and existing nodes that are leaving the collaboration. In an Ad Hoc Grid this detection has to be optimised towards the detection of frequent changes in the Grid structure. When a new node has joined the collaboration then its properties and shared resources will have to be discovered which is described by the node property assessment task. In addition to this high-level resource information some Grid environments also provide low-level resource information about the nodes. Such low-level resource information can include properties like the operating system type and available hardware resources. But depending on the abstraction level implemented by the Grid environment such low-level resource information might not be needed nor be accessible for Grid applications.

The minimum resource discovery functionality that an Ad Hoc Grid environment has to provide is the node discovery and more specifically the detection of new nodes joining the Grid structure and existing nodes that are leaving the structure.

## 3.2 ProActive

ProActive is a Grid middleware implemented in Java that supports parallel, distributed, and concurrent computing including mobility and security features within a uniform framework [15]. It is developed and maintained as an open-source project at INRIA[4] and uses the *Active Object* pattern to provide remotely accessible objects that can act as Grid services or mobile agents. Calls to such active objects can be performed asynchronous using a *future-based* synchronisation scheme known as *wait-by-necessity* for return values. A detailed documentation including programming tutorials as well as the full source code can be found at the ProActive Web site [15].

---

[4] Institut national de recherche en informatique et en automatique (National Institute for Research in Computer Science and Control)

ProActive was chosen as the Grid environment for the implementation of this project because it fulfils the specific requirements of Ad Hoc Grids as outlined in 3.1. As such it is very well suited for the dynamic and transient structure of Ad Hoc Grids and allows the setup of Grid structures with very little administrative overhead. The next few sections will briefly describe the features of ProActive that make it especially suited for Ad Hoc Grids.

## 3.2.1 Descriptor-Based Deployment

ProActive uses a deployment descriptor XML file to separate Grid applications and their source code from deployment related information. The source code of such Grid applications will only refer to virtual nodes. The actual mapping from a virtual node to real ProActive nodes is defined by the deployment descriptor file. When a Grid application is started ProActive will read the deployment descriptor file and will provide access to the actual nodes within the Grid application. The deployment descriptor file includes information about how the nodes are acquired or created. ProActive supports the creation of its nodes on physical nodes via several protocols, these include for instance *ssh*, *rsh*, *rlogin* as well as other Grid environments like *Globus Toolkit* or *glite*. Alternatively ProActive nodes can be started manually using the *startNode.sh* script provided. For the actual communication between Grid nodes, ProActive can use a variety of communication protocols like for instance *rmi*, *http* or *soap*. Even file transfer is supported as part of the deployment process. Further details about the deployment functionality provided by ProActive can be found in its documentation at the ProActive Web site [15].

## 3.2.2 Peer-to-Peer Infrastructure

ProActive provides a self-organising Peer-to-Peer functionality that can be used to discover new nodes, which are not defined within the deployment descriptor file of a Grid application. The only thing required is an entry point into an existing ProActive-based Peer-to-Peer network, for instance through a known node that is already part of that network. Further nodes from the Peer-to-Peer network can then be discovered and used by the Grid application. The Peer-to-Peer functionality of ProActive is not limited to sub-networks, it can communicate through firewalls and NAT routers and is therefore suitable for Internet-based Peer-to-Peer infrastructures. It is also self-organising which means that an existing Peer-to-Peer network tries to keep itself alive as long as there are nodes belonging to it.

### 3.2.3 Active Object Migration

In ProActive Active Objects can easily be migrated between different nodes. This can either be triggered by the Active Object itself or by an external tool. The migration functionality is based on standard Java serialisation, which is why Active Objects that need to be migrated and their private passive objects have to be serialisable. A detailed description of the migration functionality including examples can be found in the ProActive documentation.

### 3.2.4 Transparent Fault Tolerance

ProActive can provide fault tolerance to Grid applications that is fully transparent. Fault tolerance can be enabled for Grid applications just by configuring it within the deployment descriptor configuration. The only requirement is that Active Objects for which fault tolerance is to be enabled need to be serialisable.

There are currently two fault tolerance protocols provided by ProActive. Both protocols use checkpointing and are based on the standard Java serialisation functionality. Further details about how the fault tolerance works and how it is configured can be found in the ProActive documentation.

# 4 Parallel Transaction-oriented Simulation

## 4.1 Past research work

Past research performed by the author looked at the parallelisation of transaction-oriented simulation using an existing Matlab-based[5] GPSS simulator and Message-Passing for the communication [19]. It was shown that the Breathing Time Buckets algorithm, which is also known as SPEEDES algorithm (a description can be found in section 2.5.2), can be applied to transaction-oriented simulation. This algorithm uses a relatively simple communication scheme without anti-events and cancellations.

But further evaluation has revealed that the Breathing Time Buckets algorithm is not well suited for transaction-oriented simulation. The reason for this is that the Breathing Time Buckets algorithm makes use of what is know as the *event horizon* [29]. This event horizon is the time stamp of the earliest new event generated by the execution of the current events. Using this event horizon the Breathing Time Buckets algorithm can execute local events until it reaches the time of a new event that needs to be sent to another LP. At this point a GVT calculation is required because only events guaranteed by the GVT can be sent. The Breathing Time Buckets algorithm works well for discrete event models that have a large event horizon, i.e. where current events create new events that are relatively far in the future so that many local events can be executed before a GVT calculation is required. This is where the Breathing Time Buckets algorithm fails when it is applied to transaction-oriented simulation. In transaction-oriented simulation the simulation time does not change while a Transaction is moved. Whenever a Transaction moves from one LP to another this results in an event horizon of zero because the time stamp of the Transaction in the new LP will be the same like the time stamp it had in the LP from which it was sent. The validation of the parallel transaction-oriented simulator based on Breathing Time Buckets (alias SPEEDES) showed that a GVT calculation was required each time a Transaction needed to be sent to another LP.

---

[5] MATLAB is a numerical computing environment and programming language created by *The MathWorks Inc.*

The described past research comes to the conclusion that the Breathing Time Buckets algorithm does not perform well for transaction-oriented simulation but the research still provides some useful findings about the application of discrete event simulation algorithms to transaction-oriented simulation. Some of these findings that also apply to this work are outlined in the following sections.

### 4.1.1 Transactions as events

An event can be described as a change of the state at a specified time. From the simulation perspective this change of state is always caused by an action (e.g. the execution of an event procedure). Therefore an event can also be seen as an action that is performed at a specific point in time. In transaction-oriented simulation the state of the simulation system is changed by the execution of blocks through Transactions. Transactions are moved from block to block at a specific point in time as long as they are movable, i.e. not advanced and not terminated. Therefore this movement of a Transaction for a specific point in time and as long as the Transaction is movable describes an action, which is equivalent to the event describing an action in the discrete event model.

Considering this equivalence it is generally possible to apply synchronisation algorithms and other techniques for discrete event simulation also to transaction-oriented simulation. But because transaction-oriented simulation has specific properties certain algorithms are more and other less well suited for transaction-oriented simulation.

### 4.1.2 Accessing objects in other LPs

In a simulation that performs partitions of the simulation model on different LPs it is possible that the simulation of the model partition within one LP needs to access an object in another LP. For instance this could be a TEST block within one LP trying to access the properties of a STORAGE entity within another LP. The main problem for accessing objects like this in other LPs is that at a certain point of real time each LP can have a different simulation time. Figure 7 shows an example for this problem. In this example the $LP_1$ that contains object $o_1$ has already reached the simulation time 12 and the $LP_2$, which is trying to access the object $o_1$ has reached the simulation time 5. It can be seen that event $e_4$ from $LP_2$ would potentially read the wrong value for object $o_1$ because this read access should happen at the simulation time 5 which is before the event $e_2$ at $LP_1$ overwrote the value of $o_1$. Instead event $e_4$ reads the value of $o_1$ as it appears at the simulation time 12.

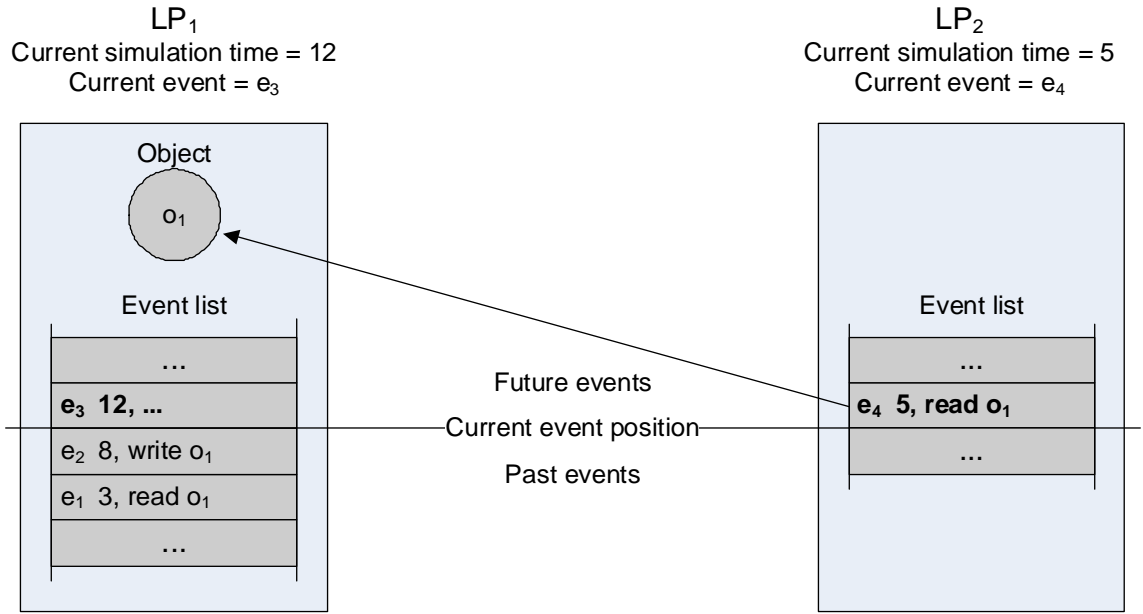

Figure 7: Accessing objects in other LPs

Because accessing an object within another LP is an action that is linked to a specific point of simulation time it can also be viewed as an event according to the description of events in section 4.1.1. Like other events they have to be executed in the correct causal order. This means that event $e_4$ that is reading the value of object o1 has to happen at the simulation time 5. Sending this event to $LP_1$ would cause $LP_1$ to roll back to the simulation time 5 so that $e_4$ would read the correct value.

Treating the access to objects as a special kind of event solves the problem mentioned above. Such a solution can also be applied to transaction-oriented simulation by implementing a simulation scheduler that besides Transactions can also handle these kinds of object access events. Alternatively the object access could be implemented as a pseudo Transaction that does not point to its next block but instead to an access method that when executed performs the object access and for a read access returns the value. Such a pseudo Transaction would send the value back to the originating LP and then be deleted. Depending on the synchronisation algorithm it can also be useful treat read and write access differently. If for instance an optimistic synchronisation algorithm is used that saves system states for possible required rollbacks then rollbacks as a result of object read access events can be avoided if the LP that contains the object has passed the time of the read access. In this case the object value for the required point in simulation time could be read from the saved system state instead of rolling back the whole LP.

Another even simpler solution to the problem of accessing objects in other LPs is to prevent the access of objects in other LPs all together, i.e. to allow only access to objects within the same LP. This sounds like a contradiction but by preventing one LP from accessing local objects in another LP the event that wants to access a particular object needs to be moved to the LP that holds that object. For the example from Figure 7 this means that instead of synchronizing the object access from event $e_4$ on $LP_2$ to object $o_1$ held by $LP_1$ the event $e_4$ is moved to $LP_1$ that holds object $o_1$ so that accessing the object can be performed as a local action. This solution reduces the problem to the general problem of moving events and synchronisation between LPs as solved by discrete event synchronisation algorithms (see section 2.5).

## 4.1.3 Analysis of GPSS language

A synchronisation strategy is a requirement for parallel discrete event simulation because LPs cannot predict the correct causal order of the events they will execute as they can receive further events from other LPs at any time. When applying discrete event synchronisation algorithms to transaction-oriented simulation based on GPSS/H it is first of interest to analyse which of the GPSS/H blocks[6] can actually cause the transfer of Transactions to another LP or which of them require access to objects that might be located at a different LP. Because Transactions usually move from one block to the next a transfer to a different LP can only be the result of a block that causes the execution of a Transaction to jump to a different block than the next following including blocks that can cause a conditional branching of the execution path. The following two tables list GPSS/H blocks that can change the execution path of a Transaction or that access other objects within the model.

---

[6] A detailed description of the GPSS/H language and its block types can be found in [26].

**Blocks that can change the execution path**

| Block | Change of execution path |
|---|---|
| TRANSFER | Jump to specified block |
| SPLIT | Jump of Transaction copy to specified block |
| GATE | Jump to specified block depending on Logic Switch |
| TEST | Jump to specified block depending on condition |
| LINK | Jump to specified block depending on condition |
| UNLINK | Jump of the unlinked Transactions to specified block and possible jump of Transaction causing the unlink operation |

Table 1: Change of Transaction execution path

**Blocks that can access objects**

| Block | Access to object |
|---|---|
| SEIZE<br>RELEASE<br>GATE | Access to Facility object |
| ENTER<br>LEAVE<br>GATE | Access to Storage object |
| QUEUE<br>DEPART | Access to Queue object |
| LOGIC<br>GATE | Access to Logic Switch |
| LINK<br>UNLINK | Access to User Chain |
| TERMINATE | Access to Termination Counter |

Table 2: Access to objects

## 4.2  Synchronisation algorithm

An important conclusion from section 4.1 is that the choice of synchronisation algorithm has a large influence on how much of the parallelism that exists in a simulation model can be utilised by the parallel simulation system. A basic overview of the classification of synchronisation algorithms for discrete event simulation was given in section 2.5. Conservative algorithms utilise the parallelism less well than optimistic algorithms because they require guarantees, which are often derived from additional knowledge about the behaviour of the simulation model, like for instance the communication topology or lookahead attributes of the model. For this reason conservative algorithms are often used to simulate very specific systems where such knowledge is given or can easily be derived from the model. For general simulation systems optimistic algorithms are better suited as they can utilise the parallelism within a model to a higher degree without requiring any guarantees or additional knowledge.

Another important aspect for choosing the right synchronisation algorithm is the relation between the performance properties of the expected parallel hardware architecture and the granularity of the parallel algorithm as outlined in section 2.2. In order for the parallel algorithm to perform well in general on the target hardware environment the granularity of the algorithm, i.e. the ratio between computation and communication has to fit the ratio of the computation performance and communication performance of the parallel hardware.

The goal of this work is to provide a basic parallel transaction-oriented simulation system for Ad Hoc Grid environments. Ad Hoc Grids can make use of special high performance hardware but more likely will be based on standard hardware machines using Intranet or Internet as the communication channel. It can therefore be expected that Ad Hoc Grids will mostly be targeted at parallel systems with reasonable computation performance but relatively poor communication performance.

### 4.2.1 Requirements

Considering the target environment of Ad Hoc Grids and the goal of designing and implementing a general parallel simulation system based on the transaction-oriented simulation language GPSS it can be concluded that the best suitable synchronisation algorithm is an optimistic or hybrid algorithm that has a coarse grained granularity. The algorithm should require only little communication compared to the amount of computation it performs. At the same time the algorithm should be flexible enough to adapt to a changing environment, as this is the case in Ad Hoc Grids. A further requirement is that the algorithm can be adapted to and is suitable for transaction-oriented simulation. Finding such an algorithm is a condition for achieving the outlined goals.

## 4.2.2 Algorithm selection

Most optimistic algorithms are based on the Time Warp algorithm but attempt to limit the optimism. As described in section 2.5.2 these algorithms can be grouped into non-adaptive algorithms, adaptive algorithms with local state and adaptive algorithms with global state. Non-adaptive algorithms usually rely on external parameters (e.g. the window size for window based algorithms) to specify how strongly the optimism is limited. Such algorithms are not ideal for a general simulation system as it can be difficult for a simulation modeller to find the optimum parameters for each simulation model. It is also common that simulation models change their behaviour during the runtime of the simulation.

As a result later research has focused more on the adaptive algorithms, which qualify for a general simulation system. They are also better suited for dynamically changing environments like Ad Hoc Grids.

Two interesting adaptive algorithms are the Elastic Time algorithm [28] and the Adaptive Memory Management algorithm [6]. The Elastic Time algorithm is based on Near Perfect State Information (NPSI). It requires a feedback system that constantly receives input state vectors from all LPs, processes these using several functions and then returns output vectors to all LPs that describe how the optimism of each LP needs to be controlled. As described in [28] for a shared memory system such a near-perfect state information feedback system can be implemented using a dedicated set of processes and processors but for a distributed memory system a high speed asynchronous reduction network would be needed. This shows that the Elastic Time algorithm is not suited for a parallel simulation system based on Grid environments where communication links between nodes might use the Internet and nodes might not be physically close to each other. Similar to the Elastic Time algorithm the Adaptive Memory Management algorithm is also best suited for shared memory systems. The Adaptive Memory Management algorithm is based on the link between optimism and memory usage in optimistic algorithms. The more over-optimistic an LP is the more memory does it use to store the executed events and state information which cannot be committed and fossil collected as they are far ahead of the GVT. It is shown that by limiting the overall memory available to the optimistic simulation artificially, the optimism can also be controlled. For this the Adaptive Memory Management algorithm uses a shared memory pool providing the memory used by all LPs. The algorithm then dynamically changes the size of the memory pool and therefore the total amount of memory available to the simulation based on several parameters like frequency of rollbacks, fossil collections and cancel backs in order to find the optimum amount of memory for the best performance. The required shared memory pool can easily be provided in a shared memory system but in a distributed memory system implementing it would require extensive synchronisation and communication between the nodes which makes this algorithm unsuitable for this work.

An algorithm that is more applicable to Grid environments as it does not need a shared memory or a high speed reduction network is the algorithm suggested in [33]. This algorithm uses a Global Progress Window (GPW) described by the GVT and the Global Furthest Time (GFT). Because the GVT is equivalent to the LVT of the slowest LP and the GFT is the LVT of the LP furthest ahead in simulation time the GPW represents the window in simulation time in which all LPs are located. This time window is then divided further into the slow zone, the fast zone and the hysteresis zone as shown in Figure 8.

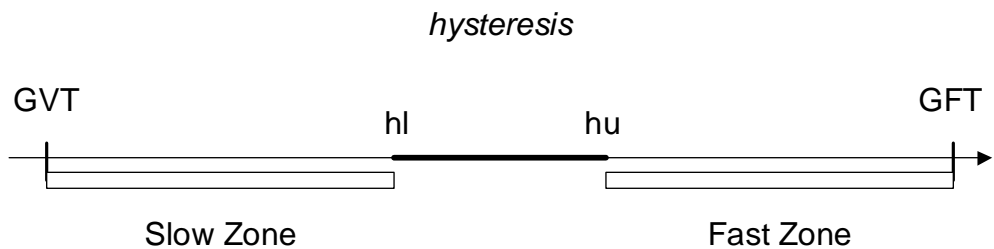

Figure 8: Global Progress Window with its zones [33]

The algorithm will slow down LPs in the fast zone and try to accelerate LPs in the slow zone with the hysteresis zone acting as a buffer between the other two zones. This algorithm could be implemented without any additional communication overhead because the GFT can be determined and passed back to the LPs by the same process that performs the GVT calculation. It is therefore well suited for very loosely coupled systems based on relatively slow communication channels. The only small disadvantage is that similar to many other algorithms the fast LPs will always be penalized even if they don't actually contribute to the majority of the cascaded rollback. In [33] the authors also explore how feasible LP migration and load balancing is for reducing the runtime of a parallel simulation.

The most promising algorithm regards the requirements outlined in 4.2.1 is the Shock Resistant Time Warp algorithm [8]. This algorithm follows similar ideas like the Elastic Time algorithm and the Adaptive Memory Management algorithm mentioned above but at the same time is very different. Similar to the Elastic Time algorithm state vectors are used to describe the current states of all LPs plus a set of functions to determine the output vector but the Shock Resistant Time Warp algorithm does not require a global state. Instead each LP tries to optimise its parameters towards the best performance. And similar to the Adaptive Memory Management algorithm the optimism is controlled indirectly be setting artificial memory limits but each LP will artificially limit its own memory instead of using an overall memory limit for the whole simulation.

The Shock Resistant Time Warp algorithm was chosen for the implementation of the parallel transaction-oriented simulator because it promises to be very adaptable and at the same time is very flexible regards changes in the environment and it does not create any additional communication overhead compared to Time Warp. The following section will describe this algorithm in more detail.

## 4.2.3  Shock resistant Time Warp Algorithm

The Shock Resistant Time Warp algorithm [8] is a fully distributed approach to controlling the optimism in Time Warp LPs that requires no additional communication between the LPs. It is based on the Time Warp algorithm but extends each LP with a control component called LPCC that constantly collects information about the current state of the LP using a set of sensors. These sets of sensor values are then translated into sets of indicator values representing state vectors for the LP. The LPCC will keep a history of such state vectors so that it can search for past state vectors that are similar to the current state vector but provide a better performance indicator. An actuator value will be derived from the most similar of such state vectors that is subsequently used to control the optimism of the LP. Figure 9 gives an overview of the interaction between LPCC and LP.

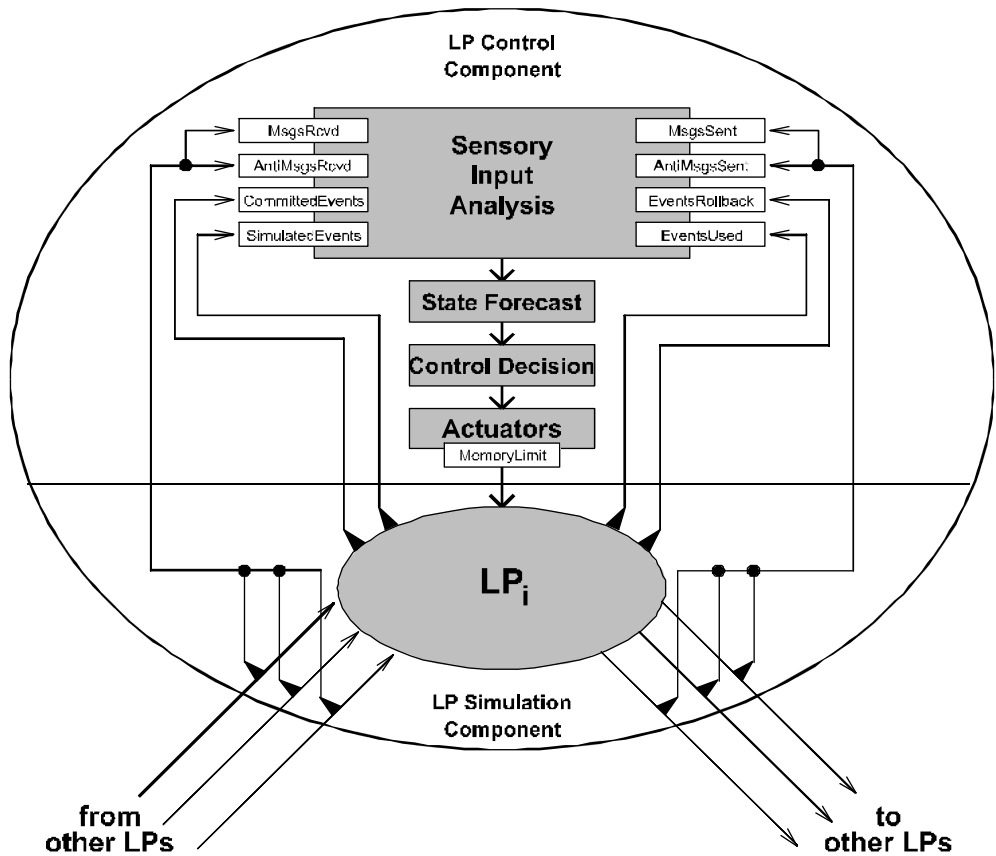

Figure 9: Overview of LP and LPCC in Shock Resistant Time Warp [8]

The specific sensors used by the LPCC are described in Table 3 but other or additional sensors could be used if appropriate. There are two types of sensors. The *point sample sensors* describe a momentary value of a performance or state metric, which can fluctuate significantly whereas the *cumulative sensors* characterise metrics that contain a sum value produced over the runtime of the simulation. The indicator for each sensor is calculated depending on which type of sensor it is. For cumulative sensors the rate of increase over a specified time period is used as the indicator values and for point sample sensors the arithmetic mean value over the same time period. Table 4 shows the corresponding indicators.

| Sensor | Type | Description |
|---|---|---|
| CommittedEvents | cumulative | total number of events committed |
| SimulatedEvents | cumulative | total number of events simulated |
| MsgsSent | cumulative | total number of (positive) messages sent |
| AntiMsgsSent | cumulative | total number of anti-messages sent |
| MsgsRcvd | cumulative | total number of (positive) messages received |
| AntiMsgsRcvd | cumulative | total number of anti-messages received |
| EventsRollback | cumulative | total number of events rolled back |
| EventsUsed | point sample | momentary number of events in use |

Table 3: Shock Resistant Time Warp sensors

| Indicator | Description |
|---|---|
| EventRate | number of events committed per second |
| SimulationRate | number of events simulated per second |
| MsgsSentRate | number of (positive) messages sent per second |
| AntiMsgsSentRate | number of anti-messages sent per second |
| MsgsRcvdRate | number of (positive) messages received per second |
| AntiMsgsRcvd | number of anti-messages received per second |
| EventsRollbackRate | number of events rolled back per second |
| MemoryConsumption | average number of events in use |

Table 4: Shock Resistant Time Warp indicators

Two of these indicators are slightly special. The EventRate indicator, which describes the number of events committed per second during a time period, is the performance indicator used to identify how much useful work has been performed. And the actuator value MemoryLimit is derived from the MemoryConsumption indicator. For a state vector with n different indicator values the LPCC will use an n-dimensional state vector space to store and compare the state vectors. The similarity of two state vectors within this state vector space is characterised by the Euclidean distance between the vectors. When searching for the most similar historic state vector that has a higher performance indicator then the Euclidean distance is calculated by ignoring the indicators EventRate and MemoryConsumption because the EventRate is the indicator that the LPCC is trying to optimise and the MemoryConsumption is directly linked to the MemoryLimit actuator controlled by the LPCC.

Keeping a full history of the past state vectors would require a large amount of memory and would create an exponentially increasing performance overhead. For these reasons the Shock Resistant Time Warp algorithm uses a clustering mechanism to cluster similar state vectors. The algorithm will keep a defined number of clusters. At first each new state is stores as a new cluster but when the cluster limit is reached then new states are added to existing clusters if the distance between the state and the cluster is smaller than any of the inter-cluster distances and otherwise the two closest clusters are merged into one and the second cluster is replaced with the new state vector. The clustering mechanism limits the total number of clusters stored and at the same time clusters will move their location within the state space to reflect the mean position of the state vectors they represent.

The Shock Resistant Time Warp algorithm as described in [8] is specific to discrete event simulation but it can also be applied to transaction-oriented simulation because of the equivalence between events in discrete event simulation and the movement of Transactions in transaction-oriented simulation as outlined in 4.1.1. Because the transaction-oriented simulation does not know events as such the names of the sensors and indicators described above need to be changed to avoid confusion when applying the Shock Resistant Time Warp algorithm to transaction-oriented simulation. The two tables below show the sensor and indicator names that will be used for this work.

| Discrete event sensor | Transaction-oriented sensor |
|---|---|
| CommittedEvents | CommittedMoves |
| EventsUsed | UncommittedMoves |
| SimulatedEvents | SimulatedMoves |
| MsgsSent | XactsSent |
| AntiMsgsSent | AntiXactsSent |
| MsgsRcvd | XactsReceived |
| AntiMsgsRcvd | AntiXactsReceived |
| EventsRollback | MovesRolledback |

Table 5: Transaction-oriented sensor names

| Discrete event indicator | Transaction-oriented indicator |
|---|---|
| EventRate | CommittedMoveRate |
| MemoryConsumption | AvgUncommittedMoves |
| SimulationRate | SimulatenRate |
| MsgsSentRate | XactSentRate |
| AntiMsgsSentRate | AntiXactSentRate |
| MsgsRcvdRate | XactReceivedRate |
| AntiMsgsRcvd | AntiXactReceivedRate |
| EventsRollbackRate | MovesRolledbackRate |

Table 6: Transaction-oriented indicator names

## 4.3 GVT Calculation

The concept of Global Virtual Time (GVT) was mentioned and briefly explained in 2.5.2. GVT is a fundamental concept of optimistic synchronisation algorithms and describes a lower bound on the simulation times of all LPs. Its main purpose is to guarantee past simulation states as being correct so that the memory for these saved states can be reclaimed through fossil collection. Another important purpose is to determine the overall progress of the simulation, which includes the detection of the simulation end. Besides these reasons optimistic parallel simulations can often run without any additional GVT calculations for long time periods or even until they reach the simulation end if enough memory for the required state saving is available. In environments with a relatively low communication performance like Computing Grids it is desirable to minimise the need for GVT calculations because the GVT calculation process is based on the exchange of messages and adds a communication overhead.

The best-known GVT calculation algorithm was suggested by Jefferson [16]. It defines the GVT as the minimum of all local simulation times and the time stamps of all events sent but not yet acknowledged as being handled by the receiving LP. The planned parallel simulator will use this algorithm for the GVT calculation because it is relatively easy to implement and well studied. Future work could also look at alternative GVT algorithms that might be suitable for Grid environments, like the one suggested in [20].

The movement of a Transaction in transaction-oriented simulation can be seen as equivalent to an event being executed in discrete event simulation as concluded in 4.1.1. But in transaction-oriented simulation the causal order is not only determined by the movement time of a Transaction but also by its priority because if several Transactions exist that have the same move time then they are moved through the system in order of their priority, i.e. Transactions with higher priority first. As a result the priority had to be included in the GVT calculation in [19] because the Breathing Time Buckets algorithm (SPEEDES algorithm) used there needs the GVT to guarantee outgoing Transactions.

For a parallel transaction-oriented simulator based on the Time Warp algorithm or the Shock Resistant Time Warp algorithm it is not necessary to include the Transaction priority in the GVT calculation because the GVT is only used to determine the progress of the overall simulation and to regain memory through fossil collection. For the Shock Resistant Time Warp algorithm one additional use of the GVT is to determine realistic values for the *CommittedEvents* sensor. Events are committed when receiving a GVT that is greater than the event's time. As a result the number of committed events during a certain period of time is only known if GVT calculations have been performed. The suggested parallel simulator based on the Shock Resistant Time Warp algorithm will therefore synchronise the processing of its LPCC with GVT calculations.

## 4.4 End of Simulation

In transaction-oriented simulation a simulation is complete when the defined end state is reached, i.e. the termination counter reaches a value less or equal to zero. When using an optimistic synchronisation algorithm for the parallelisation of transaction-oriented simulation it is crucial to consider that optimistic algorithms will first execute all local events without guarantee that the causal order is correct. They will recover from wrong states by performing a rollback if it later turns out that the causal order was violated. Therefore any local state reached by an optimistic LP has to be considered provisional until a GVT has been received that guarantees the state. In addition it needs to be considered that at any point in real time it is most likely that each of the LPs has reached a different local simulation time so that after an end state has been reached by one of the LPs that is guaranteed by a GVT it is important to synchronise the states of all LPs so that the combined end state from all model partitions is equivalent to the model end state that would have been reached in a sequential simulator.

To summarise, a parallel transaction-oriented simulation based on an optimistic algorithm is only complete when the defined end state has been reached in one of the LPs and when this state has been confirmed by a GVT. Furthermore if the confirmed end of the simulation has been reached by one of the LPs then the states of all the other LPs need to be synchronised so that they all reflect the state that would exist within the model when the Transaction causing the simulation end executed its TERMINATE block. These significant aspects regarding the simulation end of a parallel transaction-oriented simulation that had not been considered in [19].

A mechanism is suggested for this work that leads to a consistent and correct global end state of the simulation considering the problems mentioned above. For this mechanism the LP reaching a provisional end state is switched into the *provisional end mode*. In this mode the LP will stop to process any further Transactions leaving the local model partition in the same state but it will still respond to and process control messages like GVT parameter requests and it will receive Transactions from other LPs that might cause a rollback. The LP will stay in this provisional end mode until the end of the simulation is confirmed by a GVT or a received Transaction causes a rollback with a potential re-execution that is not resulting in the same end state. While the LP is in the provisional end mode additional GVT parameters are passed on for every GVT calculation denoting the fact that a provisional end state has been reached and the simulation time and priority of the Transaction that caused the provisional end. The GVT calculation process can then assess whether the earliest current provisional end state is guaranteed by the GVT. If this is the case then all other LPs are forced to synchronise to the correct end state by rolling back using the simulation time and priority of the Transaction that caused the provisional end and the simulation is stopped.

## 4.5  Cancellation Techniques

Transaction-oriented simulation has some specific properties compared to discrete event simulation. One of these properties is that Transactions do not consume simulation time while they are moving from block to block. This has an influence on which of the synchronisation algorithms are suitable for transaction-oriented simulation as described in 4.1 but also on the cancellation techniques used. If a Transaction moves from $LP_1$ to $LP_2$ then it will arrive at $LP_2$ with the same simulation time that it had at $LP_1$. A Transaction moving from one LP to another is therefore equivalent to an event in discrete event simulation that when executed creates another event for the other LP with exactly the same time stamp. Because simulation models can contain loops as it is common for the models of quality control systems where an item failing the quality control needs to loop back through the production process (see [26] for an example) this specific behaviour of transaction-oriented simulation can lead to endless rollback loops if *aggressive cancellation* is used (cancellation techniques were briefly described in 2.5.2).

The example in Figure 10 demonstrates this effect. It shows the movement of a Transaction $x_1$ from $LP_1$ to $LP_2$ but without a delay in simulation time the Transaction is transferred back to $LP_1$. As a result $LP_1$ will be rolled back to the simulation time just before $x_1$ was moved. At this point two copies of Transaction $x_1$ will exist in $LP_1$. The first one is $x_1$ itself which needs to be moved again and the second is $x_1$' which is the copy that was send back from $LP_2$. This is the point from where the execution differs between lazy cancellation and aggressive cancellation. In lazy cancellation $x_1$ would be moved again resulting in the same transfer to $LP_2$. But because $x_1$ was sent to $LP_1$ already it will not be transferred again and no anti-transaction will be sent. From here $LP_1$ just proceeds moving the Transactions in its Transaction chain according to their simulation time (Transaction priorities are ignored for this example). Apposed to that the rollback in aggressive cancellation would result in an anti-Transaction being sent out for $x_1$ immediately which would cause a second rollback in $LP_2$ and another anti-Transaction for $x_1$' being sent back to $LP_1$. At the end both LPs will end up in the same state in which they were before $x_1$ was moved by $LP_1$. The same cycle of events would start again without any actual simulation progress.

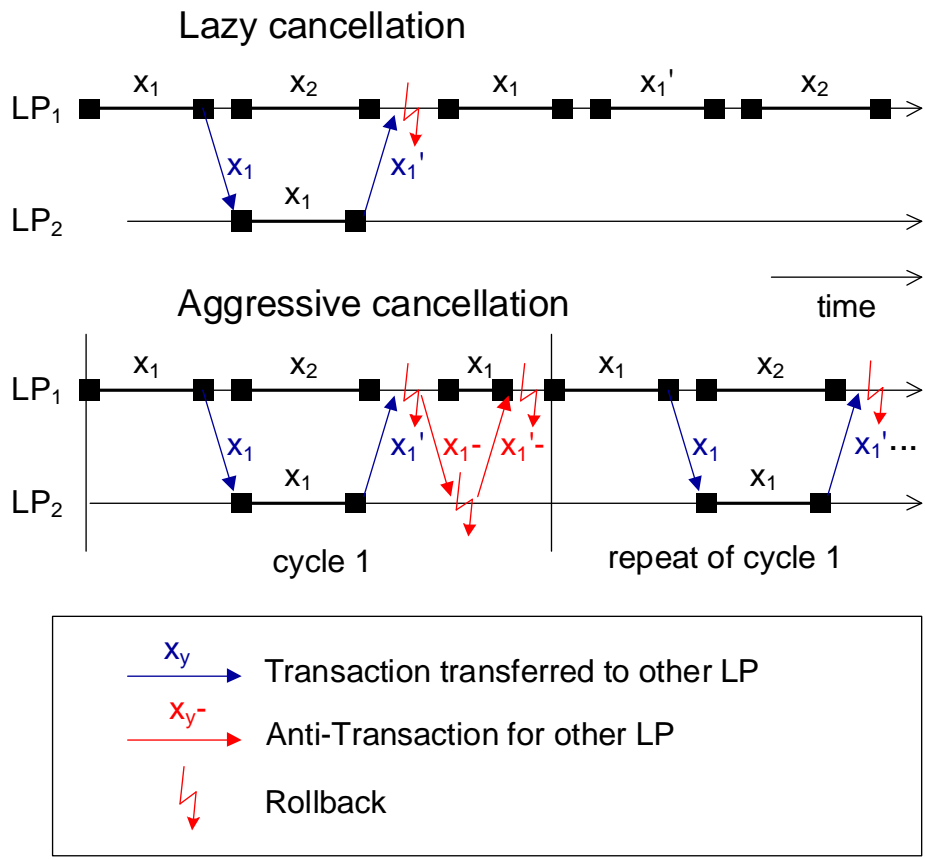

Figure 10: Cancellation in transaction-oriented simulation

It can therefore be concluded that lazy cancellation needs to be used for a parallel transaction-oriented simulation based on an optimistic algorithm in order to avoid such endless loops.

## 4.6 Load Balancing

Load balancing and the automatic migration of slow LPs to nodes or processors that have a lighter work load has been suggested in order to reduce the runtime of parallel simulations. This has also been explored by the authors of [33]. They concluded that the migration of LPs involves a "substantial amount of overheads in saving the process context, flushing the communication channels to prevent loss of messages". And especially on loosely coupled systems with relatively slow communication channels sending the full process context of the LP from one node to another can add a significant performance penalty to the overall simulation. This penalty would depend on the size of the process context as well as the communication performance between the nodes involved in the migration. The gained performance on the other hand depends on the difference in processing performance and other workload on these nodes. To determine reliably when such an automatic migration is beneficial within a loosely coupled, dynamically changing Ad Hoc Grid environment would be difficult and it is likely that the performance penalty outweighs the gains.

This work will therefore not investigate the load balancing and automatic LP migration for performance reasons but only support automatic LP migration as part of the fault tolerance functionality provided by ProActive and described in 3.1.3. Manual LP migration will be supported by the parallel simulator using ProActive tools.

## 4.7  Model Partitioning

Besides the chosen synchronisation algorithm the partitioning of the simulation model also has a large influence on the performance of the parallel simulation because the communication required between the Logical Processes depends to a large degree on how independent the partitions of a simulation model are. Looking at the requirements of a general-purpose transaction-oriented simulation system for Ad Hoc Grid environments in 4.2 the conclusion was drawn that the required communication needs to be kept to a minimum in order to reach acceptable performance results through parallelisation in such environments. The communication required for the exchange of Transactions between the Logical Processes is part of this overall communication.

A simulation model that is supposed to be run in a Grid based parallel simulation system therefore needs to be partitioned in such a way that the expected amount of Transactions moved within the partitions is significantly larger than the amount of Transactions that need to be transferred between these partitions.  This means that Grid based parallel simulation systems are best suited for the simulation of systems that contain relatively independent sub-systems.

In practice the ratio of computation performance to communication performance provided by the underlying hardware architecture of the Grid environment will have to match the ratio of computation performance to communication performance required by the parallel simulation as reasoned in 2.2. Whether a partitioned simulation model will perform well will therefore also depend on the underlying hardware architecture.

# 5 Implementation

The GPSS based parallel transaction-oriented simulator will be implemented using the Java[TM] 2 Platform Standard Edition 5.0, also known as J2SE5.0 [31] and ProActive version 3.1 [15] as the Grid environment. An object-oriented design will be applied for the implementation of the simulator and resulting classes will be grouped into a hierarchy of packages according to the functional parts of the parallel simulator and the implementation phases. The parallel simulator will use the logging library log4j [3] for all its output, which will provide very flexible means to enable or disable specific parts of the output as required. The log4j library is the same logging library that is used by ProActive so that only one configuration file will be needed to configure the logging of ProActive and the parallel simulator.

## 5.1 Implementation Considerations

### 5.1.1 Overall Architecture

Figure 11 shows the suggested architecture of the parallel simulator including its main components. The main parts of the parallel simulator will be the Simulation Controller and the Logical Processes. The Simulation Controller controls the overall simulation. It is created when the user starts the simulation and will use the Model Parser component to read the simulation model file and parse it into an in memory object structure representation of the model. After the model is parsed the Simulation Controller will create Logical Process instances, one for each model partition contained within the simulation model. The Simulation Controller and the Logical Processes will be implemented as ProActive Active Objects so that they can communicate with each other via method calls. Communication will take place between the Simulation Controller and the Logical Processes but also between the Logical Processes for instance in order to exchange Transactions. Note that the communication between the Logical Processes is not illustrated in Figure 11. After the Logical Process instances have been created, they will be initialised, they will receive the model partitions from the Simulation Controller that they are going to simulate and the simulation is started.

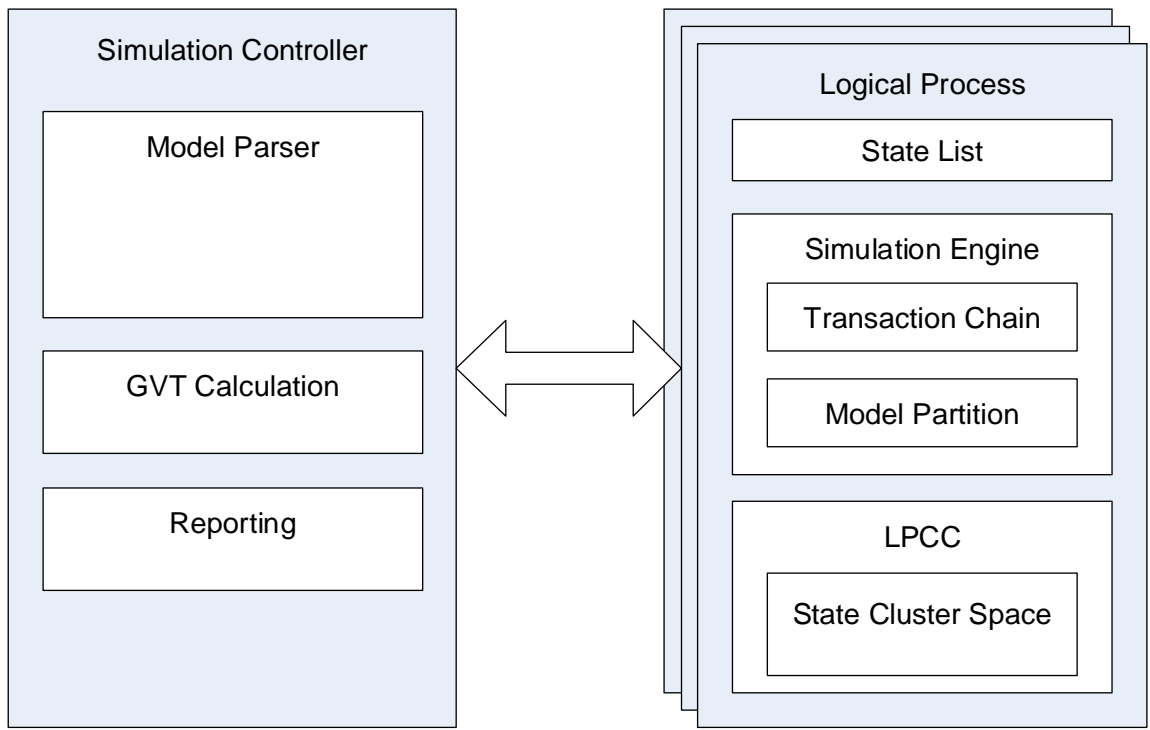

Figure 11: Architecture overview

Each Logical Process implements an LP according to the Shock Resistant Time Warp algorithm. The main component of the Logical Process is the Simulation Engine, which contains the Transaction chain and the model partition that is simulated. The Simulation Engine is the part that is performing the actual simulation. It is moving the Transactions from block to block by executing the block functionality using the Transactions. Another important part of the Logical Process is the State List. It contains historic simulation states in order to allow rollbacks as required by optimistic synchronisation algorithms. Note that there will be other lists like for instance the list of Transactions received and the list of Transactions sent to other Logical Processes, which are not shown in Figure 11. Furthermore the Logical Process will contain the Logical Process Control Component (LPCC) according to the Shock Resistant Time Warp algorithm described in 4.2.3. Using specific sensors within the Logical Process the LPCC will limit the optimism by the means of an artificial memory limit if this promises a better simulation performance at the current circumstances.

The Simulation Controller will perform GVT calculations in order to establish the overall progress of the simulation and if requested by one of the Logical Processes that needs to reclaim memory using fossil collection. GVT calculation will also be used to confirm a provisional simulation end state that might be reached by one of the Logical Processes.

When the end of the simulation is reached then the Simulation Controller will ensure that the partial models in all Logical Processes are set to the correct and consistent end state and it will collect information from all Logical Processes in order to assemble and output the post simulation report.

## 5.1.2 Transaction Chain and Scheduling

Thomas J. Schriber gives an overview in section 4. and 7. of [26] on how the Transaction chains and the scheduling work in the original GPSS/H implementation. The scheduling and the Transaction chains for the parallel transaction-oriented simulator of this work will be based on his description but with some significant differences. Only one Transaction chain is used containing Transactions for current and future simulation time in a sorted order. The Transactions within the chain are first sorted ascending by their next moving time and Transactions for the same time are also sorted descending by their priority. Transactions will be taken out of the chain before they are moved and put back into the chain after they have been moved (unless the Transaction has been terminated). The functionality to put a Transaction back into the chain will do so at the right position ensuring the correct sort order of the Transactions within the chain. The scheduling will be slightly simpler than described in [26] because no "Model's Status changed flag" will be needed.

Another difference is that in order to keep the time management as simple as possible the proposed parallel simulator will restrict the simulation time and time related parameters to integer values instead of floating point values. At first this might seem like a major restriction but it is not because decimal places can be represented using scaling. If for instance a simulation time with three decimal places is needed for a simulation then a scale of 1000:1 can be used, which means 1000 time units of the parallel simulator represent one second of simulation time or a single time unit represents 1ms. The Java integer type *long* will be used for time values providing a large value range that allows flexible scaling for different required precisions.

The actual scheduling will be implemented using a *SimulationEngine* class. This class will for instance contain the functionality for moving Transactions, updating the simulation time and also the Transaction chain itself. The *SimulationEngine* class will be implemented as part of the Basic GPSS Simulation Engine implementation phase detailed in section 5.2.2 and a description of how the scheduling was implemented can be found in 5.3.1.

## 5.1.3 Generation and Termination of Transactions

Performing transaction-oriented simulation in a parallel simulator also has an influence on how Transactions are generated and how they are terminated.

**Generating Transactions**

Transactions are generated in the GENERATE blocks of the simulation. During the generation each Transaction receives a unique numerical ID that identifies the Transaction during the rest of the simulation. In a parallel transaction-oriented simulator GENERATE blocks can exist in any of the model partitions and therefore in any of the LPs. This requires a scheme, which ensures that the Transaction IDs generated in each LP are unique across the overall parallel simulation. Ideally such a scheme requires as little communication between the LPs as possible.

The scheme used for this parallel simulator will generate unique Transaction IDs without any additional communication overhead. This is achieved by partitioning the value range of the numeric IDs according to the number of partitions in the simulation model. The only requirement of this scheme is that all LPs are aware of the total number of partitions and LPs within the simulation. This information will be passed to them during the initialisation. The used scheme is based on an offset that depends on the total number of LPs. Each LP has its own counter that is used to generate unique Transaction IDs. These counters are initialised with different starting values and the same offset is used for incrementing the counters when one of their values has been used for a Transaction ID. A simulation with n LPs (i.e. n partitions) will use the offset n to increment the local Transaction ID counters and each LP will initialise its counter with its own number in the list of LPs. In a simulation with 3 LPs, $LP_1$ would initialise its counter to the value 1, $LP_2$ to 2 and $LP_3$ to 3 and the increment offset used would be the total number of LPs which is 3. The sequence of IDs generated by these LPs would be $LP_1$: 1, 4, 7, … and by $LP_2$: 2, 5, 8, … and by $LP_3$: 3, 6, 9, …  and so forth. Further advantages of this scheme are that it partitions the possible ID value range into equally large numerical partitions independent of the total size of the value range and it also makes it possible to determine in which partition a Transaction was generated using their ID.

**Terminating Transactions**

The termination of Transactions raises similar problems like their generation. According to GPSS Transactions are terminated in TERMINATE blocks. Each time a Transaction is terminated a global Termination Counter is decremented by the decrement parameter of the TERMINATE block involved. A GPSS simulation initialises the Termination Counter with a positive value at the start of the simulation and the counter is then used to detect the end of the simulation, which is reached as soon as the Termination Counter has a value of zero or less. The required global Termination Counter could be located in one of the LPs but accessing it form other LPs would require additional synchronisation. The problem of accessing such a Termination Counter is the same like accessing other objects from different LPs as outlined in 4.1.2.

In order to avoid the additional complexity and communication overhead of implementing a global Termination Counter the parallel simulator will use a separate local Termination Counter in each LP. This solution will perform simulations that don't require a global Termination Counter without additional communication overhead. For simulations that do require a global Termination Counter the problem can be reduced to the synchronisation and the movement of Transactions between LPs as solved by the synchronisation algorithm. In this case all TERMINATE blocks of such a simulation need to be located within the same partition. This will result in additional communication and synchronisation when Transactions are moved from other LPs to the one containing the TERMINATE blocks. Figure 12 below shows how a simulation model with two partitions that needs a single Termination Counter can be converted into equivalent simulation models so that the simulation will effectively use a single Termination Counter.

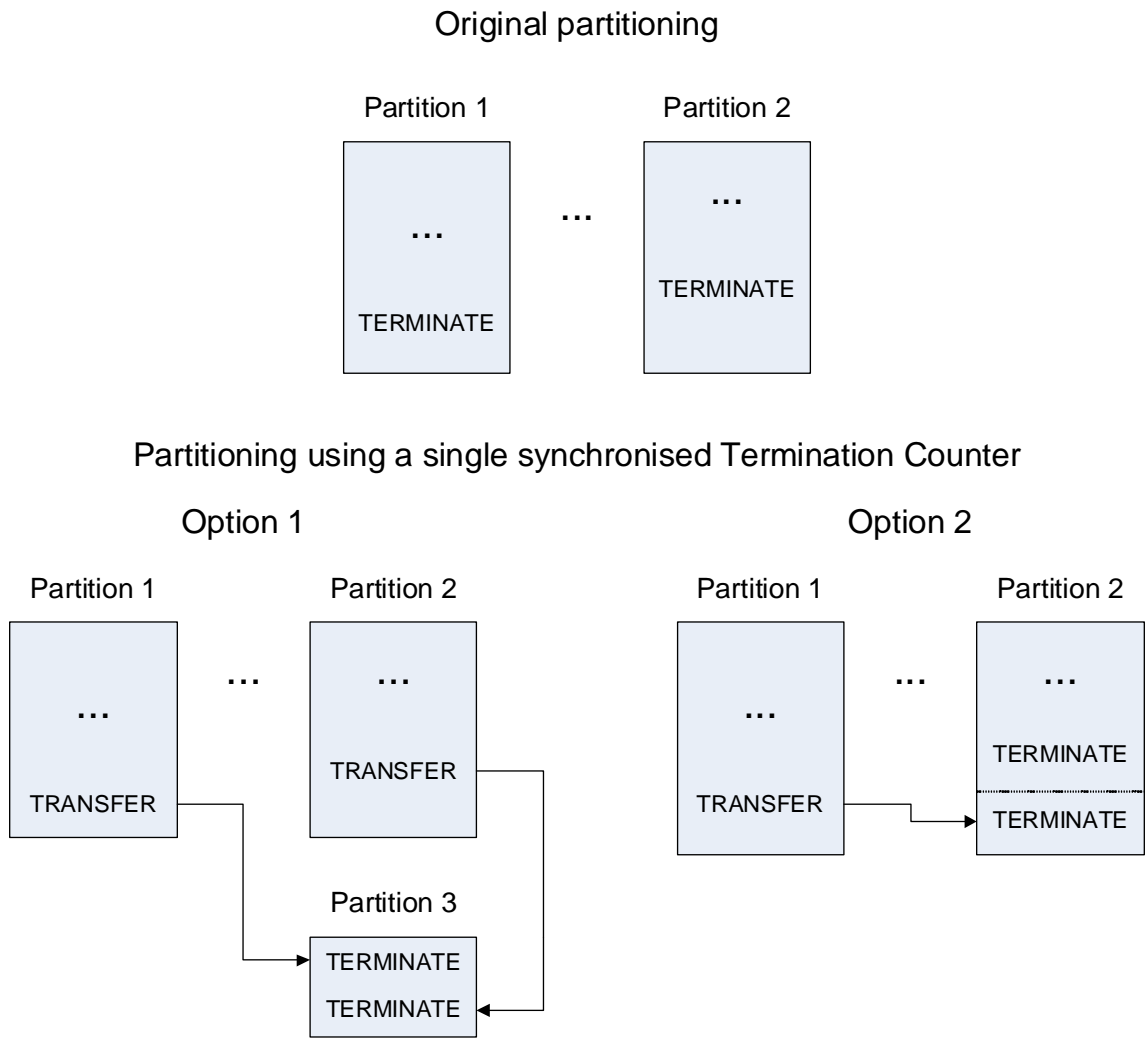

Figure 12: Single synchronised Termination Counter

### 5.1.4 Supported GPSS Syntax

In order to ease the migration of existing GPSS models to this new parallel GPSS simulator the GPSS syntax supported will be kept as close as possible to the original GPSS/H language described in [26]. At the same time only a sub-set of the full GPSS/H language will be implemented but this sub-set will include all the main GPSS functionality including functionality needed to demonstrate the parallel simulation of partitioned models on more than one LP. The simulator will not support Transaction cloning, Logic Switches, User Chains and user defined properties for Transactions but such functionality can easily be added in future if required.

A detailed description of the GPSS syntax expected by the parallel GPSS simulator can be found in Appendix A. In particular the simulator will support the generating, delay and termination of Transactions as well as their transfer to other partitions of the model. It will also support Facilities, Queues, Storages and labels. The following table gives an overview of the supported GPSS block types.

| Block type | Short description |
| --- | --- |
| GENERATE | Generate Transactions |
| TERMINATE | Terminate Transactions |
| ADVANCE | Delay the movement of Transactions |
| SEIZE | Capture Facility |
| RELEASE | Release Facility |
| ENTER | Capture Storage units |
| LEAVE | Release Storage units |
| QUEUE | Enter Queue |
| DEPART | Leave Queue |
| TRANSFER | Jump to a different block than the next following |

Table 7: Overview of supported GPSS block types

In addition to the GPSS functionality described above the new reserved word PARTITION is introduced. This reserved word marks the start of a new partition within the model. If the model does not start with such a partition definition then a default partition is created for the following blocks. All block definitions following a partition definition are automatically assigned to that partition. During the simulation each partition will be performed on a separate LP.

## 5.1.5 Simulation Termination at Specific Simulation Time

The parallel simulator will not provide any syntax or configuration options for terminating a simulation when a specific simulation time is reached but such behaviour can easily be modelled in any simulation model using an additional GENERATE and TERMINATE block. For instance if a simulation is supposed to be terminated when it reaches a simulation time of 10,000 then an additional GENERATE block is added that generates a single Transaction for the simulation time 10,000 immediately followed by a TERMINATE block that stops the simulation when that Transaction is terminated. This additional set of GENERATE and TERMINATE block can either be added to the end of an existing partition or as an additional partition. All other TERMINATE blocks in such a simulation will need to have a decrement parameter of 0. The following GPSS code shows an example model that will terminate at the simulation time 10,000.

```
PARTITION Partition1,1     sets Termination Counter to 1
…                          original model partition
GENERATE 1,0,10000         generates a Transaction for time 10000
TERMINATE 1                end of simulation after 1 Transaction
```

## 5.2  Implementation Phases

The following sections will describe the four main development phases of the parallel simulator.

### 5.2.1 Model Parsing

The classes for parsing and validating the GPSS model read from the model file can be found in the package *parallelJavaGpssSimulator.gpss.parser*. A GPSS model file is parsed by calling the method *ModelFileParser.parseFile()*. This method returns an instance of the class *Model* from the package *parallelJavaGpssSimulator.gpss* that contains the whole GPSS model as an object structure. The *Model* instance contains a list of model partitions represented by instances of the class *Partition* and each *Partition* instance contains a list of GPSS blocks and lists of other entities like labels, queues, facilities and storages that make up the model partition.

**Global GPSS block references**

GPSS simulators require a way of referencing GPSS blocks. A TRANSFER block for instance needs to reference the block it should transfer Transactions to. Sequential simulators often just use the block index within the model to refer to a specific block.

But in a parallel GPSS simulator each Logical Process only knows its own model partition. Still a block reference needs to uniquely identify a GPSS block within the whole simulation model and ideally it should also be possible to determine the target partition from a block reference. The *GlobalBlockReference* class in the package *parallelJavaGpssSimulator.gpss* implements a block reference that fulfils these criteria and it is used to represent global block references and also block labels within the runtime model object structure and other parts of the simulator.

**Parser class hierarchy**

A parallel class hierarchy and the Builder design pattern [22] is used in order to separate the code for parsing and validating the GPSS model from the code that represents the model at the runtime of the simulation. This second class hierarchy is found in the *parallelJavaGpssSimulator.gpss.parser* package and contains a builder class for all element types that can make up the model structure at runtime. Figure 13 shows the UML diagram of the two class hierarchies including some of the relevant methods. When loading and parsing a GPSS model file the instance of the *ModelFileParser* class internally creates an instance of the *ModelBuilder* class, which for each partition found in the model file holds an instance of a *PartitionBuilder* class and the *PartitionBuilder* class holds builder classes for all blocks and other entities that are found in the partition. The parsing and validation of the different model elements is delegated to these builder classes. In addition the Factory design pattern [22] is used by the *BlockBuilderFactory* class that creates the correct builder class for a GPSS block depending on the block type keyword found in the model file. All builder classes have a *build()* method that returns an instance of the corresponding simulation runtime class for that element. These *build()* methods are called recursively so that the *ModelBuilder.build()* method calls the *build()* method of the *PartitionBuilder* instances it contains and each *PartitionBuilder* instance calls the *build()* method of all builder classes it contains. This delegation of responsibility within the class hierarchy makes it possible to return an instance of the *Model* class representing the whole GPSS model just by calling the *ModelBuilder.build()* method.

As mention above the package *parallelJavaGpssSimulator.gpss.parser* is only used to load, parse and verify a GPSS model from file into the object structure used at simulation runtime. For this reason the only class from this package with public visibility is *ModelFileParser*. All other classes of this package are only visible within the package itself.

Runtime model object structure classes in
*parallelJavaGpssSimulator.gpss*

Model parsing object structure classes in
*parallelJavaGpssSimulator.gpss.parser*

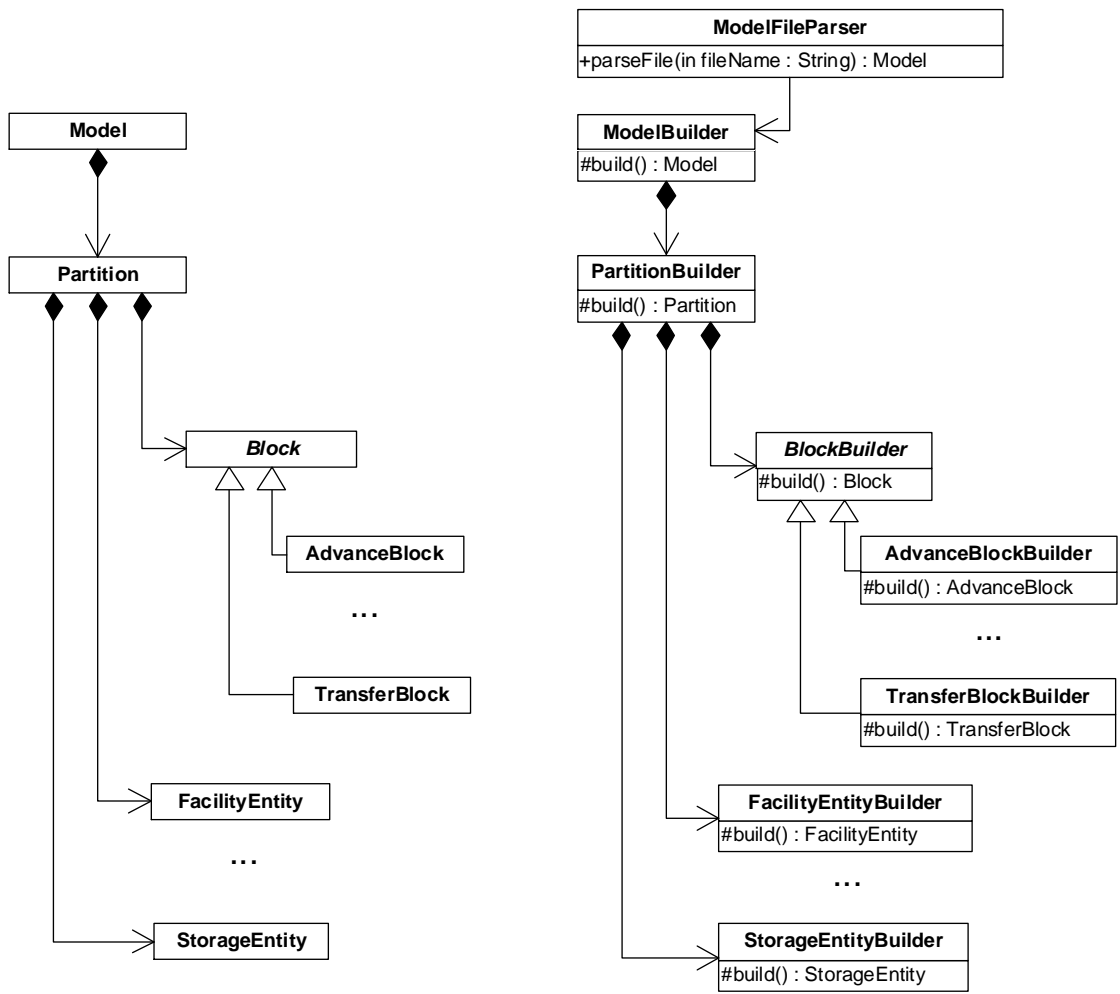

Figure 13: Simulation model class hierarchies for parsing and simulation runtime

**Test and Debugging of the Model Parsing**

In order to test and debug the parsing of the GPSS model file and the correct creation of the object structure representing the GPSS model at simulation run time, the *toString()* methods of all classes from the runtime class hierarchy were implemented to output their properties in textual form and to recursively call the *toString()* methods of any sub-element contained. A test application class with a *main()* method was implemented to load a GPSS model from a file using the *ModelFileParser.parseFile()* method and to output the whole structure of the model in textual form. Using this application different

GPSS test models were parsed that contained all of the supported GPSS entities and the textual output of the resulting object structures was checked. These tests included checking the default values of the different GPSS block types and parsing errors for invalid model files.

## 5.2.2 Basic GPSS Simulation Engine

The second implementation phase focused on the development of the basic GPSS simulation functionality. A GPSS simulation engine was implemented that can perform the sequential simulation of one model partition. This sequential simulation engine will be the basis of the parallel simulation engine implemented in the third phase.

The classes for the basic GPSS simulation functionality can be found in the package *parallelJavaGpssSimulator.gpss*. The main class in this package is the *SimulationEngine* class that encapsulates the GPSS simulation engine functionality. It uses the runtime model object structure class hierarchy mentioned in 5.2.1 to represent the model partition and the model state in memory. The runtime model object structure class hierarchy contains classes for the GPSS block types plus some additional classes to represent other GPSS entities like Facilities, Queues and Storages. Each of these classes implements the functionality that will be performed when a block of that type is executed or the GPSS entity is used by a Transaction. Two further classes in this package are the *Transaction* class representing a single Transaction and the *GlobalBlockReference* class introduced in 5.2.1.

The basic GPSS simulation scheduling is also implemented using the *SimulationEngine* class (a detailed description of the scheduling can be found in 5.3.1). It will generate Transactions and move them through the simulation model partition using the runtime model object structure. All block instances within this runtime model object structure inherit from the abstract *Block* class and therefore have to implement the *execute()* method. When a Transaction is moved through the model it will call this *execute()* method for each block it enters.

**Test and Debugging of Basic GPSS Simulation functionality**

The test application class *TestSimulationApp* was used to test and debug the basic GPSS simulation functionality implemented during this phase. This class contains a *main()* method and can therefore be run as an application. It allows the simulation of a single model partition using the *SimulationEngine* class. The exact simulation processing can be followed using the log4j logger *parallelJavaGpssSimulator.gpss* (details of the logging can be found in 5.3.5). In debug mode this logger outputs detailed steps of the simulation. Several test models where used to test the correct implementation of the basic scheduling and the GPSS blocks and other entities. They will not be described in more detail here because the same functionality will be tested again in the final version of the simulator (see validation phase in section 6).

### 5.2.3 Time Warp Parallel Simulation Engine

During the third development phase the parallel simulator was implemented based on the Time Warp synchronisation algorithm. The functionality of the parallel simulator is split into the Simulation Controller side and the Logical Process side, each found in a different package. Figure 14 shows the general architecture and the main classes involved at each side during this development phase. Classes marked with (AO) are instantiated as ProActive Active Objects. Instances of the *LogicalProcess* class also communicate with each other, for instance in order to exchange Transactions, which is not displayed in Figure 14.

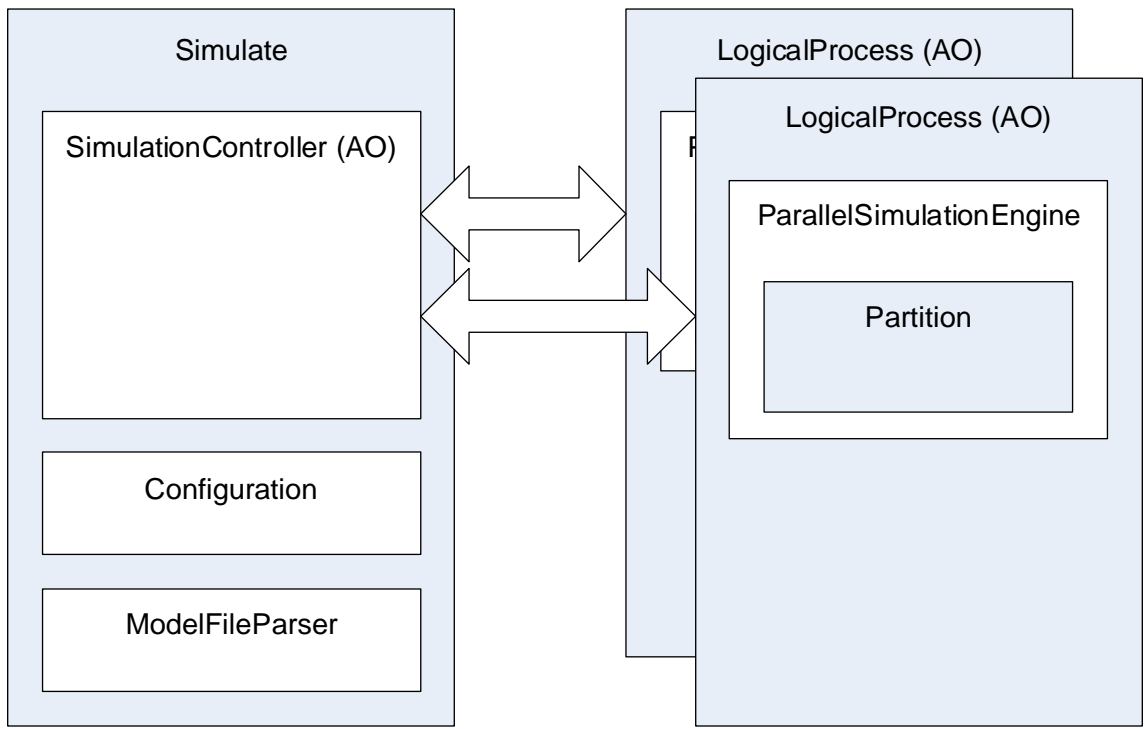

Figure 14: Parallel simulator main class overview (excluding LPCC)

**Simulation Controller side**

At the Simulation Controller side is the root package *parallelJavaGpssSimulator*. It contains the class Simulate, which is the application class that is used to start the parallel simulator. When run the Simulate class will load the configuration from command line arguments or the configuration file, it will also load and parse the simulation model file and then create an Active Object instance of the *SimulationController* class (the JavaDoc documentation for this class can be found in Appendix E). The *SimulationController* instance receives the configuration settings and the simulation model when its *simulate()* method is called. As a result of this call the *SimulationController* class will read the deployment descriptor file and create the required number of *LogicalProcess* instances at the specified nodes.

**Logical Process side**

The functionality of the Logical Processes is found in the package *parallelJavaGpssSimulator.lp*. This package contains the *LogicalProcess* class, the *ParallelSimulationEngine* class (the JavaDoc documentation for both can be found in Appendix E) and a few helper classes. The *LogicalProcess* instances are created as Active Objects by the Simulation Controller. After their creation the *LogicalProcess* instances receive the simulation model partitions and the configuration when their *initialize()* method is called. When all *LogicalProcess* instances are initialised then the Simulation Controller calls their *startSimulation()* method to start the simulation. Figure 15 illustrates the communication flow between the Simulation Controller and the Logical Processes before and at the end of the simulation. The method calls just described can be found at the start of this communication flow. When the Simulation Controller detects that a confirmed simulation end has been reached then all Logical Processes are requested to end the simulation with a consistent state matching that confirmed simulation end using the *endOfSimulationByTransaction()* method. The Logical Processes will confirm when they reached the consistent simulation state after which the Simulation Controller will request the post simulation report details from each Logical Process. Further specific details about the implementation of the parallel simulator can be found in section 5.3.

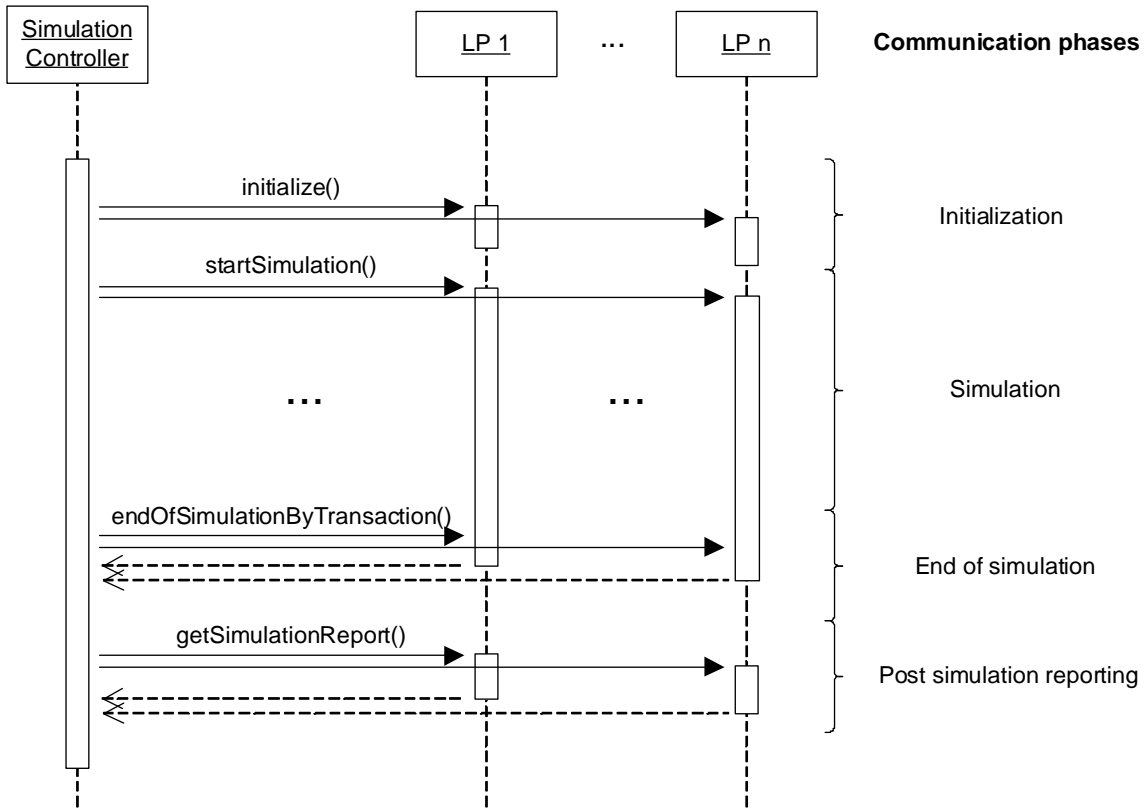

Figure 15: Main communication sequence diagram

**Test and Debugging of the Time Warp parallel simulator**

The parallel simulator resulting from this development phase was tested and debugged with extensive logging enabled and using different models. The functionality was tested again in the final version of the parallel simulator as part of the validation phase, of which details can be found in section 6.

## 5.2.4 Shock Resistant Time Warp

This development phase extended the Time Warp based parallel simulator from the former development phase to support the Shock Resistant Time Warp algorithm by adding the LPCC and the required sensor value functionality to the *LogicalProcess* class. The functionality for the Shock Resistant Time Warp algorithm is found in the package *parallelJavaGpssSimulator.lp.lpcc*. The main class in this package is the class *LPControlComponent* that implements the LPCC (see Appendix E for the JavaDoc documentation of this class). The package also contains two classes that represent the sets of sensor and indicator values and the class *StateClusterSpace* that encapsulates the functionality to store and retrieve past indicator state information using the cluster technique described in [8]. The Shock Resistant Time Warp algorithm of the parallel simulator is implemented so that it can be enabled and disabled by a configuration setting of the parallel simulator as required. If the LPCC and therefore the Shock Resistant Time Warp algorithm is disabled then the parallel simulator will simulate according to the normal Time Warp algorithm, if it is enabled then the Shock Resistant Time Warp algorithm will be used. The option to enable/disable the LPCC makes it possible to compare the performance of both algorithms for specific simulation models and hardware setup using the same parallel simulator.

**LPCC**

The Logical Process Control Component (LPCC) implemented by the *LPControlComponent* class is used by the LogicalProcess instances during a simulation according to the Shock Resistant Time Warp algorithm. It is the main component of this algorithm that attempts to steer the parameters of the LPs towards values of past states that promise better performance using an actuator that limits the number of uncommitted Transaction moves allowed. Figure 16 shows the architecture and main classes used by the final version of the parallel simulator including the *LPControlComponent* class representing the LPCC.

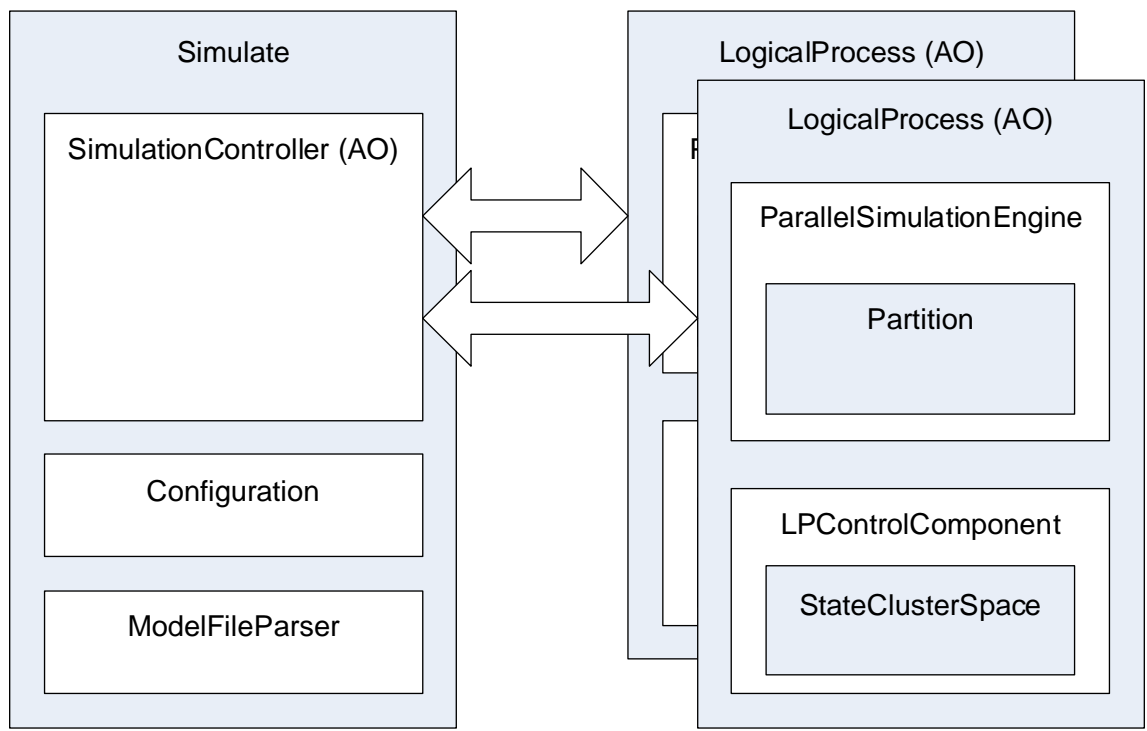

Figure 16: Parallel simulator main class overview

The LPCC receives the current sensor values with each simulation time update cycle (details of the scheduling can be found in 5.3.1) but the main processing of the LPCC is only called during specified time intervals as set in the configuration file of the simulator. When the main processing of the LPCC is called using its *processSensorValues()* method then the LPCC will create a set of indicator values for the sensor values cumulated. Using the State Cluster Space it will search for a similar indicator set that promises better performance and it will set the actuator according to the indicator set found. Finally the current indicator set will be added to the State Cluster Space. The LPCC is also used to check whether the current number of uncommitted Transaction moves exceeds the current actuator limit. Within the scheduling cycle the LP will call the *isUncommittedMovesValueWithinActuatorRange()* method of its *LPControlComponent* instance to perform this check. As a result the number of uncommitted Transaction moves passed in is compared to the maximum actuator limit determined by the mean actuator value and the standard deviation with a confidence level of 95% as described in [8]. The method will return false if the number of uncommitted Transaction moves exceeds the maximum actuator limit forcing the LP into cancelback mode (see 5.3.4).

**State Cluster Space**

The *StateClusterSpace* class encapsulates the functionality to store sets of indicator values and to return a similar indicator set for a given one. Each stored indicator set is treated as a vector in an n-dimensional vector space with n being the number of indicators per set. The similarity between two indicator sets is determined by their Euclidean vector distance. A clustering technique is used that groups similar indicator sets into clusters to limit the amount of memory required when large numbers of indicator sets are stored.

The two main public methods provided by the *StateClusterSpace* class are *addIndicatorSet()* and *getClosestIndicatorSetForHigherCommittedMoveRate()*. The first method adds a new indicator set to the State Cluster Space and the second returns the indicator set most similar to the one passed in that has a higher *CommittedMoveRate* indicator value. Note that the two indicators *AvgUncommittedMoves* and *CommittedMoveRate* are ignored when determining the similarity by calculating the Euclidean distance because *AvgUncommittedMoves* is directly linked to the actuator and *CommittedMoveRate* is the performance indicator that is hoped to be maximized.

**Test and Debugging of the Shock Resistant Time Warp and the State Cluster Space**

The State Cluster Space was tested and debugged using the test application class TestStateClusterSpaceApp, which allows for the *StateClusterSpace* class to be tested outside the parallel simulator. Using this class the detailed functionality of the State Cluster Space was tested using specific scenarios that would have been difficult to create within the parallel simulator. The test application class is left in the project so that possible future changes or enhancements to the *StateClusterSpace* class can also be tested outside the parallel simulator.

The implementation of the Shock Resistant Time Warp algorithm was tested and debugged in the final version of the parallel simulator using a selection of different models of which a significant one was chosen for validation 5 in section 6.5.

## 5.3  Specific Implementation Details

The following sections describe some specific implementation details of the parallel simulator.

### 5.3.1  Scheduling

The scheduling of the parallel simulator was implemented in two phases. The first part is the basic scheduling of the GPSS simulation that was implemented using the *SimulationEngine* class as described in section 5.2.2. This scheduling algorithm was later extended for the parallel simulation by the *LogicalProcess* class and the *ParallelSimulationEngine* class, which inherits from the *SimulationEngine* class as part of the Time Warp parallel simulator implementation phase described in 5.2.3.

**Basic GPSS Scheduling**

The basic GPSS scheduling is implemented using the functionality provided by the *SimulationEngine* class. A flowchart diagram of the scheduling algorithm is shown in Figure 17. As seen from this diagram the scheduling algorithm will first initialise the GENERATE blocks in order to create the first Transactions. Subsequent Transactions are created whenever a Transaction leaves a GENERATE block. The algorithm then updates the simulation time to the move time of the earliest movable Transaction. After the simulation time has been updated all movable Transactions with a move time of the current simulation time are moved through the model as far as possible. Unless this results in the simulation being completed the algorithm will repeat the cycle of updating the simulation time and moving the Transactions.

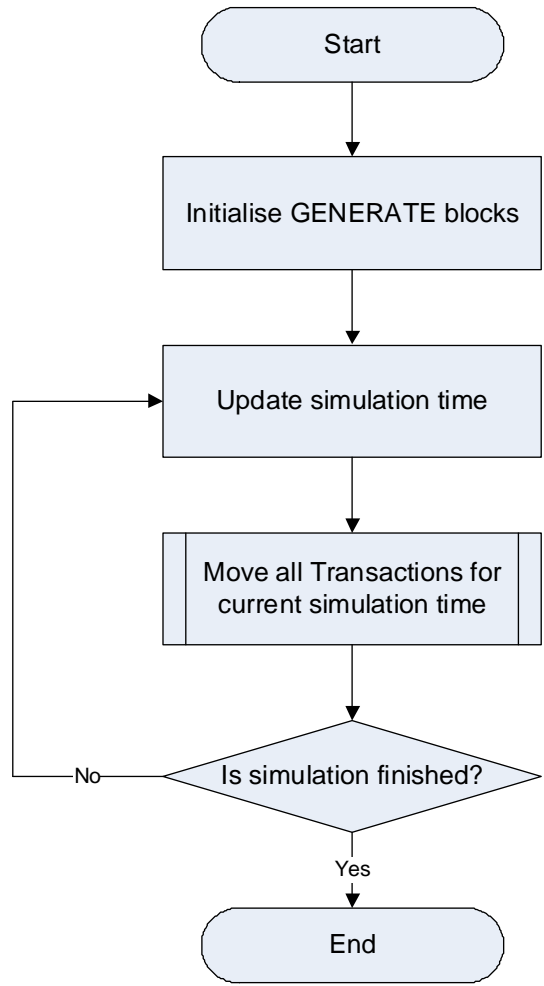

Figure 17: Scheduling flowchart - part 1

Figure 18 shows the flowchart of the *Move all Transactions for current simulation time* processing block from Figure 17. The algorithm for this block will retrieve the first movable Transaction for the current simulation time and take this Transaction out of the Transaction chain. If no such Transaction is found then the processing block is left. Otherwise the Transaction is moved through the model as far as possible. If the Transaction is not terminated as a result then it is chained back into the Transaction chain at the correct position according to its move time and priority (note that the move time and priority could have changed while the Transaction was moved).

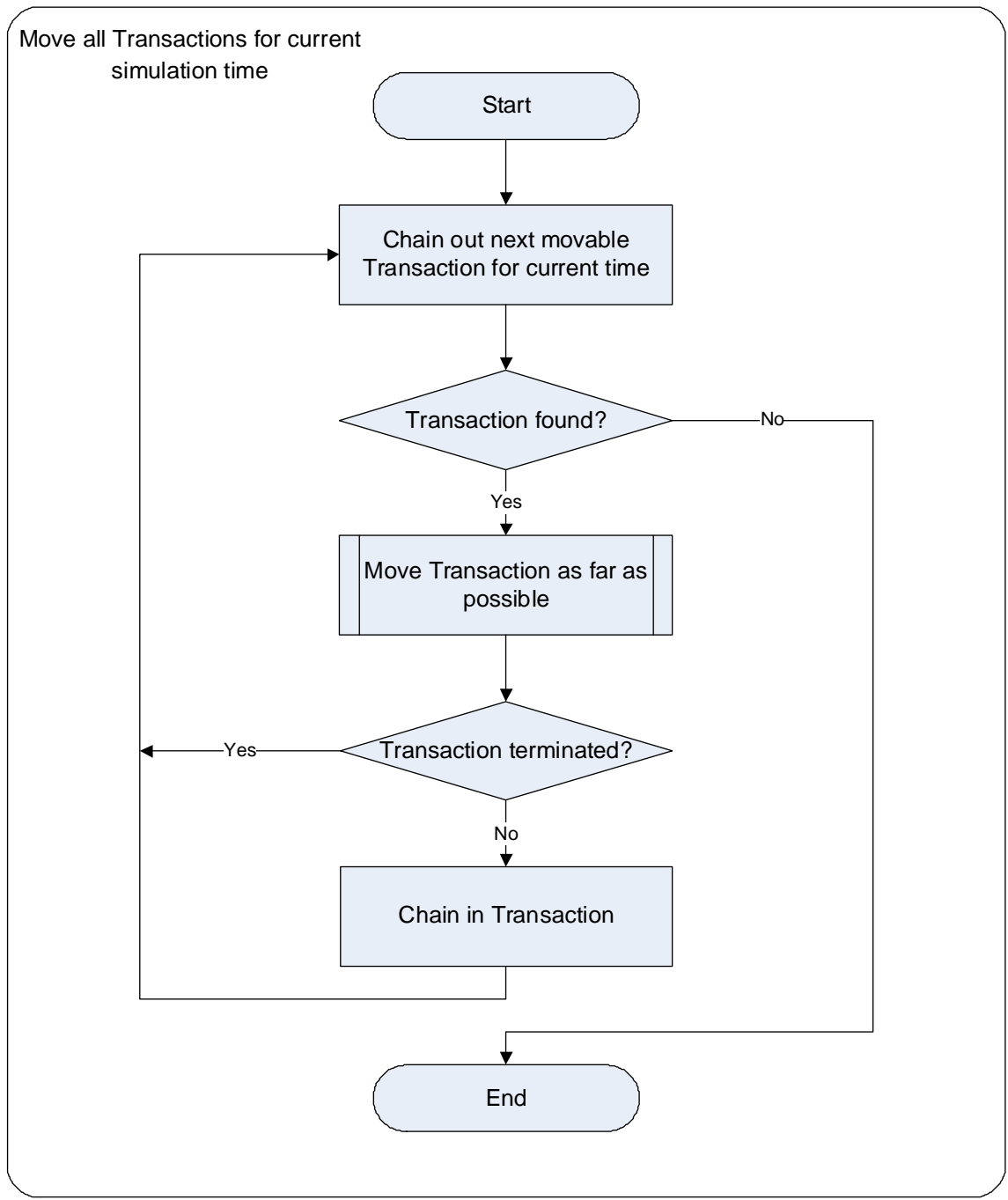

Figure 18: Scheduling flowchart - part 2

The *Move Transaction as far as possible* processing block is split down further and its algorithm illustrated in the flowchart shown in Figure 19.

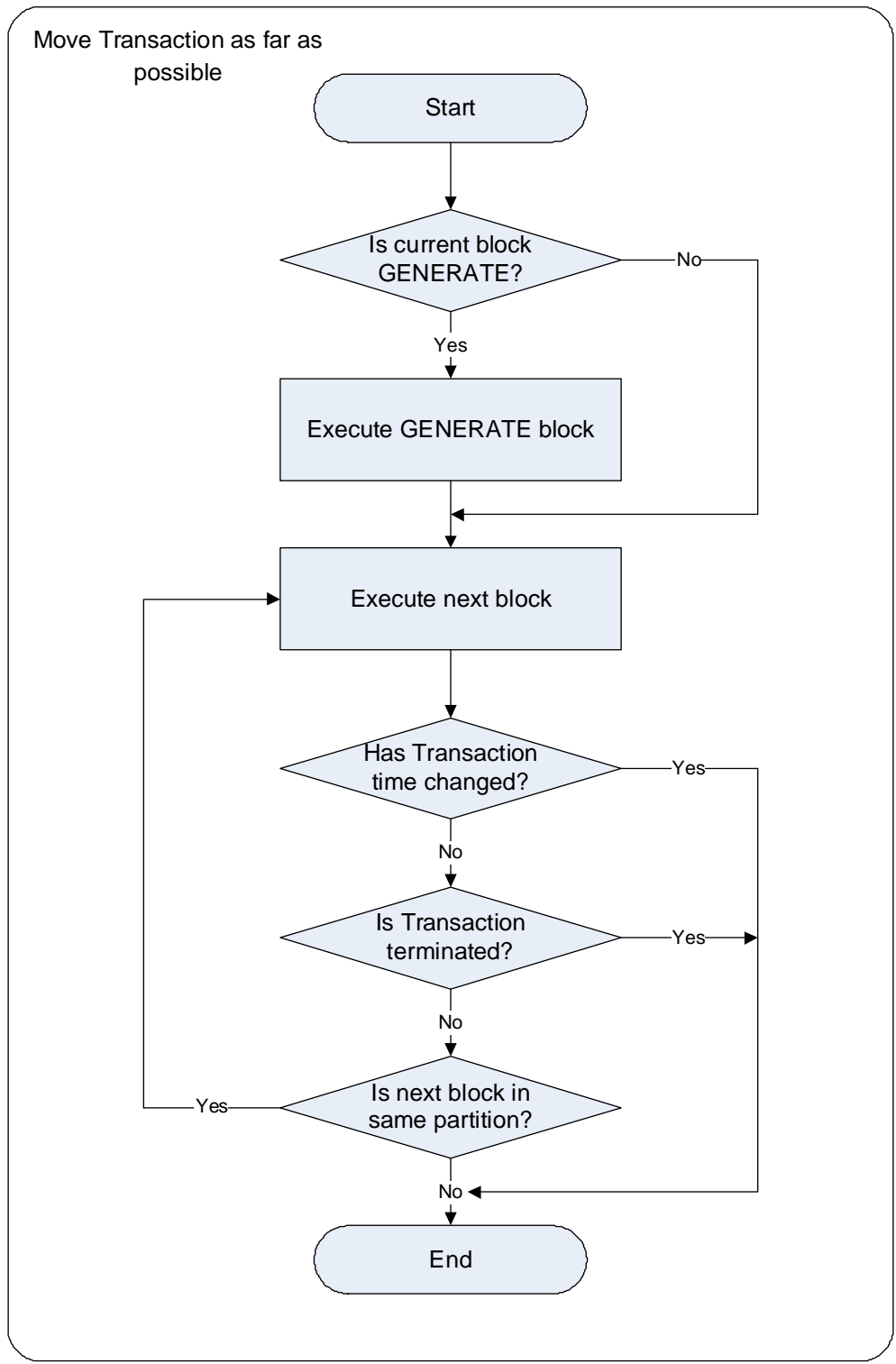

Figure 19: Scheduling flowchart - part 3

The algorithm will first check whether the Transaction is currently within a GENERATE block and if so the GENERATE block is execute. Then the Transaction is moved into the next following block by executing it. Unless the move time of the Transaction changed, the Transaction got terminated or the next block of the Transaction lays within a different partition the algorithm will repeatedly execute the next block for the Transaction in a loop and therefore move the Transaction from block to block. From this flowchart it can be seen that the execution of GENERATE blocks is treated different to the execution of other blocks. The reason is that GENERATE blocks are the only blocks that are executed when a Transaction leaves the block where as all other blocks are executed when the Transaction enters them. This allows a GENERATE block to create the next Transaction when the last one created leaves it. The table below mentions the different methods that implement the flowchart processing blocks described.

| Flowchart processing block | Method |
|---|---|
| Initialise GENERATE blocks | SimulationEngine.initializeGenerateBlocks() |
| Update simulation time | SimulationEngine.updateClock() |
| Move all Transactions for current simulation time | SimulationEngine. moveAllTransactionsAtCurrentTime() |
| Chain out next movable Transaction for current time | SimulationEngine. chainOutNextMovableTransactionForCurrentTime() |
| Move Transaction as far as possible | SimulationEngine.moveTransaction() |
| Chain in Transaction | SimulationEngine.chainIn() |
| Execute GENERATE block | GenerateBlock.execute() |
| Execute next block | Calls the execute() method of the next block instance for the Transaction |

Table 8: Methods implementing basic GPSS scheduling functionality

**Extended parallel simulation scheduling**

For the parallel simulator the simulation scheduling is implemented in the Logical Processes. It integrates the Active Object request processing of the *LogicalProcess* class and the synchronisation algorithm of the parallel simulation. This results in a scheduling algorithm that looks quite different to the one for the basic GPSS simulation. A slightly simplified flowchart of this algorithm can be found in Figure 20 (note that the darker flowchart processing blocks are blocks that already existed in the basic GPSS scheduling algorithm).

Because the *LogicalProcess* class is used as an Active Object its scheduling algorithm is implemented in the *runActivity()* method inherited from the *org.objectweb.proactive. RunActive* interface that is part of the ProActive library. The algorithm first checks whether the body of the Active Object is still active and then processes any Active Object method requests received. If the Logical Process is not in the mode SIMULATING then the algorithm will return and loop through checking the body and processing Active Object requests. If the mode is changed to SIMULATING then it will proceed to update the simulation time. This step existed already in the basic GPSS scheduling algorithm. Note that the functionality to initialize the GENERATE blocks is not part of the actual scheduling algorithm any more as it is performed when the *LogicalProcess* class is initialized using the *initialize()* method. After the simulation time has been updated the start state for the new simulation time will be saved. The state saving and rollback process is described in detail in section 5.3.3. The next step is to handle received Transactions, which includes anti-Transactions and cancelbacks. They are received via ProActive remote method calls and stored in an input list during the *Process Active Object requests* step. Normal received Transactions are handled by chaining them into the Transaction chain. This might require a rollback if the local simulation time has already passed the move time of the new Transaction. In order to handle a received anti-Transaction the matching normal Transaction has to be found and deleted. If the normal Transaction has been moved through the model already then a rollback is required as well. Cancelback requests are also handled by performing a rollback (see section 5.3.4 for details of the memory management and cancelback).

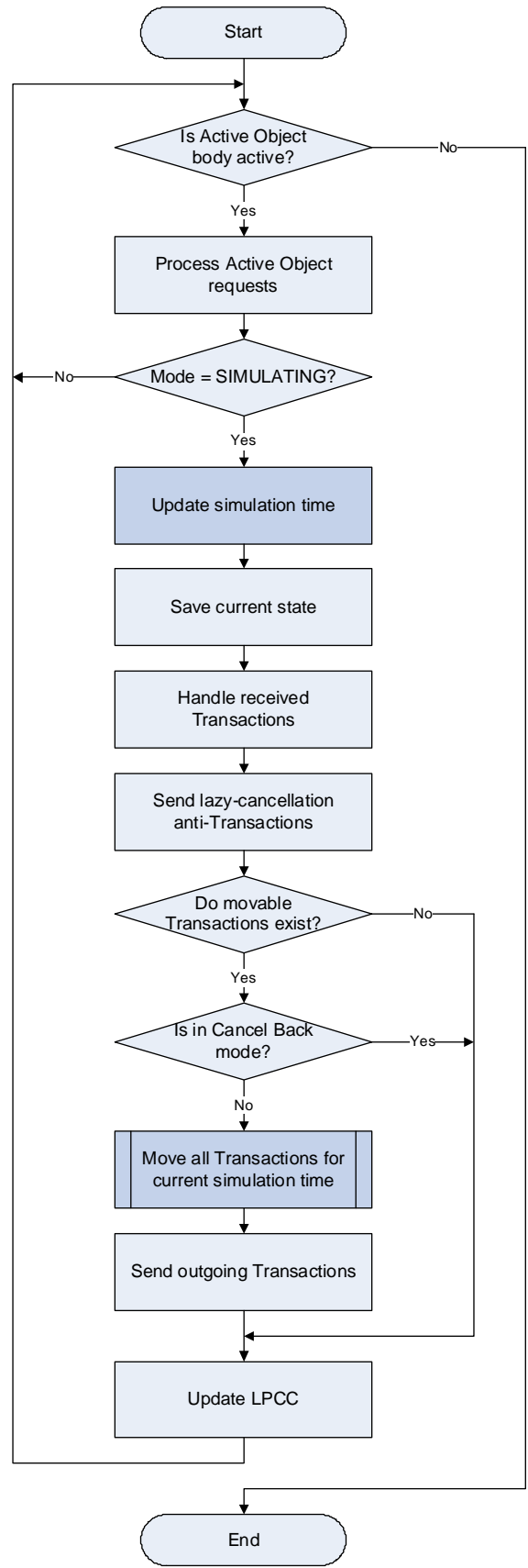

Figure 20: Extended parallel simulation scheduling flowchart

Following the handling of received Transactions and anti-Transactions the scheduling algorithm will send out any anti-Transactions required by the lazy-cancellation mechanism. It will identify all Transactions that have been sent out for an earlier simulation time and which have been rolled back and subsequently not sent again. Such Transactions need to be cancelled by sending out anti-Transactions. If following the lazy-cancellation handling the Simulation Engine has movable Transactions and is not in cancelback mode then all movable Transactions for the current simulation time are moved through the simulation model. Any outgoing Transactions are sent to their destination Logical Process and the LPCC sensors are updated. The whole scheduling algorithm will be repeated until the *LogicalProcess* instance is terminated and its Active Object body becomes inactive. The methods implementing the flowchart processing blocks described are shown below.

| Flowchart processing block | Method |
|---|---|
| Process Active Object requests | LogicalProcess**.**processActiveObjectRequests() |
| Update simulation time | SimulationEngine**.**updateClock() |
| Save current state | LogicalProcess**.**saveCurrentState() |
| Handle received Transactions | LogicalProcess**.**handleReceivedTransactions() |
| Send lazy-cancellation anti-Transactions | LogicalProcess**.** sendLazyCancellationAntiTransactions() |
| Move all Transactions for current simulation time | ParallelSimulationEngine**.** moveAllTransactionsAtCurrentTime() |
| Send outgoing Transactions | LogicalProcess**.** sendTransactionsFromSimulationEngine() |
| Update LPCC | LogicalProcess**.**updateLPControlComponent() |

Table 9: Methods implementing extended parallel simulation scheduling

## 5.3.2 GVT Calculation and End of Simulation

Details of why and how the GVT is calculated during the simulation have already been described in 4.3 but here the focus lies on the actual implementation. The GVT calculation is performed by the SimulationController class within the private method *performGvtCalculation()*. During the GVT calculation the Simulation Controller will request the required parameters from each LP, determine the GVT and pass the GVT back to the LPs so that these can perform the fossil collection. Figure 21 shows the sequence diagram of the GVT calculation process.

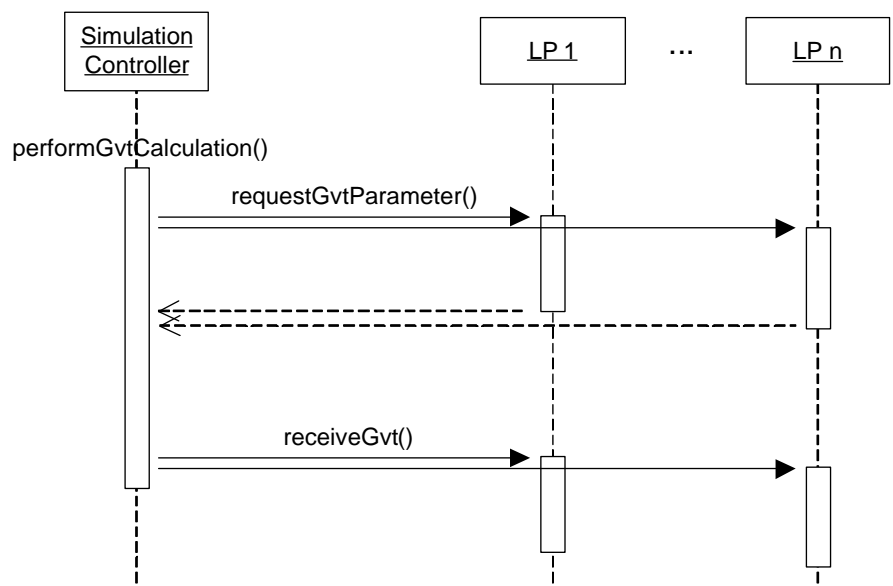

Figure 21: GVT calculation sequence diagram

There are different circumstances that can cause a GVT calculation within the parallel simulator. First LPs can request a GVT calculation from the Simulation Controller by calling its *requestGvtCalculation()* method. This happens when an LP reached certain defined memory limits (as described in 5.3.4) or when a provisional simulation end is reached by one of the LPs, which is described in more detail further below. Another reason for a GVT calculation is that the LPCC processing is required because the defined processing time interval has passed. For the Shock Resistant Time Warp algorithm the LPCC processing is linked to a GVT calculation so that the sensor and indicator for the number of committed Transaction moves have realistic values that reflect the simulation progress made during the time interval. For this reason the LPCC processing times are controlled by the Simulation Controller and linked to GVT calculations that are triggered when the next LPCC processing is needed. An additional parameter for the method *receiveGvt()* of the *LogicalProcess* class indicates to the LP that an LPCC processing is needed after the new GVT has been received. Finally the user can also trigger a GVT calculation, which is useful for simulations in normal Time Warp mode that might not require any GVT calculation for large parts of the simulation. Forcing a GVT calculation allows the user to check what progress the simulation has made so far as the GVT is an approximation for the confirmed simulation times that has been reached by all LPs.

**End of simulation**

The detection of the simulation end is closely linked to the GVT calculation because a provisional simulation end reached by one of the LPs can only be confirmed by a GVT. The background of detecting the simulation end has already been discussed in 4.4 but the actual implementation will be explained here. When an LP reaches a provisional simulation end then the parallel simulator will attempt to confirm this simulation end as soon as possible if at all possible. First the LP reaching the provisional simulation end will request a GVT calculation from the Simulation Controller. But the resulting GVT might not confirm the provisional simulation end if the LP is ahead of other LPs in respect of the simulation time. For this case a scheme is introduced in which the LP reaching the provisional simulation end tells all other LPs to request a GVT calculation themselves if they pass the simulation time of that provisional simulation end. The method *forceGvtAt()* of the *LogicalProcess* class is used to tell other LPs about the provisional simulation end time. Because it is possible for more than one LP to reach a provisional simulation end before any of them is confirmed this method will keep a list of the times at which the LPs need to request GVT calculations.

Whether or not a provisional simulation end reached by one of the LPs is confirmed, is detected by the private method *performGvtCalculation()* of the SimulationController class that also performs the calculation of the GVT. In order to make this possible the method *requestGvtCalculation()* of the *LogicProcess* class returns additional information about a possible simulation end reached by that LP. This way the GVT calculation process described above is also used to confirm a provisional simulation end. Such a simulation end is confirmed during the GVT calculation when it is found that all other LPs have reached a later simulation time than the one that reported the provisional simulation end. In this case no future rollback could occur that can undo the provisional simulation end, which is therefore guaranteed. If the GVT calculation confirms a simulation end then no GVT is send back to the LPs but instead the Simulation Controller calls the method *endOfSimulationByTransaction( )* of all LPs as shown in Figure 15 of section 5.2.3.

## 5.3.3 State Saving and Rollbacks

Optimistic synchronisation algorithms execute all local events without guarantee that additional events received later will not violate the causal order of events already execute. In order to correct such causal violations they have to provide means to restore a past state before the causal violation occurred so that the new event can be inserted into the event chain and the events be executed again in the correct order. A common technique to allow the restoration of past states is called State Saving or State Checkpointing where an LP saves the state of the simulation into a list of simulation states each time the state changes or in defined intervals.

The parallel simulator implemented employs a relatively simple state saving scheme. Each time the simulation time is incremented the LP serialises the state of the Simulation Engine and saves it together with the corresponding simulation time into a state list. Each state record therefore describes the simulation state at the beginning of that time, i.e. before any Transactions were moved. To keep the complexity of this solution low the standard object serialisation functionality provided by Java is used to serialise and deserialise the state to and from a Stream object that is then stored in the state list. The state list keeps all states sorted by their time. The saving of the state is implemented in the method *saveCurrentState( )* of the *LogicalProcess* class.

The purpose of saving the simulation state is to allow LPs to rollback to past simulation states if required. The functionality to rollback to a past simulation state is implemented in the method *rollbackState( )*. Using the example shown in Figure 22 the principle of rolling back to a past simulation state is briefly explained.

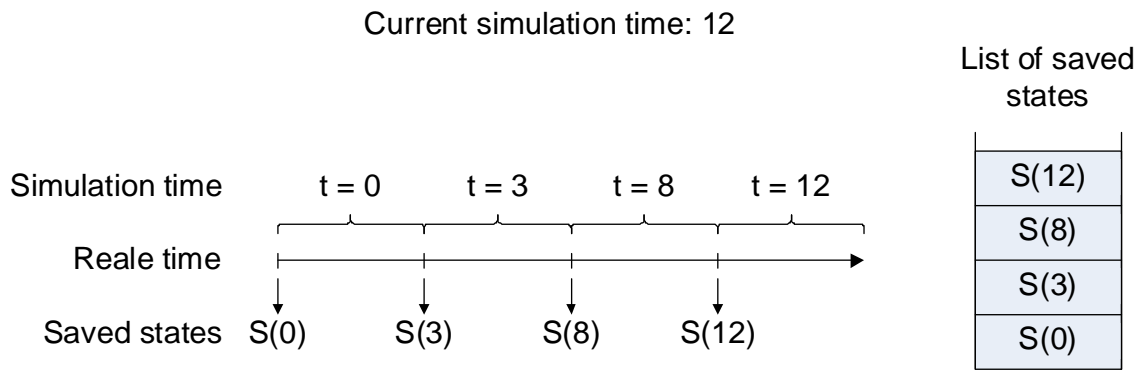

Figure 22: State saving example

Figure 22 shows the state information of an LP that has gone through the simulation times 0, 3, 8 and 12 and that has saved the simulation state at the beginning of each of these times. There are two possible options for a rollback depending on whether a state for the simulation time that needs to be restored exists in the state list or not. If for instance the LP receives a Transaction for the time 3 then the LP will just restore the state of the time 3, chain in the new Transaction and proceed moving the Transactions in their required order. But if a Transaction for the simulation time 5 is received, which implies that a rollback to the simulation time 5 is needed then the state of the time 8 is restored because this is the same state that would have existed at the simulation time 5. Recapitulating it can be said that if no saved state exists for the simulation time to that a rollback is needed then the rollback functionality will restore the state with the next higher simulation time.

In addition to the basic task of restoring the correct simulation state the *rollbackState()* method also performs a few related tasks like chaining in any Transactions that were received after that restored state was saved or marking any Transactions sent out after the rollback simulation time for the lazy-cancellation mechanism.

A further task related to the state management is performing the fossil collection which is implemented by the *commitState()* method of the *LogicalProcess* class. This method is called when the LP receives a new GVT. It will remove any past simulation states and other information, for instance about Transactions received or sent, that are not needed any more.

Because of the time scale of this project and in order to keep the complexity of the implementation low, the state saving scheme used by the parallel simulator is a relatively basic periodic state saving scheme. Future work on the simulator could look at enhancing the state saving using an adaptive periodic checkpointing scheme with variable state saving intervals as suggested in [25]. Alternatively an incremental state saving scheme could be used but this would drastically increase the complexity of the state saving because the standard Java serialisation functionality could not be used or would need to be extended. An incremental state saving scheme would also add an additional overhead for restoring a specific state so that an adaptive periodic checkpointing scheme appears to be the best option for future enhancements.

## 5.3.4 Memory Management

Optimistic synchronisation algorithms make extensive use of the available memory in order to save state information that allow the restoration and the rollback to a past simulation state required if an LP receives an event or Transaction that would violate the causal order of events or Transactions already executed. At the same time a parallel simulator has to avoid running out of available memory completely as this would mean the abortion of the simulation. The parallel simulator implemented here will therefore use a relatively simple mechanism to avoid reaching the given memory limits. It will monitor the memory available to the LP within the JVM during the simulation and perform defined actions if the available memory drops below certain limits. The first limit is defined at 5MB. If the amount of available memory goes below this limit then the LP will request a GVT calculation from the Simulation Controller in the expectation that a new GVT will confirm some of the uncommitted Transaction moves and saved simulation states so that fossil collection can free up some of the memory currently used.

In some circumstances GVT calculations will not free up any memory used by the LP or not enough. This is for instance the case when the LP is far ahead in simulation time compared to the other LPs. If none if its uncommitted Transaction moves or saved simulation states are confirmed by the new GVT then no memory will be freed by fossil collection. Otherwise it is also possible that only very few uncommitted Transaction moves and saved simulation states are confirmed by the new GVT resulting in very little memory being free. The parallel simulator defines a second memory limit of 1MB for the case that GVT calculations did not help in freeing memory. When the memory available to the LP drops below this second limit then the LP switches into cancelback mode. A cancelback strategy was already mentioned by David Jefferson [17] but the cancelback strategy used here will differ slightly from the one suggested by him. When the LP operates in cancelback mode then it will still respond to control messages and will still receive Transactions from other LPs but it will stop moving or processing any local Transaction so that no simulation progress is made by the LP and no simulation state information are saved as a result. Further the LP will attempt to cancel back Transactions that it received from other LPs in order to free memory or at least stop memory usage growing further. To cancelback a Transaction means that all local traces that a Transaction was received are removed and the Transaction is sent back to its original sender that will rollback to the move time of that Transaction. The main methods involved with the cancelback mechanism are the method *LogicalProcess.needToCancelBackTransactions()* which is called by an LP that is in cancelback mode and the method *LogicalProcess.cancelBackTransaction()* which is used to send a cancelled Transaction back to the sender LP. This cancelback mechanism of the parallel simulator is not only used for the general memory management but also when the Actuator value of the LPCC has been exceeded.

### 5.3.5 Logging

The parallel simulator uses the Java logging library log4j [3] for its logging and standard user output. It is the same logging library that is used by ProActive. The log4j library makes it possible to enable or disable parts or all of the logging or to change the detail of logging by means of a configuration file without any changes to the Java code. To utilise the same logging library for ProActive and the parallel simulator means that only a single configuration file can be used to configure the logging output for both. A hierarchical structure of loggers combined with inheritance between loggers makes it very easy and fast to configure the logging of the simulator. A detailed description of the log4j library and its configuration can be found at [3]. The specific loggers used by the parallel simulator are described in Appendix C.

As mentioned above the parallel simulator will use the same log4j configuration file like ProActive. By default this is the file proactive-log4j but a different file can be specified as described in the ProActive documentation [15]. The log4j root logger for all output from the parallel simulator is *parallelJavaGpssSimulator* (in the log4j configuration file all loggers have to be prefixed with "log4j.logger." so that this logger would appear as

*log4j.logger.parallelJavaGpssSimulator*). A hierarchy of lower level loggers allow the configuration of which information will be output or logged by the parallel simulator. The log4j logging library supports the inheritance of logger properties, which means that a lower level logger that is not specifically configured will inherit the configuration from a logger at a higher level within the same name space. For example if only the logger *parallelJavaGpssSimulator* is configured then all other loggers of the parallel simulator would inherit the same configuration settings from it.

## 5.4  Running the Parallel Simulator

### 5.4.1  Prerequisites

The parallel GPSS simulator was implemented using the Java$^{TM}$ 2 Platform Standard Edition 5.0, also known as J2SE5.0 or Java 1.5 [31] and ProActive version 3.1 [15] as the Grid environment. J2SE5.0 or the JRE of the same version plus ProActive 3.1 need to be installed on all nodes that are supposed to be used by the parallel GPSS simulator. The parallel simulator might also work with later versions of the Java Runtime Environment and ProActive as long as these are backwards compatible but the author of this work can give no guarantees in this respect.

Because the parallel simulator and the libraries it uses are written in Java it can be run on many different platforms. But the main target platforms of this work are Unix based systems because the scripts that are part of the parallel simulator are only provided as Unix shell scripts. These relatively basic scripts will need to be rewritten before the parallel simulator can be used on Windows or other non-Unix based platforms.

### 5.4.2  Files

The following files are required or are optionally needed in order to run the parallel simulator. They can be found in the folder /ParallelJavaGpssSimulator/ on the attached CD and will briefly be described here.

**deploymentDescriptor.xml**

This is the ProActive deployment descriptor file mentioned in 3.2.1. It is read by ProActive to determine which nodes the parallel simulator should use and how these need to be accessed. A detailed description of this file and the deployment configuration of ProActive can be found at the ProActive project Web site [15]. ProActive uses the concept of virtual nodes for its deployment. For the parallel simulator the ProActive deployment descriptor file needs to contain the virtual node *ParallelJavaGpssSimulator*. If this virtual node is not found then the parallel simulator will abort with an error message. In addition the deployment descriptor file needs to define enough JVM nodes linked to this virtual node for the number of partitions within the simulation model to be simulated.

**DescriptorSchema.xsd**

This is the XML schema file that describes the structure of the deployment descriptor XML file. It is used by ProActive to verify that the XML structure of the file *deploymentDescriptor.xml* mentioned above is correct.

**env.sh**

This Unix shell script is part of ProActive and is only included because it is needed by the file *startNode.sh* described further down. It can also be found in the ProActive installation. Together with the file *startNode.sh* it is used to start ProActive nodes directly from this folder. But first the environment variable *PROACTIVE* defined in the beginning of this file might have to be changed to point to the installation location of the ProActive library.

**ParallelJavaGpssSimulator.jar**

This is the JAR file (Java archive) that contains the Java class files, which make up the parallel simulator. It is required by the script *simulate.sh* described further down in order to start and run the parallel simulator.

**proactive.java.policy**

This is a copy of the default security policy file provided by ProActive. It can also be found in the ProActive installation and is provided here so that the parallel simulator can be run straight from this folder. This security policy file basically disables any access restrictions by granting all permissions. It should only be used when no security and access restrictions are needed. Please refer to the ProActive documentation [15] regards defining a proper security policy for a ProActive Grid application like the parallel simulator.

**proactive-log4j**

This is the log4j logging configuration file used by the parallel simulator and ProActive. A description of this file and how logging is configured for the parallel simulator can be found in section 5.3.5.

**simulate.config**

This is the default configuration file for the parallel simulator. The configuration of the parallel simulator is explained in detail in 5.4.3.

**simulate.sh**

This Unix shell script is used to start the parallel simulator. It defines the two environment variables *PROACTIVE* and *SIMULATOR*. Both might need to be changed before the parallel simulator can be run so that *PROACTIVE* points to the ProActive installation directory and *SIMULATOR* points to the directory containing the parallel simulator JAR file *ParallelJavaGpssSimulator.jar*. Further details about how to run the parallel simulator can be found in 5.4.4.

**startNode.sh**

This Unix shell script is part of ProActive and is used to start a ProActive node. It is a copy of the of the same file found in the ProActive installation and is only provided here so that ProActive nodes for the LPs of the parallel simulator can be started straight from the same directory. The file *env.sh* is called be this script to setup all environment variables needed by ProActive.

## 5.4.3 Configuration

The parallel simulator can be configured using command line arguments or by a configuration file. The reading of the configuration settings from the command line arguments or from the configuration file is handled by the *Configuration* class in the root package. If the parallel simulator is started with no further command line arguments after the simulation model file name then the default configuration file *simulate.config* is used for the configuration. If the next command line argument after the simulation model file name has the format *ConfigFile=...* then the specified configuration file is used. Otherwise the configuration is read from the existing command line arguments and default values are used for any configuration settings not specified.

Configuration settings have the format *=<value>* and Boolean configuration settings can be specified without value and equal sign in which case they are set to true. This is useful when specifying configuration settings as command line arguments. For instance to get the parallel simulator to output the parsed simulation model it is enough to add the command line argument *ParseModelOnly* instead of *ParseModelOnly=true*. A detailed description of the configuration settings can be found in Appendix B.

## 5.4.4 Starting a Simulation

Before a simulation model can be simulated using the parallel simulator the *deploymentDescriptor.xml* needs to contain enough node definitions linked to the virtual node *ParallelJavaGpssSimulator* for the number of partitions within the simulation model. If the deployment descriptor file does not define how ProActive can automatically start the required nodes then the ProActive nodes have to be created manually on the relevant machines using the *startNode.sh* script before the parallel simulator can be started.

The parallel simulator is started using the shell script *simulate.sh*. The exact syntax is:

**simulate.sh <simulation model file> [<command line argument>] […]**

The configuration of the parallel simulator and possible command line arguments are described in 5.4.3 and the files required to run the parallel simulator and their meaning are explained in 5.4.2.

## 5.4.5 Increasing Memory Provided by JVM

By default the JVM of J2SE5.0 provides only a maximum of 64MB of memory to the Java applications that run inside it (Maximum Memory Allocation Pool Size). Considering that at the time of this paper standard PCs already come with a physical memory of around 1GB and dedicated server machines even more, the Maximum Memory Allocation Pool Size of the JVM does not seem appropriate. Therefore in order to make the best possible use of the memory provided by the Grid nodes the Maximum

Memory Allocation Pool Size of the JVM needs to be increased to the amount of memory available. This is especially important for long running simulations and complex simulation models.

The Maximum Memory Allocation Pool Size of the JVM can be set using the command line argument –Xmx*n* of the java command (see Java documentation for more details [31]). If the ProActive nodes running the LPs of the parallel simulator are started using the *startNode.sh* script then this command line argument with the appropriate memory size can be added to this script, otherwise if the nodes are started via the deployment descriptor file then the command line argument has to be added there. The following example shows how the *startNode.sh* script needs to be changed in order to increase the Maximum Memory Allocation Pool Size from its default value to 512MB.

```
…
$JAVACMD  org.objectweb.proactive.core.node.StartNode  $1 $2 $3 $4
$5 $6 $7
…
```

Extract of the *startNode.sh* script with default memory pool size

```
…
$JAVACMD  –Xmx512m  org.objectweb.proactive.core.node.StartNode  $1
$2 $3 $4 $5 $6 $7
…
```

Extract of the *startNode.sh* script with memory pool size of 512MB

# 6    Validation of the Parallel Simulator

The functionality of the parallel simulator was validated using a set of example simulation models. These simulation models were deliberately kept very simple in order to evaluate specific aspects of the parallel simulator as complex models would possibly hide some of the findings and would make the analysis of the results more difficult. Each of the validations evaluates a particular part of the overall functionality and the example simulation models were specifically chosen for that evaluation. They therefore don't represent any real live systems. Of course it cannot be expected that this validation using example simulation models will prove the absolute correctness of the implemented functionality. But instead the different validation runs performed provide a sufficient level of confidence that the functionality of the parallel simulator is correct.

All files required to perform these validations including the specific configuration files and the resulting validation output log files can be found in specific sub folders of the attached CD. For further details about the CD see Appendix D. The relevant output log files of the validation runs performed are also included in Appendix F. Line numbers in brackets were added to all lines of the output log files in order to make it possible to refer to a particular line. The log4j logging system [3] was specifically configured for each validation run to include certain details or exclude details that were not relevant to that particular validation. The Termination Counters for the validation runs were chosen so that the simulation runs were long enough to evaluate the specific aspects but also kept as short as possible in order to avoid unnecessary long output log files. Nevertheless some of the validations still resulted in long output log files. In these cases some of the lines that were not relevant to the validation have been removed from the output logs listed in Appendix F. The complete output log files can still be found on the attached CD.

The validation runs were performed on a standard PC with a single CPU (Intel Pentium 4 with 3.2GHz, 1GB RAM) running SuSE Linux 10.0. As the validation was performed only on a single CPU it should be noted that it does not represent a detailed investigation into the performance of the parallel simulator. Such an investigation would exceed the expected time scale of this project because the performance of a parallel simulation depends on a lot of different factors besides the simulation system (e.g.

simulation model, computation and communication performance of the hardware used) and would need to be analysed using a variety of simulation models and on different systems in order to draw any reliable conclusions. Nevertheless some basic performance conclusions where made as part of Validation 5 and 6.

## 6.1  Validation 1

The first validation checks the correct parsing of the supported GPSS syntax elements. Two models are used to evaluate the parser component of the parallel simulator. Both include examples of all GPSS block types and other GPSS entities but in the first model all possible parameters of the blocks and entities are used whereas in the second model all optional parameters are left out in order to test the correct defaulting by the parser. For both models the simulator was started using the *ParseModelOnly* command line argument option. When this option is specified then the simulator will not actually perform the simulation but instead parse the specified simulation model and either output the parsed in memory object structure representation of the simulation model or parsing errors if found.

**Validation 1.1**

The first simulation model used is shown below:

```
PARTITION Partition1,5
STORAGE Storage1,2
GENERATE 1,0,100,50,5
ENTER Storage1,1
ADVANCE 5,3
LEAVE Storage1,1
TRANSFER 0.5,Label1
TERMINATE 1
PARTITION Partition2,10
Label1 QUEUE Queue1
DEPART Queue1
SEIZE Facility1
RELEASE Facility1
TERMINATE 1
```

Simulation model file model_validation1.1.gps

The output log for this simulation model can be found in Appendix F. A comparison of the original simulation model file and the in memory object structure representation that was output by the simulator shows that they are equivalent and that the parser correctly parsed all lines of the simulation model.

**Validation 1.2**

The simulation model for this validation is based on the earlier simulation model but all optional elements of the model were removed.

```
STORAGE Storage1
GENERATE
ENTER Storage1
ADVANCE
LEAVE Storage1
TRANSFER Label1
TERMINATE
PARTITION Partition2
Label1 QUEUE Queue1
DEPART Queue1
SEIZE Facility1
RELEASE Facility1
TERMINATE
```

Simulation model file model_validation1.2.gps

As described this simulation model tests the parser regards setting default values for optional elements and parameters. Comparing the simulation model file to corresponding output log in Appendix F it can be found that the parser automatically created a new partition before parsing the first line of the model so that the in memory representation of the model contains two partitions (see line 2 of output log). The default name given to this partition by the parser is 'Partition 1' (see line 6 of output log). Line 9 shows that the Storage size was set to its maximum value of 2147483647. The GENERATE block at line 10 was parsed with all its parameters set to its default values as described in Appendix A. This also applies to the ADVANCED block at line 12 and the TERMINATE blocks at line 15 and 27 of the output log. The usage count of the ENTER and LEAVE block at the lines 11 and 13 were set to the expected default value of 1 and the TRANSFER block at line 14 of the output log also has the default transfer probability of 1 so that all Transactions would be transferred to the specified label. It can be seen that all the missing parameters were set to their expected default values.

## 6.2  Validation 2

This validation evaluates the basic GPSS functionality of the parallel simulator. This includes the basic scheduling and the movement of Transactions as well as the correct processing of the GPSS blocks. The simulation model used for this contains only a single partition but otherwise all possible GPSS block types and entities. There is even a TRANSFER block that transfers Transactions with a probability of 0.5. The model is shown below:

```
PARTITION Partition1,4
STORAGE Storage1,2
GENERATE 3,2
QUEUE Queue1
ENTER Storage1,1
ADVANCE 5,3
LEAVE Storage1,1
DEPART Queue1
TRANSFER 0.5,Label1
SEIZE Facility1
RELEASE Facility1
Label1 TERMINATE 1
```

Simulation model file model_validation2.gps

The model is simulated with the log4j loggers *parallelJavaGpssSimulator.simulation* and *parallelJavaGpssSimulator.gpss* set to DEBUG (see configuration file proactive-log4j at the corresponding sub folder on the attached CD). The last of these two loggers will result in a very detailed logging of the GPSS processing and Transaction movement. For this reason the Termination Counter is kept very small, i.e. set to 4 so that the simulation is stopped after 4 Transactions have been terminated. Otherwise the output log would be too long to be useful. The deployment descriptor XML file is set to a single ProActive node as the model contains exactly one partition and the simulation will require only one LP.

The interesting output for this validation is the output log of the LP. Following this output log the simulation starts with initialising the GENERATE block (line 4 to 6). This results in a new Transaction with the ID 1 being chained in for the move time 4. The model above shows the GENERATE block with an average interarrival time of 3 and a half range of 2. This means that the interarrival times of the generated Transactions will lie in the open interval (1,5) with possible values of 2, 3 or 4. The current block of the new Transaction is (1,1) which is the GENERATE block itself as this Transaction has not been moved yet (in the logging of the parallel simulator a block reference is shown as a comma separated set of the partition number and the block number within that partition). The next step of the simulator found in the log is the updating of the simulation time to the value 4 at line 7 because the first movable Transaction (the one just generated) has a move time of 4. The lines 8 to 16 show how this Transaction is moved through the model until it reaches the ADVANCED block where it is delayed. The first block to be executed by the Transaction is the GENERATE block which results in a second Transaction being created when the first one is leaving this block as shown in line 10 and 11. The lines 12 and 13 show the first Transaction executing the QUEUE and ENTER block until it reaches the ADVANCE block at line 14. The ADVANE block changes the move time of the Transaction from 4 to 9 (delay by a value of 5), which means that, this Transaction is no longer movable at the simulation time of 4. At line 16 the Transaction is therefore chain back into the Transaction chain and because there is no other movable Transaction for the time of 4 the current simulation time is updated to the move time of the next movable Transaction, which is the one with an ID of 2 and a move time of 7. In the lines 18 to 26 the second Transaction is going through the same move process like the first Transaction before and when it is leaving the GENERATE block this results in a third Transaction with a move time of 10 being created and chained in. When the ADVANCE block changes the move time of the second Transaction from 7 to 13 as shown in line 24 the current simulation time is updated to the value of 9 and the first Transaction starts moving again (see line 27 to 35). It will execute the LEAVE and DEPART block before reaching the TRANSFER block at line 34. Here it is transferred directly to the TERMINATE block which can be seen from the next block property of the Transaction jumping from the block (1,7) to block (1,10). After executing the TERMINATE block the Transaction stops moving but is not chained back into the Transaction chain as it has been terminated (see line 35 and 36). The simulator proceeds with updating the simulation time and moving the next Transaction. The rest of the output log can be followed analogue to before.

The output log of the Simulate process is also found in Appendix F. This log contains the post simulation report and shows the interesting fact that all of the 4 Transactions that executed the TRANSFER block were transferred directly to the TERMINATE block. There should have been a ratio of 50% of the Transactions transferred but because the number of Transactions is very low this results in a large statistical error. Nevertheless Validation 3 will show that for a large number of Transactions the statistical behaviour of the TRANSFER block is correct.

The validation has shown that the Transaction scheduling and movement as well as the processing of the blocks is performed by the simulator as expected.

## 6.3  Validation 3

The third validation focuses on the exchange of Transaction between LPs. It evaluates that the sending of Transactions from one LP to another works correctly including the correct functioning of the TRANSFER block. In addition it shows that an LP can correctly handle the situation where it has no movable Transactions left. In a sequential simulator this would lead to an error and the abortion of the simulation but in a parallel simulator this is a valid state as long as at least one of the LPs has movable Transactions. Further this validation shows the correct processing of the simulation end by the Simulation Controller and the LPs.

**Validation 3.1**

This validation run uses a very simple model with two partitions. The first partition contains a GENERATE block and a TRANSFER block and the second partition the TERMINATE block. When run, the model will generate Transactions in the first partition and then transfer them to the second partition where they are terminated. Detailed GPSS logging is used again in order to follow the Transaction processing. The loggers enabled for debug output are shown below.

```
…
log4j.logger.parallelJavaGpssSimulator.gpss=DEBUG
log4j.logger.parallelJavaGpssSimulator.lp=DEBUG
…
log4j.logger.parallelJavaGpssSimulator.lp.rollback=DEBUG
…
log4j.logger.parallelJavaGpssSimulator.simulation=DEBUG
…
```

Extract of the used log4j configuration file proactive-log4j

To avoid unnecessary long output log files the Termination Counter for the partitions is set to 4 again so that the simulation will be finished after 4 Transactions have been terminated. In addition the GENERATE block has a limit count parameter of 10 so that it will only create a maximum of 10 Transactions. This limit is used because otherwise LP1 simulating the first partition would create more Transactions before the Simulation Controller has established the confirmed end of the simulation resulting in a longer output log with details that are not relevant to the simulation. The whole simulation model is shown below.

```
PARTITION Partition1,4
GENERATE 3,2,,10
TRANSFER Label1
PARTITION Partition2,4
Label1 TERMINATE 1
```

Simulation model file model_validation3.1.gps

Looking at the output log of LP1 the start of the Transaction processing is similar to the one of Validation 2. The initialisation of the GENERATE block creates the first Transaction (see line 5 to 7) and when the Transaction leaves the GENERATE block it creates the next Transaction and so forth. After the first Transaction with the ID 1 executed the TRANSFER block at line 15 it stops moving but is not chained back into the Transaction chain because it has been transferred to LP2 that simulates partition 2. The next block property of the Transaction shown in that line now points to the first block of partition two, i.e. has the value (2,1). Swapping to the output log of LP2 it can be seen at line 10 that LP2 just received the Transaction with the ID 1 and that this Transaction is chained in. Because of the communication delay and LP1 being ahead in simulation time LP2 already receives the next few Transactions as well as shown in line 12 and 13. In the lines 14 to 16 the first Transaction received is now chained out and moved into the TERMINATE block where it is terminated. The processing of any subsequent Transactions within LP1 and LP2 follows the same pattern.

The correct handling of the situation when an LP has no movable Transaction can be seen at line 9 of the output log of LP2. The LP will just stay in a waiting position not moving any Transaction until it either receives a Transaction from another LP or until the Simulation Controller establishes that the end of the simulation has been reached.

Using all three output log files from LP1, LP2 and the simulate process the correct processing of the simulation end can be followed. The first step of the simulation end is the forth Transaction being terminated in LP2 (see line 32 of the output log of LP2). The LP detects that a provisional simulation end has been reached (line 33) and requests a GVT calculation from the Simulation Controller (line 34). Subsequently it is still receiving Transactions from LP1 (e.g. line 35, 38 and 41) but no Transactions are moved because the LP is in the provisional simulation end mode. The output log of the simulate process shows at the lines 14 to 16 that the Simulation Controller performs a GVT calculation and receives the information that LP1 has reached the simulation time 26 and LP2 has reached a provisional simulation end at the time 11. Because all other LPs except LP2 have passed the provisional simulation end time the Simulation Controller concludes that the simulation end is confirmed. It now informs the LPs of the simulation end which can be seen in line 55 of the output log of LP2 and line 83 of the output log of LP1. Because LP1 is ahead of the simulation end time this information causes it to rollback to the state at the start of simulation time 11 (line 84). The rollback leads to the Transaction with ID 7 being moved to the TRANSFER block again to reach the exact state needed for the simulation end. It can be seen from the output log of LP1 that the lines 32 to 37 are identical to the lines 85 to 90 which is a result of the rollback and re-execution in order to reach a state that is consistent with the simulation end in LP2. Both LPs confirm to the Simulation Controller that they reached the consistent simulation end state, which then outputs the combined post simulation report showing the correct counts as seen in line 23 to 30 of the simulate process output log. This post simulation report confirms that four Transactions were moved through all blocks and a 5[th] is already waiting in the GENERATE. The GENERATE block has not yet been executed by the 5[th] Transaction because GENERATE blocks are executed when a Transaction leaves them.

The output logs confirm that the transfer of Transactions between LPs and the handling of the simulation end reached by one of the LPs works correctly as expected.

**Validation 3.2**

The second validation of this validation group looks at the correct statistical behaviour of the TRANSFER block when it is used with a transfer probability parameter. The model used for the validation run is similar to the model used by validation 3.1 but differs in the fact that this time the partition 1 has its own TERMINATE block and that the TRANSFER block only transfers 25% of the Transactions to partition 2. Below is the complete simulation model used for this run.

```
PARTITION Partition1,750
GENERATE 3,2
TRANSFER 0.25,Label1
TERMINATE 1
PARTITION Partition2,750
Label1 TERMINATE 1
```

Simulation model file model_validation3.2.gps

Another difference is that larger Termination Counter values are used in order to get reliable values for the statistical behaviour. With such large Termination Counter values the output logs of the LPs would be very long and not of much use, which is why, they are not included in Appendix F. The interesting output log for this validation run is the one of the simulate process and specifically the post simulation report. The expected simulation behaviour from the model shown above would be that LP1 reaches the simulation end after 750 Transactions have been terminated in it's TERMINATE block. Because the TRANSFER block transfers 25% of the Transactions to LP2 this means that about 1000 Transactions would need to be generated in LP1 of which around 250 should end up in LP2. The output log of the simulate process confirms in the lines 27 to 31 that this is the case. In fact the number of Transactions that reached LP2 and the overall count is only short of two Transactions. This proves that the statistical behaviour of the TRANSFER block is correct.

## 6.4  Validation 4

Evaluating the memory management of the parallel simulator is the subject of this validation. It will show that the simulator performs the correct actions when the two different defined memory limits are reached. The simulation model used for this validation is shown below.

```
PARTITION Partition1,2000
GENERATE 1,0
Label1 QUEUE Queue1
DEPART Queue1
TERMINATE 1
PARTITION Partition2
GENERATE 1,0,2000
TRANSFER Label1
```

Simulation model file model_validation4.gps

The simulation model contains two partitions. The first partition has a GENERATE block that will create a Transaction for each time unit. All Transactions will be terminated in the TERMINATE block of the first partition. The only purpose of the QUEUE and DEPART block in between is to slightly slow down the processing of the Transactions by the LP. The second partition also generates Transactions for each time unit but with an offset of 2000 so that its Transactions will start from the simulation time 2000. The second partition will therefore always be ahead in simulation time compared to the first partition and all its Transactions are transferred to the QUEUE block within the first partition. From the design of this model it can be seen that the first partition will become a bottleneck because on top of its own Transactions it will also receive Transactions from the second partition. The number of Transactions in its Transaction chain, i.e. Transactions that still need to be moved will constantly grow.

In order to reach the memory limits of the parallel simulator more quickly the script startnode.sh, which is used to start the ProActive nodes for the LPs, is changed so that the command line argument –Xmx12m is passed to the Java Virtual Machine. This instructs the JVM to make only 12MB of memory available to its Java programs. The LPs for this validation are therefore run with a memory limit of 12MB. To avoid any memory management side effects introduced by the LPCC, the LPCC is switched off in the config file *simulate.config*. The Termination Counter for the simulation model was chosen as small as possible but just about large enough for the simulation to reach the desired effects on the hardware used. The logging configuration was changed to include the current time and the debug output was enabled only for the loggers shown below.

```
…
log4j.logger.parallelJavaGpssSimulator.lp=DEBUG
log4j.logger.parallelJavaGpssSimulator.lp.commit=DEBUG
log4j.logger.parallelJavaGpssSimulator.lp.rollback=DEBUG
…
log4j.logger.parallelJavaGpssSimulator.simulation=DEBUG
…
```

Extract of the used log4j configuration file proactive-log4j

The output log of LP1 shows at line 7 that the memory limit 1 of 5MB available memory left is already reached after around 2 minutes of simulating. As expected the LP requests a GVT calculation so that some of its past states can be confirmed and fossil collected in order to free up memory. The GVT is received from the Simulation Controller and possible states are committed and fossil collection at line 9 and 10. Because Java uses garbage collection the memory freed by the LP does not become available immediately. As a result the LP is still operating pass the memory limit 1 and requests a few further GVT calculations until between line 65 and 66 there is more than one minute of simulation without GVT calculation because the garbage collector has freed enough memory for the LP to be out of memory limit 1. This pattern repeats itself several time as the memory used by the LP keeps growing until at line 376 of the output log LP1 reaches the memory limit 2 of 1MB of available memory left. At this point LP1 turns into cancelback mode and cancels back 25 of the Transactions it received from LP2 in order to free up memory. This can be seen at line 377. After the cancelback of these Transactions and another GVT calculation the memory available to LP1 raises above the memory limit 2 and the LP changes back from cancelback mode into normal simulation mode as shown in line 381. The effects of the Transactions cancelled back on LP2 can be seen in line 294, 295 and the following lines of the output log of LP2. Because Transactions are cancelled back one by one as they might have been received from different LPs they do not all arrive at LP2 at once. The log shows that the 25 Transactions are cancelled back by LP2 in groups of 9, 9, 5 and 2 Transactions. It also shows that LP2 has reached the memory limit 2 even earlier than LP1. This fact can be explained by looking at the output log of the simulate process. The GVT calculation shown at the lines 311 to 313 indicates that LP2 has a lot more saved simulation states then LP1. LP1 hast started simulating at the time 1, has created one Transaction for every time unit and has reached a simulation time of 1520. That makes it 1520 saved simulation states of which some will have been fossil collected already. LP2 has started simulating from the time 2000 and has reached a simulation time of 4481 which means 2481 saves simulation states of which non will have been fossil collected as the GVT has not yet reached 2000. Saved simulation states require more memory than outstanding Transactions. This explains why LP2 had reached the memory limit 2 and the cancelback mode before LP1. But because LP2 does not receive Transactions from other LPs it has no Transactions that it can cancelback.

The validation showed that the memory management of the parallel simulator works as expected. The LPs perform the required actions when they reach the defined memory limits and avoided Out Of Memory Exceptions by not running completely out of memory.

## 6.5  Validation 5

The fifth validation evaluates the correct functioning of the Shock Resistant Time Warp algorithm and its main component, the LPCC. It will show that the LPCC within the LPs is able to steer the simulation towards an actuator value that results in less rolled back Transaction moves compared to normal Time Warp.

The simulation model used for this validation contains two partitions. Both partitions have a GENERATE block and a TERMINATE block but in addition partition 1 also contains a TRANSFER block that with a very small probability of 0.001 sends some of its Transactions to partition 2. The whole model is constructed so that partition 2 is usually ahead in simulation time compared to partition 1, achieved through the different configuration of the GENERATE blocks, and that occasionally partition 2 receives a Transaction from the first partition. Because partition 2 is usually ahead in simulation time this will lead to rollbacks in this partition. The simulation stops when 20000 Transactions have been terminated in partition 2. This model attempts to emulate the common scenario where a distributed simulation uses nodes with different performance parameters or partitions that create different loads so that during the simulation the LPs drift apart and some of them are further ahead in simulation time than others leading to rollbacks and re-execution. The details of the model used can be seen below.

```
PARTITION Partition1,20000
GENERATE 1,0
TRANSFER 0.001,Label1
TERMINATE 0
PARTITION Partition2,20000
GENERATE 4,0,5000
Label1 TERMINATE 1
```

Simulation model file model_validation5.gps

In order to reduce the influence of the general memory management on this validation the amount of memory available to the LPs was increase from the default value of 64MB to 128MB by adding the JVM command line argument -Xmx128m in the *startNode.sh* script used to start the ProActive nodes for the LPs. The logging configuration was extended to get additional debug output relevant to the processing of the LPCC and some additional LP statistics at the end of the simulation. The loggers for which debug logging was enabled are shown below.

```
…
log4j.logger.parallelJavaGpssSimulator.lp=DEBUG
log4j.logger.parallelJavaGpssSimulator.lp.commit=DEBUG
log4j.logger.parallelJavaGpssSimulator.lp.rollback=DEBUG
…
log4j.logger.parallelJavaGpssSimulator.lp.stats=DEBUG
log4j.logger.parallelJavaGpssSimulator.lp.lpcc=DEBUG
log4j.logger.parallelJavaGpssSimulator.lp.lpcc.statespace=DEBUG
log4j.logger.parallelJavaGpssSimulator.simulation=DEBUG
…
```

Extract of the used log4j configuration file proactive-log4j

**Validation 5.1**

The first validation run of this model was performed with the LPCC enabled and the LPCC processing time interval set to 5 seconds. The extract of the simulation configuration file below shows all the configuration settings relevant to the LPCC.

```
…
LpccEnabled=true
LpccClusterNumber=500
LpccUpdateInterval=5
```

Extract of the configuration file simulate.config for validation 5.1

From the output log of LP2 it can be seen that LP2 constantly has to rollback to an earlier simulation time because of Transactions it receives from LP1. For instance in line 6 of this output log LP2 has to roll back from simulation time 13332 to the time 1133 and in line 7 from time 6288 to 1439. The LPCC is processing the indicator set around every 5 seconds. Such a processing step is shown for instance in the lines 18 to 25 and lines 37 to 44. During these first two LPCC processing steps no actuator is being set (9223372036854775807 is the Java value of Long.MAX_VALUE meaning no actuator is set) because no past indicator set that promises a better performance indicator could be found. But at the third LPCC indicator processing a better indicator set is found as shown from line 55 to 62 and the actuator was set to 4967 as a result. At this point the number of uncommitted Transaction moves does not reach this limit but slightly later during the simulation, when the actuator limit is 4388, the LP reaches a number of uncommitted Transaction moves that exceeds the current actuator limit forcing the LP into cancelback mode. This can be found in the output log from line 109. While in cancelback mode LP2 is not cancelling back any Transactions received from LP1 as these are earlier than any Transaction generated within LP2 and therefore have been executed and terminated already but being in cancelback mode also means that no local Transactions are processed reducing the lead in simulation time of LP2 compared to LP1. LP2 stays in cancelback mode until Transaction moves are committed during the next GVT calculation reducing the number of uncommitted Transaction moves below the actuator limit. The following table shows the Actuator values set by the LPCC during the simulation and whether the Actuator limit was exceeded resulting in the cancelback mode.

| LPCC processing step | Time | Actuator limit | Limit exceeded |
|---:|---|---:|---|
| 1 | 19:37:10 | no limit | No |
| 2 | 19:37:15 | no limit | No |
| 3 | 19:37:20 | 4967 | No |
| 4 | 19:37:25 | 4267 | No |
| 5 | 19:37:30 | 4388 | Yes |
| 6 | 19:37:35 | 4396 | No |
| 7 | 19:37:40 | 3135 | Yes |
| 8 | 19:37:45 | 3146 | Yes |
| 9 | 19:37:50 | 3762 | No |
| 10 | 19:37:55 | 2817 | No |

Table 10: Validation 5.1 LPCC Actuator values

From this table it is possible to see that the LPCC is limiting the number of uncommitted Transactions and therefore the progress of LP2 in order to reduce the number of rolled back Transaction moves and increase the number of committed Transaction moves per second, which is the performance indicator. The graph below shows the same Actuator values in graphical form.

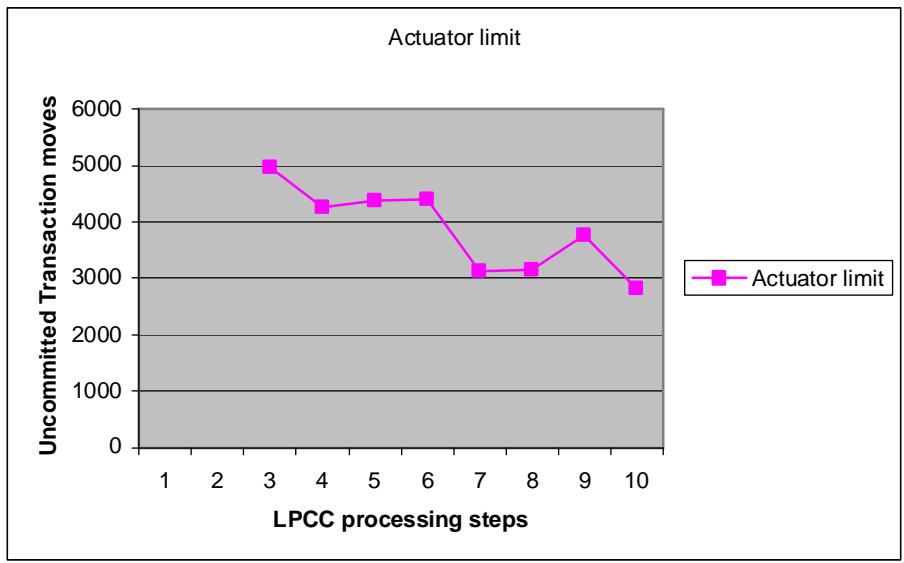

Figure 23: Validation 5.1 Actuator value graph

**Validation 5.2**

Exactly the same simulation model and logging was used for the second simulation run but this time the LPCC was switched off so that the normal Time Warp algorithm was used to simulate the model. The only configuration setting changed for this simulation is the following:

```
…
LpccEnabled=false
…
```

Extract of the configuration file simulate.config for validation 5.2

Regards rollbacks the output log for LP2 looks similar compared to Validation 5.1 but does not contain any logging from the LPCC as this was switched off. Therefore the LP does not reach any Actuator limit and does not switch into cancelback mode.

**Comparison of validation 5.1 and 5.2**

Comparing the output logs of validation 5.1 and 5.2 it becomes visible that performing the given simulation model using the Shock Resistant Time Warp algorithm reduces the number of Transaction moves rolled back. This can be seen especially when comparing the statistic information output by LP2 at the end of both simulation runs as found in the output log files for LP2 or in the table below.

| LP statistic item | Validation 5.1 | Validation 5.2 |
|---|---|---|
| Total committed Transaction moves | 19639 | 19953 |
| Total Transaction moves rolled back | 70331 | 77726 |
| Total simulated Transaction moves | 90330 | 97725 |

Table 11: LP2 processing statistics of validation 5

Table 11 shows that the simulation run of validation 5.1 using the Shock Resistant Time Warp algorithm required around 7400 less rolled back Transaction moves, which is about 10% less compared to the simulation run of validation 5.2 using the Time Warp algorithm. As a result the total number of Transaction moves performed by the simulation was reduced as well. The simulation run using the Shock Resistant Time Warp algorithm also performed slightly better than the simulation run using the Time Warp algorithm. This can be seen from the simulation performance information output as part of the post simulation report found in the output logs of the simulate process for both simulation runs. For validation 5.1 the simulation performance was 1640 time units per second real time and for validation 5.2 1607 time units per second real time. The performance difference is quite small which suggests that for the example model used the processing saved on rolled back Transaction moves just about out weights the extra processing required for the LPCC, additional GVT calculations and the extra logging for the LPCC (the LP2 output log size of validation 5.2 is only around 3% of the one of validation 5.1). But for more complex simulation models where rollbacks in one LP lead to cascaded rollbacks in other LPs a much larger saving on the number of rolled back Transaction moves can be expected. It also needs to be considered that the hardware setup used (i.e. all nodes being run on a single CPU machine) is not ideal for a performance evaluation as the main purpose of this validation is to evaluate the functionality of the parallel simulator.

## 6.6  Validation 6

During the testing of the parallel simulator it became apparent that in same cases the normal Time Warp algorithm can outperform the Shock Resistant Time Warp algorithm. This last validation is showing this in an example. The simulation model used is very similar to the one used for validation 5. It contains two partitions with the first partition transferring some of its Transactions to the second partition but this time the GENERATE blocks are configured so that the first partition is ahead in simulation time compared to the second. The simulation is finished when 3000 Transactions have been terminated in one of the partitions. The complete simulation model can be seen here:

```
PARTITION Partition1,3000
GENERATE 1,0,2000
TRANSFER 0.3,Label1
TERMINATE 1
PARTITION Partition2,3000
GENERATE 1,0
Label1 TERMINATE 1
```

Simulation model file model_validation6.gps

As a result of the changed GENERATE block configuration and the first partition being ahead of the second partition in simulation time, all Transactions received by partition 2 from partition 1 are in the future for partition 2 and no rollbacks will be caused. But it will lead to an increase of the number of outstanding Transactions within partition 2 pushing up the number of uncommitted Transaction moves during the simulation.

The logging configuration for this validation is also similar to the one used for validation 5 except that the LP statistic is not needed this time and is therefore switched off. The extract below shows the loggers for which debug output was enabled.

```
…
log4j.logger.parallelJavaGpssSimulator.lp=DEBUG
log4j.logger.parallelJavaGpssSimulator.lp.commit=DEBUG
log4j.logger.parallelJavaGpssSimulator.lp.rollback=DEBUG
…
log4j.logger.parallelJavaGpssSimulator.lp.lpcc=DEBUG
log4j.logger.parallelJavaGpssSimulator.lp.lpcc.statespace=DEBUG
log4j.logger.parallelJavaGpssSimulator.simulation=DEBUG
…
```

Extract of the used log4j configuration file proactive-log4j

Like for validation 5 the script startNode.sh used to run the LP nodes is changed to extend the memory limit of the JVM to 128MB.

**Validation 6.1**

For the first validation run the LPCC was enabled using the same configuration like for validation 5.1 resulting in the model being simulated using the Shock Resistant Time Warp algorithm. The significant effect of the simulation run is that the LPCC in LP2 starts setting actuator values in order to steer the local simulation processing towards a state that promises better performance but because the number of uncommitted Transaction moves within the second partition increases as a result of the Transactions received from partition 1 the actuator limits set by the LPCC are reached and the LP is switched into cancelback mode leading to its simulation progress being slowed down. In addition LP1 is also slowed down by the Transactions cancelled back from LP 2 as indicated by the output log of LP1. The actuator being set for LP2 can for instance be seen in line 25 to 28 of the output log of LP2. Subsequently the actuator limit is reached as shown in line 32, 36 and 42 of the output log and the LP turned into cancelback mode in the lines below. LP2 keeps switching into cancel back mode and keeps cancelling back Transactions to LP1 for large parts of the simulation resulting in a significant slowdown of the overall simulation progress.

**Validation 6.2**

The output logs for validation 6.2 are very short compared to the former simulation run because the simulation is performed using the normal Time Warp algorithm with the LPCC being switched off. Therefore no actuator values are set and none of the LPs is switching into cancelback mode. There are also no rollbacks so that the simulation progresses with the optimum performance for the model and setup used.

**Comparison of validation 6.1 and 6.2**

The simulation model for this validation is processed with the optimum simulation performance by the normal Time Warp algorithm. As a result no performance increase can be expected from the Shock Resistant Time Warp algorithm. But the Shock Resistant Time Warp Algorithm performs significantly worse than the normal Time Warp algorithm. This is caused by the Shock Resistant Time Warp algorithm slowing down the LP that is already behind in simulation time, i.e. the slowest LP. The validation shows that the approach of the Shock Resistant Time Warp algorithm to optimise the parameters of each LP by only considering local status information within these LPs does not always work.

## 6.7  Validation Analysis

The first few validations evaluate basic functional aspects of the parallel simulator. For instance validation 1 focuses on the GPSS parser component of the simulator and validation 2 on the basic GPSS simulation engine functionality. The transfer of Transactions between LPs is the main subject matter of validation 3 and the memory management is evaluated by validation 4. Validation 5 examines the correct functioning of the LPCC as the main component of the Shock Resistant Time Warp algorithm. Using specific simulation models and configurations these validations demonstrate with a certain degree of confidence that the parallel simulator is functionally correct and working as expected.

In addition validation 5 and 6 give some basic idea about the performance of the parallel transaction-oriented simulation based on the Shock Resistant Time Warp algorithm and about the performance of the parallel simulator even so the validations performed here cannot be seen as proper performance validations. Validation 5 shows that the Shock Resistant Time Warp algorithm can successfully reduce the number of rolled back Transaction moves resulting in more useful processing during the simulation and possibly better performance. But validation 6 revealed that the Shock Resistant Time Warp algorithm can also perform significantly worse than normal Time Warp. This is usually the case when the LPCC of the LP that is already furthest behind in simulation time decides to set an actuator value that limits the simulation progress resulting in the LP and the overall simulation progress being slowed down further.

The problem of the Shock Resistant Time Warp algorithm found is a direct result of the fact that if implemented as described in [8] the control decisions of each LPCC are only based on local sensor values within each LP and not on an overall picture of the simulation progress. Such problems could possibly be avoided by combining the Shock Resistant Time Warp algorithm with ideas from the adaptive throttle scheme suggested in [33] which is also briefly described in section 4.2.2. The GFT needed by this algorithm in order to describe the Global Progress Window could easily be determined and passed back to the LPs together with the GVT after the GVT calculation without much additional processing being required or communication overhead being created. Using the information of such a global progress window one option to improve the Shock Resistant Time Warp algorithm would be to add another sensor to the LPCC that describes the position of the current LP within the Global Progress Window. But another option that promises greater influence of the global progress information on the local LPCC control decisions would be to change the function that determines the new actuator value in a way that makes direct use of the global progress information. Such a function could for instance ignore the actuator value resulting from the best past state found if the position of the LP within the Global Progress Window is very close towards the GVT. It could also increase or decrease the actuator value returned by the original function depending on whether the LP is located within the slow or the fast zone of the Global Progress Window (see Figure 8 in 4.2.2). And a finer influence of the position within the Global Progress Window could be reached by dividing this window into more than 2 zones, each resulting in a slightly different effect on the actuator value and the future progress of the LP. Future work on this parallel simulator could investigate and compare these options with a prospect of creating a synchronisation algorithm that combines the advantages of both these algorithms.

# 7    Conclusions

Even so the performance of modern computer systems has steadily increased during that last decades the ever growing demand for the simulation of more and more complex systems can still lead to long runtimes. The runtime of a simulation can often be reduced by performing its parts distributed across several processors or nodes of a parallel computer system. Purpose-build parallel computer systems are usually very expensive. This is where Computing Grids provide a cost-saving alternative by allowing several organisations to share their computing resources. A certain type of Computing Grids called Ad Hoc Grid offers a dynamic and transient resource-sharing infrastructure, suitable for short-term collaborations and with a very small administrative overhead that makes it even for small organisations or individual users possible to form Computing Grids.

In the first part of this paper the requirements of Ad Hoc Grids are outlined and the Grid framework ProActive [15] is identified as a Grid environment that fulfils these requirements. The second part analyses the possibilities of performing parallel transaction-oriented simulation with a special focus on the space-parallel approach and synchronisation algorithms for discrete event simulation. From the algorithms considered the Shock Resistant Time Warp algorithm [8] was chosen as the most suitable for transaction-oriented simulation as well as the target environment of Ad Hoc Grids. This algorithm was subsequently applied to transaction-oriented simulation, considering further aspects and properties of this simulation type. These additional considerations included the GVT calculation, detection of the simulation end, cancellation techniques suitable for transaction-oriented simulation and the influence of the model partitioning. Following the theoretical decisions a Grid-based parallel transaction-oriented simulator was then implemented in order to demonstrate the decisions made. Finally the functionality of the simulator was evaluated using different simulation models in several validation runs in order to show the correctness of the implementation.

The main contribution of this work is to provide a Grid-based parallel transaction-oriented simulator that can be used for further research, for educational purpose or even for real live simulations. The chosen Grid framework ProActive ensures its suitability for Ad Hoc Grids. The parallel simulator can operate according to the normal Time Warp or the Shock Resistant Time Warp algorithm allowing large-scale performance comparisons of these two synchronisation algorithms using different simulation models and on different hardware environments. It was shown that the Shock Resistant Time Warp algorithm can successfully reduce the number of rolled back Transaction moves, which for simulations with many or long cascaded rollbacks will lead to a better simulation performance. But this work also revealed a problem of the Shock Resistant Time Warp algorithm, implemented as described in [8]. Because according to this algorithm all LPs try to optimise their properties based only on local information it is possible for the Shock Resistant Time Warp algorithm to perform significantly worse than the normal Time Warp algorithm. Future work on this simulator could improve the Shock Resistant Time Warp algorithm by making the LPs aware of their position within the GPW as suggested in [33]. Combining these two synchronisation algorithms would create an algorithm that has the advantages of both without any major additional communication and processing overhead.

Future work on improving this parallel transaction-oriented simulator could also look at employing different GVT algorithms and state saving schemes. Possible options were suggested in 4.3 and 5.3.3. This work also discussed the aspect of accessing objects in a different LP including a single shared Termination Counter. As mentioned in the report these options were not implemented in the parallel simulator and could be considered for future enhancements. Finally the simulator does not support the full GPSS/H language but only a large sub-set of the most important entities, which leaves further room for improvements.

# Appendix A: Detailed GPSS Syntax

This is a detailed description of the GPSS syntax supported by the parallel GPSS simulator. The correct syntax of the simulation models loaded into the simulator is validated by the GPSS parser within the simulator (see section 5.2.1).

GPSS simulation model files are line-based text files. Each line contains a block definition, a Storage definition or a partition definition. Comment lines starting with the sign * or empty lines are ignored by the parser.

A block definition line starts with an optional label followed by the reserved word of a block type and an optional comma separated list of parameters. The label is used to reference the block definition for jumps or a branched execution path from other blocks. The label, the block type and the parameter list need to be separated by at least one space. Note that a comma separated parameter list cannot contain any spaces. Any other characters following the comma separated parameter list are considered to be comments and are ignored. Labels as well as other entity names (i.e. for Facilities, Queues and Storages) can contain any alphanumerical characters but no spaces and they must be different to the defined reserved words. Labels for two different block definitions within the same model cannot be same. But labels are case sensitive so that two labels that only differ in the case of their characters are considered different. Apposed to that reserved words are not case sensitive but is an accepted convention in GPSS to use upper case letters for GPSS reserved words. Storage definitions and partition definitions cannot start with a label. They therefore start with the reserved word STORAGE or PARTITION. Below the syntax of the different GPSS definitions is explained in more detail. For a detailed description of the actual GPSS functionality see [26].

## Partition definition:

**Reserved word:** PARTITION

**Syntax:** PARTITION [<partition name>[,<termination counter>]]

**Description:**

The partition definition declares the beginning of a new partition. If the optional *partition name* parameter is not specified then it will default to 'Partition x' with x being the number of the partition within the model. The *partition name* cannot contain any spaces. If the optional *termination counter* parameter is not specified then the default termination counter value from the simulator configuration will be used. If specified then the *termination counter* parameter has to be a positive integer value.

## Storage definition:

**Reserved word:** STORAGE

**Syntax:** STORAGE <Storage name>[,<Storage capacity>]

**Description:**

The Storage definition declares a Storage entity. The *Storage name* parameter is required. If the optional *Storage capacity* parameter is not specified then it will default to the maximum value of the Java *int* type. If specified then the *Storage capacity* parameter needs to be a positive integer value. The Storage definition has to appear in the simulation model before any block that uses the specific Storage.

## Block definitions:

**Reserved word:** ADVANCE

**Syntax:** [<label>] ADVANCE [<average holding time>[,<half range>]]

**Description:**

The ADVANCE block delays Transactions by a fixed or random amount of time. The optional *average holding time* parameter describes the average time by which the Transaction is delayed defaulting to zero. Together with the *average holding time* the second parameter which is the *half range* parameter describe the uniformly distributed random value range from which the actually delay time is drawn. If the half range parameter is not specified then the delay time will always have the deterministic value of the *average holding time* parameter. Both parameters can either be a positive integer value or zero. In addition the half range parameter cannot be greater than the average holding time parameter.

**Reserved word:** DEPART

**Syntax:** [<label>] DEPART <Queue name>

**Description:**

The DEPART block causes the Transaction to leave the specified Queue entity. The required *Queue name* parameter describes the Queue entity that the Transaction will leave.

**Reserved word:** ENTER

**Syntax:** [<label>] ENTER <Storage name>[,<usage count>]

**Description:**

Through the ENTER block a Transaction will capture a certain number of units from the specified Storage entity. The *Storage name* parameter defines the Storage entity used and the second optional *usage count* parameter specifies how many Storage units will be captured by the Transaction. If not specified this parameter will default to 1. Otherwise this parameter has to have a positive integer value that is less or equal to the size of the Storage. If the specified number of units are not available for that Storage then the Transaction will be blocked until they become available.

**Reserved word:** GENERATE

**Syntax:** [<label>] GENERATE [<average interarrival time>[,<half range>
          [,<time offset>[,<limit count>[,<priority>]]]]]

**Description:**

The GENERATE block generates new Transactions that enter the simulation. The first two parameter *average interarrival time* and *half range* describe the uniformly distributed random value range from which the interarrival time is drawn. The interarrival time is the time between the last Transaction entering the simulation through this block and the next one. Both these parameters default to zero if not specified. The next parameter is the *time offset* parameter that describes the arrival time of the first Transaction. If it is not specified then the arrival time of the first Transaction will be determined via a uniformly distributed random sample using the first two parameters. The *limit count* parameter specifies the total count of Transactions that will enter the simulation through this GENERATE block. If the count is reached then no further Transactions are generated. If this parameter is not specified then no limit applies. The *priority* parameter specifies the priority value assigned to the generated Transactions and will default to 0. If specified all these parameters are required to have a positive integer value or zero. In addition the *half range* parameter cannot be greater than the average *interarrival time* parameter.

**Reserved word:** LEAVE

**Syntax:** [<label>] LEAVE <Storage name>[,<usage count>]

**Description:**

The LEAVE block will release the specified number of Storage units held by the Transaction. The *Storage name* parameter defines the Storage entity from which units will be released and the second optional *usage count* parameter specifies how many Storage units will be released by the Transaction. If not specified this parameter will default to 1. Otherwise this parameter has to have a positive integer value that is less or equal to the size of the Storage. If the specified number of units is greater than the number of units currently held by the Transaction then a runtime error will occur.

**Reserved word:** QUEUE

**Syntax:** [<label>] QUEUE <Queue name>

**Description:**

The QUEUE block causes the Transaction to enter the specified Queue entity. The required *Queue name* parameter describes the Queue entity that the Transaction will enter.

**Reserved word:** RELEASE

**Syntax:** [<label>] RELEASE <Facility name>

**Description:**

The RELEASE block will release the specified Facility entity held by the Transaction. The *Facility name* parameter defines the Facility entity that will be released. If the Facility entity is not currently held by the Transaction then a runtime error will occur.

**Reserved word:** SEIZE

**Syntax:** [<label>] SEIZE <Facility name>

**Description:**

Through the SEIZE block a Transaction will capture the specified Facility entity. The *Facility name* parameter defines the Facility entity that will be captured. If the Facility entity is already held by another Transaction then the Transaction will be blocked until it becomes available.

**Reserved word:** TERMINATE

**Syntax:** [<label>] TERMINATE [<Termination Counter decrement>]

**Description:**

TERMINATE blocks are used to destroy Transactions. When a Transaction enters a terminate block then the Transaction is removed from the model and not chained back into the Transaction chain. Each time a Transaction is destroyed by a TERMINATE block the local Termination counter is decremented by the decrement specified for that TERMINATE block. The *Termination Counter decrement* parameter is optional and will default to zero if it is not specified. TERMINATE blocks with a zero decrement parameter will not change the Termination Counter when they destroy a Transaction.

**Reserved word:** TRANSFER

**Syntax:** [<label>] TRANSFER [<transfer probability>,]<destination label>

**Description:**

A TRANSFER block changes the execution path of a Transaction based on the specified probability. Normally a Transaction is moved from one block to the next but when it executes a TRANSFER block the Transaction can be transferred to a different block than the next following. This destination can even be located in a different partition of the model. For the decision of whether a Transaction is transferred or not a random value is drawn and compared to the specified probability. If the random value is less than or equal to the probability then the Transaction is transferred. The *transfer probability* parameter needs to be a floating point number between 0 and 1 (inclusive). It is optional and will default to 1, which will transfer all Transactions, if not specified. The *destination label* parameter has to be a valid block label within the model.

# Appendix B: Simulator configuration settings

This appendix describes the configuration settings that can be used for the parallel simulator. Most of these settings can be applied as command line arguments or as settings within the *simulate.config* file. A general description of the simulator configuration can be found in 5.4.3.

**Setting:** ConfigFile

**Default value:** simulate.config

**Description:**

This configuration setting can only be used as a command line argument and it has to follow straight after the simulation model file name. It specifies the name of the configuration file used by the parallel simulator.

**Setting:** DefaultTC

**Default value:** none

**Description:**

This is a configuration setting that defines the default Termination Counter used for partitions that do not have a Termination Counter defined in the simulation model file. When specified then it needs to have a non-negative numeric value.

**Setting:** DeploymentDescriptor

**Default value:** ./deploymentDescriptor.xml

**Description:**

The ProActive deployment descriptor file used by the parallel simulator is specified using this configuration setting.

**Setting:** LogConfigDetails

**Default value:** false

**Description:**

If this Boolean configuration setting is switched on then the parallel simulator will always output the current configuration setting used including default ones at the start of a simulation. This can be useful for debugging purposes.

**Setting:** LpccClusterNumber

**Default value:** 1000

**Description:**

This numeric configuration setting sets the maximum number of clusters stored in the State Cluster Space used by the LPCC. If a new indicator set is added to the State Cluster Space and the maximum number of clusters has been reached already then two clusters or a cluster and the new indicator set are merged. The larger the value of this setting the more distinct state indicator sets can be stored and used by the Shock Resistant Time Warp algorithm but the more memory is also need to store such information.

**Setting:** LpccEnabled

**Default value:** true

**Description:**

If this Boolean configuration setting is set to true which is also the default of this setting then the LPCC is enabled and the simulation is performed according to the Shock Resistant Time Warp algorithm. Otherwise the LPCC is switched off and the normal Time Warp algorithm is used for the parallel simulation.

**Setting:** LpccUpdateInterval

**Default value:** 10

**Description:**

This is a configuration setting that defines the LPCC processing time interval. Its value has to be a positive number greater than zero and describes the number of seconds between LPCC processing steps. It also specifies how often the LPCC tries to find and set a new actuator value. For long simulation runs on systems with large memory pools larger LPCC processing intervals can be beneficial because less GVT calculation are needed. On the other hand if the systems used have frequently changing additional loads and the simulation model is known to have a frequently changing behaviour pattern then small values might give better results.

**Setting:** ParseModelOnly

**Default value:** false

**Description:**

If this configuration setting is enabled then the parallel simulator will parse the simulation model file and output the in memory representation of the model but no simulation will be performed. This setting can therefore be used to evaluate whether the simulation model was parsed correctly and to check which defaults have been set for optional GPSS parameter.

# Appendix C: Simulator log4j loggers

The following loggers of the parallel simulator can be configured in its log4j configuration file. In this section each logger is briefly described and its supported log levels are mentioned.

**Logger:** parallelJavaGpssSimulator.gpss

**Log levels used:** debug

**Description:**

This logger is used to output debug information of the GPSS block and Transaction processing during the simulation. It creates a detailed log of when a Transaction is moved, which blocks it executes and when it is chained in or out. It is also the root logger for any logging related to the GPSS simulation processing.

**Logger:** parallelJavaGpssSimulator.gpss.facility

**Log levels used:** debug

**Description:**

Whenever a Transaction releases a Facility entity this logger outputs detailed information about the Transaction and when it captured and released the Facility.

**Logger:** parallelJavaGpssSimulator.gpss.queue

**Log levels used:** debug

**Description:**

Whenever a Transaction leaves a Queue entity this logger outputs detailed information about the Transaction and when it entered and left the Queue.

**Logger:** parallelJavaGpssSimulator.gpss.storage

**Log levels used:** debug

**Description:**

Whenever a Transaction releases a Storage entity this logger outputs detailed information about the Transaction and when it captured and released the Storage.

**Logger:** parallelJavaGpssSimulator.lp

**Log levels used:** debug, info, error, fatal

**Description:**

This logger is the root logger for any output of the LPs. The default log level is *info* which outputs some basic information about what partition was assigned to the LP and when the simulation is completed. Errors within the LP are also output using this logger and if the *debug* log level is enabled then it outputs detailed information about the communication and the processing of the LP related to the synchronisation algorithm.

**Logger:** parallelJavaGpssSimulator.lp.commit

**Log levels used:** debug

**Description:**

This logger outputs information about when simulation states are committed and for which simulation time.

**Logger:** parallelJavaGpssSimulator.lp.rollback

**Log levels used:** debug

**Description:**

A logger that outputs information about the rollbacks performed.

**Logger:** parallelJavaGpssSimulator.lp.memory

**Log levels used:** debug

**Description:**

This logger outputs detailed information about the current memory usage of the LP and the amount of available memory within the JVM. It is called with each scheduling cycle and can therefore create very large logs if enabled.

**Logger:** parallelJavaGpssSimulator.lp.stats

**Log levels used:** debug

**Description:**

This logger outputs the values of the sensor counters that are also user by the LPCC as a statistic of the overall LP processing at the end of the simulation.

**Logger:** parallelJavaGpssSimulator.lp.lpcc

**Log levels used:** debug

**Description:**

A logger that outputs detailed information about the processing of the LPCC, including for instance any actuator values set or when an actuator limit has been exceeded.

**Logger:** parallelJavaGpssSimulator.lp.lpcc.statespace

**Log levels used:** debug

**Description:**

This logger outputs information about the processing of the State Cluster Space. This includes details of new indicator sets added or possible past indicator sets found that promises better performance.

**Logger:** parallelJavaGpssSimulator.simulation

**Log levels used:** debug, info, error, fatal

**Description:**

This is the root logger for all general output about the simulation, the Simulation Controller and the simulate process. The default log level is *info,* which outputs the standard information about the simulation. The logger also outputs errors thrown during the simulation and in *debug* mode gives detailed information about the processing of the Simulation Controller.

**Logger:** parallelJavaGpssSimulator.simulation.gvt

**Log levels used:** debug, info

**Description:**

A logger that outputs detailed information about GVT calculations if in *debug* level. If the logger is set to the *info* log level then only basic information about the GVT reached is logged.

**Logger:** parallelJavaGpssSimulator.simulation.report

**Log levels used:** info

**Description:**

This logger is the root logger for the post simulation report. It can be used to switch off the output of the post simulation report by setting the log level to *off*.

**Logger:** parallelJavaGpssSimulator.simulation.report.block

**Log levels used:** info

**Description:**

This is the logger that is used for the block section of the post simulation report. It allows this section to be switched off if required.

**Logger:** parallelJavaGpssSimulator.simulation.report.summary

**Log levels used:** info

**Description:**

This is the logger that is used for the summary section of the post simulation report. It allows this section to be switched off if required.

**Logger:** parallelJavaGpssSimulator.simulation.report.chain

**Log levels used:** info

**Description:**

This is the logger that is used for the Transaction chain section of the post simulation report. It allows this section to be switched off if required.

# Appendix D: Structure of the attached CD

The folder structure of the attached CD is briefly explained in this section. The root folder of the CD also contains this report as a Microsoft Word and PDF document.

**/ParallelJavaGpssSimulator**

This is the main folder of the parallel simulator. It contains all the files required to run the simulator as described in 5.4.2. It also contains some of the folders mentioned below.

**/ParallelJavaGpssSimulator/bin**

The folder structure within this folder contains all the binary Java class files of the parallel simulator. The same Java class files are also included in the main JAR file of the simulator.

**/ParallelJavaGpssSimulator/doc**

This folder contains the full JavaDoc documentation of the parallel simulator. The JavaDoc documentation can be viewed by opening the index.html file within this folder in a Web browser. It describes the source code of the parallel simulator and is generated from comments within the source code using the JavaDoc tool.

**/ParallelJavaGpssSimulator/src**

The src folder contains the actual source code of the parallel simulator, i.e. all the java files.

**/ParallelJavaGpssSimulator/validation**

This folder contains a sub-folder for each validation. All files required to repeat the validation runs can be found in these sub-folders, including simulation models, configuration and all the output log files of the validation runs described in section 6. The validation runs can be performed directly from these folders.

**/ProActive**

This additional folder contains the compressed archive of the ProActive version used.

# Appendix E: Documentation of selected classes

This section contains the JavaDoc documentation of the following selected classes.

- parallelJavaGpssSimulator.**SimulationController**

- parallelJavaGpssSimulator.lp.**LogicalProcess**

- parallelJavaGpssSimulator.lp.**ParallelSimulationEngine**

- parallelJavaGpssSimulator.lp.lpcc.**LPControlComponent**

The full JavaDoc documentation of all classes can be found on the attached CD, see Appendix D for further details.

**parallelJavaGpssSimulator**

# Class SimulationController

java.lang.Object

  └─ **parallelJavaGpssSimulator.SimulationController**

**All Implemented Interfaces:**

>   java.io.Serializable,
>
>   org.objectweb.proactive.Active,
>
>   org.objectweb.proactive.RunActive

---

public class **SimulationController**

extends java.lang.Object

implements org.objectweb.proactive.RunActive, java.io.Serializable

The SimulationController class creates, initializes and controls the Logical Processes and the overall simulation. It is created by the main simulation application class as a ProActive Active Object and communicates with the LogicProcess instances via ProActive remote method calls.

**Author:**

>   Gerald Krafft

**See Also:**

>   Serialized Form

---

## Field Summary

| static java.lang.String | **VIRTUAL_NODE_NAME**<br>        Name of the virtual node that needs to be defined in the deployment descriptor file. |
| --- | --- |

# Constructor Summary

**SimulationController**()

    Main constructor

# Method Summary

| | |
|---|---|
| static SimulationController | **createActiveInstance**(org.objectweb.proactive.core.node.Node node) <br><br> Static method that creates an Active Object SimulationController instance on the specified node |
| SimulationState | **getSimulationState**() <br><br> Returns the state of the simulation |
| void | **reportException**(java.lang.Exception e, int logicalProcessIndex) <br><br> Called by logical process instances to report exceptions thrown by the simulation. |
| void | **requestGvtCalculation**() <br><br> Called by LPs to request a GVT calculation by the SimulationController. |
| void | **runActivity**(org.objectweb.proactive.Body body) <br><br> Implements the main activity loop of the Active Object |
| void | **simulate**(Model model, Configuration configuration) <br><br> Starts parallel simulation of the specified model and using the specified configuration |
| void | **terminateLPs**() <br><br> This method terminates all LPs. |

| **Methods inherited from class java.lang.Object** |
|---|
| clone, equals, finalize, getClass, hashCode, notify, notifyAll, toString, wait, wait, wait |

# Field Detail

## VIRTUAL_NODE_NAME

public static final java.lang.String **VIRTUAL_NODE_NAME**

Name of the virtual node that needs to be defined in the deployment descriptor file. Its value is "ParallelJavaGpssSimulator".

**See Also:**

Constant Field Values

# Constructor Detail

## SimulationController

public **SimulationController**()

Main constructor

# Method Detail

## runActivity

public void **runActivity**(org.objectweb.proactive.Body body)

Implements the main activity loop of the Active Object

**Specified by:**

runActivity in interface org.objectweb.proactive.RunActive

**Parameters:**

`body` - body of the Active Object

**See Also:**

RunActive.runActivity(org.objectweb.proactive.Body)

---

## createActiveInstance

public static SimulationController
**createActiveInstance**(org.objectweb.proactive.core.node.Node node)
                          throws org.objectweb.proactive.ActiveObjectCreationException,
                                 org.objectweb.proactive.core.node.NodeException

Static method that creates an Active Object SimulationController instance on the specified node

**Parameters:**

`node` - node at which the instance will be created or within current JVM if null

**Returns:**

active instance of SimulationController

**Throws:**

org.objectweb.proactive.ActiveObjectCreationException

org.objectweb.proactive.core.node.NodeException

---

## simulate

public void **simulate**(Model model, Configuration configuration)
        throws org.objectweb.proactive.core.ProActiveException,
            org.objectweb.proactive.core.node.NodeException,
            CriticalSimulatorException,
            org.objectweb.proactive.core.mop.ClassNotReifiableException,

java.lang.ClassNotFoundException

Starts parallel simulation of the specified model and using the specified configuration

**Parameters:**

model - GPSS model that will be simulated

configuration - configuration settings

**Throws:**

org.objectweb.proactive.core.ProActiveException - can be thrown by ProActive

org.objectweb.proactive.core.node.NodeException - can be thrown by ProActive

CriticalSimulatorException - Critical error that makes simulation impossible

org.objectweb.proactive.core.mop.ClassNotReifiableException - can be thrown by ProActive

java.lang.ClassNotFoundException - can be thrown by ProActive

---

# terminateLPs

public void **terminateLPs**()

       throws java.io.IOException

This method terminates all LPs. It is called by the main application Simulate class. It returns an exception if terminating the LPs fails which automatically forces calls to this method to be synchronous.

**Throws:**

java.io.IOException

---

## reportException

**public void reportException(java.lang.Exception e, int logicalProcessIndex)**

Called by logical process instances to report exceptions thrown by the simulation. This method is used for exceptions that occur within runActivity() of these instances and not within remote method calls. Exceptions thrown within remote method calls are automatically passed back by ProActive.

**Parameters:**

**e** - Exception that was thrown in LP

**logicalProcessIndex** - index of the LP that reports the exception

---

## getSimulationState

**public SimulationState getSimulationState()**

Returns the state of the simulation

**Returns:**

state of the simulation

---

## requestGvtCalculation

**public void requestGvtCalculation()**

Called by LPs to request a GVT calculation by the SimulationController.

parallelJavaGpssSimulator.lp

# Class LogicalProcess

java.lang.Object

   └ **parallelJavaGpssSimulator.lp.LogicalProcess**

**All Implemented Interfaces:**

>java.io.Serializable,
>
>org.objectweb.proactive.Active,
>
>org.objectweb.proactive.RunActive

---

public class **LogicalProcess**

extends java.lang.Object

implements org.objectweb.proactive.RunActive, java.io.Serializable

This class implements a Logical Process (LP) according to the parallel simulation synchronisation algorithm. Each Logical Process simulates a specific partition of the simulation model. Logical Processes are instantiated as ProActive Active Objects. They communicate with each other in order to exchange Transactions and with the SimulationController instance for the overall control of the simulation.

**Author:**

>Gerald Krafft

**See Also:**

>Serialized Form

---

## Constructor Summary

**LogicalProcess**()

>Main constructor (also used for serialization purpose)

## Method Summary

| | |
|---|---|
| void | **cancelBackTransaction**(Transaction xact)<br><br>Called by other Logical Processes to force a cancel back of the specified Transaction sent by this LP. |
| protected void | **commitState**(long gvt)<br><br>Performs fossil collection for changes earlier than the GVT. |
| static LogicalProcess | **createActiveInstance**(org.objectweb.proactive.core.node.Node node)<br><br>This static method is called by the Simulation Controller in order to create a ProActive Active Object instance of the LogicalProcess class. |
| org.objectweb.proactive.core.util.wrapper.BooleanWrapper | **endOfSimulationByTransaction**(Transaction xact)<br><br>Requests the Logical Process to end the simulation at the specified Transaction. |
| void | **forceGvtAt**(long time)<br><br>Calling this method will force the Logical Process to request a GVT calculation as soon as it passes the specified simulation time. |
| SimulationReportSet | **getSimulationReport**(boolean includeChainReport)<br><br>Returns the simulation report. |
| protected void | **handleReceivedTransactions**()<br><br>Goes through the list of received Transactions and anti-Transactions and either chains the new Transaction in or undoes the original Transaction for received anti-Transaction. |

| void | **initialize**(Partition partition, |
|---|---|
| | org.objectweb.proactive.core.group.Group logicalProcessGroup, |
| | SimulationController simulationController, |
| | Configuration configuration) |
| | Initializes the Logical Process. |
| protected void | **needToCancelBackTransactions**(long count) |
| | Cancel back a certain number of Transactions. |
| void | **receiveGvt**(long gvt, boolean lpccProcessingNeeded) |
| | Called by SimulationController to send the calculated GVT (global virtual time). |
| org.objectweb.proactive.core.util. wrapper.BooleanWrapper | **receiveTransaction**(Transaction xact, boolean undo) |
| | Public method that is used by other Logical Processes to send a Transaction or anti-Transaction to this Logical Process. |
| LocalGvtParameter | **requestGvtParameter**() |
| | Returns the parameters of this Logical Process required for the GVT calculation. |
| protected void | **rollbackState**(long time) |
| | Rolls the state of the simulation engine back to the state for the given time or the next later state. |
| void | **runActivity**(org.objectweb.proactive.Body body) |
| | Implements the main activity loop of the Active Object |
| protected void | **saveCurrentState**() |
| | Saves the current state of the simulation engine into the local state list (unless an unconfirmed end of simulation has |

| | been reached by this LP or the LP is in Cancelback mode, in both cases the local simulation time would have the value of Long.MAX_VALUE). |
|---|---|
| protected void | **sendLazyCancellationAntiTransactions**()<br><br>    This method performs the main lazy-cancellation for Transactions that have been sent and subsequently rolled back. |
| void | **startSimulation**()<br><br>    Tells the LP to start simulating the local partition of the simulation model. |

| **Methods inherited from class java.lang.Object** |
|---|
| clone, equals, finalize, getClass, hashCode, notify, notifyAll, toString, wait, wait, wait |

# Constructor Detail

## LogicalProcess

public **LogicalProcess**()

    Main constructor (also used for serialization purpose)

# Method Detail

## createActiveInstance

public static LogicalProcess

**createActiveInstance**(org.objectweb.proactive.core.node.Node node)

                throws org.objectweb.proactive.ActiveObjectCreationException,

                    org.objectweb.proactive.core.node.NodeException

This static method is called by the Simulation Controller in order to create a ProActive Active Object instance of the LogicalProcess class. The Active Object instance is created at the specified node.

**Parameters:**

`node` - node at which the Active Object LogicalProcess instance will be created

**Returns:**

the Active Object LogicalProcess instance (i.e. a stub of the LogicalProcess instance)

**Throws:**

`org.objectweb.proactive.ActiveObjectCreationException`

`org.objectweb.proactive.core.node.NodeException`

---

## initialize

public void **initialize**([Partition](#) partition,

           org.objectweb.proactive.core.group.Group logicalProcessGroup,

           [SimulationController](#) simulationController,

           [Configuration](#) configuration)

Initializes the Logical Process. This method is called by the Simulation Controller. The initialization is done outside the constructor because it requires the group of all Logical Process active objects to be passed in. The LogicalProcess instance cannot be used before it is initialized.

**Parameters:**

`partition` - the simulation model partition that this LP will process

`logicalProcessGroup` - a group containing all LPs (i.e. stubs to all LPs)

---

## runActivity

public void **runActivity**(org.objectweb.proactive.Body body)

      Implements the main activity loop of the Active Object

      **Specified by:**

      runActivity in interface **org.objectweb.proactive.RunActive**

      **Parameters:**

      body - body of the Active Object

      **See Also:**

      RunActive.runActivity(org.objectweb.proactive.Body)

## startSimulation

public void **startSimulation**()

      Tells the LP to start simulating the local partition of the simulation model. The LP needs to be initialized by calling initialize() before the simulation can be started. This method is called by the Simulation Controller after it created and initialized all LogicalProcess instances.

## receiveTransaction

public org.objectweb.proactive.core.util.wrapper.BooleanWrapper
**receiveTransaction**(Transaction xact, boolean undo)

      Public method that is used by other Logical Processes to send a Transaction or anti-Transaction to this Logical Process.

      **Parameters:**

      xact - Transaction received from other LP

      undo - true if an anti-Transaction has been received

**Returns:**

Returns a Future object which allows the send to verify that is has been received

---

# cancelBackTransaction

public void **cancelBackTransaction**(<span style="color:blue">Transaction</span> xact)

> Called by other Logical Processes to force a cancel back of the specified
> Transaction sent by this LP.
>
> **Parameters:**
>
> xact - Transaction that needs to be cancelled back

---

# handleReceivedTransactions

protected void **handleReceivedTransactions**()

> Goes through the list of received Transactions and anti-Transactions and either
> chains the new Transaction in or undoes the original Transaction for received
> anti-Transaction. This method also handles received cancelbacks.

---

# needToCancelBackTransactions

protected void **needToCancelBackTransactions**(long count)

> Cancel back a certain number of Transactions. This method is called by the
> Logical Process if it is in CancelBack mode and it will attempt to cancel back
> the specified number of received Transactions from the end of the Transaction
> chain, i.e. the Transactions that are furthest ahead in simulation time and that
> where received from other LPs.
>
> **Parameters:**
>
> count - number of Transactions to cancel back

---

## sendLazyCancellationAntiTransactions

protected void **sendLazyCancellationAntiTransactions**()

> This method performs the main lazy-cancellation for Transactions that have been sent and subsequently rolled back. The method is called after the simulation time has been updated (increased). It looks for any past sent and rolled back Transactions that still exist in rolledBackSentHistoryList (i.e. that had not been re-sent in identical form after the rollback) and sends out anti-Transactions for these.

## commitState

protected void **commitState**(long gvt)

> Performs fossil collection for changes earlier than the GVT. This will remove any saved state information and any records in the sent and received history lists that are not needed any more.
>
> **Parameters:**
>
> guaranteedTime - time until which all Transaction movements are guarantied, this means there cannot be any rollback to a time before this time

## rollbackState

protected void **rollbackState**(long time)

> Rolls the state of the simulation engine back to the state for the given time or the next later state. This also changes some of the information within the Logical Process back to what it was at the time to which the simulation engine is rolled back.
>
> **Parameters:**
>
> time - time to which the simulation state will be rolled back

## saveCurrentState

**protected void saveCurrentState**()

Saves the current state of the simulation engine into the local state list (unless an unconfirmed end of simulation has been reached by this LP or the LP is in Cancelback mode, in both cases the local simulation time would have the value of Long.MAX_VALUE).

## requestGvtParameter

**public LocalGvtParameter requestGvtParameter**()

Returns the parameters of this Logical Process required for the GVT calculation. This method is called by the Simulation Controller when it performs a GVT calculation. The parameters include the minimum time of all received and not executed Transactions (i.e. either in receivedList or in the simulation engine queue) and the minimum time of any Transaction in transit (i.e. sent but not yet received).

**Returns:**

GVT parameter object

## receiveGvt

**public void receiveGvt**(long gvt, boolean lpccProcessingNeeded)

Called by SimulationController to send the calculated GVT (global virtual time). This time guarantees all executed Transactions and state changes with a time smaller than the GVT and as a result the Logical Process can perform fossil collection by committing any changes that happened before the GVT.

**Parameters:**

**gvt** - GVT (global virtual time)

## forceGvtAt

**public void forceGvtAt**(long time)

Calling this method will force the Logical Process to request a GVT calculation as soon as it passes the specified simulation time. If the specified time has been passed already then a GVT calculation is requested at the next simulation scheduling cycle. This method is called by a Logical Process that reached an unconfirmed End of Simulation in order to force other LPs to request a GVT calculation when they pass the provisional End of Simulation time.

**Parameters:**

time - simulation time after which a GVT calculation should be requested

---

## endOfSimulationByTransaction

**public org.objectweb.proactive.core.util.wrapper.BooleanWrapper**
**endOfSimulationByTransaction**(Transaction xact)

Requests the Logical Process to end the simulation at the specified Transaction. This method is called by the Simulation Controller when a GVT calculation confirms a provisional End of Simulation reached by one of the LPs. If this is the LP that reported the unconfirmed End of Simulation by this Transaction then it will have stopped simulating already. All other LPs will be rolled back to the time of this Transaction and then they will simulate any Transactions for the same time that in a sequential simulator would have been executed before the specified Transaction. Afterwards the simulation is stopped and completion is reported back to the SimulationController.

**Parameters:**

xact - Transaction that finished the simulation

**Returns:**

BooleanWrapper to indicate to the SimulationController that the LP completed the simulation at the specified end

## getSimulationReport

public SimulationReportSet **getSimulationReport**(boolean includeChainReport)

> Returns the simulation report. This method is called by the Simulation
> Controller after the simulation has finished in order to output the combined
> simulation report from all LPs. The simulation report can optionally contain the
> Transaction chain report section. This additional section is optional because it
> can be very large. It is therefore only returned if needed, i.e. requested by the
> user. This method will be called by the Simulation Controller after the
> simulation was completed in order to output the combined reports from all LPs.

**Parameters:**

includeChainReport - include Transaction chain report section

**Returns:**

populated instance of SimulationReportSet

**parallelJavaGpssSimulator.lp**

# Class ParallelSimulationEngine

java.lang.Object

└ parallelJavaGpssSimulator.gpss.SimulationEngine

   └ **parallelJavaGpssSimulator.lp.ParallelSimulationEngine**

**All Implemented Interfaces:**

      java.io.Serializable

---

public class **ParallelSimulationEngine**

extends SimulationEngine

implements java.io.Serializable

The parallel simulation engine used by the simulator. This class extends the basic GPSS simulation engine in order to provide additional functionality required by the parallel simulator.

**Author:**

      Gerald Krafft

**See Also:**

      SimulationEngine, Serialized Form

---

## Constructor Summary

| |
|---|
| **ParallelSimulationEngine**() |
|     Constructor for serialization purpose |
| **ParallelSimulationEngine**(Partition partition) |
|     Main constructor |

## Method Summary

| | |
|---|---|
| protected void | **chainIn**(Transaction inXact)<br><br>Overrides chainIn() from class parallelJavaGpssSimulator.gpss.SimulationEngine. |
| protected void | **deleteLaterTransactions**(Transaction xact)<br><br>Removes all Transactions from the local chain that would be executed/moved after the specified Transaction, i.e. all Transactions that have a move time later than the specified Transaction or with the same move time but a lower priority. |
| protected boolean | **deleteTransaction**(Transaction xact)<br><br>Removes the specified Transaction from the Transaction chain. |
| protected long | **getMinChainTime**()<br><br>Returns the minimum time of all movable Transactions in the Transaction chain. |
| protected long | **getNoOfTransactionsInChain**()<br><br>Returns the number of Transactions currently in the chain |
| long | **getTotalTransactionMoves**()<br><br>Returns the total number of Transaction moves performed since the start of the simulation. |
| protected<br>java.util.ArrayList<Transaction> | **getTransactionChain**()<br><br>Gives access to the Transaction chain for classes that inherit from SimulationEngine and makes this visible |
| | within the current package. |
| java.util.ArrayList<Transaction> | **getTransactionToSendList**()<br><br>Returns the out list of Transactions that need to be sent to other LPs. |
| protected  Transaction | **getUnconfirmedEndOfSimulationXact**()<br><br>Returns the Transaction that caused an unconfirmed end of simulation within this simulation engine. |
| void | **moveAllTransactionsAtCurrentTime**()<br><br>Moves all Transactions that are movable at the current simulation time. |
| protected  void | **moveTransaction**(Transaction xact)<br><br>Overrides the inherited method in order to add some sensor information used by the LP Control Component. |
| protected  void | **setCurrentSimulationTime**(long currentSimulationTime)<br><br>Sets the simulation time to the specified value |
| boolean | **unconfirmedEndOfSimulationReached**()<br><br>Returns whether an unconfirmed end of simulation has been reached by this engine |
| boolean | **updateClock**()<br><br>Overrides the inherited method. |

---

**Methods inherited from class parallelJavaGpssSimulator.gpss.SimulationEngine**

chainOutNextMovableTransactionForCurrentTime, getBlockForBlockReference,

getBlockReferenceForLocalBlock, getBlockReport, getChainReport, getCurrentSimulationTime,

getFacilitySummaryReport, getNextTransactionId, getNoOfTransactionsAtBlock, getPartition,

getQueueSummaryReport, getStorageSummaryReport, initializeGenerateBlocks, isTransactionBlocked,

setBlockReferenceToLocalBlock

---

**Methods inherited from class java.lang.Object**

clone, equals, finalize, getClass, hashCode, notify, notifyAll, toString, wait, wait, wait

---

# Constructor Detail

## ParallelSimulationEngine

public **ParallelSimulationEngine**()

> Constructor for serialization purpose

---

## ParallelSimulationEngine

public **ParallelSimulationEngine**(Partition partition)

> Main constructor
>
> **Parameters:**
>
> partition - model partition that will be simulated by this simulation engine

# Method Detail

## chainIn

protected void **chainIn**(Transaction inXact)

> Overrides chainIn() from class
>
> parallelJavaGpssSimulator.gpss.SimulationEngine. If the next block of the
>
> Transaction to be chained in lies in a different partition then the Transaction is stored in the out list so that it can later be sent to the LP of that partition, otherwise the inherited chainIn() method is called.

**Overrides:**

chainIn in class SimulationEngine

**Parameters:**

inXact - Transaction to be added to the chain

**See Also:**

SimulationEngine.chainIn(parallelJavaGpssSimulator.gpss.Transaction)

---

# moveAllTransactionsAtCurrentTime

public void **moveAllTransactionsAtCurrentTime**()

                          throws InvalidBlockReferenceException

Moves all Transactions that are movable at the current simulation time. This method overrides the same method from class parallelJavaGpssSimulator.gpss.SimulationEngine in order to implement end of simulation detection for parallel simulation.

**Overrides:**

moveAllTransactionsAtCurrentTime in class SimulationEngine

**Throws:**

InvalidBlockReferenceException

**See Also:**

SimulationEngine.moveAllTransactionsAtCurrentTime()

---

# moveTransaction

protected void **moveTransaction**(Transaction xact)

Overrides the inherited method in order to add some sensor information used by the LP Control Component. In addition it calls the inherited method to perform the actual movement of the Transaction.

**Overrides:**

moveTransaction in class SimulationEngine

**Parameters:**

**xact** - Transaction to move

**See Also:**

SimulationEngine.moveTransaction(Transaction xact)

---

# updateClock

public boolean **updateClock**()

Overrides the inherited method. The inherited method is only called and the simulation time updated if no provisional End of Simulation has been reached.

**Overrides:**

updateClock in class SimulationEngine

**Returns:**

true if a movable Transaction was found, otherwise false

**See Also:**

SimulationEngine.updateClock()

---

# deleteTransaction

protected boolean **deleteTransaction**(Transaction xact)

Removes the specified Transaction from the Transaction chain.

**Parameters:**

`xactId` - Id of the Transaction

**Returns:**

true if the Transaction was found and removed, otherwise false

---

# deleteLaterTransactions

protected void **deleteLaterTransactions**(Transaction xact)

> Removes all Transactions from the local chain that would be executed/moved after the specified Transaction, i.e. all Transactions that have a move time later than the specified Transaction or with the same move time but a lower priority.
>
> **Parameters:**
>
> `xact` - Transaction for which any later Transactions will be removed

---

# getTransactionChain

protected java.util.ArrayList<Transaction> **getTransactionChain**()

> Gives access to the Transaction chain for classes that inherit from SimulationEngine and makes this visible within the current package.
>
> **Overrides:**
>
> getTransactionChain in class SimulationEngine
>
> **Returns:**
>
> Returns the Transaction chain.

---

# getMinChainTime

protected long **getMinChainTime**()

Returns the minimum time of all movable Transactions in the Transaction chain. This is the current simulation time unless there are no Transactions in the chain or an unconfirmed end of simulation has been reached in which case Long.MAX_VALUE is returned. This method is used by the LP to determine the local time that will be sent to the GVT calculation.

**Returns:**

minimum local chain time

---

## getNoOfTransactionsInChain

protected long **getNoOfTransactionsInChain**()

Returns the number of Transactions currently in the chain

**Returns:**

number of Transactions in the chain

---

## setCurrentSimulationTime

protected void **setCurrentSimulationTime**(long currentSimulationTime)

**Description copied from class: SimulationEngine**

Sets the simulation time to the specified value

**Overrides:**

setCurrentSimulationTime in class SimulationEngine

**Parameters:**

currentSimulationTime - new current simulation time

---

## getUnconfirmedEndOfSimulationXact

protected Transaction **getUnconfirmedEndOfSimulationXact**()

Returns the Transaction that caused an unconfirmed end of simulation within this simulation engine.

**Returns:**

Returns Transaction that caused an unconfirmed end of simulation

---

## unconfirmedEndOfSimulationReached

public boolean **unconfirmedEndOfSimulationReached**()

Returns whether an unconfirmed end of simulation has been reached by this engine

**Returns:**

true if unconfirmed end of simulation has been reached

---

## getTotalTransactionMoves

public long **getTotalTransactionMoves**()

Returns the total number of Transaction moves performed since the start of the simulation. This information is required by the LPCC as a sensor value.

**Returns:**

total number of Transaction moves performed

---

## getTransactionToSendList

public java.util.ArrayList<Transaction> **getTransactionToSendList**()

Returns the out list of Transactions that need to be sent to other LPs.

**Returns:**

outgoing list of Transactions

**parallelJavaGpssSimulator.lp.lpcc**

# Class LPControlComponent

java.lang.Object

  └ **parallelJavaGpssSimulator.lp.lpcc.LPControlComponent**

**All Implemented Interfaces:**

     java.io.Serializable

---

public class **LPControlComponent**

extends java.lang.Object

implements java.io.Serializable

This class implements the LP Control Component, which is a major part of the Shock Resistant Time Warp algorithm. It is used to control the optimism of Time Warp by comparing the current set of sensor indicator values to past states and steering the simulation towards a state that promises a better performance.

**Author:**

    Gerald Krafft

**See Also:**

    Serialized Form

---

## Constructor Summary

**LPControlComponent**()

    Constructor for serialization

---

**LPControlComponent**(int clusterCount)

    Main constructor, initializes the cluster space with the maximum number of clusters to be held.

| | |
|---|---|
| **Method Summary** | |

| long | **getCurrentUncommittedMovesMeanLimit**() |
|---|---|
| | Returns the mean limit for uncommitted Transaction moves (AvgUncommittedMoves) that is based on the indicators passed in the last time processSensorValues() was called. |

| long | **getCurrentUncommittedMovesUpperLimit**() |
|---|---|
| | Returns the current upper limit for uncommitted Transaction moves as determined by the LPCC. |

| long | **getLastSensorProcessingTime**() |
|---|---|
| | Returns the last time the sensor values were processed in milliseconds. |

| SensorSet | **getSensorSet**() |
|---|---|
| | Returns the sensor set with the current sensor values. |

| boolean | **isUncommittedMovesValueWithinActuatorRange**(long uncommittedMoves) |
|---|---|
| | Returns true if the value is within the actuator limit using the UncommittedMoves standard deviation and a confidence level of 95%. |

| void | **processSensorValues**() |
|---|---|
| | This method performs the main processing of the sensor values which will result in a new actuator value. |

| |
|---|
| **Methods inherited from class java.lang.Object** |
| clone, equals, finalize, getClass, hashCode, notify, notifyAll, toString, wait, wait, wait |

## Constructor Detail

### LPControlComponent

public **LPControlComponent**()

> Constructor for serialization

---

### LPControlComponent

public **LPControlComponent**(int clusterCount)

> Main constructor, initializes the cluster space with the maximum number of
> clusters to be held.
>
> **Parameters:**
>
> clusterCount - maximum number of clusters to be held

## Method Detail

### getSensorSet

public **SensorSet** **getSensorSet**()

> Returns the sensor set with the current sensor values.
>
> **Returns:**
>
> sensor set

---

### getLastSensorProcessingTime

public long **getLastSensorProcessingTime**()

> Returns the last time the sensor values were processed in milliseconds.
>
> **Returns:**
>
> the last time the sensor values were processed.

## processSensorValues

**public void processSensorValues()**

> This method performs the main processing of the sensor values which will result in a new actuator value. It generates an indicator set for the sensor values, determines the closest (most similar) past indicator set with a higher performance indicator (CommittedMoveRate) using a state cluster space and then adds the current indicator set to the state cluster space.

## getCurrentUncommittedMovesMeanLimit

**public long getCurrentUncommittedMovesMeanLimit()**

> Returns the mean limit for uncommitted Transaction moves (AvgUncommittedMoves) that is based on the indicators passed in the last time processSensorValues() was called.
>
> **Returns:**
>
> mean actuator limit

## getCurrentUncommittedMovesUpperLimit

**public long getCurrentUncommittedMovesUpperLimit()**

> Returns the current upper limit for uncommitted Transaction moves as determined by the LPCC. This is the upper limit based on the average uncommitted moves limit, the standard deviation and a confidence level of 95%.
>
> **Returns:**
>
> upper actuator value

## isUncommittedMovesValueWithinActuatorRange

**public boolean isUncommittedMovesValueWithinActuatorRange**(long uncommittedMoves)

Returns true if the value is within the actuator limit using the UncommittedMoves standard deviation and a confidence level of 95%.

**Parameters:**

**uncommittedMoves** - current UncommittedMoves sample value

**Returns:**

true if UncommittedMoves sample value is within actuator limits, otherwise false

# Appendix F: Validation output logs

This appendix contains the relevant output log files resulting from the validation runs performed as part of the validation in section 6. Line numbers in brackets were added to all lines of the output log files in order to make it possible to refer to a specific line. For very long output log files non-relevant lines where removed and replaced with "...". But the complete output log files can still be found on the attached CD.

**Validation 1.1, output of simulate:**

```
(1)  Simulation model file read and parsed successfully.
(2)  2 partition(s) found in simulation model file.
(3)
(4)  Model details:
(5)    Partition details:
(6)      Name: Partition1
(7)      Partition No: 1
(8)      Partition's termination counter: 5
(9)      Storage: Storage1, capacity: 2
(10)     (1) Block: GENERATE 1,0,100,50,5
(11)     (2) Block: ENTER (Storage: Storage1, capacity: 2),1
(12)     (3) Block: ADVANCE 5,3
(13)     (4) Block: LEAVE (Storage: Storage1, capacity: 2),1
(14)     (5) Block: TRANSFER 0.5,(Label: Label1, partition: 2, block: 1)
(15)     (6) Block: TERMINATE 1
(16)   Partition details:
(17)     Name: Partition2
(18)     Partition No: 2
(19)     Partition's termination counter: 10
(20)     Label: Label1, partition: 2, block: 1
(21)     Queue: Queue1
(22)     Facility: Facility1
(23)     (1) Block: QUEUE (Queue: Queue1)
(24)     (2) Block: DEPART (Queue: Queue1)
(25)     (3) Block: SEIZE (Facility: Facility1)
(26)     (4) Block: RELEASE (Facility: Facility1)
(27)     (5) Block: TERMINATE 1
```

**Validation 1.2, output of simulate:**

(1)   Simulation model file read and parsed successfully.
(2)   2 partition(s) found in simulation model file.
(3)
(4)   Model details:
(5)     Partition details:
(6)       Name: Partition 1
(7)       Partition No: 1
(8)       Partition's termination counter: unspecified
(9)       Storage: Storage1, capacity: 2147483647
(10)      (1) Block: GENERATE 0.0,(no offset),(no limit),0
(11)      (2) Block: ENTER (Storage: Storage1, capacity: 2147483647),1
(12)      (3) Block: ADVANCE 0,0
(13)      (4) Block: LEAVE (Storage: Storage1, capacity: 2147483647),1
(14)      (5) Block: TRANSFER 1.0,(Label: Label1, partition: 2, block: 1)
(15)      (6) Block: TERMINATE 0
(16)    Partition details:
(17)      Name: Partition2
(18)      Partition No: 2
(19)      Partition's termination counter: unspecified
(20)      Label: Label1, partition: 2, block: 1
(21)      Queue: Queue1
(22)      Facility: Facility1
(23)      (1) Block: QUEUE (Queue: Queue1)
(24)      (2) Block: DEPART (Queue: Queue1)
(25)      (3) Block: SEIZE (Facility: Facility1)
(26)      (4) Block: RELEASE (Facility: Facility1)
(27)      (5) Block: TERMINATE 0

**Validation 2, output of LP:**

(1)
(2)    --- StartNode ----------------------------------
(3)   Simulation engine created for partition: Partition1
(4)   Initialize GENERATE blocks
(5)   xact(Id: 1, move time: 4, current block: (1,1), next block: (1,2)) chained in
(6)   GENERATE blocks initialized
(7)   Local simulation clock updated to 4
(8)   xact(Id: 1, move time: 4, current block: (1,1), next block: (1,2)) chained out
(9)   Move xact(Id: 1, move time: 4, current block: (1,1), next block: (1,2))
(10)  xact(Id: 2, move time: 7, current block: (1,1), next block: (1,2)) chained in
(11)  xact(Id: 1, move time: 4, current block: (1,1), next block: (1,2)) executed block (1,1) Block: GENERATE
(12)  xact(Id: 1, move time: 4, current block: (1,2), next block: (1,3)) executed block (1,2) Block: QUEUE
3,2,(no offset),(no limit),0
Queue: Queue1
(13)  xact(Id: 1, move time: 4, current block: (1,3), next block: (1,4)) executed block (1,3) Block: ENTER
(Storage: Storage1, capacity: 2),1
(14)  xact(Id: 1, move time: 9, current block: (1,4), next block: (1,5)) executed block (1,4) Block: ADVANCE
5,3
(15)  Finished moving xact(Id: 1, move time: 9, current block: (1,4), next block: (1,5))
(16)  xact(Id: 1, move time: 9, current block: (1,4), next block: (1,5)) chained in
(17)  Local simulation clock updated to 7
(18)  xact(Id: 2, move time: 7, current block: (1,1), next block: (1,2)) chained out
(19)  Move xact(Id: 2, move time: 7, current block: (1,1), next block: (1,2))
(20)  xact(Id: 3, move time: 10, current block: (1,1), next block: (1,2)) chained in
(21)  xact(Id: 2, move time: 7, current block: (1,1), next block: (1,2)) executed block (1,1) Block: GENERATE
3,2,(no offset),(no limit),0
Queue: Queue1
(22)  xact(Id: 2, move time: 7, current block: (1,2), next block: (1,3)) executed block (1,2) Block: QUEUE
(23)  xact(Id: 2, move time: 7, current block: (1,3), next block: (1,4)) executed block (1,3) Block: ENTER
(Storage: Storage1, capacity: 2),1
(24)  xact(Id: 2, move time: 13, current block: (1,4), next block: (1,5)) executed block (1,4) Block: ADVANCE
5,3

```
(25) Finished moving xact(Id: 2, move time: 13, current block: (1,4), next block: (1,5))
(26) xact(Id: 2, move time: 13, current block: (1,4), next block: (1,5)) chained in
(27) Local simulation clock updated to 9
(28) xact(Id: 1, move time: 9, current block: (1,4), next block: (1,5)) chained out
(29) Move xact(Id: 1, move time: 9, current block: (1,4), next block: (1,5))
(30) Storage: Storage1, Xact id: 1, entered at time: 4, left at time: 9, used capacity: 1
(31) xact(Id: 1, move time: 9, current block: (1,5), next block: (1,6)) executed block (1,5) Block: LEAVE
(Storage: Storage1, capacity: 2),1
(32) Queue: name, Xact id: 1, entered at time: 4, left at time: 9
(33) xact(Id: 1, move time: 9, current block: (1,6), next block: (1,7)) executed block (1,6) Block: DEPART
(Queue: Queue1)
(34) xact(Id: 1, move time: 9, current block: (1,7), next block: (1,10)) executed block (1,7) Block: TRANSFER
0.5,(Label: Label1, partition: 1, block: 10)
(35) xact(Id: 1, move time: 9, current block: (1,10), next block: (1,10)) executed block (1,10) Block:
TERMINATE 1
(36) Finished moving xact(Id: 1, move time: 9, current block: (1,10), next block: (1,10))
(37) Local simulation clock updated to 10
(38) xact(Id: 3, move time: 10, current block: (1,1), next block: (1,2)) chained out
(39) Move xact(Id: 3, move time: 10, current block: (1,1), next block: (1,2))
(40) xact(Id: 4, move time: 14, current block: (1,1), next block: (1,2)) chained in
(41) xact(Id: 3, move time: 10, current block: (1,1), next block: (1,2)) executed block (1,1) Block: GENERATE
3,2,(no offset),(no limit),0
(42) xact(Id: 3, move time: 10, current block: (1,2), next block: (1,3)) executed block (1,2) Block: QUEUE
(Queue: Queue1)
(43) xact(Id: 3, move time: 10, current block: (1,3), next block: (1,4)) executed block (1,3) Block: ENTER
(Storage: Storage1, capacity: 2),1
(44) xact(Id: 3, move time: 15, current block: (1,4), next block: (1,5)) executed block (1,4) Block: ADVANCE
5,3
(45) Finished moving xact(Id: 3, move time: 15, current block: (1,4), next block: (1,5))
(46) xact(Id: 3, move time: 15, current block: (1,4), next block: (1,5)) chained in
(47) Local simulation clock updated to 13
(48) xact(Id: 2, move time: 13, current block: (1,4), next block: (1,5)) chained out
(49) Move xact(Id: 2, move time: 13, current block: (1,4), next block: (1,5))
(50) Storage: Storage1, Xact id: 2, entered at time: 7, left at time: 13, used capacity: 1
(51) xact(Id: 2, move time: 13, current block: (1,5), next block: (1,6)) executed block (1,5) Block: LEAVE
```

(Storage: Storage1, capacity: 2),1
(52) Queue: name, Xact id: 2, entered at time: 7, left at time: 13
(53) xact(Id: 2, move time: 13, current block: (1,6), next block: (1,7)) executed block (1,6) Block: DEPART
(Queue: Queue1)
(54) xact(Id: 2, move time: 13, current block: (1,7), next block: (1,10)) executed block (1,7) Block:
TRANSFER 0.5,(Label: Label1, partition: 1, block: 10)
(55) xact(Id: 2, move time: 13, current block: (1,10), next block: (1,10)) executed block (1,10) Block:
TERMINATE 1
(56) Finished moving xact(Id: 2, move time: 13, current block: (1,10), next block: (1,10))
(57) Local simulation clock updated to 14
(58) xact(Id: 4, move time: 14, current block: (1,1), next block: (1,2)) chained out
(59) Move xact(Id: 4, move time: 14, current block: (1,1), next block: (1,2))
(60) xact(Id: 5, move time: 17, current block: (1,1), next block: (1,2)) chained in
(61) xact(Id: 4, move time: 14, current block: (1,1), next block: (1,2)) executed block (1,1) Block: GENERATE
3,2,(no offset),(no limit),0
(62) xact(Id: 4, move time: 14, current block: (1,2), next block: (1,3)) executed block (1,2) Block: QUEUE
(Queue: Queue1)
(63) xact(Id: 4, move time: 14, current block: (1,3), next block: (1,4)) executed block (1,3) Block: ENTER
(Storage: Storage1, capacity: 2),1
(64) xact(Id: 4, move time: 18, current block: (1,4), next block: (1,5)) executed block (1,4) Block: ADVANCE
5,3
(65) Finished moving xact(Id: 4, move time: 18, current block: (1,4), next block: (1,5))
(66) xact(Id: 4, move time: 18, current block: (1,4), next block: (1,5)) chained in
(67) Local simulation clock updated to 15
(68) xact(Id: 3, move time: 15, current block: (1,4), next block: (1,5)) chained out
(69) Move xact(Id: 3, move time: 15, current block: (1,4), next block: (1,5))
(70) Storage: Storage1, Xact id: 3, entered at time: 10, left at time: 15, used capacity: 1
(71) xact(Id: 3, move time: 15, current block: (1,5), next block: (1,6)) executed block (1,5) Block: LEAVE
(Storage: Storage1, capacity: 2),1
(72) Queue: name, Xact id: 3, entered at time: 10, left at time: 15
(73) xact(Id: 3, move time: 15, current block: (1,6), next block: (1,7)) executed block (1,6) Block: DEPART
(Queue: Queue1)
(74) xact(Id: 3, move time: 15, current block: (1,7), next block: (1,10)) executed block (1,7) Block:
TRANSFER 0.5,(Label: Label1, partition: 1, block: 10)
(75) xact(Id: 3, move time: 15, current block: (1,10), next block: (1,10)) executed block (1,10) Block:

```
TERMINATE 1
(76) Finished moving xact(Id: 3, move time: 15, current block: (1,10), next block: (1,10))
(77) Local simulation clock updated to 17
(78) xact(Id: 5, move time: 17, current block: (1,1), next block: (1,2)) chained out
(79) Move xact(Id: 5, move time: 17, current block: (1,1), next block: (1,2))
(80) xact(Id: 6, move time: 19, current block: (1,1), next block: (1,2)) chained in
(81) xact(Id: 5, move time: 17, current block: (1,1), next block: (1,2)) executed block (1,1) Block: GENERATE
3,2,(no offset),(no limit),0
(82) xact(Id: 5, move time: 17, current block: (1,2), next block: (1,3)) executed block (1,2) Block: QUEUE
Queue: Queue1
(83) xact(Id: 5, move time: 17, current block: (1,3), next block: (1,4)) executed block (1,3) Block: ENTER
(Storage: Storage1, capacity: 2),1
(84) xact(Id: 5, move time: 22, current block: (1,4), next block: (1,5)) executed block (1,4) Block: ADVANCE
5,3
(85) Finished moving xact(Id: 5, move time: 22, current block: (1,4), next block: (1,5))
(86) xact(Id: 5, move time: 22, current block: (1,4), next block: (1,5)) chained in
(87) Local simulation clock updated to 18
(88) xact(Id: 4, move time: 18, current block: (1,4), next block: (1,5)) chained out
(89) Move xact(Id: 4, move time: 18, current block: (1,4), next block: (1,5))
(90) Storage: Storage1, Xact id: 4, entered at time: 14, left at time: 18, used capacity: 1
(91) xact(Id: 4, move time: 18, current block: (1,5), next block: (1,6)) executed block (1,5) Block: LEAVE
(Storage: Storage1, capacity: 2),1
(92) Queue: name, Xact id: 4, entered at time: 14, left at time: 18
(93) xact(Id: 4, move time: 18, current block: (1,6), next block: (1,7)) executed block (1,6) Block: DEPART
(Queue: Queue1)
(94) xact(Id: 4, move time: 18, current block: (1,7), next block: (1,10)) executed block (1,7) Block:
TRANSFER 0.5,(Label: Label1, partition: 1, block: 10)
(95) xact(Id: 4, move time: 18, current block: (1,10), next block: (1,10)) executed block (1,10) Block:
TERMINATE 1
(96) GVT parameter requested by SimulationController, min. time: Infinite (unconfirmed end of simulation for
time 18)
```

**Validation 2, output of simulate:**

```
(1)  Simulation model file read and parsed successfully.
(2)  1 partition(s) found in simulation model file.
(3)
(4)  Creating SimulationController instance
(5)  parallelJavaGpssSimulator.SimulationController
(6)  SimulationController.runActivity() started
(7)  Creating LPs and starting simulation
(8)  Simulation started
(9)
(10) Please press:
(11) G + <Enter> to force GVT calculation
(12) X + <Enter> to terminate the simulation
(13)
(14) Initiated GVT calculation
(15) Min time received from LP1: Infinite (no movable transaction)
(16) Simulation finished
(17) ***************    Simulation report    ****************
(18) The simulation was completed at the simulation time: 18
(19) by the transaction xact(id: 4, move time: 18, current block: (1,10), next block: (1,10))
(20) The average simulation perfomance in simulation time per second real time was: 67,164177 (time units/s)
(21)
(22) Block report section:
(23) Block          current          total
(24)                xacts            xacts
(25) Partition: Partition1
(26) (1,1)              1               5      Block: GENERATE 3,2,(no offset),(no limit),0
(27) (1,2)              0               5      Block: QUEUE (Queue: Queue1)
(28) (1,3)              0               5      Block: ENTER (Storage: Storage1, capacity: 2),1
(29) (1,4)              1               5      Block: ADVANCE 5,3
(30) (1,5)              0               4      Block: LEAVE (Storage: Storage1, capacity: 2),1
(31) (1,6)              0               4      Block: DEPART (Queue: Queue1)
(32) (1,7)              0               4      Block: TRANSFER 0.5,(Label: Label1, partition: 1, block: 10)
```

```
(33) (1,8)           0   Block: SEIZE (Facility: Facility1)
(34) (1,9)           0   Block: RELEASE (Facility: Facility1)
(35) (1,10)          4   Block: TERMINATE 1
(36)
(37) Summary entity report section:
(38) Facility    average     total    average    current                current
(39)             usage       entries  time/unit  xact ID                content
(40) Facility1   0.0         0
(41) Queue       maximum     average  total    zero     percent  average    current   maximum
(42)             content     content  entries  entries  zeros    time/unit  content   content
(43) Queue1      2           1.1578947  5        0        0.0      4.4        1         1
(44) Storage     average     total    average  capacity  average    current   maximum
(45)             usage       captures time/unit          content    content   content
(46) Storage1    0.57894737  5        4.4      2         1.1578947  1         2
(47)
(48) SimulationController.terminateLPs() called
```

**Validation 3.1, output of LP1:**

```
(1)   --- StartNode ----------------------------
(2)   LogicalProcess.runActivity() started
(3)   Simulation engine created for partition: Partition1
(4)   Initialize GENERATE blocks
(5)   GENERATE blocks initialized
(6)   xact(Id: 1, move time: 2, current block: (1,1), next block: (1,2)) chained in
(7)   GENERATE blocks initialized
(8)   LP1 with partition 'Partition1' initialized
(9)   LP simulation state changed from INITIALIZED to SIMULATING
(10)  Local simulation clock updated to 2
(11)  xact(Id: 1, move time: 2, current block: (1,1), next block: (1,2)) chained out
(12)  Move xact(Id: 1, move time: 2, current block: (1,1), next block: (1,2))
(13)  xact(Id: 3, move time: 5, current block: (1,1), next block: (1,2)) chained in
(14)  xact(Id: 1, move time: 2, current block: (1,1), next block: (1,2)) executed block (1,1) Block: GENERATE
3,2,(no offset),10,0
(15)  xact(Id: 1, move time: 2, current block: (1,2), next block: (2,1)) executed block (1,2) Block: TRANSFER
1.0,(Label1, partition: 2, block: 1)
(16)  Finished moving xact(Id: 1, move time: 2, current block: (1,2), next block: (2,1))
(17)  Local simulation clock updated to 5
(18)  xact(Id: 3, move time: 5, current block: (1,1), next block: (1,2)) chained out
(19)  Move xact(Id: 3, move time: 5, current block: (1,1), next block: (1,2))
(20)  xact(Id: 5, move time: 7, current block: (1,1), next block: (1,2)) chained in
(21)  xact(Id: 3, move time: 5, current block: (1,1), next block: (1,2)) executed block (1,1) Block: GENERATE
3,2,(no offset),10,0
(22)  xact(Id: 3, move time: 5, current block: (1,2), next block: (2,1)) executed block (1,2) Block: TRANSFER
1.0,(Label1, partition: 2, block: 1)
(23)  Finished moving xact(Id: 3, move time: 5, current block: (1,2), next block: (2,1))
(24)  Local simulation clock updated to 7
(25)  xact(Id: 5, move time: 7, current block: (1,1), next block: (1,2)) chained out
(26)  Move xact(Id: 5, move time: 7, current block: (1,1), next block: (1,2))
(27)  xact(Id: 7, move time: 11, current block: (1,1), next block: (1,2)) chained in
(28)  xact(Id: 5, move time: 7, current block: (1,1), next block: (1,2)) executed block (1,1) Block: GENERATE
```

```
3,2,(no offset),10,0
(29) xact(Id: 5, move time: 7, current block: (1,2), next block: (2,1)) executed block (1,2) Block: TRANSFER
1,0,(Label: Label1, partition: 2, block: 1)
(30) Finished moving xact(Id: 5, move time: 7, current block: (1,2), next block: (2,1))
(31) Local simulation clock updated to 11
(32) xact(Id: 7, move time: 11, current block: (1,1), next block: (1,2)) chained out
(33) Move xact(Id: 7, move time: 11, current block: (1,1), next block: (1,2))
(34) xact(Id: 9, move time: 14, current block: (1,1), next block: (1,2)) chained in
(35) xact(Id: 7, move time: 11, current block: (1,1), next block: (1,2)) executed block (1,1) Block: GENERATE
3,2,(no offset),10,0
(36) xact(Id: 7, move time: 11, current block: (1,2), next block: (2,1)) executed block (1,2) Block: TRANSFER
1,0,(Label: Label1, partition: 2, block: 1)
(37) Finished moving xact(Id: 7, move time: 11, current block: (1,2), next block: (2,1))
(38) Local simulation clock updated to 14
(39) xact(Id: 9, move time: 14, current block: (1,1), next block: (1,2)) chained out
(40) Move xact(Id: 9, move time: 14, current block: (1,1), next block: (1,2))
(41) xact(Id: 11, move time: 17, current block: (1,1), next block: (1,2)) chained in
(42) xact(Id: 9, move time: 14, current block: (1,1), next block: (1,2)) executed block (1,1) Block: GENERATE
3,2,(no offset),10,0
(43) xact(Id: 9, move time: 14, current block: (1,2), next block: (2,1)) executed block (1,2) Block: TRANSFER
1,0,(Label: Label1, partition: 2, block: 1)
(44) Finished moving xact(Id: 9, move time: 14, current block: (1,2), next block: (2,1))
(45) Local simulation clock updated to 17
(46) xact(Id: 11, move time: 17, current block: (1,1), next block: (1,2)) chained out
(47) Move xact(Id: 11, move time: 17, current block: (1,1), next block: (1,2))
(48) xact(Id: 13, move time: 19, current block: (1,1), next block: (1,2)) chained in
(49) xact(Id: 11, move time: 17, current block: (1,1), next block: (1,2)) executed block (1,1) Block:
GENERATE 3,2,(no offset),10,0
(50) xact(Id: 11, move time: 17, current block: (1,2), next block: (2,1)) executed block (1,2) Block:
TRANSFER 1,0,(Label: Label1, partition: 2, block: 1)
(51) Finished moving xact(Id: 11, move time: 17, current block: (1,2), next block: (2,1))
(52) Local simulation clock updated to 19
(53) xact(Id: 13, move time: 19, current block: (1,1), next block: (1,2)) chained out
(54) Move xact(Id: 13, move time: 19, current block: (1,1), next block: (1,2))
(55) xact(Id: 15, move time: 22, current block: (1,1), next block: (1,2)) chained in
```

```
(56) xact(Id: 13, move time: 19, current block: (1,1), next block: (1,2), executed block (1,1) Block:
GENERATE 3,2,(no offset),10,0
(57) xact(Id: 13, move time: 19, current block: (1,2), next block: (2,1)) executed block (1,2) Block:
TRANSFER 1.0,(Label: Label1, partition: 2, block: 1)
(58) Finished moving xact(Id: 13, move time: 19, current block: (1,2), next block: (2,1))
(59) Received a request from other LP for local LP to initiate GVT calculation when it passed the simulation
time 12
(60) Local simulation clock updated to 22
(61) xact(Id: 15, move time: 22, current block: (1,1), next block: (1,2)) chained out
(62) Move xact(Id: 15, move time: 22, current block: (1,1), next block: (1,2))
(63) xact(Id: 17, move time: 24, current block: (1,1), next block: (1,2)) chained in
(64) xact(Id: 15, move time: 22, current block: (1,1), next block: (1,2)) executed block (1,1) Block:
GENERATE 3,2,(no offset),10,0
(65) xact(Id: 15, move time: 22, current block: (1,2), next block: (2,1)) executed block (1,2) Block:
TRANSFER 1.0,(Label: Label1, partition: 2, block: 1)
(66) Finished moving xact(Id: 15, move time: 22, current block: (1,2), next block: (2,1))
(67) Sent GVT calculation request to SimulationController
(68) Local simulation clock updated to 24
(69) xact(Id: 17, move time: 24, current block: (1,1), next block: (1,2)) chained out
(70) Move xact(Id: 17, move time: 24, current block: (1,1), next block: (1,2))
(71) xact(Id: 19, move time: 26, current block: (1,1), next block: (1,2)) chained in
(72) xact(Id: 17, move time: 24, current block: (1,1), next block: (1,2)) executed block (1,1) Block:
GENERATE 3,2,(no offset),10,0
(73) xact(Id: 17, move time: 24, current block: (1,2), next block: (2,1)) executed block (1,2) Block:
TRANSFER 1.0,(Label: Label1, partition: 2, block: 1)
(74) Finished moving xact(Id: 17, move time: 24, current block: (1,2), next block: (2,1))
(75) Local simulation clock updated to 26
(76) xact(Id: 19, move time: 26, current block: (1,1), next block: (1,2)) chained out
(77) Move xact(Id: 19, move time: 26, current block: (1,1), next block: (1,2))
(78) xact(Id: 19, move time: 26, current block: (1,1), next block: (1,2)) executed block (1,1) Block:
GENERATE 3,2,(no offset),10,0
(79) xact(Id: 19, move time: 26, current block: (1,2), next block: (2,1)) executed block (1,2) Block:
TRANSFER 1.0,(Label: Label1, partition: 2, block: 1)
(80) Finished moving xact(Id: 19, move time: 26, current block: (1,2), next block: (2,1))
(81) GVT parameter requested by SimulationController, min. time: 26
```

```
(82) No transactions moved in moveTransactionsAtCurrentTime()
(83) SimulationController reported confirmed End of Simulation by xact: xact(Id: 7, move time: 11, current
block: (1,1), next block: (2,1))
(84) Rollback for time 11, state restored for time 11, time before rollback 26
(85) xact(Id: 7, move time: 11, current block: (1,1), next block: (1,2)) chained out
(86) Move xact(Id: 7, move time: 11, current block: (1,1), next block: (1,2))
(87) xact(Id: 9, move time: 14, current block: (1,1), next block: (1,2)) chained in
(88) xact(Id: 7, move time: 11, current block: (1,1), next block: (1,2)) executed block (1,1) Block: GENERATE
3,2,(no offset),10,0
(89) xact(Id: 7, move time: 11, current block: (1,2), next block: (2,1)) executed block (1,2) Block: TRANSFER
1.0,(Label: Label1, partition: 2, block: 1)
(90) Finished moving xact(Id: 7, move time: 11, current block: (1,2), next block: (2,1))
(91) Simulation stopped and simulation state changed to TERMINATED
(92) LogicalProcess.runActivity() left (ActiveObject stopped)
```

**Validation 3.1, output of LP2:**

```
(1)
(2)  --- StartNode ---------------------------------------
(3)  LogicalProcess.runActivity() started
(4)  Simulation engine created for partition: Partition2
(5)  Initialize GENERATE blocks
(6)  GENERATE blocks initialized
(7)  LP2 with partition 'Partition2' initialized
(8)  LP simulation state changed from INITIALIZED to SIMULATING
(9)  No transactions moved in moveTransactionsAtCurrentTime()
(10) xact(Id: 1, move time: 2, current block: (1,2), next block: (2,1)) chained in
(11) Local simulation clock updated to 2
(12) xact(Id: 3, move time: 5, current block: (1,2), next block: (2,1)) chained in
(13) xact(Id: 5, move time: 7, current block: (1,2), next block: (2,1)) chained in
(14) xact(Id: 1, move time: 2, current block: (1,2), next block: (2,1)) chained out
(15) Move xact(Id: 1, move time: 2, current block: (1,2), next block: (2,1))
(16) xact(Id: 1, move time: 2, current block: (1,1), next block: (2,1)) executed block (2,1) Block: TERMINATE
1
(17) Finished moving xact(Id: 1, move time: 2, current block: (1,1), next block: (2,1))
(18) Local simulation clock updated to 5
(19) xact(Id: 3, move time: 5, current block: (1,2), next block: (2,1)) chained out
(20) Move xact(Id: 3, move time: 5, current block: (1,2), next block: (2,1))
(21) xact(Id: 3, move time: 5, current block: (1,1), next block: (2,1)) executed block (2,1) Block: TERMINATE
1
(22) Finished moving xact(Id: 3, move time: 5, current block: (1,1), next block: (2,1))
(23) Local simulation clock updated to 7
(24) xact(Id: 7, move time: 11, current block: (1,2), next block: (2,1)) chained in
(25) xact(Id: 5, move time: 7, current block: (1,2), next block: (2,1)) chained out
(26) Move xact(Id: 5, move time: 7, current block: (1,2), next block: (2,1))
(27) xact(Id: 5, move time: 7, current block: (1,1), next block: (2,1)) executed block (2,1) Block: TERMINATE
1
(28) Finished moving xact(Id: 5, move time: 7, current block: (1,1), next block: (2,1))
(29) Local simulation clock updated to 11
```

```
(30) xact(Id: 7, move time: 11, current block: (1,2), next block: (2,1)) chained out
(31) Move xact(Id: 7, move time: 11, current block: (1,2), next block: (2,1))
(32) xact(Id: 7, move time: 11, current block: (1,1), next block: (2,1)) executed block (2,1)) Block:
TERMINATE 1
(33) Unconfirmed End Of Simulation reached by xact: xact(Id: 7, move time: 11, current block: (1,1), next
block: (2,1))
(34) Sent GVT calculation request to SimulationController
(35) xact(Id: 9, move time: 14, current block: (1,2), next block: (2,1)) chained in
(36) No transactions moved in moveTransactionsAtCurrentTime()
(37) Local simulation clock updated to 14
(38) xact(Id: 11, move time: 17, current block: (1,2), next block: (2,1)) chained in
(39) No transactions moved in moveTransactionsAtCurrentTime()
(40) Local simulation clock updated to 14
(41) xact(Id: 13, move time: 19, current block: (1,2), next block: (2,1)) chained in
(42) No transactions moved in moveTransactionsAtCurrentTime()
(43) Local simulation clock updated to 14
(44) xact(Id: 15, move time: 22, current block: (1,2), next block: (2,1)) chained in
(45) No transactions moved in moveTransactionsAtCurrentTime()
(46) Local simulation clock updated to 14
(47) xact(Id: 17, move time: 24, current block: (1,2), next block: (2,1)) chained in
(48) No transactions moved in moveTransactionsAtCurrentTime()
(49) GVT parameter requested by SimulationController, min. time: Infinite (unconfirmed end of simulation for
time 11)
(50) Local simulation clock updated to 14
(51) No transactions moved in moveTransactionsAtCurrentTime()
(52) Local simulation clock updated to 14
(53) xact(Id: 19, move time: 26, current block: (1,2), next block: (2,1)) chained in
(54) No transactions moved in moveTransactionsAtCurrentTime()
(55) SimulationController reported confirmed End of Simulation by xact: xact(Id: 7, move time: 11, current
block: (1,1), next block: (2,1))
(56) Simulation stopped and simulation state changed to TERMINATED
(57) LogicalProcess.runActivity() left (ActiveObject stopped)
```

**Validation 3.1, output of simulate:**

```
(1)   Simulation model file read and parsed successfully.
(2)   2 partition(s) found in simulation model file.
(3)
(4)   Creating SimulationController instance
(5)   parallelJavaGpssSimulator.SimulationController
(6)   SimulationController.runActivity() started
(7)   Creating LPs and starting simulation
(8)   Simulation started
(9)
(10)  Please press:
(11)  G + <Enter> to force GVT calculation
(12)  X + <Enter> to terminate the simulation
(13)
(14)  Initiated GVT calculation
(15)  Min time received from LP1: 26
(16)  Min time received from LP2: Infinite (unconfirmed end of simulation for time 11)
(17)  Simulation finished
(18)  **************   Simulation report   ****************
(19)  The simulation was completed at the simulation time: 11
(20)  by the transaction xact(Id: 7, move time: 11, current block: (1,1), next block: (2,1))
(21)  The average simulation perfomance in simulation time per second real time was: 15,151515 (time units/s)
(22)
(23)  Block report section:
(24)  Block              current      total
(25)                     xacts        xacts
(26)  Partition: Partition1
(27)  (1,1)                 1          4      Block: GENERATE 3,2,(no offset),10,0
(28)  (1,2)                 0          4      Block: TRANSFER 1.0,(Label: Label1, partition: 2, block: 1)
(29)  Partition: Partition2
(30)  (2,1)                 0          4      Block: TERMINATE 1
(31)
(32)  Summary entity report section:
```

```
(33)
(34) SimulationController.terminateLPs() called
```

**Validation 4, output of LP1:**

```
(1)
(2)   --- StartNode ----------------------------------
(3)   20:08:02,537 LogicalProcess.runActivity() started
(4)   20:08:03,136 LP1 with partition 'Partition1' initialized
(5)   20:08:03,440 LP simulation state changed from INITIALIZED to SIMULATING
(6)   20:08:05,321 Sent GVT calculation request to SimulationController
(7)   20:08:05,321 Requested GVT calculation because memory limit 1 (free memory for JVM below 5242880 byte)
reached
(8)   20:08:05,693 GVT parameter requested by SimulationController, min. time: 841
(9)   20:08:05,771 Received new GVT of 841 from SimulationController, LPCC processing needed = false
(10)  20:08:05,785 Simulation state committed for time 841
(11)  20:08:05,831 Sent GVT calculation request to SimulationController
(12)  20:08:05,831 Requested GVT calculation because memory limit 1 (free memory for JVM below 5242880 byte)
reached
(13)  20:08:05,882 GVT parameter requested by SimulationController, min. time: 848
(14)  20:08:05,899 Received new GVT of 848 from SimulationController, LPCC processing needed = false
(15)  20:08:05,899 Simulation state committed for time 848
(16)  20:08:05,921 Sent GVT calculation request to SimulationController
(17)  20:08:05,922 Requested GVT calculation because memory limit 1 (free memory for JVM below 5242880 byte)
reached
(18)  20:08:05,944 GVT parameter requested by SimulationController, min. time: 855
(19)  20:08:05,978 Received new GVT of 855 from SimulationController, LPCC processing needed = false
(20)  20:08:05,982 Simulation state committed for time 855
(21)  20:08:05,997 Sent GVT calculation request to SimulationController
(22)  20:08:05,997 Requested GVT calculation because memory limit 1 (free memory for JVM below 5242880 byte)
reached
(23)  20:08:06,023 GVT parameter requested by SimulationController, min. time: 862
(24)  20:08:06,049 Received new GVT of 862 from SimulationController, LPCC processing needed = false
(25)  20:08:06,058 Simulation state committed for time 862
(26)  20:08:06,071 Sent GVT calculation request to SimulationController
(27)  20:08:06,071 Requested GVT calculation because memory limit 1 (free memory for JVM below 5242880 byte)
reached
```

```
(28) 20:08:06,110 GVT parameter requested by SimulationControler, min. time: 867
(29) 20:08:06,156 Received new GVT of 867 from SimulationController, LPCC processing needed = false
(30) 20:08:06,157 Simulation state committed for time 867
(31) 20:08:06,173 Sent GVT calculation request to SimulationController
(32) 20:08:06,174 Requested GVT calculation because memory limit 1 (free memory for JVM below 5242880 byte)
reached
(33) 20:08:06,211 GVT parameter requested by SimulationControler, min. time: 877
(34) 20:08:06,237 Received new GVT of 877 from SimulationController, LPCC processing needed = false
(35) 20:08:06,238 Simulation state committed for time 877
(36) 20:08:06,252 Sent GVT calculation request to SimulationController
(37) 20:08:06,252 Requested GVT calculation because memory limit 1 (free memory for JVM below 5242880 byte)
reached
(38) 20:08:06,261 GVT parameter requested by SimulationControler, min. time: 878
(39) 20:08:06,291 Received new GVT of 878 from SimulationController, LPCC processing needed = false
(40) 20:08:06,291 Simulation state committed for time 878
(41) 20:08:06,303 Sent GVT calculation request to SimulationController
(42) 20:08:06,304 Requested GVT calculation because memory limit 1 (free memory for JVM below 5242880 byte)
reached
(43) 20:08:06,346 GVT parameter requested by SimulationControler, min. time: 886
(44) 20:08:06,462 Received new GVT of 886 from SimulationController, LPCC processing needed = false
(45) 20:08:06,462 Simulation state committed for time 886
(46) 20:08:06,471 Sent GVT calculation request to SimulationController
(47) 20:08:06,471 Requested GVT calculation because memory limit 1 (free memory for JVM below 5242880 byte)
reached
(48) 20:08:06,484 GVT parameter requested by SimulationControler, min. time: 896
(49) 20:08:06,505 Received new GVT of 896 from SimulationController, LPCC processing needed = false
(50) 20:08:06,507 Simulation state committed for time 896
(51) 20:08:06,539 Sent GVT calculation request to SimulationController
(52) 20:08:06,540 Requested GVT calculation because memory limit 1 (free memory for JVM below 5242880 byte)
reached
(53) 20:08:06,557 GVT parameter requested by SimulationControler, min. time: 898
(54) 20:08:06,573 Received new GVT of 898 from SimulationController, LPCC processing needed = false
(55) 20:08:06,574 Simulation state committed for time 898
(56) 20:08:06,588 Sent GVT calculation request to SimulationController
(57) 20:08:06,588 Requested GVT calculation because memory limit 1 (free memory for JVM below 5242880 byte)
```

```
reached
(58) 20:08:06,618  GVT parameter requested by SimulationController, min. time: 903
(59) 20:08:06,647  Received new GVT of 903 from SimulationController, LPCC processing needed = false
(60) 20:08:06,648  Simulation state committed for time 903
(61) 20:08:06,658  Sent GVT calculation request to SimulationController
(62) 20:08:06,658  Requested GVT calculation because memory limit 1 (free memory for JVM below 5242880 byte)
reached
(63) 20:08:06,739  GVT parameter requested by SimulationController, min. time: 909
(64) 20:08:06,805  Received new GVT of 909 from SimulationController, LPCC processing needed = false
(65) 20:08:06,806  Simulation state committed for time 909
(66) 20:08:08,073  Sent GVT calculation request to SimulationController
(67) 20:08:08,073  Requested GVT calculation because memory limit 1 (free memory for JVM below 5242880 byte)
reached
(68) 20:08:08,099  GVT parameter requested by SimulationController, min. time: 1011
(69) 20:08:08,119  Received new GVT of 1011 from SimulationController, LPCC processing needed = false
(70) 20:08:08,126  Simulation state committed for time 1011
(71) 20:08:08,164  Sent GVT calculation request to SimulationController
(72) 20:08:08,164  Requested GVT calculation because memory limit 1 (free memory for JVM below 5242880 byte)
reached
(73) 20:08:08,187  GVT parameter requested by SimulationController, min. time: 1013
(74) 20:08:08,214  Received new GVT of 1013 from SimulationController, LPCC processing needed = false
(75) 20:08:08,214  Simulation state committed for time 1013
(76) 20:08:08,232  Sent GVT calculation request to SimulationController
(77) 20:08:08,232  Requested GVT calculation because memory limit 1 (free memory for JVM below 5242880 byte)
reached
(78) 20:08:08,285  GVT parameter requested by SimulationController, min. time: 1017
(79) 20:08:08,305  Received new GVT of 1017 from SimulationController, LPCC processing needed = false
(80) 20:08:08,305  Simulation state committed for time 1017
(81) 20:08:08,328  Sent GVT calculation request to SimulationController
(82) 20:08:08,328  Requested GVT calculation because memory limit 1 (free memory for JVM below 5242880 byte)
reached
(83) 20:08:08,351  GVT parameter requested by SimulationController, min. time: 1019
(84) 20:08:08,378  Received new GVT of 1019 from SimulationController, LPCC processing needed = false
(85) 20:08:08,378  Simulation state committed for time 1019
(86) 20:08:08,398  Sent GVT calculation request to SimulationController
```

```
(87) 20:08:08,399 Requested GVT calculation because memory limit 1 (free memory for JVM below 5242880 byte)
reached
(88) 20:08:08,421 GVT parameter requested by SimulationControler, min. time: 1021
(89) 20:08:08,438 Received new GVT of 1021 from SimulationControler, LPCC processing needed = false
(90) 20:08:08,438 Simulation state committed for time 1021
(91) 20:08:08,459 Sent GVT calculation request to SimulationControler
(92) 20:08:08,459 Requested GVT calculation because memory limit 1 (free memory for JVM below 5242880 byte)
reached
(93) 20:08:08,477 GVT parameter requested by SimulationControler, min. time: 1023
(94) 20:08:08,506 Received new GVT of 1023 from SimulationControler, LPCC processing needed = false
(95) 20:08:08,506 Simulation state committed for time 1023
(96) 20:08:08,531 Sent GVT calculation request to SimulationControler
(97) 20:08:08,531 Requested GVT calculation because memory limit 1 (free memory for JVM below 5242880 byte)
reached
(98) 20:08:08,549 GVT parameter requested by SimulationControler, min. time: 1025
(99) 20:08:08,585 Received new GVT of 1025 from SimulationControler, LPCC processing needed = false
(100) 20:08:08,586 Simulation state committed for time 1025
...
(359) 20:08:26,728 Sent GVT calculation request to SimulationControler
(360) 20:08:26,728 Requested GVT calculation because memory limit 1 (free memory for JVM below 5242880 byte)
reached
(361) 20:08:27,440 GVT parameter requested by SimulationControler, min. time: 1482
(362) 20:08:27,516 Received new GVT of 1482 from SimulationControler, LPCC processing needed = false
(363) 20:08:27,517 Simulation state committed for time 1482
(364) 20:08:27,556 Sent GVT calculation request to SimulationControler
(365) 20:08:27,556 Requested GVT calculation because memory limit 1 (free memory for JVM below 5242880 byte)
reached
(366) 20:08:27,683 GVT parameter requested by SimulationControler, min. time: 1487
(367) 20:08:27,727 Received new GVT of 1487 from SimulationControler, LPCC processing needed = false
(368) 20:08:27,728 Simulation state committed for time 1487
(369) 20:08:28,080 Sent GVT calculation request to SimulationControler
(370) 20:08:28,080 Requested GVT calculation because memory limit 1 (free memory for JVM below 5242880 byte)
reached
(371) 20:08:28,160 GVT parameter requested by SimulationControler, min. time: 1500
(372) 20:08:28,627 Received new GVT of 1500 from SimulationControler, LPCC processing needed = false
```

```
(373) 20:08:28,628 Simulation state committed for time 1500
(374) 20:08:28,662 Sent GVT calculation request to SimulationController
(375) 20:08:28,662 Requested GVT calculation because memory limit 1 (free memory for JVM below 5242880 byte)
reached
(376) 20:08:28,923 Changed Cancelback mode to ON because memory limit 2 (free memory for JVM below 1048576
byte) reached
(377) 20:08:29,470 Cancelback mode - cancelled back 25 received transactions
(378) 20:08:29,470 GVT parameter requested by SimulationController, min. time: 1520
(379) 20:08:29,561 Received new GVT of 1520 from SimulationController, LPCC processing needed = false
(380) 20:08:29,563 Simulation state committed for time 1520
(381) 20:08:29,563 Changed Cancelback mode to OFF
(382) 20:08:29,612 Sent GVT calculation request to SimulationController
(383) 20:08:29,612 Requested GVT calculation because memory limit 1 (free memory for JVM below 5242880 byte)
reached
(384) 20:08:30,106 GVT parameter requested by SimulationController, min. time: 1533
(385) 20:08:30,247 Received new GVT of 1533 from SimulationController, LPCC processing needed = false
(386) 20:08:30,248 Simulation state committed for time 1533
(387) 20:08:30,420 Sent GVT calculation request to SimulationController
(388) 20:08:30,421 Requested GVT calculation because memory limit 1 (free memory for JVM below 5242880 byte)
reached
(389) 20:08:30,664 GVT parameter requested by SimulationController, min. time: 1544
(390) 20:08:30,794 Received new GVT of 1544 from SimulationController, LPCC processing needed = false
(391) 20:08:30,794 Simulation state committed for time 1544
(392) 20:08:31,329 Sent GVT calculation request to SimulationController
(393) 20:08:31,329 Requested GVT calculation because memory limit 1 (free memory for JVM below 5242880 byte)
reached
(394) 20:08:31,460 GVT parameter requested by SimulationController, min. time: 1563
(395) 20:08:31,729 Changed Cancelback mode to ON because memory limit 2 (free memory for JVM below 1048576
byte) reached
(396) 20:08:32,228 Cancelback mode - cancelled back 24 received transactions
(397) 20:08:32,237 Received new GVT of 1563 from SimulationController, LPCC processing needed = false
(398) 20:08:32,238 Simulation state committed for time 1563
(399) 20:08:32,312 Changed Cancelback mode to OFF
...
(599) 20:08:53,537 Sent GVT calculation request to SimulationController
```

```
(600) 20:08:53,537 Requested GVT calculation because memory limit 1 (free memory for JVM below 5242880 byte)
reached
(601) 20:08:53,626 GVT parameter requested by SimulationController, min. time: 1946
(602) 20:08:53,699 Received new GVT of 1946 from SimulationController, LPCC processing needed = false
(603) 20:08:53,699 Simulation state committed for time 1946
(604) 20:08:53,737 Sent GVT calculation request to SimulationController
(605) 20:08:53,738 Requested GVT calculation because memory limit 1 (free memory for JVM below 5242880 byte)
reached
(606) 20:08:54,060 GVT parameter requested by SimulationController, min. time: 1956
(607) 20:08:54,103 Received new GVT of 1956 from SimulationController, LPCC processing needed = false
(608) 20:08:54,103 Simulation state committed for time 1956
(609) 20:08:54,672 Sent GVT calculation request to SimulationController
(610) 20:08:54,672 Requested GVT calculation because memory limit 1 (free memory for JVM below 5242880 byte)
reached
(611) 20:08:54,918 GVT parameter requested by SimulationController, min. time: 1973
(612) 20:08:55,266 Received new GVT of 1973 from SimulationController, LPCC processing needed = false
(613) 20:08:55,267 Simulation state committed for time 1973
(614) 20:08:55,311 Sent GVT calculation request to SimulationController
(615) 20:08:55,312 Requested GVT calculation because memory limit 1 (free memory for JVM below 5242880 byte)
reached
(616) 20:08:55,744 GVT parameter requested by SimulationController, min. time: 1990
(617) 20:08:55,935 Received new GVT of 1990 from SimulationController, LPCC processing needed = false
(618) 20:08:55,936 Simulation state committed for time 1990
(619) 20:08:55,974 Sent GVT calculation request to SimulationController
(620) 20:08:55,975 Requested GVT calculation because memory limit 1 (free memory for JVM below 5242880 byte)
reached
(621) 20:08:56,063 GVT parameter requested by SimulationController, min. time: 1998
(622) 20:08:56,136 Unconfirmed End Of Simulation reached by xact: xact(Id: 2, move time: 2000, current block:
(2,4), next block: (1,4))
(623) 20:08:56,292 Received new GVT of 1998 from SimulationController, LPCC processing needed = false
(624) 20:08:56,293 Simulation state committed for time 1998
(625) 20:08:56,304 Sent GVT calculation request to SimulationController
(626) 20:08:56,304 Requested GVT calculation because memory limit 1 (free memory for JVM below 5242880 byte)
reached
(627) 20:08:56,313 GVT parameter requested by SimulationController, min. time: Infinite (unconfirmed end of
```

```
simulation for time 2000)
(628) 20:08:56,438 SimulationController reported confirmed End of Simulation by xact: xact(Id: 2, move time:
2000, current block: (2,4), next block: (1,4))
(629) 20:08:56,439 Simulation stopped and simulation state changed to TERMINATED
(630) 20:08:56,810 LogicalProcess.runActivity() left (ActiveObject stopped)
```

**Validation 4, output of LP2:**

```
(1)
(2)   --- StartNode -------------------------------------
(3)   20:08:02,780 LogicalProcess.runActivity() started
(4)   20:08:03,514 LP2 with partition 'Partition2' initialized
(5)   20:08:03,517 LP simulation state changed from INITIALIZED to SIMULATING
(6)   20:08:05,596 GVT parameter requested by SimulationController, min. time: 2154
(7)   20:08:05,729 Received new GVT of 841 from SimulationController, LPCC processing needed = false
(8)   20:08:05,879 GVT parameter requested by SimulationController, min. time: 2221
(9)   20:08:05,915 Received new GVT of 848 from SimulationController, LPCC processing needed = false
(10)  20:08:05,948 GVT parameter requested by SimulationController, min. time: 2228
(11)  20:08:05,982 Received new GVT of 855 from SimulationController, LPCC processing needed = false
(12)  20:08:06,026 GVT parameter requested by SimulationController, min. time: 2238
(13)  20:08:06,050 Received new GVT of 862 from SimulationController, LPCC processing needed = false
(14)  20:08:06,141 GVT parameter requested by SimulationController, min. time: 2258
(15)  20:08:06,172 Received new GVT of 867 from SimulationController, LPCC processing needed = false
·::·
(68)  20:08:11,370 GVT parameter requested by SimulationController, min. time: 3166
(69)  20:08:11,390 Received new GVT of 1153 from SimulationController, LPCC processing needed = false
(70)  20:08:11,441 GVT parameter requested by SimulationController, min. time: 3175
(71)  20:08:11,462 Received new GVT of 1155 from SimulationController, LPCC processing needed = false
(72)  20:08:11,525 GVT parameter requested by SimulationController, min. time: 3194
(73)  20:08:11,543 Sent GVT calculation request to SimulationController
(74)  20:08:11,543 Requested GVT calculation because memory limit 1 (free memory for JVM below 5242880 byte)
reached
(75)  20:08:11,551 Received new GVT of 1157 from SimulationController, LPCC processing needed = false
(76)  20:08:11,573 GVT parameter requested by SimulationController, min. time: 3201
(77)  20:08:11,655 Received new GVT of 1158 from SimulationController, LPCC processing needed = false
(78)  20:08:11,658 GVT parameter requested by SimulationController, min. time: 3205
(79)  20:08:11,683 Received new GVT of 1159 from SimulationController, LPCC processing needed = false
(80)  20:08:12,520 GVT parameter requested by SimulationController, min. time: 3388
(81)  20:08:12,548 Received new GVT of 1189 from SimulationController, LPCC processing needed = false
(82)  20:08:12,630 GVT parameter requested by SimulationController, min. time: 3403
```

```
(83)  20:08:12,658  Received new GVT of 1192 from SimulationController, LPCC processing needed = false
(84)  20:08:12,705  GVT parameter requested by SimulationController, min. time: 3413
(85)  20:08:12,734  Received new GVT of 1193 from SimulationController, LPCC processing needed = false
(86)  20:08:12,844  GVT parameter requested by SimulationController, min. time: 3434
(87)  20:08:12,869  Received new GVT of 1195 from SimulationController, LPCC processing needed = false
(88)  20:08:13,926  GVT parameter requested by SimulationController, min. time: 3574
(89)  20:08:13,953  Received new GVT of 1228 from SimulationController, LPCC processing needed = false
(90)  20:08:14,034  GVT parameter requested by SimulationController, min. time: 3590
(91)  20:08:14,068  Received new GVT of 1230 from SimulationController, LPCC processing needed = false
(92)  20:08:14,142  GVT parameter requested by SimulationController, min. time: 3612
(93)  20:08:14,233  Received new GVT of 1232 from SimulationController, LPCC processing needed = false
(94)  20:08:15,185  GVT parameter requested by SimulationController, min. time: 3812
(95)  20:08:15,218  Received new GVT of 1260 from SimulationController, LPCC processing needed = false
(96)  20:08:15,303  GVT parameter requested by SimulationController, min. time: 3842
(97)  20:08:15,332  Received new GVT of 1262 from SimulationController, LPCC processing needed = false
(98)  20:08:15,400  GVT parameter requested by SimulationController, min. time: 3858
(99)  20:08:15,428  Received new GVT of 1263 from SimulationController, LPCC processing needed = false
(100) 20:08:15,548  GVT parameter requested by SimulationController, min. time: 3867
(101) 20:08:15,576  Received new GVT of 1265 from SimulationController, LPCC processing needed = false
(102) 20:08:16,120  GVT parameter requested by SimulationController, min. time: 3995
(103) 20:08:16,146  Received new GVT of 1276 from SimulationController, LPCC processing needed = false
(104) 20:08:16,710  GVT parameter requested by SimulationController, min. time: 4101
(105) 20:08:16,732  Received new GVT of 1287 from SimulationController, LPCC processing needed = false
(106) 20:08:16,995  Changed Cancelback mode to ON because memory limit 2 (free memory for JVM below 1048576
byte) reached
(107) 20:08:16,997  Changed Cancelback mode to OFF
(108) 20:08:17,033  Changed Cancelback mode to ON because memory limit 2 (free memory for JVM below 1048576
byte) reached
(109) 20:08:17,037  Changed Cancelback mode to OFF
(110) 20:08:17,070  Changed Cancelback mode to ON because memory limit 2 (free memory for JVM below 1048576
byte) reached
(111) 20:08:17,078  Changed Cancelback mode to OFF
(112) 20:08:17,140  Changed Cancelback mode to ON because memory limit 2 (free memory for JVM below 1048576
byte) reached
(113) 20:08:17,143  Changed Cancelback mode to OFF
```

```
(114) 20:08:17,171 Changed Cancelback mode to ON because memory limit 2 (free memory for JVM below 1048576
byte) reached
(115) 20:08:17,253 Changed Cancelback mode to OFF
(116) 20:08:17,299 Changed Cancelback mode to ON because memory limit 2 (free memory for JVM below 1048576
byte) reached
(117) 20:08:17,378 Changed Cancelback mode to OFF
(118) 20:08:17,471 GVT parameter requested by SimulationController, min. time: 4212
(119) 20:08:17,489 Received new GVT of 1299 from SimulationController, LPCC processing needed = false
(120) 20:08:17,631 GVT parameter requested by SimulationController, min. time: 4223
(121) 20:08:17,656 Received new GVT of 1301 from SimulationController, LPCC processing needed = false
(122) 20:08:18,016 Changed Cancelback mode to ON because memory limit 2 (free memory for JVM below 1048576
byte) reached
(123) 20:08:18,018 Changed Cancelback mode to OFF
...
(265) 20:08:24,947 Changed Cancelback mode to ON because memory limit 2 (free memory for JVM below 1048576
byte) reached
(266) 20:08:25,032 Received new GVT of 1422 from SimulationController, LPCC processing needed = false
(267) 20:08:25,165 Changed Cancelback mode to OFF
(268) 20:08:25,235 GVT parameter requested by SimulationController, min. time: 4476
(269) 20:08:25,239 Changed Cancelback mode to ON because memory limit 2 (free memory for JVM below 1048576
byte) reached
(270) 20:08:25,245 Received new GVT of 1427 from SimulationController, LPCC processing needed = false
(271) 20:08:25,644 Changed Cancelback mode to OFF
(272) 20:08:25,655 GVT parameter requested by SimulationController, min. time: 4477
(273) 20:08:25,659 Changed Cancelback mode to ON because memory limit 2 (free memory for JVM below 1048576
byte) reached
(274) 20:08:25,670 Received new GVT of 1437 from SimulationController, LPCC processing needed = false
(275) 20:08:26,196 GVT parameter requested by SimulationController, min. time: 4478
(276) 20:08:26,318 Received new GVT of 1451 from SimulationController, LPCC processing needed = false
(277) 20:08:26,585 Changed Cancelback mode to OFF
(278) 20:08:26,596 GVT parameter requested by SimulationController, min. time: 4478
(279) 20:08:26,601 Changed Cancelback mode to ON because memory limit 2 (free memory for JVM below 1048576
byte) reached
(280) 20:08:26,610 Received new GVT of 1460 from SimulationController, LPCC processing needed = false
(281) 20:08:27,478 GVT parameter requested by SimulationController, min. time: 4479
```

```
(282)  20:08:27,488  Received new GVT of 1482 from SimulationController, LPCC processing needed = false
(283)  20:08:27,619  Changed Cancelback mode to OFF
(284)  20:08:27,688  GVT parameter requested by SimulationControler, min. time: 4479
(285)  20:08:27,690  Changed Cancelback mode to ON because memory limit 2 (free memory for JVM below 1048576
byte) reached
(286)  20:08:27,698  Received new GVT of 1487 from SimulationController, LPCC processing needed = false
(287)  20:08:28,233  Changed Cancelback mode to OFF
(288)  20:08:28,582  GVT parameter requested by SimulationController, min. time: 4480
(289)  20:08:28,589  Changed Cancelback mode to ON because memory limit 2 (free memory for JVM below 1048576
byte) reached
(290)  20:08:28,872  Changed Cancelback mode to OFF
(291)  20:08:29,161  Received new GVT of 1500 from SimulationController, LPCC processing needed = false
(292)  20:08:29,171  GVT parameter requested by SimulationControler, min. time: 4481
(293)  20:08:29,251  Changed Cancelback mode to ON because memory limit 2 (free memory for JVM below 1048576
byte) reached
(294)  20:08:29,254  9 received cancelbacks require rollback to time 4472
(295)  20:08:29,368  Rollback for time 4472, state restored for time 4472, time before rollback 4482
(296)  20:08:29,374  9 received cancelbacks require rollback to time 4463
(297)  20:08:29,460  Rollback for time 4463, state restored for time 4463, time before rollback 4472
(298)  20:08:29,463  5 received cancelbacks require rollback to time 4458
(299)  20:08:29,467  Rollback for time 4458, state restored for time 4458, time before rollback 4463
(300)  20:08:29,469  2 received cancelbacks require rollback to time 4456
(301)  20:08:29,471  Rollback for time 4456, state restored for time 4456, time before rollback 4458
(302)  20:08:29,842  Changed Cancelback mode to OFF
(303)  20:08:30,010  Received new GVT of 1520 from SimulationController, LPCC processing needed = false
(304)  20:08:30,015  Changed Cancelback mode to ON because memory limit 2 (free memory for JVM below 1048576
byte) reached
(305)  20:08:30,103  GVT parameter requested by SimulationController, min. time: 4456
(306)  20:08:30,394  Changed Cancelback mode to OFF
(307)  20:08:30,566  Received new GVT of 1533 from SimulationController, LPCC processing needed = false
(308)  20:08:30,572  Changed Cancelback mode to ON because memory limit 2 (free memory for JVM below 1048576
byte) reached
(309)  20:08:30,657  GVT parameter requested by SimulationController, min. time: 4457
(310)  20:08:30,943  Changed Cancelback mode to OFF
(311)  20:08:31,142  Received new GVT of 1544 from SimulationController, LPCC processing needed = false
```

```
(312)  20:08:31,146  Changed Cancelback mode to ON because memory limit 2 (free memory for JVM below 1048576
byte)
(313)  20:08:31,491  Changed Cancelback mode to OFF
(314)  20:08:31,787  GVT parameter requested by SimulationController, min. time: 4459
(315)  20:08:31,801  Changed Cancelback mode to ON because memory limit 2 (free memory for JVM below 1048576
reached
(316)  20:08:31,801  6 received cancelbacks require rollback to time 4454
(317)  20:08:31,805  Rollback for time 4454, state restored for time 4454, time before rollback 4460
(318)  20:08:31,924  1 received cancelbacks require rollback to time 4453
(319)  20:08:31,936  Rollback for time 4453, state restored for time 4453, time before rollback 4454
(320)  20:08:31,937  Received new GVT of 1563 from SimulationController, LPCC processing needed = false
(321)  20:08:32,024  Changed Cancelback mode to OFF
(322)  20:08:32,029  3 received cancelbacks require rollback to time 4450
(323)  20:08:32,034  Rollback for time 4450, state restored for time 4450, time before rollback 4453
(324)  20:08:32,048  Changed Cancelback mode to ON because memory limit 2 (free memory for JVM below 1048576
byte)
reached
(325)  20:08:32,049  5 received cancelbacks require rollback to time 4445
(326)  20:08:32,051  Rollback for time 4445, state restored for time 4445, time before rollback 4451
(327)  20:08:32,053  1 received cancelbacks require rollback to time 4444
(328)  20:08:32,127  Rollback for time 4444, state restored for time 4444, time before rollback 4445
(329)  20:08:32,129  1 received cancelbacks require rollback to time 4443
(330)  20:08:32,134  Rollback for time 4443, state restored for time 4443, time before rollback 4444
(331)  20:08:32,135  1 received cancelbacks require rollback to time 4442
(332)  20:08:32,137  Rollback for time 4442, state restored for time 4442, time before rollback 4443
(333)  20:08:32,138  1 received cancelbacks require rollback to time 4441
(334)  20:08:32,141  Rollback for time 4441, state restored for time 4441, time before rollback 4442
(335)  20:08:32,142  1 received cancelbacks require rollback to time 4440
(336)  20:08:32,144  Rollback for time 4440, state restored for time 4440, time before rollback 4441
(337)  20:08:32,145  1 received cancelbacks require rollback to time 4439
(338)  20:08:32,220  Rollback for time 4439, state restored for time 4439, time before rollback 4440
(339)  20:08:32,221  1 received cancelbacks require rollback to time 4438
(340)  20:08:32,223  Rollback for time 4438, state restored for time 4438, time before rollback 4439
(341)  20:08:32,224  1 received cancelbacks require rollback to time 4437
(342)  20:08:32,230  Rollback for time 4437, state restored for time 4437, time before rollback 4438
(343)  20:08:32,231  1 received cancelbacks require rollback to time 4436
```

```
(344)   20:08:32,233   Rollback for time 4436, state restored for time 4436, time before rollback 4437
(345)   20:08:32,308   Changed Cancelback mode to OFF
(346)   20:08:32,314   Changed Cancelback mode to ON because memory limit 2 (free memory for JVM below 1048576
byte)   reached
(347)   20:08:32,413   Changed Cancelback mode to OFF
:::
(700)   20:08:51,809   GVT parameter requested by SimulationControler, min. time: 4487
(701)   20:08:51,811   Changed Cancelback mode to ON because memory limit 2 (free memory for JVM below 1048576
byte)   reached
(702)   20:08:51,955   Changed Cancelback mode to OFF
(703)   20:08:51,961   Received new GVT of 1904 from SimulationController, LPCC processing needed = false
(704)   20:08:51,965   Changed Cancelback mode to ON because memory limit 2 (free memory for JVM below 1048576
byte)   reached
(705)   20:08:52,717   GVT parameter requested by SimulationControler, min. time: 4489
(706)   20:08:52,724   Received new GVT of 1911 from SimulationController, LPCC processing needed = false
(707)   20:08:52,877   GVT parameter requested by SimulationControler, min. time: 4489
(708)   20:08:53,025   Received new GVT of 1930 from SimulationController, LPCC processing needed = false
(709)   20:08:53,183   GVT parameter requested by SimulationControler, min. time: 4489
(710)   20:08:53,350   Received new GVT of 1936 from SimulationController, LPCC processing needed = false
(711)   20:08:53,487   Changed Cancelback mode to OFF
(712)   20:08:53,533   Changed Cancelback mode to ON because memory limit 2 (free memory for JVM below 1048576
byte)   reached
(713)   20:08:53,624   GVT parameter requested by SimulationControler, min. time: 4489
(714)   20:08:53,926   Changed Cancelback mode to OFF
(715)   20:08:53,972   Received new GVT of 1946 from SimulationController, LPCC processing needed = false
(716)   20:08:54,063   GVT parameter requested by SimulationControler, min. time: 4490
(717)   20:08:54,065   Changed Cancelback mode to ON because memory limit 2 (free memory for JVM below 1048576
byte)   reached
(718)   20:08:54,074   Received new GVT of 1956 from SimulationController, LPCC processing needed = false
(719)   20:08:54,365   Changed Cancelback mode to OFF
(720)   20:08:54,725   Changed Cancelback mode to ON because memory limit 2 (free memory for JVM below 1048576
byte)   reached
(721)   20:08:54,917   Changed Cancelback mode to OFF
(722)   20:08:55,142   GVT parameter requested by SimulationControler, min. time: 4492
(723)   20:08:55,226   Changed Cancelback mode to ON because memory limit 2 (free memory for JVM below 1048576
```

```
byte) reached
(724) 20:08:55,526  Changed Cancelback mode to OFF
(725) 20:08:55,877  Received new GVT of 1973 from SimulationController, LPCC processing needed = false
(726) 20:08:55,887  GVT parameter requested by SimulationController, min. time: 4493
(727) 20:08:55,895  Changed Cancelback mode to ON because memory limit 2 (free memory for JVM below 1048576
byte) reached
(728) 20:08:56,047  Received new GVT of 1990 from SimulationController, LPCC processing needed = false
(729) 20:08:56,167  Changed Cancelback mode to OFF
(730) 20:08:56,277  GVT parameter requested by SimulationControler, min. time: 4494
(731) 20:08:56,285  Received a request from other LP for local LP to initiate GVT calculation when it passed
the simulation time 2001
(732) 20:08:56,285  Changed Cancelback mode to ON because memory limit 2 (free memory for JVM below 1048576
byte) reached
(733) 20:08:56,289  Received new GVT of 1998 from SimulationController, LPCC processing needed = false
(734) 20:08:56,392  GVT parameter requested by SimulationControler, min. time: 4495
(735) 20:08:56,511  SimulationController reported confirmed End of Simulation by xact: xact(Id: 2, move time:
2000, current block: (2,4), next block: (1,4))
(736) 20:08:56,622  Rollback for time 2000, state restored for time 2000, time before rollback 4495
(737) 20:08:56,622  Simulation stopped and simulation state changed to TERMINATED
(738) 20:08:56,820  LogicalProcess.runActivity() left (ActiveObject stopped)
```

**Validation 4, output of simulate:**

```
(1)   20:07:59,838  Simulation model file read and parsed successfully.
(2)   20:07:59,840  2 partition(s) found in simulation model file.
(3)
(4)   20:07:59,848  Creating SimulationController instance
(5)   20:07:59,849  parallelJavaGpssSimulator.SimulationController
(6)   20:08:01,352  SimulationController.runActivity() started
(7)   20:08:01,416  Creating LPs and starting simulation
(8)   20:08:03,453  Simulation started
(9)
(10)  20:08:03,454  Please press:
(11)  G + <Enter>  to force GVT calculation
(12)  X + <Enter>  to terminate the simulation
(13)
(14)  20:08:05,320  Initiated GVT calculation
(15)  20:08:05,699  Min time received from LP1: 841
(16)  20:08:05,700  Min time received from LP2: 2154
(17)  20:08:05,728  Simulation reached Global Virtual Time: 841
(18)  20:08:05,830  Initiated GVT calculation
(19)  20:08:05,889  Min time received from LP1: 848
(20)  20:08:05,890  Min time received from LP2: 2221
(21)  20:08:05,908  Simulation reached Global Virtual Time: 848
(22)  20:08:05,923  Initiated GVT calculation
(23)  20:08:05,959  Min time received from LP1: 855
(24)  20:08:05,959  Min time received from LP2: 2228
(25)  20:08:05,986  Simulation reached Global Virtual Time: 855
(26)  20:08:06,003  Initiated GVT calculation
(27)  20:08:06,030  Min time received from LP1: 862
(28)  20:08:06,030  Min time received from LP2: 2238
(29)  20:08:06,055  Simulation reached Global Virtual Time: 862
(30)  20:08:06,076  Initiated GVT calculation
(31)  20:08:06,144  Min time received from LP1: 867
(32)  20:08:06,145  Min time received from LP2: 2258
```

```
(33)  20:08:06,178  Simulation reached Global Virtual Time: 867
(34)  20:08:06,184  Initiated GVT calculation
(35)  20:08:06,223  Min time received from LP1: 877
(36)  20:08:06,223  Min time received from LP2: 2268
(37)  20:08:06,234  Simulation reached Global Virtual Time: 877
(38)  20:08:06,252  Initiated GVT calculation
(39)  20:08:06,283  Min time received from LP1: 878
(40)  20:08:06,284  Min time received from LP2: 2276
(41)  20:08:06,329  Simulation reached Global Virtual Time: 878
(42)  20:08:06,329  Initiated GVT calculation
(43)  20:08:06,365  Min time received from LP1: 886
(44)  20:08:06,374  Min time received from LP2: 2286
(45)  20:08:06,393  Simulation reached Global Virtual Time: 886
(46)  20:08:06,470  Initiated GVT calculation
(47)  20:08:06,491  Min time received from LP1: 896
(48)  20:08:06,491  Min time received from LP2: 2294
(49)  20:08:06,508  Simulation reached Global Virtual Time: 896
...
(298) 20:08:26,737  Initiated GVT calculation
(299) 20:08:27,482  Min time received from LP1: 1482
(300) 20:08:27,483  Min time received from LP2: 4479
(301) 20:08:27,487  Simulation reached Global Virtual Time: 1482
(302) 20:08:27,555  Initiated GVT calculation
(303) 20:08:27,690  Min time received from LP1: 1487
(304) 20:08:27,690  Min time received from LP2: 4479
(305) 20:08:27,698  Simulation reached Global Virtual Time: 1487
(306) 20:08:28,079  Initiated GVT calculation
(307) 20:08:28,596  Min time received from LP1: 1500
(308) 20:08:28,596  Min time received from LP2: 4480
(309) 20:08:28,999  Simulation reached Global Virtual Time: 1500
(310) 20:08:29,000  Initiated GVT calculation
(311) 20:08:29,476  Min time received from LP1: 1520
(312) 20:08:29,476  Min time received from LP2: 4481
(313) 20:08:30,006  Simulation reached Global Virtual Time: 1520
(314) 20:08:30,007  Initiated GVT calculation
```

```
(315) 20:08:30,119 Min time received from LP1: 1533
(316) 20:08:30,119 Min time received from LP2: 4456
(317) 20:08:30,550 Simulation reached Global Virtual Time: 1533
(318) 20:08:30,550 Initiated GVT calculation
(319) 20:08:30,674 Min time received from LP1: 1544
(320) 20:08:30,776 Min time received from LP2: 4457
(321) 20:08:31,121 Simulation reached Global Virtual Time: 1544
(322) 20:08:31,328 Initiated GVT calculation
(323) 20:08:31,801 Min time received from LP1: 1563
(324) 20:08:31,801 Min time received from LP2: 4459
(325) 20:08:31,936 Simulation reached Global Virtual Time: 1563
(326) 20:08:32,358 Initiated GVT calculation
(327) 20:08:32,429 Min time received from LP1: 1572
(328) 20:08:32,430 Min time received from LP2: 4437
(329) 20:08:32,558 Simulation reached Global Virtual Time: 1572
(330) 20:08:32,558 Initiated GVT calculation
(331) 20:08:32,751 Min time received from LP1: 1577
(332) 20:08:32,751 Min time received from LP2: 4439
(333) 20:08:32,760 Simulation reached Global Virtual Time: 1577
(334) 20:08:33,955 Initiated GVT calculation
(335) 20:08:34,242 Min time received from LP1: 1589
(336) 20:08:34,242 Min time received from LP2: 4446
(337) 20:08:34,249 Simulation reached Global Virtual Time: 1589
(338) 20:08:34,386 Initiated GVT calculation
(339) 20:08:34,469 Min time received from LP1: 1595
(340) 20:08:34,469 Min time received from LP2: 4450
(341) 20:08:34,603 Simulation reached Global Virtual Time: 1595
(342) 20:08:34,603 Initiated GVT calculation
(343) 20:08:34,690 Min time received from LP1: 1599
(344) 20:08:34,691 Min time received from LP2: 4452
(345) 20:08:34,814 Simulation reached Global Virtual Time: 1599
(346) 20:08:35,210 Initiated GVT calculation
(347) 20:08:35,426 Min time received from LP1: 1610
(348) 20:08:35,427 Min time received from LP2: 4462
(349) 20:08:35,440 Simulation reached Global Virtual Time: 1610
```

```
(494)  ...
(495)  20:08:56,048  Initiated GVT calculation
(496)  20:08:56,283  Min time received from LP1: 1998
(497)  20:08:56,283  Min time received from LP2: 4494
(498)  20:08:56,308  Simulation reached Global Virtual Time: 1998
(499)  20:08:56,308  Initiated GVT calculation
(500)  20:08:56,395  Min time received from LP1: Infinite (unconfirmed end of simulation for time 2000)
(501)  20:08:56,395  Min time received from LP2: 4495
(502)  20:08:56,630  Simulation finished
(503)  20:08:56,774  *****************     Simulation report    ******************
(504)  20:08:56,781  The simulation was completed at the simulation time: 2000
(505)  20:08:56,781  by the transaction xact(Id: 2, move time: 2000, current block: (2,4), next block: (1,4))
(506)  20:08:56,783  The average simulation perfomance in simulation time per second real time was: 37,498127
       (time units/s)
(507)
(508)  20:08:56,784  Block report section:
(509)  Block              current      total
(510)                     xacts        xacts
(511)  Partition: Partition1
(512)  (1,1)                1            1999      Block: GENERATE 1,0,(no offset),(no limit),0
(513)  (1,2)                0            2000      Block: QUEUE (Queue: Queue1)
(514)  (1,3)                0            2000      Block: DEPART (Queue: Queue1)
(515)  (1,4)                0            2000      Block: TERMINATE 1
(516)  Partition: Partition2
(517)  (2,1)                1            1         Block: GENERATE 1,0,2000,(no limit),0
(518)  (2,2)                0            1         Block: TRANSFER 1.0,(Label: Label1, partition: 1, block: 2)
(519)
(520)  20:08:56,785  Summary entity report section:
(521)  Queue    maximum    average    total     zero      percent     average     current
(522)           content    content    entries   entries   zeros       time/unit   content
(523)  Queue1      1        0.0        2000      2000      100.0       0.0         0
(524)
(525)  20:08:56,795  SimulationController.terminateLPs() called
```

**Validation 5.1, output of LP1:**

```
(1)
(2)   --- StartNode ---------------------------
(3)   19:37:04,033 LogicalProcess.runActivity() started
(4)   19:37:04,710 LP1 with partition 'Partition1' initialized
(5)   19:37:05,006 LP simulation state changed from INITIALIZED to SIMULATING
(6)   19:37:10,601 GVT parameter requested by SimulationController, min. time: 7252
(7)   19:37:10,637 Received new GVT of 7220 from SimulationController, LPCC processing needed = true
(8)   19:37:10,685 Simulation state committed for time 7220
(9)   19:37:10,696 CommittedMoves now: 7219, last time: 0, time diff 6029ms
(10)  19:37:10,697 Uncommitted moves avg: 3654, sum: 26699866, sqr sum: 130072426410, count: 7307
(11)  19:37:10,700 Set new std derivation: 2109,472934
(12)  19:37:10,701 Max uncommitted moves: 7307
(13)  19:37:10,701 Get better indicator set: non found
(14)  19:37:10,718 Current indicator set [1197,3654,1211,1,0,0,0,0] added to Clustered State Space
(15)  19:37:10,718 LPCC processed, total committed moves: 7219
(16)  19:37:10,719 Actuator upper limit 9223372036854775807, mean limit 9223372036854775807
(17)  19:37:15,229 GVT parameter requested by SimulationController, min. time: 16409
(18)  19:37:15,255 Received new GVT of 16409 from SimulationController, LPCC processing needed = true
(19)  19:37:15,287 Simulation state committed for time 16409
(20)  19:37:15,289 CommittedMoves now: 16408, last time: 7219, time diff 4592ms
(21)  19:37:15,289 Uncommitted moves avg: 4651, sum: 42458742, sqr sum: 260876559204, count: 9127
(22)  19:37:15,290 Set new std derivation: 2634,894132
(23)  19:37:15,290 Max uncommitted moves: 9215
(24)  19:37:15,291 Get better indicator set: non found
(25)  19:37:15,297 Current indicator set [2001,4651,1987,1,0,0,0,0] added to Clustered State Space
(26)  19:37:15,297 LPCC processed, total committed moves: 16408
(27)  19:37:15,298 Actuator upper limit 9223372036854775807, mean limit 9223372036854775807
(28)  19:37:20,270 GVT parameter requested by SimulationController, min. time: 25718
(29)  19:37:20,297 Received new GVT of 25718 from SimulationController, LPCC processing needed = true
(30)  19:37:20,326 Simulation state committed for time 25718
(31)  19:37:20,327 CommittedMoves now: 25717, last time: 16408, time diff 5039ms
(32)  19:37:20,328 Uncommitted moves avg: 4675, sum: 43482114, sqr sum: 270330540434, count: 9299
```

```
(33) 19:37:20,329  Set new std derivation: 2684.536275
(34) 19:37:20,329  Max uncommitted moves: 9325
(35) 19:37:20,329  Get better indicator set: [2001,4651,1987,1,0,0,0]
(36) 19:37:20,330  Current indicator set [1847,4675,1845,1,0,0,0] added to Clustered State Space
(37) 19:37:20,330  LPCC processed, total committed moves: 25717
(38) 19:37:20,330  Actuator upper limit 9914, mean limit 4651
(39) 19:37:25,296  GVT parameter requested by SimulationController, min. time: 34517
(40) 19:37:25,319  Received new GVT of 34517 from SimulationController, LPCC processing needed = true
(41) 19:37:25,348  Simulation state committed for time 34517
(42) 19:37:25,350  CommittedMoves now: 34516, last time: 25717, time diff 5022ms
(43) 19:37:25,351  Uncommitted moves avg: 4410, sum: 38768269, sqr sum: 227583517119, count: 8789
(44) 19:37:25,352  Set new std derivation: 2537,312064
(45) 19:37:25,352  Max uncommitted moves: 8805
(46) 19:37:25,352  Get better indicator set: [1847,4675,1845,1,0,0,0]
(47) 19:37:25,353  Current indicator set [1752,4410,1750,2,0,0,0] added to Clustered State Space
(48) 19:37:25,353  LPCC processed, total committed moves: 34516
(49) 19:37:25,353  Actuator upper limit 9649, mean limit 4675
(50) 19:37:30,320  GVT parameter requested by SimulationController, min. time: 42689
(51) 19:37:30,341  Received new GVT of 42689 from SimulationController, LPCC processing needed = true
(52) 19:37:30,370  Simulation state committed for time 42689
(53) 19:37:30,371  CommittedMoves now: 42688, last time: 34516, time diff 5022ms
(54) 19:37:30,371  Uncommitted moves avg: 4097, sum: 33525755, sqr sum: 183017156743, count: 8182
(55) 19:37:30,372  Set new std derivation: 2362,082170
(56) 19:37:30,372  Max uncommitted moves: 8188
(57) 19:37:30,373  Get better indicator set: [1752,4410,1750,2,0,0,0]
(58) 19:37:30,373  Current indicator set [1627,4097,1629,1,0,0,0] added to Clustered State Space
(59) 19:37:30,373  LPCC processed, total committed moves: 42688
(60) 19:37:30,374  Actuator upper limit 9041, mean limit 4410
(61) 19:37:35,405  GVT parameter requested by SimulationController, min. time: 51021
(62) 19:37:35,424  Received new GVT of 51021 from SimulationController, LPCC processing needed = true
(63) 19:37:35,453  Simulation state committed for time 51021
(64) 19:37:35,454  CommittedMoves now: 51020, last time: 42688, time diff 5083ms
(65) 19:37:35,455  Uncommitted moves avg: 4176, sum: 34756806, sqr sum: 193190728676, count: 8321
(66) 19:37:35,455  Set new std derivation: 2402,212421
(67) 19:37:35,456  Max uncommitted moves: 8337
```

```
(68)  19:37:35,457  Get better indicator set: [1752,4410,1750,2,0,0,0]
(69)  19:37:35,457  Current indicator set [1639,4176,1637,1,0,0,0] added to Clustered State Space
(70)  19:37:35,457  LPCC processed, total committed moves: 51020
(71)  19:37:35,458  Actuator upper limit 9119, mean limit 4410
(72)  19:37:40,439  GVT parameter requested by SimulationController, min. time: 59249
(73)  19:37:40,462  Received new GVT of 59249 from SimulationController, LPCC processing needed = true
(74)  19:37:40,488  Simulation state committed for time 59249
(75)  19:37:40,489  CommittedMoves now: 59248, last time: 51020, time diff 5035ms
(76)  19:37:40,489  Uncommitted moves avg: 4128, sum: 34035377, sqr sum: 187205908063, count: 8245
(77)  19:37:40,490  Set new std derivation: 2380,267254
(78)  19:37:40,490  Max uncommitted moves: 8250
(79)  19:37:40,491  Get better indicator set: [1639,4176,1637,1,0,0,0]
(80)  19:37:40,491  Current indicator set [1634,4128,1637,1,0,0,0] added to Clustered State Space
(81)  19:37:40,491  LPCC processed, total committed moves: 59248
(82)  19:37:40,492  Actuator upper limit 8842, mean limit 4176
(83)  19:37:45,463  GVT parameter requested by SimulationController, min. time: 67139
(84)  19:37:45,497  Received new GVT of 67139 from SimulationController, LPCC processing needed = true
(85)  19:37:45,528  Simulation state committed for time 67139
(86)  19:37:45,531  CommittedMoves now: 67138, last time: 59248, time diff 5042ms
(87)  19:37:45,532  Uncommitted moves avg: 3966, sum: 31287750, sqr sum: 165002505152, count: 7888
(88)  19:37:45,533  Set new std derivation: 2277,214234
(89)  19:37:45,533  Max uncommitted moves: 7910
(90)  19:37:45,534  Get better indicator set: [1627,4097,1629,1,0,0,0]
(91)  19:37:45,534  Current indicator set [1564,3966,1564,2,0,0,0] added to Clustered State Space
(92)  19:37:45,534  LPCC processed, total committed moves: 67138
(93)  19:37:45,535  Actuator upper limit 8561, mean limit 4097
(94)  19:37:50,495  GVT parameter requested by SimulationController, min. time: 74915
(95)  19:37:50,519  Received new GVT of 74915 from SimulationController, LPCC processing needed = true
(96)  19:37:50,547  Simulation state committed for time 74915
(97)  19:37:50,548  CommittedMoves now: 74914, last time: 67138, time diff 5017ms
(98)  19:37:50,549  Uncommitted moves avg: 3904, sum: 30330148, sqr sum: 157485367252, count: 7768
(99)  19:37:50,551  Set new std derivation: 2242,574564
(100) 19:37:50,552  Max uncommitted moves: 7788
(101) 19:37:50,553  Get better indicator set: [1564,3966,1564,2,0,0,0]
(102) 19:37:50,554  Current indicator set [1549,3904,1548,0,0,0,0] added to Clustered State Space
```

```
(103) 19:37:50,555 LPCC processed, total committed moves: 74914
(104) 19:37:50,555 Actuator upper limit 8362, mean limit 3966
(105) 19:37:55,193 Received a request from other LP for local LP to initiate GVT calculation when it passed
the simulation time 84653
(106) 19:37:55,235 GVT parameter requested by SimulationController, min. time: 82106
(107) 19:37:55,249 Received new GVT of 82106 from SimulationController, LPCC processing needed = false
(108) 19:37:55,267 Simulation state committed for time 82106
(109) 19:37:55,634 GVT parameter requested by SimulationController, min. time: 82840
(110) 19:37:55,692 Received new GVT of 82840 from SimulationController, LPCC processing needed = true
(111) 19:37:55,695 Simulation state committed for time 82840
(112) 19:37:55,697 CommittedMoves now: 82839, last time: 74914, time diff 5148ms
(113) 19:37:55,698 Uncommitted moves avg: 3296, sum: 26348733, sqr sum: 125297270181, count: 7993
(114) 19:37:55,699 Set new std derivation: 2193,1079980
(115) 19:37:55,699 Max uncommitted moves: 7213
(116) 19:37:55,700 Get better indicator set: [1549,3904,1548,0,0,0,0]
(117) 19:37:55,703 Current indicator set [1539,3296,1552,2,0,0,0,0] added to Clustered State Space
(118) 19:37:55,704 LPCC processed, total committed moves: 82839
(119) 19:37:55,704 Actuator upper limit 8203, mean limit 3904
(120) 19:37:55,881 Received a request from other LP for local LP to initiate GVT calculation when it passed
the simulation time 84645
(121) 19:37:55,898 GVT parameter requested by SimulationController, min. time: 83206
(122) 19:37:55,924 Received new GVT of 83206 from SimulationController, LPCC processing needed = false
(123) 19:37:55,926 Simulation state committed for time 83206
(124) 19:37:56,397 Sent GVT calculation request to SimulationController
(125) 19:37:56,405 GVT parameter requested by SimulationController, min. time: 84663
(126) 19:37:56,464 SimulationController reported confirmed End of Simulation by xact: xact(Id: 39824, move
time: 84644, current block: (2,2), next block: (2,2))
(127) 19:37:56,469 Rollback for time 84644, state restored for time 84644, time before rollback 84747
(128) 19:37:56,471 Simulation stopped and simulation state changed to TERMINATED
(129) 19:37:56,472 Statistics about the simulation processing of this LP:
(130) 19:37:56,475  Total committed transaction moves: 83205
(131) 19:37:56,476  Total transaction moves rolled back: 104
(132) 19:37:56,476  Total simulated transaction moves: 84748
(133) 19:37:56,476  Total transactions sent: 88
(134) 19:37:56,477  Total anti-transactions sent: 0
```

```
(135) 19:37:56,477  Total transactions received: 0
(136) 19:37:56,478  Total anti-transactions received: 0
(137) 19:37:56,647 LogicalProcess.runActivity() left (ActiveObject stopped)
```

**Validation 5.1, output of LP2:**

```
(1)   --- StartNode ------------------------------
(2)   19:37:04,283 LogicalProcess.runActivity() started
(3)   19:37:05,073 LP2 with partition 'Partition2' initialized
(4)   19:37:05,075 LP simulation state changed from INITIALIZED to SIMULATING
(5)   19:37:06,948 Rollback for time 1133, state restored for time 5000, time before rollback 13332
(6)   19:37:07,113 Rollback for time 1439, state restored for time 5000, time before rollback 6288
(7)   19:37:07,285 Rollback for time 1615, state restored for time 5000, time before rollback 6604
(8)   19:37:07,762 Rollback for time 1975, state restored for time 5000, time before rollback 8356
(9)   19:37:08,284 Rollback for time 3058, state restored for time 5000, time before rollback 9804
(10)  19:37:08,998 Rollback for time 4089, state restored for time 5000, time before rollback 11264
(11)  19:37:09,763 Rollback for time 5672, state restored for time 5672, time before rollback 12272
(12)  19:37:10,536 Rollback for time 7109, state restored for time 7112, time before rollback 12276
(13)  19:37:10,542 Rollback for time 7111, state restored for time 7112, time before rollback 7112
(14)  19:37:10,608 GVT parameter requested by SimulationController, min. time: 7220
(15)  19:37:10,648 Received new GVT of 7220 from SimulationController, LPCC processing needed = true
(16)  19:37:10,652 Simulation state committed for time 7220
(17)  19:37:10,659 CommittedMoves now: 564, last time: 0, time diff 5628ms
(18)  19:37:10,659 Uncommitted moves avg: 802, sum: 7965700, sqr sum: 9204719892, count: 9921
(19)  19:37:10,665 Set new std derivation: 532,128610
(20)  19:37:10,665 Max uncommitted moves: 2084
(21)  19:37:10,666 Get better indicator set: non found
(22)  19:37:10,676 Current indicator set [100,802,1762,0,0,1,0,1661] added to Clustered State Space
(23)  19:37:10,677 LPCC processed, total committed moves: 564
(24)  19:37:10,677 Actuator upper limit 9223372036854775807, mean limit 9223372036854775807
(25)  19:37:11,055 Rollback for time 8001, state restored for time 8004, time before rollback 10540
(26)  19:37:11,345 Rollback for time 8641, state restored for time 8644, time before rollback 10504
(27)  19:37:12,894 Rollback for time 11752, state restored for time 11752, time before rollback 21460
(28)  19:37:13,654 Rollback for time 13301, state restored for time 13304, time before rollback 18160
(29)  19:37:14,309 Rollback for time 14634, state restored for time 14636, time before rollback 18204
(30)  19:37:14,368 Rollback for time 14724, state restored for time 14724, time before rollback 15064
(31)  19:37:14,699 Rollback for time 15361, state restored for time 15364, time before rollback 17688
(32)  
```

```
(33)  19:37:15,156  Rollback for time 16248, state restored for time 16248, time before rollback 19340
(34)  19:37:15,235  GVT parameter requested by SimulationController, min. time: 16764
(35)  19:37:15,260  Received new GVT of 16409 from SimulationController, LPCC processing needed = true
(36)  19:37:15,273  Simulation state committed for time 16409
(37)  19:37:15,274  CommittedMoves now: 2869, last time: 564, time diff 4616ms
(38)  19:37:15,275  Uncommitted moves avg: 1838, sum: 17402692, sqr sum: 39014243410, count: 9466
(39)  19:37:15,276  Set new std derivation: 861,233128
(40)  19:37:15,276  Max uncommitted moves: 3563
(41)  19:37:15,277  Get better indicator set: non found
(42)  19:37:15,285  Current indicator set [499,1838,2051,0,0,1,0,1531] added to Clustered State Space
(43)  19:37:15,285  LPCC processed, total committed moves: 2869
(44)  19:37:15,286  Actuator upper limit 92233720368547755807, mean limit 92233720368547775807
(45)  19:37:15,745  Rollback for time 17321, state restored for time 17324, time before rollback 20564
(46)  19:37:16,305  Rollback for time 18244, state restored for time 18244, time before rollback 21368
(47)  19:37:16,367  Rollback for time 18357, state restored for time 18360, time before rollback 18684
(48)  19:37:17,031  Rollback for time 19628, state restored for time 19628, time before rollback 24184
(49)  19:37:17,567  Rollback for time 20668, state restored for time 20668, time before rollback 24204
(50)  19:37:19,930  Rollback for time 25189, state restored for time 25192, time before rollback 40212
(51)  19:37:20,114  Rollback for time 25425, state restored for time 25428, time before rollback 26148
(52)  19:37:20,274  GVT parameter requested by SimulationController, min. time: 26708
(53)  19:37:20,297  Received new GVT of 25718 from SimulationController, LPCC processing needed = true
(54)  19:37:20,312  Simulation state committed for time 25718
(55)  19:37:20,313  CommittedMoves now: 5203, last time: 2869, time diff 5039ms
(56)  19:37:20,313  Uncommitted moves avg: 2295, sum: 23238716, sqr sum: 79118752802, count: 10125
(57)  19:37:20,314  Set new std derivation: 1595,803780
(58)  19:37:20,314  Max uncommitted moves: 5957
(59)  19:37:20,314  Get better indicator set: [499,1838,2051,0,0,1,0,1531]
(60)  19:37:20,315  Current indicator set [463,2295,2009,0,0,1,0,1514] added to Clustered State Space
(61)  19:37:20,315  LPCC processed, total committed moves: 5203
(62)  19:37:20,315  Actuator upper limit 4967, mean limit 1838
(63)  19:37:20,372  Rollback for time 25770, state restored for time 25772, time before rollback 27104
(64)  19:37:20,427  Rollback for time 25852, state restored for time 25852, time before rollback 26188
(65)  19:37:20,963  Rollback for time 26829, state restored for time 26832, time before rollback 30308
(66)  19:37:21,819  Rollback for time 28219, state restored for time 28220, time before rollback 33976
(67)  19:37:21,840  Rollback for time 28245, state restored for time 28248, time before rollback 28340
```

```
(68)  19:37:21,859  Rollback for time 28269, state restored for time 28272, time before rollback 28348
(69)  19:37:22,454  Rollback for time 29378, state restored for time 29380, time before rollback 32380
(70)  19:37:22,867  Rollback for time 30108, state restored for time 30108, time before rollback 32708
(71)  19:37:23,651  Rollback for time 31542, state restored for time 31544, time before rollback 36440
(72)  19:37:23,758  Rollback for time 31734, state restored for time 31736, time before rollback 32372
(73)  19:37:23,841  Rollback for time 31878, state restored for time 31880, time before rollback 32372
(74)  19:37:25,308  GVT parameter requested by SimulationController, min. time: 42904
(75)  19:37:25,319  Received new GVT of 34517 from SimulationController, LPCC processing needed = true
(76)  19:37:25,332  Simulation state committed for time 34517
(77)  19:37:25,337  CommittedMoves now: 7414, last time: 5203, time diff 5024ms
(78)  19:37:25,338  Uncommitted moves avg: 1730, sum: 16821636, sqr sum: 3893818210,0 count: 9720
(79)  19:37:25,338  Set new std derivation: 1005,505259
(80)  19:37:25,338  Max uncommitted moves: 4311
(81)  19:37:25,339  Get better indicator set: [463,2295,2009,0,0,1,0,1514]
(82)  19:37:25,339  Current indicator set [440,1730,1935,0,0,2,0,1129] added to Clustered State Space
(83)  19:37:25,339  LPCC processed, total committed moves: 7414
(84)  19:37:25,340  Actuator upper limit 4267, mean limit 2295
(85)  19:37:25,547  Rollback for time 34836, state restored for time 34836, time before rollback 44540
(86)  19:37:25,714  Rollback for time 35082, state restored for time 35084, time before rollback 35996
(87)  19:37:25,756  Rollback for time 35163, state restored for time 35164, time before rollback 35448
(88)  19:37:26,188  Rollback for time 35725, state restored for time 35728, time before rollback 38728
(89)  19:37:26,572  Rollback for time 36403, state restored for time 36404, time before rollback 38768
(90)  19:37:26,760  Rollback for time 36690, state restored for time 36692, time before rollback 37860
(91)  19:37:26,810  Rollback for time 36764, state restored for time 36764, time before rollback 37028
(92)  19:37:28,334  Rollback for time 39432, state restored for time 39432, time before rollback 48328
(93)  19:37:28,751  Rollback for time 40153, state restored for time 40156, time before rollback 42716
(94)  19:37:28,814  Rollback for time 40251, state restored for time 40252, time before rollback 40544
(95)  19:37:30,324  GVT parameter requested by SimulationController, min. time: 51212
(96)  19:37:30,342  Received new GVT of 42689 from SimulationController, LPCC processing needed = true
(97)  19:37:30,352  Simulation state committed for time 42689
(98)  19:37:30,353  CommittedMoves now: 9467, last time: 7414, time diff 5016ms
(99)  19:37:30,354  Uncommitted moves avg: 1847, sum: 17465420, sqr sum: 4303471448,0 count: 9453
(100) 19:37:30,355  Set new std derivation: 1067,223113
(101) 19:37:30,355  Max uncommitted moves: 4195
(102) 19:37:30,355  Get better indicator set: [463,2295,2009,0,0,1,0,1514]
```

```
(103)  19:37:30,356  Current indicator set [409,1847,1885,0,0,1,0,1467] added to Clustered State Space
(104)  19:37:30,356  LPCC processed, total committed moves: 9467
(105)  19:37:30,356  Actuator upper limit 4388, mean limit 2295
(106)  19:37:30,468  Rollback for time 42804, state restored for time 42804, time before rollback 52012
(107)  19:37:30,499  Rollback for time 42838, state restored for time 42840, time before rollback 43008
(108)  19:37:31,624  Rollback for time 44789, state restored for time 44792, time before rollback 51596
(109)  19:37:33,694  Actuator limit (limit of uncommitted transaction moves) exceeded, current uncommitted
moves: 4389, limit: 4388
(110)  19:37:33,695  Changed Cancelback mode to ON because of actuator limit reached
(111)  19:37:33,695  Actuator limit (limit of uncommitted transaction moves) exceeded, current uncommitted
moves: 4389, limit: 4388
(112)  19:37:33,696  Actuator limit (limit of uncommitted transaction moves) exceeded, current uncommitted
moves: 4389, limit: 4388
(113)  19:37:33,696  Actuator limit (limit of uncommitted transaction moves) exceeded, current uncommitted
moves: 4389, limit: 4388
(114)  19:37:33,697  Actuator limit (limit of uncommitted transaction moves) exceeded, current uncommitted
moves: 4389, limit: 4388
...
(466)  19:37:33,987  Actuator limit (limit of uncommitted transaction moves) exceeded, current uncommitted
moves: 4389, limit: 4388
(467)  19:37:34,006  Actuator limit (limit of uncommitted transaction moves) exceeded, current uncommitted
moves: 4389, limit: 4388
(468)  19:37:34,011  Rollback for time 48566, state restored for time 48568, time before rollback 60228
(469)  19:37:34,012  Changed Cancelback mode to OFF
(470)  19:37:34,263  Rollback for time 49049, state restored for time 49052, time before rollback 50276
(471)  19:37:34,504  Rollback for time 49464, state restored for time 49464, time before rollback 50664
(472)  19:37:34,988  Rollback for time 50310, state restored for time 50312, time before rollback 53232
(473)  19:37:35,378  GVT parameter requested by SimulationController, min. time: 53220
(474)  19:37:35,421  Received new GVT of 51021 from SimulationController, LPCC processing needed = true
(475)  19:37:35,429  Simulation state committed for time 51021
(476)  19:37:35,431  CommittedMoves now: 11557, last time: 9467, time diff 5078ms
(477)  19:37:35,432  Uncommitted moves avg: 2093, sum: 19259674, sqr sum: 5087957926, count: 9200
(478)  19:37:35,432  Set new std derivation: 1071.452725
(479)  19:37:35,432  Max uncommitted moves: 4389
(480)  19:37:35,433  Get better indicator set: [463,2295,2009,0,0,1,0,1514]
```

```
(481) 19:37:35,433  Current indicator set [411,2093,1741,0,0,1,0,1633] added to Clustered State Space
(482) 19:37:35,433  LPCC processed, total committed moves: 11557
(483) 19:37:35,433  Actuator upper limit 4396, mean limit 2295
(484) 19:37:36,155  Rollback for time 52203, state restored for time 52204, time before rollback 58192
(485) 19:37:36,458  Rollback for time 52741, state restored for time 52744, time before rollback 54408
(486) 19:37:36,742  Rollback for time 53215, state restored for time 53216, time before rollback 54696
(487) 19:37:37,489  Rollback for time 54501, state restored for time 54504, time before rollback 58704
(488) 19:37:37,627  Rollback for time 54753, state restored for time 54756, time before rollback 55496
(489) 19:37:38,708  Rollback for time 56328, state restored for time 56328, time before rollback 62968
(490) 19:37:39,767  Rollback for time 58130, state restored for time 58132, time before rollback 63500
(491) 19:37:39,944  Rollback for time 58421, state restored for time 58424, time before rollback 59440
(492) 19:37:40,433  GVT parameter requested by SimulationController, min. time: 62096
(493) 19:37:40,466  Received new GVT of 59249 from SimulationController, LPCC processing needed = true
(494) 19:37:40,476  Simulation state committed for time 59249
(495) 19:37:40,477  CommittedMoves now: 13622, last time: 11557, time diff 5046ms
(496) 19:37:40,477  Uncommitted moves avg: 1659, sum: 14868795, sqr sum: 2929603113, count: 8962
(497) 19:37:40,477  Set new std derivation: 716,543205
(498) 19:37:40,478  Max uncommitted moves: 3127
(499) 19:37:40,478  Get better indicator set: [440,1730,1935,0,0,2,0,1129]
(500) 19:37:40,478  Current indicator set [409,1659,1776,0,0,1,0,1342] added to Clustered State Space
(501) 19:37:40,483  LPCC processed, total committed moves: 13622
(502) 19:37:40,484  Actuator upper limit 3135, mean limit 1730
(503) 19:37:40,534  Rollback for time 59322, state restored for time 59324, time before rollback 62480
(504) 19:37:40,656  Rollback for time 59516, state restored for time 59516, time before rollback 60184
(505) 19:37:40,962  Rollback for time 60020, state restored for time 60020, time before rollback 61800
(506) 19:37:42,622  Actuator limit (limit of uncommitted transaction moves) exceeded, current uncommitted
moves: 3136, limit: 3135
(507) 19:37:42,623  Changed Cancelback mode to ON because of actuator limit reached
(508) 19:37:42,624  Actuator limit (limit of uncommitted transaction moves) exceeded, current uncommitted
moves: 3136, limit: 3135
(509) 19:37:42,624  Actuator limit (limit of uncommitted transaction moves) exceeded, current uncommitted
moves: 3136, limit: 3135
(510) 19:37:42,625  Actuator limit (limit of uncommitted transaction moves) exceeded, current uncommitted
moves: 3136, limit: 3135
...
```

```
(700)  19:37:42,719  Actuator limit (limit of uncommitted transaction moves) exceeded, current uncommitted
moves: 3136, limit: 3135
(701)  19:37:42,722  Rollback for time 62924, state restored for time 62924, time before rollback 71776
(702)  19:37:42,723  Changed Cancelback mode to OFF
(703)  19:37:43,182  Rollback for time 63661, state restored for time 63664, time before rollback 66336
(704)  19:37:43,408  Rollback for time 64007, state restored for time 64008, time before rollback 65252
(705)  19:37:43,915  Rollback for time 64610, state restored for time 64612, time before rollback 67748
(706)  19:37:43,992  Rollback for time 64733, state restored for time 64736, time before rollback 65132
(707)  19:37:44,062  Rollback for time 64836, state restored for time 64836, time before rollback 65208
(708)  19:37:44,844  Rollback for time 66130, state restored for time 66132, time before rollback 70592
(709)  19:37:45,137  Rollback for time 66592, state restored for time 66592, time before rollback 68180
(710)  19:37:45,465  GVT parameter requested by SimulationControler, min. time: 68208
(711)  19:37:45,490  Received new GVT of 67139 from SimulationController, LPCC processing needed = true
(712)  19:37:45,494  Simulation state committed for time 67139
(713)  19:37:45,495  CommittedMoves now: 15605, last time: 13622, time diff 5018ms
(714)  19:37:45,496  Uncommitted moves avg: 1625, sum: 14300262, sqr sum: 28295825508, count: 8799
(715)  19:37:45,497  Set new std derivation: 757,986422
(716)  19:37:45,497  Max uncommitted moves: 3136
(717)  19:37:45,498  Get better indicator set: [409,1659,1776,0,0,1,0,1342]
(718)  19:37:45,498  Current indicator set [395,1625,1715,0,2,0,1411] added to Clustered State Space
(719)  19:37:45,499  LPCC processed, total committed moves: 15605
(720)  19:37:45,499  Actuator upper limit 3146, mean limit 1659
(721)  19:37:46,292  Rollback for time 68368, state restored for time 68368, time before rollback 74028
(722)  19:37:46,702  Rollback for time 69018, state restored for time 69020, time before rollback 71308
(723)  19:37:47,139  Rollback for time 69710, state restored for time 69712, time before rollback 72100
(724)  19:37:47,441  Rollback for time 70197, state restored for time 70200, time before rollback 71868
(725)  19:37:47,802  Rollback for time 70787, state restored for time 70788, time before rollback 72836
(726)  19:37:49,116  Actuator limit (limit of uncommitted transaction moves) exceeded, current uncommitted
moves: 3147, limit: 3146
(727)  19:37:49,117  Changed Cancelback mode to ON because of actuator limit reached
(728)  19:37:49,117  Actuator limit (limit of uncommitted transaction moves) exceeded, current uncommitted
moves: 3147, limit: 3146
(729)  19:37:49,117  Actuator limit (limit of uncommitted transaction moves) exceeded, current uncommitted
moves: 3147, limit: 3146
(730)  19:37:49,118  Actuator limit (limit of uncommitted transaction moves) exceeded, current uncommitted
```

```
moves: 3147, limit: 3146
...
(3758) 19:37:50,489 Actuator limit (limit of uncommitted transaction moves) exceeded, current uncommitted
moves: 3147, limit: 3146
(3759) 19:37:50,489 Actuator limit (limit of uncommitted transaction moves) exceeded, current uncommitted
moves: 3147, limit: 3146
(3760) 19:37:50,494 Actuator limit (limit of uncommitted transaction moves) exceeded, current uncommitted
moves: 3147, limit: 3146
(3761) 19:37:50,495 GVT parameter requested by SimulationController, min. time: 79700
(3762) 19:37:50,506 Actuator limit (limit of uncommitted transaction moves) exceeded, current uncommitted
moves: 3147, limit: 3146
...
(3798) 19:37:50,517 Actuator limit (limit of uncommitted transaction moves) exceeded, current uncommitted
moves: 3147, limit: 3146
(3799) 19:37:50,518 Received new GVT of 74915 from SimulationController, LPCC processing needed = true
(3800) 19:37:50,529 Simulation state committed for time 74915
(3801) 19:37:50,529 Changed Cancelback mode to OFF
(3802) 19:37:50,530 CommittedMoves now: 17554, last time: 15605, time diff 5034ms
(3803) 19:37:50,530 Uncommitted moves avg: 1905, sum: 18029686, sqr sum: 44528826786, count: 9460
(3804) 19:37:50,530 Set new std derivation: 1036,407354
(3805) 19:37:50,530 Max uncommitted moves: 3147
(3806) 19:37:50,531 Get better indicator set: [440,1730,1935,0,0,2,0,1129]
(3807) 19:37:50,531 Current indicator set [387,1905,1269,0,0,0,697] added to Clustered State Space
(3808) 19:37:50,531 LPCC processed, total committed moves: 17554
(3809) 19:37:50,531 Actuator upper limit 3762, mean limit 1730
(3810) 19:37:51,137 Rollback for time 75829, state restored for time 75832, time before rollback 84116
(3811) 19:37:51,529 Rollback for time 76447, state restored for time 76448, time before rollback 78616
(3812) 19:37:51,888 Rollback for time 77021, state restored for time 77024, time before rollback 79016
(3813) 19:37:52,000 Rollback for time 77172, state restored for time 77172, time before rollback 77736
(3814) 19:37:52,748 Rollback for time 78374, state restored for time 78376, time before rollback 82576
(3815) 19:37:52,953 Rollback for time 78708, state restored for time 78708, time before rollback 78996
(3816) 19:37:53,331 Rollback for time 79296, state restored for time 79296, time before rollback 81444
(3817) 19:37:53,431 Rollback for time 79454, state restored for time 79456, time before rollback 79932
(3818) 19:37:54,093 Rollback for time 80472, state restored for time 80472, time before rollback 84188
(3819) 19:37:54,661 Rollback for time 81152, state restored for time 81152, time before rollback 84532
```

```
(3820)  19:37:55,164  Unconfirmed End Of Simulation reached by xact: xact(Id: 39828, move time: 84652, current
block: (2,2), next block: (2,2))
(3821)  19:37:55,220  Sent GVT calculation request to SimulationController
(3822)  19:37:55,220  GVT parameter requested by SimulationController, min. time: Infinite (unconfirmed end of
simulation for time 84652)
(3823)  19:37:55,254  Received new GVT of 82106 from SimulationController, LPCC processing needed = false
(3824)  19:37:55,263  Simulation state committed for time 82106
(3825)  19:37:55,384  Rollback for time 82435, state restored for time 82436, time before rollback 84656
(3826)  19:37:55,553  Rollback for time 82697, state restored for time 82700, time before rollback 83584
(3827)  19:37:55,641  GVT parameter requested by SimulationController, min. time: 83196
(3828)  19:37:55,701  Received new GVT of 82840 from SimulationController, LPCC processing needed = true
(3829)  19:37:55,703  Simulation state committed for time 82840
(3830)  19:37:55,704  CommittedMoves now: 19548, last time: 17554, time diff 5175ms
(3831)  19:37:55,705  Uncommitted moves avg: 1368, sum: 11682582, sqr sum: 1913840332, count: 8539
(3832)  19:37:55,705  Set new std derivation: 607,883839
(3833)  19:37:55,705  Max uncommitted moves: 2446
(3834)  19:37:55,705  Get better indicator set: [395,1625,1715,0,0,2,0,1411]
(3835)  19:37:55,708  Current indicator set [385,1368,1650,0,0,2,0,1464] added to Clustered State Space
(3836)  19:37:55,708  LPCC processed, total committed moves: 19548
(3837)  19:37:55,708  Actuator upper limit 2817, mean limit 1625
(3838)  19:37:55,870  Unconfirmed End Of Simulation reached by xact: xact(Id: 39824, move time: 84644, current
block: (2,2), next block: (2,2))
(3839)  19:37:55,886  Sent GVT calculation request to SimulationController
(3840)  19:37:55,898  GVT parameter requested by SimulationController, min. time: Infinite (unconfirmed end of
simulation for time 84644)
(3841)  19:37:55,916  Received new GVT of 83206 from SimulationController, LPCC processing needed = false
(3842)  19:37:55,917  Simulation state committed for time 83206
(3843)  19:37:56,411  GVT parameter requested by SimulationController, min. time: Infinite (unconfirmed end of
simulation for time 84644)
(3844)  19:37:56,470  SimulationController reported confirmed End of Simulation by xact: xact(Id: 39824, move
time: 84644, current block: (2,2), next block: (2,2))
(3845)  19:37:56,470  Simulation stopped and simulation state changed to TERMINATED
(3846)  19:37:56,470  Statistics about the simulation processing of this LP:
(3847)  19:37:56,470      Total committed transaction moves: 19639
(3848)  19:37:56,470      Total transaction moves rolled back: 70331
```

```
(3849) 19:37:56,470   Total simulated transaction moves: 90330
(3850) 19:37:56,470   Total transactions sent: 0
(3851) 19:37:56,471   Total anti-transactions sent: 0
(3852) 19:37:56,471   Total transactions received: 88
(3853) 19:37:56,471   Total anti-transactions received: 0
(3854) 19:37:56,651 LogicalProcess.runActivity() left (ActiveObject stopped)
```

**Validation 5.1, output of simulate:**

```
(1)  19:37:01,371  Simulation model file read and parsed successfully.
(2)  19:37:01,373  2 partition(s) found in simulation model file.
(3)
(4)  19:37:01,382  Creating SimulationController instance
(5)  19:37:01,382  parallelJavaGpssSimulator.SimulationController.SimulationController
(6)  19:37:02,853  SimulationController.runActivity() started
(7)  19:37:02,912  Creating LPs and starting simulation
(8)  19:37:05,029  Simulation started
(9)
(10) 19:37:05,029  Please press:
(11) G + <Enter>  to force GVT calculation
(12) X + <Enter>  to terminate the simulation
(13)
(14) 19:37:10,185  Initiated GVT calculation
(15) 19:37:10,615  Min time received from LP1: 7252
(16) 19:37:10,615  Min time received from LP2: 7220
(17) 19:37:10,647  Simulation reached Global Virtual Time: 7220
(18) 19:37:15,218  Initiated GVT calculation
(19) 19:37:15,241  Min time received from LP1: 16409
(20) 19:37:15,242  Min time received from LP2: 16764
(21) 19:37:15,261  Simulation reached Global Virtual Time: 16409
(22) 19:37:20,259  Initiated GVT calculation
(23) 19:37:20,278  Min time received from LP1: 25718
(24) 19:37:20,279  Min time received from LP2: 26708
(25) 19:37:20,296  Simulation reached Global Virtual Time: 25718
(26) 19:37:25,282  Initiated GVT calculation
(27) 19:37:25,311  Min time received from LP1: 34517
(28) 19:37:25,312  Min time received from LP2: 42904
(29) 19:37:25,319  Simulation reached Global Virtual Time: 34517
(30) 19:37:30,310  Initiated GVT calculation
(31) 19:37:30,328  Min time received from LP1: 42689
(32) 19:37:30,329  Min time received from LP2: 51212
```

```
(33)  19:37:30,341  Simulation reached Global Virtual Time: 42689
(34)  19:37:35,365  Initiated GVT calculation
(35)  19:37:35,414  Min time received from LP1: 51021
(36)  19:37:35,414  Min time received from LP2: 53220
(37)  19:37:35,424  Simulation reached Global Virtual Time: 51021
(38)  19:37:40,423  Initiated GVT calculation
(39)  19:37:40,443  Min time received from LP1: 59249
(40)  19:37:40,443  Min time received from LP2: 62096
(41)  19:37:40,462  Simulation reached Global Virtual Time: 59249
(42)  19:37:45,452  Initiated GVT calculation
(43)  19:37:45,476  Min time received from LP1: 67139
(44)  19:37:45,477  Min time received from LP2: 68208
(45)  19:37:45,496  Simulation reached Global Virtual Time: 67139
(46)  19:37:50,487  Initiated GVT calculation
(47)  19:37:50,505  Min time received from LP1: 74915
(48)  19:37:50,506  Min time received from LP2: 79700
(49)  19:37:50,518  Simulation reached Global Virtual Time: 74915
(50)  19:37:55,210  Initiated GVT calculation
(51)  19:37:55,237  Min time received from LP1: 82106
(52)  19:37:55,237  Min time received from LP2: Infinite (unconfirmed end of simulation for time 84652)
(53)  19:37:55,253  Simulation reached Global Virtual Time: 82106
(54)  19:37:55,549  Initiated GVT calculation
(55)  19:37:55,644  Min time received from LP1: 82840
(56)  19:37:55,644  Min time received from LP2: 83196
(57)  19:37:55,702  Simulation reached Global Virtual Time: 82840
(58)  19:37:55,886  Initiated GVT calculation
(59)  19:37:55,910  Min time received from LP1: 83206
(60)  19:37:55,911  Min time received from LP2: Infinite (unconfirmed end of simulation for time 84644)
(61)  19:37:55,924  Simulation reached Global Virtual Time: 83206
(62)  19:37:56,396  Initiated GVT calculation
(63)  19:37:56,414  Min time received from LP1: 84663
(64)  19:37:56,414  Min time received from LP2: Infinite (unconfirmed end of simulation for time 84644)
(65)  19:37:56,480  Simulation finished
(66)  19:37:56,612  ************* Simulation report  *************
(67)  19:37:56,618  The simulation was completed at the simulation time: 84644
```

```
(68) 19:37:56,619 by the transaction xact(Id: 39824, move time: 84644, current block: (2,2), next block:
(2,2))
(69) 19:37:56,622 The average simulation perfomance in simulation time per second real time was: 1640,260498
(time units/s)
(70)
(71) 19:37:56,623 Block report section:
(72) Block           current    total
(73)                 xacts      xacts
(74) Partition: Partition1
(75) (1,1)              1        84644    Block: GENERATE 1,0,(no offset),(no limit),0
(76) (1,2)              0        84644    Block: TRANSFER 0.0010,(Label: Label1, partition: 2, block: 2)
(77) (1,3)              0        84556    Block: TERMINATE 0
(78) Partition: Partition2
(79) (2,1)              1        19912    Block: GENERATE 4,0,5000,(no limit),0
(80) (2,2)              0        20000    Block: TERMINATE 1
(81)
(82) 19:37:56,624 Summary entity report section:
(83)
(84) 19:37:56,625 SimulationController.terminateLPs() called
```

**Validation 5.2, output of LP1:**

```
(1)
(2)    --- StartNode -------------------------------------------
(3)    19:41:38,243 LogicalProcess.runActivity() started
(4)    19:41:38,924 LP1 with partition 'Partition1' initialized
(5)    19:41:39,222 LP simulation state changed from INITIALIZED to SIMULATING
(6)    19:41:58,924 Sent GVT calculation request to SimulationController
(7)    19:41:58,925 Requested GVT calculation because memory limit 1 (free memory for JVM below 5242880 byte)
reached
(8)    19:41:59,094 GVT parameter requested by SimulationController, min. time: 32250
(9)    19:41:59,155 Received new GVT of 32250 from SimulationController, LPCC processing needed = false
(10)   19:41:59,320 Simulation state committed for time 32250
(11)   19:41:59,377 Sent GVT calculation request to SimulationController
(12)   19:41:59,377 Requested GVT calculation because memory limit 1 (free memory for JVM below 5242880 byte)
reached
(13)   19:41:59,388 GVT parameter requested by SimulationController, min. time: 32444
(14)   19:41:59,532 Received new GVT of 32444 from SimulationController, LPCC processing needed = false
(15)   19:41:59,539 Simulation state committed for time 32444
(16)   19:42:19,316 Sent GVT calculation request to SimulationController
(17)   19:42:19,316 Requested GVT calculation because memory limit 1 (free memory for JVM below 5242880 byte)
reached
(18)   19:42:19,497 GVT parameter requested by SimulationController, min. time: 64728
(19)   19:42:19,516 Received new GVT of 64728 from SimulationController, LPCC processing needed = false
(20)   19:42:19,647 Simulation state committed for time 64728
(21)   19:42:19,756 Sent GVT calculation request to SimulationController
(22)   19:42:19,757 Requested GVT calculation because memory limit 1 (free memory for JVM below 5242880 byte)
reached
(23)   19:42:19,767 GVT parameter requested by SimulationController, min. time: 64911
(24)   19:42:19,780 Received new GVT of 64911 from SimulationController, LPCC processing needed = false
(25)   19:42:19,781 Simulation state committed for time 64911
(26)   19:42:19,791 Sent GVT calculation request to SimulationController
(27)   19:42:19,792 Requested GVT calculation because memory limit 1 (free memory for JVM below 5242880 byte)
reached
```

```
(28) 19:42:19,940 GVT parameter requested by SimulationControler, min. time: 64954
(29) 19:42:19,961 Received new GVT of 64954 from SimulationControler, LPCC processing needed = false
(30) 19:42:19,961 Simulation state committed for time 64954
(31) 19:42:25,685 Received a request from other LP for local LP to initiate GVT calculation when it passed
the simulation time 84669
(32) 19:42:25,715 GVT parameter requested by SimulationControler, min. time: 74461
(33) 19:42:25,746 Received new GVT of 74461 from SimulationControler, LPCC processing needed = false
(34) 19:42:25,768 Simulation state committed for time 74461
(35) 19:42:29,145 Received a request from other LP for local LP to initiate GVT calculation when it passed
the simulation time 84637
(36) 19:42:29,168 GVT parameter requested by SimulationControler, min. time: 79884
(37) 19:42:29,220 Received new GVT of 79884 from SimulationControler, LPCC processing needed = false
(38) 19:42:29,228 Simulation state committed for time 79884
(39) 19:42:30,653 Received a request from other LP for local LP to initiate GVT calculation when it passed
the simulation time 84621
(40) 19:42:30,676 GVT parameter requested by SimulationControler, min. time: 82417
(41) 19:42:30,692 Received new GVT of 82417 from SimulationControler, LPCC processing needed = false
(42) 19:42:30,699 Simulation state committed for time 82417
(43) 19:42:31,018 Received a request from other LP for local LP to initiate GVT calculation when it passed
the simulation time 84617
(44) 19:42:31,040 GVT parameter requested by SimulationControler, min. time: 82978
(45) 19:42:31,050 Received new GVT of 82978 from SimulationControler, LPCC processing needed = false
(46) 19:42:31,052 Simulation state committed for time 82978
(47) 19:42:31,541 Received a request from other LP for local LP to initiate GVT calculation when it passed
the simulation time 84613
(48) 19:42:31,557 GVT parameter requested by SimulationControler, min. time: 84428
(49) 19:42:31,575 Received new GVT of 84428 from SimulationControler, LPCC processing needed = false
(50) 19:42:31,577 Simulation state committed for time 84428
(51) 19:42:31,643 Sent GVT calculation request to SimulationControler
(52) 19:42:31,652 GVT parameter requested by SimulationControler, min. time: 84614
(53) 19:42:31,709 SimulationControler reported confirmed End of Simulation by xact: xact(Id: 39808, move
time: 84612, current block: (2,2), next block: (2,2))
(54) 19:42:31,712 Rollback for time 84612, state restored for time 84612, time before rollback 84697
(55) 19:42:31,713 Simulation stopped and simulation state changed to TERMINATED
(56) 19:42:31,713 Statistics about the simulation processing of this LP:
```

```
(57) 19:42:31,714    Total committed transaction moves: 84427
(58) 19:42:31,714    Total transaction moves rolled back: 86
(59) 19:42:31,714    Total simulated transaction moves: 84698
(60) 19:42:31,714    Total transactions sent: 96
(61) 19:42:31,715    Total anti-transactions sent: 0
(62) 19:42:31,715    Total transactions received: 0
(63) 19:42:31,715    Total anti-transactions received: 0
(64) 19:42:32,080 LogicalProcess.runActivity() left (ActiveObject stopped)
```

**Validation 5.2, output of LP2:**

```
(1)    ---------------------------------------------------------
(2)    --- StartNode ---------------------------------------------
(3)    19:41:38,489 LogicalProcess.runActivity() started
(4)    19:41:39,247 LP2 with partition 'Partition2' initialized
(5)    19:41:39,288 LP simulation state changed from INITIALIZED to SIMULATING
(6)    19:41:40,982 Rollback for time 894, state restored for time 5000, time before rollback 13060
(7)    19:41:41,394 Rollback for time 1417, state restored for time 5000, time before rollback 7316
(8)    19:41:41,991 Rollback for time 2097, state restored for time 5000, time before rollback 10416
(9)    19:41:42,160 Rollback for time 2495, state restored for time 5000, time before rollback 5804
(10)   19:41:42,348 Rollback for time 2888, state restored for time 5000, time before rollback 6652
(11)   19:41:42,415 Rollback for time 3022, state restored for time 5000, time before rollback 5524
(12)   19:41:43,115 Rollback for time 3996, state restored for time 5000, time before rollback 10916
(13)   19:41:43,318 Rollback for time 4418, state restored for time 5000, time before rollback 6876
(14)   19:41:44,106 Rollback for time 6093, state restored for time 6096, time before rollback 11788
(15)   19:41:45,541 Rollback for time 8955, state restored for time 8956, time before rollback 18292
(16)   19:41:45,630 Rollback for time 9109, state restored for time 9112, time before rollback 9792
(17)   19:41:45,742 Rollback for time 9305, state restored for time 9308, time before rollback 10084
(18)   19:41:46,238 Rollback for time 10215, state restored for time 10216, time before rollback 12884
(19)   19:41:47,592 Rollback for time 12810, state restored for time 12812, time before rollback 21596
(20)   19:41:47,867 Rollback for time 13365, state restored for time 13368, time before rollback 15200
(21)   19:41:48,282 Rollback for time 14177, state restored for time 14180, time before rollback 16992
(22)   19:41:49,063 Rollback for time 15737, state restored for time 15740, time before rollback 20088
(23)   19:41:49,145 Rollback for time 15869, state restored for time 15872, time before rollback 16364
(24)   19:41:49,291 Rollback for time 16154, state restored for time 16156, time before rollback 17016
(25)   19:41:49,505 Rollback for time 16524, state restored for time 16524, time before rollback 17832
(26)   19:41:49,992 Rollback for time 17457, state restored for time 17460, time before rollback 20648
(27)   19:41:50,186 Rollback for time 17801, state restored for time 17804, time before rollback 19116
(28)   19:41:50,327 Rollback for time 18067, state restored for time 18068, time before rollback 19000
(29)   19:41:50,819 Rollback for time 18738, state restored for time 18740, time before rollback 21704
(30)   19:41:51,287 Rollback for time 19607, state restored for time 19608, time before rollback 22824
(31)   19:41:51,419 Rollback for time 19855, state restored for time 19856, time before rollback 20764
(32)   19:41:51,549 Rollback for time 20080, state restored for time 20080, time before rollback 20884
```

```
(33)  19:41:52,566  Rollback for time 22020, state restored for time 22020, time before rollback 27784
(34)  19:41:53,139  Rollback for time 23099, state restored for time 23100, time before rollback 26984
(35)  19:41:53,585  Rollback for time 23955, state restored for time 23956, time before rollback 26844
(36)  19:41:53,661  Rollback for time 24082, state restored for time 24084, time before rollback 24528
(37)  19:41:53,752  Rollback for time 24236, state restored for time 24236, time before rollback 24776
(38)  19:41:54,088  Rollback for time 24860, state restored for time 24860, time before rollback 26976
(39)  19:41:54,460  Rollback for time 25552, state restored for time 25552, time before rollback 27892
(40)  19:41:54,909  Rollback for time 26357, state restored for time 26360, time before rollback 28544
(41)  19:41:55,075  Rollback for time 26657, state restored for time 26660, time before rollback 27748
(42)  19:41:55,889  Rollback for time 28104, state restored for time 28104, time before rollback 33344
(43)  19:41:56,358  Rollback for time 28968, state restored for time 28968, time before rollback 31956
(44)  19:41:56,826  Rollback for time 29762, state restored for time 29764, time before rollback 32440
(45)  19:41:57,326  Rollback for time 30613, state restored for time 30616, time before rollback 33588
(46)  19:41:58,571  Rollback for time 31908, state restored for time 31908, time before rollback 40708
(47)  19:41:58,984  GVT parameter requested by SimulationControler, min. time: 35276
(48)  19:41:59,163  Received new GVT of 32250 from SimulationControler, LPCC processing needed = false
(49)  19:41:59,196  Simulation state committed for time 32250
(50)  19:41:59,199  Rollback for time 32319, state restored for time 32320, time before rollback 36352
(51)  19:41:59,391  GVT parameter requested by SimulationControler, min. time: 33796
(52)  19:41:59,414  Received new GVT of 32444 from SimulationControler, LPCC processing needed = false
(53)  19:41:59,425  Simulation state committed for time 32444
(54)  19:41:59,677  Rollback for time 32716, state restored for time 32716, time before rollback 35832
(55)  19:41:59,915  Rollback for time 33146, state restored for time 33148, time before rollback 34572
(56)  19:42:01,563  Rollback for time 36138, state restored for time 36140, time before rollback 46204
(57)  19:42:02,280  Rollback for time 37435, state restored for time 37436, time before rollback 41912
(58)  19:42:03,061  Rollback for time 38798, state restored for time 38800, time before rollback 43812
(59)  19:42:03,578  Rollback for time 39738, state restored for time 39740, time before rollback 42980
(60)  19:42:04,062  Rollback for time 40588, state restored for time 40588, time before rollback 43596
(61)  19:42:04,149  Rollback for time 40722, state restored for time 40724, time before rollback 41204
(62)  19:42:04,267  Rollback for time 40910, state restored for time 40912, time before rollback 41520
(63)  19:42:04,849  Rollback for time 41960, state restored for time 41960, time before rollback 45536
(64)  19:42:06,074  Rollback for time 44003, state restored for time 44004, time before rollback 51060
(65)  19:42:07,223  Rollback for time 45998, state restored for time 46000, time before rollback 53196
(66)  19:42:07,243  Rollback for time 46021, state restored for time 46024, time before rollback 46100
(67)  19:42:07,297  Rollback for time 46094, state restored for time 46096, time before rollback 46412
```

```
(68)  19:42:08,110 Rollback for time 47510, state restored for time 47512, time before rollback 52536
(69)  19:42:08,738 Rollback for time 48599, state restored for time 48600, time before rollback 52492
(70)  19:42:09,935 Rollback for time 50696, state restored for time 50696, time before rollback 57196
(71)  19:42:09,952 Rollback for time 50719, state restored for time 50720, time before rollback 50788
(72)  19:42:12,159 Rollback for time 54473, state restored for time 54476, time before rollback 88212
(73)  19:42:12,558 Rollback for time 55147, state restored for time 55148, time before rollback 57540
(74)  19:42:12,565 Rollback for time 55151, state restored for time 55152, time before rollback 55156
(75)  19:42:12,921 Rollback for time 55659, state restored for time 55660, time before rollback 57084
(76)  19:42:13,505 Rollback for time 56651, state restored for time 56652, time before rollback 60220
(77)  19:42:13,675 Rollback for time 56928, state restored for time 56928, time before rollback 57916
(78)  19:42:14,280 Rollback for time 57932, state restored for time 57932, time before rollback 61564
(79)  19:42:15,124 Rollback for time 59339, state restored for time 59340, time before rollback 64388
(80)  19:42:15,176 Rollback for time 59411, state restored for time 59412, time before rollback 59628
(81)  19:42:15,747 Rollback for time 60366, state restored for time 60368, time before rollback 63712
(82)  19:42:16,388 Rollback for time 61466, state restored for time 61468, time before rollback 64348
(83)  19:42:16,807 Rollback for time 62169, state restored for time 62172, time before rollback 64680
(84)  19:42:17,522 Rollback for time 63339, state restored for time 63340, time before rollback 67672
(85)  19:42:19,342 GVT parameter requested by SimulationController, min. time: 77988
(86)  19:42:19,539 Received new GVT of 64728 from SimulationController, LPCC processing needed = false
(87)  19:42:19,576 Simulation state committed for time 64728
(88)  19:42:19,743 Rollback for time 64884, state restored for time 64884, time before rollback 79672
(89)  19:42:19,766 GVT parameter requested by SimulationController, min. time: 65008
(90)  19:42:19,786 Received new GVT of 64911 from SimulationController, LPCC processing needed = false
(91)  19:42:19,786 Simulation state committed for time 64911
(92)  19:42:19,940 GVT parameter requested by SimulationController, min. time: 66208
(93)  19:42:19,959 Received new GVT of 64954 from SimulationController, LPCC processing needed = false
(94)  19:42:19,960 Simulation state committed for time 64954
(95)  19:42:20,043 Rollback for time 65082, state restored for time 65084, time before rollback 66776
(96)  19:42:21,112 Rollback for time 66864, state restored for time 66864, time before rollback 73188
(97)  19:42:21,159 Rollback for time 66932, state restored for time 66932, time before rollback 67172
(98)  19:42:21,232 Rollback for time 67053, state restored for time 67056, time before rollback 67468
(99)  19:42:22,422 Rollback for time 69025, state restored for time 69028, time before rollback 76012
(100) 19:42:23,275 Rollback for time 70413, state restored for time 70416, time before rollback 75320
(101) 19:42:23,706 Rollback for time 71145, state restored for time 71148, time before rollback 73580
(102) 19:42:23,869 Rollback for time 71411, state restored for time 71412, time before rollback 72324
```

```
(103) 19:42:25,651 Unconfirmed End Of Simulation reached by xact: xact(Id: 39836, move time: 84668, current
block: (2,2), next block: (2,2))
(104) 19:42:25,715 Sent GVT calculation request to SimulationController
(105) 19:42:25,716 GVT parameter requested by SimulationControler, min. time: Infinite (unconfirmed end of
simulation for time 84668)
(106) 19:42:25,735 Received new GVT of 74461 from SimulationController, LPCC processing needed = false
(107) 19:42:25,754 Simulation state committed for time 74461
(108) 19:42:25,871 Rollback for time 74755, state restored for time 74756, time before rollback 84672
(109) 19:42:26,393 Rollback for time 75466, state restored for time 75468, time before rollback 78172
(110) 19:42:27,502 Rollback for time 77259, state restored for time 77260, time before rollback 83884
(111) 19:42:27,603 Rollback for time 77400, state restored for time 77400, time before rollback 78028
(112) 19:42:27,794 Rollback for time 77687, state restored for time 77688, time before rollback 78764
(113) 19:42:28,080 Rollback for time 78139, state restored for time 78140, time before rollback 79740
(114) 19:42:28,296 Rollback for time 78487, state restored for time 78488, time before rollback 79728
(115) 19:42:28,320 Rollback for time 78522, state restored for time 78524, time before rollback 78636
(116) 19:42:29,134 Unconfirmed End Of Simulation reached by xact: xact(Id: 39820, move time: 84636, current
block: (2,2), next block: (2,2))
(117) 19:42:29,160 Sent GVT calculation request to SimulationController
(118) 19:42:29,197 GVT parameter requested by SimulationControler, min. time: Infinite (unconfirmed end of
simulation for time 84636)
(119) 19:42:29,208 Received new GVT of 79884 from SimulationController, LPCC processing needed = false
(120) 19:42:29,214 Simulation state committed for time 79884
(121) 19:42:29,384 Rollback for time 80396, state restored for time 80396, time before rollback 84640
(122) 19:42:29,792 Rollback for time 81022, state restored for time 81024, time before rollback 83428
(123) 19:42:30,200 Rollback for time 81659, state restored for time 81660, time before rollback 84040
(124) 19:42:30,255 Rollback for time 81745, state restored for time 81748, time before rollback 82060
(125) 19:42:30,642 Unconfirmed End Of Simulation reached by xact: xact(Id: 39812, move time: 84620, current
block: (2,2), next block: (2,2))
(126) 19:42:30,659 Sent GVT calculation request to SimulationController
(127) 19:42:30,674 GVT parameter requested by SimulationControler, min. time: Infinite (unconfirmed end of
simulation for time 84620)
(128) 19:42:30,691 Received new GVT of 82417 from SimulationController, LPCC processing needed = false
(129) 19:42:30,693 Simulation state committed for time 82417
(130) 19:42:30,734 Rollback for time 82513, state restored for time 82516, time before rollback 84624
(131) 19:42:31,010 Unconfirmed End Of Simulation reached by xact: xact(Id: 39810, move time: 84616, current
```

```
block: (2,2), next block: (2,2))
(132) 19:42:31,026 Sent GVT calculation request to SimulationController
(133) 19:42:31,034 GVT parameter requested by SimulationControler, min. time: Infinite (unconfirmed end of
simulation for time 84616)
(134) 19:42:31,048 Received new GVT of 82978 from SimulationController, LPCC processing needed = false
(135) 19:42:31,049 Simulation state committed for time 82978
(136) 19:42:31,507 Rollback for time 84370, state restored for time 84372, time before rollback 84620
(137) 19:42:31,536 Unconfirmed End Of Simulation reached by xact: xact(Id: 39808, move time: 84612, current
block: (2,2), next block: (2,2))
(138) 19:42:31,545 Sent GVT calculation request to SimulationController
(139) 19:42:31,555 GVT parameter requested by SimulationControler, min. time: Infinite (unconfirmed end of
simulation for time 84612)
(140) 19:42:31,574 Received new GVT of 84428 from SimulationController, LPCC processing needed = false
(141) 19:42:31,575 Simulation state committed for time 84428
(142) 19:42:31,648 GVT parameter requested by SimulationController, min. time: Infinite (unconfirmed end of
simulation for time 84612)
(143) 19:42:31,697 SimulationController reported confirmed End of Simulation by xact: xact(Id: 39808, move
time: 84612, current block: (2,2), next block: (2,2))
(144) 19:42:31,697 Simulation stopped and simulation state changed to TERMINATED
(145) 19:42:31,698 Statistics about the simulation processing of this LP:
(146) 19:42:31,698   Total committed transaction moves: 19953
(147) 19:42:31,698   Total transaction moves rolled back: 77726
(148) 19:42:31,698   Total simulated transaction moves: 97725
(149) 19:42:31,699   Total transactions sent: 0
(150) 19:42:31,699   Total anti-transactions sent: 0
(151) 19:42:31,700   Total transactions received: 96
(152) 19:42:31,705   Total anti-transactions received: 0
(153) 19:42:32,077 LogicalProcess.runActivity() left (ActiveObject stopped)
```

**Validation 5.2, output of simulate:**

```
(1)  19:41:35,569  Simulation model file read and parsed successfully.
(2)  19:41:35,571  2 partition(s) found in simulation model file.
(3)
(4)  19:41:35,580  Creating SimulationController instance
(5)  19:41:35,581  parallelJavaGpssSimulator.SimulationController
(6)  19:41:37,071  SimulationController.runActivity() started
(7)  19:41:37,129  Creating LPs and starting simulation
(8)  19:41:39,230  Simulation started
(9)
(10) 19:41:39,287  Please press:
(11) G + <Enter> to force GVT calculation
(12) X + <Enter> to terminate the simulation
(13)
(14) 19:41:58,909  Initiated GVT calculation
(15) 19:41:59,101  Min time received from LP1: 32250
(16) 19:41:59,102  Min time received from LP2: 35276
(17) 19:41:59,157  Simulation reached Global Virtual Time: 32250
(18) 19:41:59,376  Initiated GVT calculation
(19) 19:41:59,394  Min time received from LP1: 32444
(20) 19:41:59,395  Min time received from LP2: 33796
(21) 19:41:59,531  Simulation reached Global Virtual Time: 32444
(22) 19:42:19,314  Initiated GVT calculation
(23) 19:42:19,504  Min time received from LP1: 64728
(24) 19:42:19,505  Min time received from LP2: 77988
(25) 19:42:19,537  Simulation reached Global Virtual Time: 64728
(26) 19:42:19,755  Initiated GVT calculation
(27) 19:42:19,771  Min time received from LP1: 64911
(28) 19:42:19,771  Min time received from LP2: 65008
(29) 19:42:19,785  Simulation reached Global Virtual Time: 64911
(30) 19:42:19,790  Initiated GVT calculation
(31) 19:42:19,946  Min time received from LP1: 64954
(32) 19:42:19,946  Min time received from LP2: 66208
```

```
(33) 19:42:19,959  Simulation reached Global Virtual Time: 64954
(34) 19:42:25,703  Initiated GVT calculation
(35) 19:42:25,725  Min time received from LP1: 74461
(36) 19:42:25,726  Min time received from LP2: Infinite (unconfirmed end of simulation for time 84668)
(37) 19:42:25,746  Simulation reached Global Virtual Time: 74461
(38) 19:42:29,157  Initiated GVT calculation
(39) 19:42:29,201  Min time received from LP1: 79884
(40) 19:42:29,201  Min time received from LP2: Infinite (unconfirmed end of simulation for time 84636)
(41) 19:42:29,220  Simulation reached Global Virtual Time: 79884
(42) 19:42:30,657  Initiated GVT calculation
(43) 19:42:30,685  Min time received from LP1: 82417
(44) 19:42:30,686  Min time received from LP2: Infinite (unconfirmed end of simulation for time 84620)
(45) 19:42:30,694  Simulation reached Global Virtual Time: 82417
(46) 19:42:31,025  Initiated GVT calculation
(47) 19:42:31,042  Min time received from LP1: 82978
(48) 19:42:31,043  Min time received from LP2: Infinite (unconfirmed end of simulation for time 84616)
(49) 19:42:31,050  Simulation reached Global Virtual Time: 82978
(50) 19:42:31,544  Initiated GVT calculation
(51) 19:42:31,560  Min time received from LP1: 84428
(52) 19:42:31,563  Min time received from LP2: Infinite (unconfirmed end of simulation for time 84612)
(53) 19:42:31,574  Simulation reached Global Virtual Time: 84428
(54) 19:42:31,642  Initiated GVT calculation
(55) 19:42:31,654  Min time received from LP1: 84614
(56) 19:42:31,654  Min time received from LP2: Infinite (unconfirmed end of simulation for time 84612)
(57) 19:42:31,718  Simulation finished
(58) 19:42:31,843  *************    Simulation report    *************
(59) 19:42:31,850  The simulation was completed at the simulation time: 84612
(60) 19:42:31,851  by the transaction xact(Id: 39808, move time: 84612, current block: (2,2), next block:
(2,2))
(61) 19:42:31,853  The average simulation perfomance in simulation time per second real time was: 1607,706787
(time units/s)
(62)
(63) 19:42:31,854  Block report section:
(64) Block                   current         total
(65)                         xacts           xacts
```

```
(66) Partition: Partition1
(67) (1,1)           1      84612  Block: GENERATE 1,0,(no offset),(no limit),0
(68) (1,2)           0      84612  Block: TRANSFER 0.0010,(Label: Label1, partition: 2, block: 2)
(69) (1,3)           0      84516  Block: TERMINATE 0
(70) Partition: Partition2
(71) (2,1)           1      19904  Block: GENERATE 4,0,5000,(no limit),0
(72) (2,2)           0      20000  Block: TERMINATE 1
(73)
(74) 19:42:31,854 Summary entity report section:
(75)
(76) 19:42:32,059 SimulationController.terminateLPs() called
```

**Validation 6.1, output of LP1:**

```
(1)  ------ StartNode ------
(2)  --- StartNode ------
(3)  14:24:54,614 LogicalProcess.runActivity() started
(4)  14:24:55,257 LP1 with partition 'Partition1' initialized
(5)  14:24:55,552 LP simulation state changed from INITIALIZED to SIMULATING
(6)  14:25:00,787 GVT parameter requested by SimulationControler, min. time: 3667
(7)  14:25:00,822 Received new GVT of 1624 from SimulationControler, LPCC processing needed = true
(8)  14:25:00,856 CommittedMoves now: 0, last time: 0, time diff 561ms
(9)  14:25:00,856 Uncommitted moves avg: 838, sum: 1401974, sqr sum: 1565071424, count: 1673
(10) 14:25:00,859 Set new std derivation: 483,097816
(11) 14:25:00,859 Max uncommitted moves: 1674
(12) 14:25:00,860 Get better indicator set: non found
(13) 14:25:00,869 Current indicator set [0,838,297,90,0,0,0,0] added to Clustered State Space
(14) 14:25:00,870 LPCC processed, total committed moves: 0
(15) 14:25:00,871 Actuator upper limit 9223372036854775807, mean limit 9223372036854775807
(16) 14:25:05,721 Unconfirmed End Of Simulation reached by xact: xact(Id: 8759, move time: 6379, current
block: (1,3), next block: (1,3))
(17) 14:25:05,802 Sent GVT calculation request to SimulationControler
(18) 14:25:05,813 GVT parameter requested by SimulationControler, min. time: Infinite (unconfirmed end of
simulation for time 6379)
(19) 14:25:05,843 Received new GVT of 1906 from SimulationControler, LPCC processing needed = true
(20) 14:25:05,859 CommittedMoves now: 0, last time: 0, time diff 504ms
(21) 14:25:05,860 Uncommitted moves avg: 3028, sum: 8205555, sqr sum: 26511298705, count: 2709
(22) 14:25:05,860 Set new std derivation: 782,161434
(23) 14:25:05,860 Max uncommitted moves: 4380
(24) 14:25:05,861 Get better indicator set: non found
(25) 14:25:05,885 Current indicator set [0,3028,540,174,0,0,0,0] added to Clustered State Space
(26) 14:25:05,886 LPCC processed, total committed moves: 0
(27) 14:25:05,887 Actuator upper limit 9223372036854775807, mean limit 9223372036854775807
(28) 14:25:05,895 GVT parameter requested by SimulationControler, min. time: Infinite (unconfirmed end of
simulation for time 6379)
(29) 14:25:05,907 Received new GVT of 1910 from SimulationControler, LPCC processing needed = false
```

```
(30) 14:25:07,367  1 received cancelbacks require rollback to time 6373
(31) 14:25:07,372  Rollback for time 6373, state restored for time 6373, time before rollback 6380
(32) 14:25:07,382  2 received cancelbacks require rollback to time 6368
(33) 14:25:07,398  Rollback for time 6368, state restored for time 6368, time before rollback 6374
(34) 14:25:07,406  6 received cancelbacks require rollback to time 6358
(35) 14:25:07,410  Rollback for time 6358, state restored for time 6358, time before rollback 6369
(36) 14:25:07,415  2 received cancelbacks require rollback to time 6347
(37) 14:25:07,421  Rollback for time 6347, state restored for time 6347, time before rollback 6359
(38) 14:25:07,428  4 received cancelbacks require rollback to time 6338
(39) 14:25:07,432  Rollback for time 6338, state restored for time 6338, time before rollback 6348
(40) 14:25:07,499  Unconfirmed End Of Simulation reached by xact: xact(Id: 8759, move time: 6379, current
block: (1,3), next block: (1,3))
(41) 14:25:07,519  1 received cancelbacks require rollback to time 6373
(42) 14:25:07,523  Rollback for time 6373, state restored for time 6373, time before rollback 6380
(43) 14:25:07,527  2 received cancelbacks require rollback to time 6368
(44) 14:25:07,532  Rollback for time 6368, state restored for time 6368, time before rollback 6374
(45) 14:25:07,537  3 received cancelbacks require rollback to time 6364
(46) 14:25:07,541  Rollback for time 6364, state restored for time 6364, time before rollback 6369
(47) 14:25:07,545  3 received cancelbacks require rollback to time 6358
(48) 14:25:07,549  Rollback for time 6358, state restored for time 6358, time before rollback 6365
(49) 14:25:07,554  2 received cancelbacks require rollback to time 6347
(50) 14:25:07,558  Rollback for time 6347, state restored for time 6347, time before rollback 6359
(51) 14:25:07,568  2 received cancelbacks require rollback to time 6344
(52) 14:25:07,573  Rollback for time 6344, state restored for time 6344, time before rollback 6348
(53) 14:25:07,578  2 received cancelbacks require rollback to time 6338
(54) 14:25:07,582  Rollback for time 6338, state restored for time 6338, time before rollback 6345
(55) 14:25:07,635  Unconfirmed End Of Simulation reached by xact: xact(Id: 8759, move time: 6379, current
block: (1,3), next block: (1,3))
(56) 14:25:07,647  1 received cancelbacks require rollback to time 6373
(57) 14:25:07,650  Rollback for time 6373, state restored for time 6373, time before rollback 6380
(58) 14:25:07,660  3 received cancelbacks require rollback to time 6367
(59) 14:25:07,666  Rollback for time 6367, state restored for time 6367, time before rollback 6374
(60) 14:25:07,672  3 received cancelbacks require rollback to time 6362
(61) 14:25:07,676  Rollback for time 6362, state restored for time 6362, time before rollback 6368
(62) 14:25:07,681  3 received cancelbacks require rollback to time 6357
```

```
(63)  14:25:07,686  Rollback for time 6357, state restored for time 6357, time before rollback 6363
(64)  14:25:07,691  2 received cancelbacks require rollback to time 6345
(65)  14:25:07,696  Rollback for time 6345, state restored for time 6345, time before rollback 6358
(66)  14:25:07,701  3 received cancelbacks require rollback to time 6338
(67)  14:25:07,705  Rollback for time 6338, state restored for time 6338, time before rollback 6346
(68)  14:25:07,761  Unconfirmed End Of Simulation reached by xact(id: 8759, move time: 6379, current
block: (1,3), next block: (1,3))
(69)  14:25:07,775  1 received cancelbacks require rollback to time 6373
(70)  14:25:07,780  Rollback for time 6373, state restored for time 6373, time before rollback 6380
(71)  14:25:07,785  3 received cancelbacks require rollback to time 6367
(72)  14:25:07,789  Rollback for time 6367, state restored for time 6367, time before rollback 6374
(73)  14:25:07,797  2 received cancelbacks require rollback to time 6364
(74)  14:25:07,800  Rollback for time 6364, state restored for time 6364, time before rollback 6368
...
(401)  14:25:10,784  2 received cancelbacks require rollback to time 6160
(402)  14:25:10,788  Rollback for time 6160, state restored for time 6160, time before rollback 6169
(403)  14:25:10,795  GVT parameter requested by SimulationControler, min. time: 6160
(404)  14:25:10,819  1 received cancelbacks require rollback to time 6158
(405)  14:25:10,822  Rollback for time 6158, state restored for time 6158, time before rollback 6161
(406)  14:25:10,876  Received new GVT of 2147 from SimulationController, LPCC processing needed = true
(407)  14:25:10,884  Simulation state committed for time 2147
(408)  14:25:10,885  CommittedMoves now: 147, last time: 0, time diff 5026ms
(409)  14:25:10,886  Uncommitted moves avg: 4273, sum: 6838298, sqr sum: 2923096534, count: 1600
(410)  14:25:10,886  Set new std derivation: 53,148661
(411)  14:25:10,887  Max uncommitted moves: 4380
(412)  14:25:10,887  Get better indicator set: non found
(413)  14:25:10,888  Current indicator set [29.4273,317.82,0,0,0,352] added to Clustered State Space
(414)  14:25:10,889  LPCC processed, total committed moves: 147
(415)  14:25:10,890  Actuator upper limit 92233720368547758071, mean limit 92233720368547758071
(416)  14:25:10,908  1 received cancelbacks require rollback to time 6202
(417)  14:25:10,912  Rollback for time 6202, state restored for time 6202, time before rollback 6222
(418)  14:25:10,916  2 received cancelbacks require rollback to time 6200
(419)  14:25:10,920  Rollback for time 6200, state restored for time 6200, time before rollback 6203
...
(1136)  14:25:15,960  2 received cancelbacks require rollback to time 5276
```

```
(1137) 14:25:15,965 Rollback for time 5276, state restored for time 5276, time before rollback 5285
(1138) 14:25:15,971 GVT parameter requested by SimulationController, min. time: 5276
(1139) 14:25:15,977 1 received cancelbacks require rollback to time 5274
(1140) 14:25:15,980 Rollback for time 5274, state restored for time 5274, time before rollback 5277
(1141) 14:25:16,009 1 received cancelbacks require rollback to time 5284
(1142) 14:25:16,011 Rollback for time 5284, state restored for time 5284, time before rollback 5287
(1143) 14:25:16,015 2 received cancelbacks require rollback to time 5274
(1144) 14:25:16,019 Rollback for time 5274, state restored for time 5274, time before rollback 5285
(1145) 14:25:16,025 3 received cancelbacks require rollback to time 5267
(1146) 14:25:16,029 Rollback for time 5267, state restored for time 5267, time before rollback 5275
(1147) 14:25:16,063 1 received cancelbacks require rollback to time 5286
(1148) 14:25:16,065 Rollback for time 5286, state restored for time 5286, time before rollback 5290
(1149) 14:25:16,069 2 received cancelbacks require rollback to time 5284
(1150) 14:25:16,072 Rollback for time 5284, state restored for time 5284, time before rollback 5287
(1151) 14:25:16,079 1 received cancelbacks require rollback to time 5283
(1152) 14:25:16,082 Rollback for time 5283, state restored for time 5283, time before rollback 5285
(1153) 14:25:16,086 2 received cancelbacks require rollback to time 5274
(1154) 14:25:16,089 Rollback for time 5274, state restored for time 5274, time before rollback 5284
(1155) 14:25:16,094 2 received cancelbacks require rollback to time 5268
(1156) 14:25:16,097 Rollback for time 5268, state restored for time 5268, time before rollback 5275
(1157) 14:25:16,103 2 received cancelbacks require rollback to time 5265
(1158) 14:25:16,109 Rollback for time 5265, state restored for time 5265, time before rollback 5269
(1159) 14:25:16,114 Received new GVT of 2187 from SimulationController, LPCC processing needed = true
(1160) 14:25:16,121 Simulation state committed for time 2187
(1161) 14:25:16,123 CommittedMoves now: 187, last time: 147, time diff 5237ms
(1162) 14:25:16,124 Uncommitted moves avg: 3434, sum: 744678l, sqr sum: 2576303431l, count: 2168
(1163) 14:25:16,124 Set new std derivation: 291,684951
(1164) 14:25:16,124 Max uncommitted moves: 4076
(1165) 14:25:16,124 Get better indicator set: [29,4273,317,82,0,0,352]
(1166) 14:25:16,125 Current indicator set [7,3434,413,127,0,0,592] added to Clustered State Space
(1167) 14:25:16,125 LPCC processed, total committed moves: 187
(1168) 14:25:16,125 Actuator upper limit 4846, mean limit 4273
(1169) 14:25:18,160 Unconfirmed End Of Simulation reached by xact: xact(Id: 8759, move time: 6379, current block: (1,3), next block: (1,3))
(1170) 14:25:18,176 Sent GVT calculation request to SimulationController
```

```
(1171) 14:25:18,188 GVT parameter requested by SimulationControler, min. time: Infinite (unconfirmed end of
simulation for time 6379)
(1172) 14:25:18,221 Received new GVT of 2270 from SimulationControler, LPCC processing needed = false
(1173) 14:25:18,230 Simulation state committed for time 2270
(1174) 14:25:20,842 1 received cancelbacks require rollback to time 6373
(1175) 14:25:20,844 Rollback for time 6373, state restored for time 6373, time before rollback 6380
(1176) 14:25:20,851 2 received cancelbacks require rollback to time 6368
(1177) 14:25:20,853 Rollback for time 6368, state restored for time 6368, time before rollback 6374
(1178) 14:25:20,857 1 received cancelbacks require rollback to time 6367
(1179) 14:25:20,860 Rollback for time 6367, state restored for time 6367, time before rollback 6369
(1180) 14:25:20,864 2 received cancelbacks require rollback to time 6364
(1181) 14:25:20,867 Rollback for time 6364, state restored for time 6364, time before rollback 6368
(1182) 14:25:20,871 2 received cancelbacks require rollback to time 6360
(1183) 14:25:20,876 Rollback for time 6360, state restored for time 6360, time before rollback 6365
(1184) 14:25:20,881 3 received cancelbacks require rollback to time 6347
(1185) 14:25:20,884 Rollback for time 6347, state restored for time 6347, time before rollback 6361
(1186) 14:25:20,888 2 received cancelbacks require rollback to time 6344
(1187) 14:25:20,891 Rollback for time 6344, state restored for time 6344, time before rollback 6348
(1188) 14:25:20,895 2 received cancelbacks require rollback to time 6338
(1189) 14:25:20,898 Rollback for time 6338, state restored for time 6338, time before rollback 6345
(1190) 14:25:20,903 1 received cancelbacks require rollback to time 6334
(1191) 14:25:20,906 Rollback for time 6334, state restored for time 6334, time before rollback 6339
(1192) 14:25:20,964 Unconfirmed End of Simulation reached by xact: xact(Id: 8759, move time: 6379, current
block: (1,3), next block: (1,3))
(1193) 14:25:20,973 Sent GVT calculation request to SimulationControler
(1194) 14:25:20,983 GVT parameter requested by SimulationControler, min. time: Infinite (unconfirmed end of
simulation for time 6379)
(1195) 14:25:20,991 1 received cancelbacks require rollback to time 6373
(1196) 14:25:20,994 Rollback for time 6373, state restored for time 6373, time before rollback 6380
(1197) 14:25:20,998 5 received cancelbacks require rollback to time 6364
(1198) 14:25:21,003 Rollback for time 6364, state restored for time 6364, time before rollback 6374
(1199) 14:25:21,011 2 received cancelbacks require rollback to time 6360
(1200) 14:25:21,014 Rollback for time 6360, state restored for time 6360, time before rollback 6365
(1201) 14:25:21,018 3 received cancelbacks require rollback to time 6347
(1202) 14:25:21,022 Rollback for time 6347, state restored for time 6347, time before rollback 6361
```

```
(1203) 14:25:21,026 2 received cancelbacks require rollback to time 6344
(1204) 14:25:21,029 Rollback for time 6344, state restored for time 6344, time before rollback 6348
(1205) 14:25:21,034 3 received cancelbacks require rollback to time 6334
(1206) 14:25:21,037 Rollback for time 6334, state restored for time 6334, time before rollback 6345
(1207) 14:25:21,076 Received new GVT of 2623 from SimulationController, LPCC processing needed = true
(1208) 14:25:21,089 Simulation state committed for time 2623
(1209) 14:25:21,091 CommittedMoves now: 623, last time: 187, time diff 4968ms
(1210) 14:25:21,091 Uncommitted moves avg: 3666, sum: 4355324, sqr sum: 1609601296, count: 1188
(1211) 14:25:21,092 Set new std derivation: 329,623032
(1212) 14:25:21,092 Max uncommitted moves: 4193
(1213) 14:25:21,092 Get better indicator set: non found
(1214) 14:25:21,092 Current indicator set [87,3666,238,71,0,0,20] added to Clustered State Space
(1215) 14:25:21,093 LPCC processed, total committed moves: 623
(1216) 14:25:21,093 Actuator upper limit 9223372036854775807, mean limit 9223372036854775807
(1217) 14:25:21,134 Unconfirmed End Of Simulation reached by xact: xact(Id: 8759, move time: 6379, current
block: (1,3), next block: (1,3))
(1218) 14:25:21,151 Sent GVT calculation request to SimulationController
(1219) 14:25:21,161 GVT parameter requested by SimulationControler, min. time: Infinite (unconfirmed end of
simulation for time 6379)
(1220) 14:25:21,189 Received new GVT of 2628 from SimulationController, LPCC processing needed = false
(1221) 14:25:21,195 Simulation state committed for time 2628
(1222) 14:25:21,420 1 received cancelbacks require rollback to time 6373
(1223) 14:25:21,423 Rollback for time 6373, state restored for time 6373, time before rollback 6380
(1224) 14:25:21,427 1 received cancelbacks require rollback to time 6369
(1225) 14:25:21,430 Rollback for time 6369, state restored for time 6369, time before rollback 6374
(1226) 14:25:21,434 3 received cancelbacks require rollback to time 6366
(1227) 14:25:21,437 Rollback for time 6366, state restored for time 6366, time before rollback 6370
(1228) 14:25:21,445 2 received cancelbacks require rollback to time 6362
(1229) 14:25:21,448 Rollback for time 6362, state restored for time 6362, time before rollback 6367
(1230) 14:25:21,452 2 received cancelbacks require rollback to time 6358
(1231) 14:25:21,455 Rollback for time 6358, state restored for time 6358, time before rollback 6363
(1232) 14:25:21,459 3 received cancelbacks require rollback to time 6345
(1233) 14:25:21,463 Rollback for time 6345, state restored for time 6345, time before rollback 6359
(1234) 14:25:21,511 Unconfirmed End Of Simulation reached by xact: xact(Id: 8759, move time: 6379, current
block: (1,3), next block: (1,3))
```

```
(1235) 14:25:21,519  Sent GVT calculation request to SimulationController
(1236) 14:25:21,525  GVT parameter requested by SimulationControler, min. time: Infinite (unconfirmed end of
simulation for time 6379)
(1237) 14:25:21,537  1 received cancelbacks require rollback to time 6373
(1238) 14:25:21,539  Rollback for time 6373, state restored for time 6373, time before rollback 6380
(1239) 14:25:21,543  1 received cancelbacks require rollback to time 6369
(1240) 14:25:21,546  Rollback for time 6369, state restored for time 6369, time before rollback 6375
(1241) 14:25:21,552  2 received cancelbacks require rollback to time 6367
(1242) 14:25:21,555  Rollback for time 6367, state restored for time 6367, time before rollback 6370
(1243) 14:25:21,559  2 received cancelbacks require rollback to time 6364
(1244) 14:25:21,562  Rollback for time 6364, state restored for time 6364, time before rollback 6368
(1245) 14:25:21,567  3 received cancelbacks require rollback to time 6358
(1246) 14:25:21,570  Rollback for time 6358, state restored for time 6358, time before rollback 6365
(1247) 14:25:21,575  3 received cancelbacks require rollback to time 6345
(1248) 14:25:21,579  Rollback for time 6345, state restored for time 6345, time before rollback 6359
(1249) 14:25:21,591  Received new GVT of 2663 from SimulationController, LPCC processing needed = false
(1250) 14:25:21,597  Simulation state committed for time 2663
(1251) 14:25:21,636  Unconfirmed End Of Simulation reached by xact: xact(Id: 8759, move time: 6379, current
block: (1,3), next block: (1,3))
(1252) 14:25:21,658  Sent GVT calculation request to SimulationController
(1253) 14:25:21,664  GVT parameter requested by SimulationControler, min. time: Infinite (unconfirmed end of
simulation for time 6379)
(1254) 14:25:21,687  Received new GVT of 2667 from SimulationController, LPCC processing needed = false
(1255) 14:25:21,694  Simulation state committed for time 2667
(1256) 14:25:21,992  1 received cancelbacks require rollback to time 6373
(1257) 14:25:21,995  Rollback for time 6373, state restored for time 6373, time before rollback 6380
(1258) 14:25:22,001  2 received cancelbacks require rollback to time 6368
(1259) 14:25:22,004  Rollback for time 6368, state restored for time 6368, time before rollback 6374
(1260) 14:25:22,008  1 received cancelbacks require rollback to time 6367
(1261) 14:25:22,012  Rollback for time 6367, state restored for time 6367, time before rollback 6369
(1262) 14:25:22,016  2 received cancelbacks require rollback to time 6364
(1263) 14:25:22,019  Rollback for time 6364, state restored for time 6364, time before rollback 6368
(1264) 14:25:22,032  2 received cancelbacks require rollback to time 6360
(1265) 14:25:22,037  Rollback for time 6360, state restored for time 6360, time before rollback 6365
(1266) 14:25:22,042  2 received cancelbacks require rollback to time 6357
```

```
(1267) 14:25:22,045  Rollback for time 6357, state restored for time 6357, time before rollback 6361
(1268) 14:25:22,057  2 received cancelbacks require rollback to time 6345
(1269) 14:25:22,060  Rollback for time 6345, state restored for time 6345, time before rollback 6359
(1270) 14:25:22,097  Unconfirmed End Of Simulation reached by xact: xact(Id: 8759, move time: 6379, current
block: (1,3), next block: (1,3))
(1271) 14:25:22,108  Sent GVT calculation request to SimulationController
(1272) 14:25:22,115  GVT parameter requested by SimulationController, min. time: Infinite (unconfirmed end of
simulation for time 6379)
(1273) 14:25:22,122  1 received cancelbacks require rollback to time 6373
(1274) 14:25:22,125  Rollback for time 6373, state restored for time 6373, time before rollback 6380
(1275) 14:25:22,132  4 received cancelbacks require rollback to time 6366
(1276) 14:25:22,135  Rollback for time 6366, state restored for time 6366, time before rollback 6374
(1277) 14:25:22,139  3 received cancelbacks require rollback to time 6360
(1278) 14:25:22,143  Rollback for time 6360, state restored for time 6360, time before rollback 6367
(1279) 14:25:22,150  2 received cancelbacks require rollback to time 6357
(1280) 14:25:22,153  Rollback for time 6357, state restored for time 6357, time before rollback 6361
(1281) 14:25:22,157  2 received cancelbacks require rollback to time 6345
(1282) 14:25:22,165  Rollback for time 6345, state restored for time 6345, time before rollback 6358
(1283) 14:25:22,177  Received new GVT of 2711 from SimulationController, LPCC processing needed = false
(1284) 14:25:22,183  Simulation state committed for time 2711
(1285) 14:25:22,232  Unconfirmed End Of Simulation reached by xact: xact(Id: 8759, move time: 6379, current
block: (1,3), next block: (1,3))
(1286) 14:25:22,240  Sent GVT calculation request to SimulationController
(1287) 14:25:22,252  GVT parameter requested by SimulationController, min. time: Infinite (unconfirmed end of
simulation for time 6379)
(1288) 14:25:22,261  Received new GVT of 2714 from SimulationController, LPCC processing needed = false
(1289) 14:25:22,268  Simulation state committed for time 2714
(1290) 14:25:22,686  Received a request from other LP for local LP to initiate GVT calculation when it passed
the simulation time 2767
(1291) 14:25:22,694  Sent GVT calculation request to SimulationController
(1292) 14:25:22,700  GVT parameter requested by SimulationController, min. time: Infinite (unconfirmed end of
simulation for time 6379)
(1293) 14:25:22,732  SimulationController reported confirmed End of Simulation by xact: xact(Id: 5532, move
time: 2766, current block: (2,2), next block: (2,2))
(1294) 14:25:22,739  Rollback for time 2766, state restored for time 2766, time before rollback 6380
```

```
(1295) 14:25:22,740 Simulation stopped and simulation state changed to TERMINATED
(1296) 14:25:23,189 LogicalProcess.runActivity() left (ActiveObject stopped)
```

**Validation 6.1, output of LP2:**

```
(1)
(2)  --- StartNode ------------------------------------
(3)  14:24:54,858 LogicalProcess.runActivity() started
(4)  14:24:55,612 LP2 with partition 'Partition2' initialized
(5)  14:24:55,614 LP simulation state changed from INITIALIZED to SIMULATING
(6)  14:25:00,782 GVT parameter requested by SimulationController, min. time: 1624
(7)  14:25:00,851 Received new GVT of 1624 from SimulationController, LPCC processing needed = true
(8)  14:25:00,879 Simulation state committed for time 1624
(9)  14:25:00,908 CommittedMoves now: 1623, last time: 0, time diff 5339ms
(10) 14:25:00,909 Uncommitted moves avg: 915, sum: 1493530, sqr sum: 1931525766, count: 1631
(11) 14:25:00,912 Set new std derivation: 588,164979
(12) 14:25:00,912 Max uncommitted moves: 2138
(13) 14:25:00,912 Get better indicator set: non found
(14) 14:25:00,921 Current indicator set [303,915,305,0,0,95,0,0] added to Clustered State Space
(15) 14:25:00,921 LPCC processed, total committed moves: 1623
(16) 14:25:00,922 Actuator upper limit 9223372036854775807, mean limit 9223372036854775807
(17) 14:25:05,793 Received a request from other LP for local LP to initiate GVT calculation when it passed
the simulation time 6380
(18) 14:25:05,793 GVT parameter requested by SimulationController, min. time: 1906
(19) 14:25:05,843 Received new GVT of 1906 from SimulationController, LPCC processing needed = true
(20) 14:25:05,853 Simulation state committed for time 1906
(21) 14:25:05,886 CommittedMoves now: 1905, last time: 1623, time diff 4978ms
(22) 14:25:05,886 Uncommitted moves avg: 1045, sum: 291669, sqr sum: 335528473, count: 279
(23) 14:25:05,887 Set new std derivation: 331,853263
(24) 14:25:05,887 Max uncommitted moves: 1667
(25) 14:25:05,888 Get better indicator set: [303,915,305,0,0,95,0,0]
(26) 14:25:05,896 Current indicator set [56,1045,56,0,0,175,0,0] added to Clustered State Space
(27) 14:25:05,896 LPCC processed, total committed moves: 1905
(28) 14:25:05,897 Actuator upper limit 1566, mean limit 915
(29) 14:25:05,897 GVT parameter requested by SimulationController, min. time: 1910
(30) 14:25:05,929 Received new GVT of 1910 from SimulationController, LPCC processing needed = false
(31) 14:25:05,930 Simulation state committed for time 1910
```

```
(32) 14:25:07,331 Actuator limit (limit of uncommited transaction moves) exceeded, current uncommitted moves:
1567, limit: 1566
(33) 14:25:07,331 Changed Cancelback mode to ON because of actuator limit reached
(34) 14:25:07,427 Cancelback mode - cancelled back 15 received transactions
(35) 14:25:07,449 Changed Cancelback mode to OFF
(36) 14:25:07,512 Actuator limit (limit of uncommited transaction moves) exceeded, current uncommitted moves:
1569, limit: 1566
(37) 14:25:07,513 Changed Cancelback mode to ON because of actuator limit reached
(38) 14:25:07,573 Cancelback mode - cancelled back 15 received transactions
(39) 14:25:07,574 Received a request from other LP for local LP to initiate GVT calculation when it passed
the simulation time 6380
(40) 14:25:07,583 Changed Cancelback mode to OFF
(41) 14:25:07,644 Received a request from other LP for local LP to initiate GVT calculation when it passed
the simulation time 6380
(42) 14:25:07,644 Actuator limit (limit of uncommited transaction moves) exceeded, current uncommitted moves:
1573, limit: 1566
(43) 14:25:07,644 Changed Cancelback mode to ON because of actuator limit reached
(44) 14:25:07,701 Cancelback mode - cancelled back 15 received transactions
(45) 14:25:07,714 Changed Cancelback mode to OFF
(46) 14:25:07,769 Received a request from other LP for local LP to initiate GVT calculation when it passed
the simulation time 6380
(47) 14:25:07,770 Actuator limit (limit of uncommited transaction moves) exceeded, current uncommitted moves:
1571, limit: 1566
(48) 14:25:07,770 Changed Cancelback mode to ON because of actuator limit reached
(49) 14:25:07,825 Cancelback mode - cancelled back 15 received transactions
(50) 14:25:07,837 Changed Cancelback mode to OFF
(51) 14:25:07,875 Actuator limit (limit of uncommited transaction moves) exceeded, current uncommitted moves:
1567, limit: 1566
(52) 14:25:07,875 Changed Cancelback mode to ON because of actuator limit reached
(53) 14:25:07,928 Cancelback mode - cancelled back 15 received transactions
(54) 14:25:07,947 Changed Cancelback mode to OFF
...
(155) 14:25:10,735 Actuator limit (limit of uncommited transaction moves) exceeded, current uncommitted
moves: 1568, limit: 1566
(156) 14:25:10,735 Changed Cancelback mode to ON because of actuator limit reached
```

```
(157) 14:25:10,794 Cancelback mode – cancelled back 15 received transactions
(158) 14:25:10,804 Changed Cancelback mode to OFF
(159) 14:25:10,818 GVT parameter requested by SimulationControler, min. time: 2147
(160) 14:25:10,874 Received new GVT of 2147 from SimulationController, LPCC processing needed = true
(161) 14:25:10,875 Simulation state committed for time 2147
(162) 14:25:10,898 CommittedMoves now: 2196, last time: 1905, time diff 501ms
(163) 14:25:10,899 Uncommitted moves avg: 1502, sum: 407059, sqr sum: 612458611, count: 271
(164) 14:25:10,899 Set new std derivation: 61,777916
(165) 14:25:10,900 Max uncommitted moves: 1573
(166) 14:25:10,900 Get better indicator set: [303,915,305,0,0,95,0,0]
(167) 14:25:10,900 Current indicator set [58,1502,58,0,0,83,0,0] added to Clustered State Space
(168) 14:25:10,901 LPCC processed, total committed moves: 2196
(169) 14:25:10,901 Actuator upper limit 1037, mean limit 915
(170) 14:25:10,904 Actuator limit (limit of uncommitted transaction moves) exceeded, current uncommitted moves: 1286, limit: 1037
(171) 14:25:10,904 Changed Cancelback mode to ON because of actuator limit reached
(172) 14:25:10,947 Cancelback mode – cancelled back 12 received transactions
(173) 14:25:10,955 Actuator limit (limit of uncommitted transaction moves) exceeded, current uncommitted moves: 1275, limit: 1037
(174) 14:25:10,997 Cancelback mode – cancelled back 12 received transactions
(175) 14:25:11,006 Actuator limit (limit of uncommitted transaction moves) exceeded, current uncommitted moves: 1269, limit: 1037
(176) 14:25:11,052 Cancelback mode – cancelled back 12 received transactions
(177) 14:25:11,068 Actuator limit (limit of uncommitted transaction moves) exceeded, current uncommitted moves: 1263, limit: 1037
(178) 14:25:11,111 Cancelback mode – cancelled back 12 received transactions
(179) 14:25:11,121 Actuator limit (limit of uncommitted transaction moves) exceeded, current uncommitted moves: 1260, limit: 1037
(180) 14:25:11,166 Cancelback mode – cancelled back 12 received transactions
(181) 14:25:11,176 Actuator limit (limit of uncommitted transaction moves) exceeded, current uncommitted moves: 1255, limit: 1037
(182) 14:25:11,224 Cancelback mode – cancelled back 12 received transactions
(183) 14:25:11,237 Actuator limit (limit of uncommitted transaction moves) exceeded, current uncommitted moves: 1250, limit: 1037
(184) 14:25:11,282 Cancelback mode – cancelled back 12 received transactions
```

```
(185) 14:25:11,290 Actuator limit (limit of uncommitted transaction moves) exceeded, current uncommitted
moves: 1247, limit: 1037
(186) 14:25:11,335 Cancelback mode - cancelled back 12 received transactions
...
(404) 14:25:15,920 Actuator limit (limit of uncommitted transaction moves) exceeded, current uncommitted
moves: 1041, limit: 1037
(405) 14:25:15,920 Changed Cancelback mode to ON because of actuator limit reached
(406) 14:25:15,961 Cancelback mode - cancelled back 10 received transactions
(407) 14:25:15,968 Changed Cancelback mode to OFF
(408) 14:25:15,985 Actuator limit (limit of uncommitted transaction moves) exceeded, current uncommitted
moves: 1041, limit: 1037
(409) 14:25:15,985 Changed Cancelback mode to ON because of actuator limit reached
(410) 14:25:16,025 Cancelback mode - cancelled back 10 received transactions
(411) 14:25:16,026 GVT parameter requested by SimulationController, min. time: 2187
(412) 14:25:16,038 Changed Cancelback mode to OFF
(413) 14:25:16,056 Actuator limit (limit of uncommitted transaction moves) exceeded, current uncommitted
moves: 1040, limit: 1037
(414) 14:25:16,056 Changed Cancelback mode to ON because of actuator limit reached
(415) 14:25:16,101 Cancelback mode - cancelled back 10 received transactions
(416) 14:25:16,106 Received new GVT of 2187 from SimulationController, LPCC processing needed = true
(417) 14:25:16,107 Simulation state committed for time 2187
(418) 14:25:16,114 Changed Cancelback mode to OFF
(419) 14:25:16,127 CommittedMoves now: 2247, last time: 2196, time diff 5230ms
(420) 14:25:16,128 Uncommitted moves avg: 1086, sum: 137946, sqr sum: 150527724, count: 127
(421) 14:25:16,128 Set new std derivation: 74,124491
(422) 14:25:16,129 Max uncommitted moves: 1275
(423) 14:25:16,129 Get better indicator set: [58,1502,58,0,0,83,0,0]
(424) 14:25:16,129 Current indicator set [9,1086,9,0,0,127,0,0] added to Clustered State Space
(425) 14:25:16,129 LPCC processed, total committed moves: 2247
(426) 14:25:16,130 Actuator upper limit 1648, mean limit 1502
(427) 14:25:18,174 Received a request from other LP for local LP to initiate GVT calculation when it passed
the simulation time 6380
(428) 14:25:18,203 GVT parameter requested by SimulationController, min. time: 2270
(429) 14:25:18,223 Received new GVT of 2270 from SimulationController, LPCC processing needed = false
(430) 14:25:18,223 Simulation state committed for time 2270
```

```
(431)  14:25:20,840  Actuator limit (limit of uncommitted transaction moves) exceeded, current uncommitted
moves: 1649, limit: 1648
(432)  14:25:20,840  Changed Cancelback mode to ON because of actuator limit reached
(433)  14:25:20,896  Cancelback mode - cancelled back 16 received transactions
(434)  14:25:20,907  Changed Cancelback mode to OFF
(435)  14:25:20,976  Received a request from other LP for local LP to initiate GVT calculation when it passed
the simulation time 6380
(436)  14:25:20,976  Actuator limit (limit of uncommitted transaction moves) exceeded, current uncommitted
moves: 1651, limit: 1648
(437)  14:25:20,977  Changed Cancelback mode to ON because of actuator limit reached
(438)  14:25:21,033  Cancelback mode - cancelled back 16 received transactions
(439)  14:25:21,033  GVT parameter requested by SimulationControler, min. time: 2623
(440)  14:25:21,046  Changed Cancelback mode to OFF
(441)  14:25:21,073  Received new GVT of 2623 from SimulationControler, LPCC processing needed = true
(442)  14:25:21,074  Simulation state committed for time 2623
(443)  14:25:21,087  CommittedMoves now: 2814, last time: 2247, time diff 4960ms
(444)  14:25:21,087  Uncommitted moves avg: 1423, sum: 623552, sqr sum: 898197966, count: 438
(445)  14:25:21,087  Set new std derivation: 154,917115
(446)  14:25:21,088  Max uncommitted moves: 1651
(447)  14:25:21,088  Get better indicator set: [303,915,305,0,0,95,0,0]
(448)  14:25:21,088  Current indicator set [114,1423,114,0,0,71,0,0] added to Clustered State Space
(449)  14:25:21,088  LPCC processed, total committed moves: 2814
(450)  14:25:21,088  Actuator upper limit 1220, mean limit 915
(451)  14:25:21,152  Received a request from other LP for local LP to initiate GVT calculation when it passed
the simulation time 6380
(452)  14:25:21,168  GVT parameter requested by SimulationControler, min. time: 2628
(453)  14:25:21,193  Received new GVT of 2628 from SimulationControler, LPCC processing needed = false
(454)  14:25:21,194  Simulation state committed for time 2628
(455)  14:25:21,417  Actuator limit (limit of uncommitted transaction moves) exceeded, current uncommitted
moves: 1221, limit: 1220
(456)  14:25:21,417  Changed Cancelback mode to ON because of actuator limit reached
(457)  14:25:21,459  Cancelback mode - cancelled back 12 received transactions
(458)  14:25:21,469  Changed Cancelback mode to OFF
(459)  14:25:21,515  Received a request from other LP for local LP to initiate GVT calculation when it passed
the simulation time 6380
```

```
(460)  14:25:21,531  Actuator  limit  (limit  of  uncommited  transaction  moves)  exceeded,  current  uncommitted
moves: 1224, limit: 1220
(461)  14:25:21,532  Changed Cancelback mode to ON because of actuator limit reached
(462)  14:25:21,575  Cancelback mode - cancelled back 12 received transactions
(463)  14:25:21,575  GVT parameter requested by SimulationControler, min. time: 2663
(464)  14:25:21,597  Changed Cancelback mode to OFF
(465)  14:25:21,616  Received new GVT of 2663 from SimulationControler, LPCC processing needed = false
(466)  14:25:21,617  Simulation state committed for time 2663
(467)  14:25:21,657  Received a request from other LP for local LP to initiate GVT calculation when it passed
the simulation time 6380
(468)  14:25:21,681  GVT parameter requested by SimulationControler, min. time: 2667
(469)  14:25:21,700  Received new GVT of 2667 from SimulationControler, LPCC processing needed = false
(470)  14:25:21,702  Simulation state committed for time 2667
(471)  14:25:21,988  Actuator  limit  (limit  of  uncommited  transaction  moves)  exceeded,  current  uncommitted
moves: 1221, limit: 1220
(472)  14:25:21,989  Changed Cancelback mode to ON because of actuator limit reached
(473)  14:25:22,054  Cancelback mode - cancelled back 12 received transactions
(474)  14:25:22,066  Changed Cancelback mode to OFF
(475)  14:25:22,110  Received a request from other LP for local LP to initiate GVT calculation when it passed
the simulation time 6380
(476)  14:25:22,110  Actuator  limit  (limit  of  uncommited  transaction  moves)  exceeded,  current  uncommitted
moves: 1222, limit: 1220
(477)  14:25:22,111  Changed Cancelback mode to ON because of actuator limit reached
(478)  14:25:22,158  Cancelback mode - cancelled back 12 received transactions
(479)  14:25:22,159  GVT parameter requested by SimulationControler, min. time: 2711
(480)  14:25:22,168  Changed Cancelback mode to OFF
(481)  14:25:22,183  Received new GVT of 2711 from SimulationControler, LPCC processing needed = false
(482)  14:25:22,184  Simulation state committed for time 2711
(483)  14:25:22,238  Received a request from other LP for local LP to initiate GVT calculation when it passed
the simulation time 6380
(484)  14:25:22,255  GVT parameter requested by SimulationControler, min. time: 2714
(485)  14:25:22,275  Received new GVT of 2714 from SimulationControler, LPCC processing needed = false
(486)  14:25:22,275  Simulation state committed for time 2714
(487)  14:25:22,683  Unconfirmed End Of Simulation reached by xact: xact(Id: 5532, move time: 2766, current
block: (2,2), next block: (2,2))
```

(488) 14:25:22,693 Sent GVT calculation request to SimulationController
(489) 14:25:22,699 GVT parameter requested by SimulationController, min. time: Infinite (unconfirmed end of simulation for time 2766)
(490) 14:25:22,737 SimulationController reported confirmed End of Simulation by xact: xact(Id: 5532, move time: 2766, current block: (2,2), next block: (2,2))
(491) 14:25:22,737 Simulation stopped and simulation state changed to TERMINATED
(492) 14:25:23,193 LogicalProcess.runActivity() left (ActiveObject stopped)

**Validation 6.1, output of simulate:**

```
(1)  14:24:51,886  Simulation model file read and parsed successfully.
(2)  14:24:51,888  2 partition(s) found in simulation model file.
(3)
(4)  14:24:51,898  Creating SimulationController instance
(5)  14:24:51,898  parallelJavaGpssSimulator.SimulationController
(6)  14:24:53,435  SimulationController.runActivity() started
(7)  14:24:53,492  Creating LPs and starting simulation
(8)  14:24:55,563  Simulation started
(9)
(10) 14:24:55,563  Please press:
(11) G + <Enter> to force GVT calculation
(12) X + <Enter> to terminate the simulation
(13)
(14) 14:25:00,612  Initiated GVT calculation
(15) 14:25:00,800  Min time received from LP1: 3667
(16) 14:25:00,801  Min time received from LP2: 1624
(17) 14:25:00,863  Simulation reached Global Virtual Time: 1624
(18) 14:25:05,688  Initiated GVT calculation
(19) 14:25:05,825  Min time received from LP1: Infinite (unconfirmed end of simulation for time 6379)
(20) 14:25:05,826  Min time received from LP2: 1906
(21) 14:25:05,844  Simulation reached Global Virtual Time: 1906
(22) 14:25:05,845  Initiated GVT calculation
(23) 14:25:05,902  Min time received from LP1: Infinite (unconfirmed end of simulation for time 6379)
(24) 14:25:05,902  Min time received from LP2: 1910
(25) 14:25:05,922  Simulation reached Global Virtual Time: 1910
(26) 14:25:10,786  Initiated GVT calculation
(27) 14:25:10,842  Min time received from LP1: 6160
(28) 14:25:10,843  Min time received from LP2: 2147
(29) 14:25:10,872  Simulation reached Global Virtual Time: 2147
(30) 14:25:15,918  Initiated GVT calculation
(31) 14:25:16,034  Min time received from LP1: 5276
(32) 14:25:16,035  Min time received from LP2: 2187
```

```
(33) 14:25:16,108  Simulation reached Global Virtual Time: 2187
(34) 14:25:18,175  Initiated GVT calculation
(35) 14:25:18,207  Min time received from LP1: Infinite (unconfirmed end of simulation for time 6379)
(36) 14:25:18,207  Min time received from LP2: 2270
(37) 14:25:18,231  Simulation reached Global Virtual Time: 2270
(38) 14:25:20,974  Initiated GVT calculation
(39) 14:25:21,036  Min time received from LP1: Infinite (unconfirmed end of simulation for time 6379)
(40) 14:25:21,038  Min time received from LP2: 2623
(41) 14:25:21,109  Simulation reached Global Virtual Time: 2623
(42) 14:25:21,149  Initiated GVT calculation
(43) 14:25:21,172  Min time received from LP1: Infinite (unconfirmed end of simulation for time 6379)
(44) 14:25:21,176  Min time received from LP2: 2628
(45) 14:25:21,186  Simulation reached Global Virtual Time: 2628
(46) 14:25:21,519  Initiated GVT calculation
(47) 14:25:21,581  Min time received from LP1: Infinite (unconfirmed end of simulation for time 6379)
(48) 14:25:21,581  Min time received from LP2: 2663
(49) 14:25:21,634  Simulation reached Global Virtual Time: 2663
(50) 14:25:21,656  Initiated GVT calculation
(51) 14:25:21,683  Min time received from LP1: Infinite (unconfirmed end of simulation for time 6379)
(52) 14:25:21,683  Min time received from LP2: 2667
(53) 14:25:21,693  Simulation reached Global Virtual Time: 2667
(54) 14:25:22,107  Initiated GVT calculation
(55) 14:25:22,161  Min time received from LP1: Infinite (unconfirmed end of simulation for time 6379)
(56) 14:25:22,162  Min time received from LP2: 2711
(57) 14:25:22,169  Simulation reached Global Virtual Time: 2711
(58) 14:25:22,239  Initiated GVT calculation
(59) 14:25:22,257  Min time received from LP1: Infinite (unconfirmed end of simulation for time 6379)
(60) 14:25:22,257  Min time received from LP2: 2714
(61) 14:25:22,266  Simulation reached Global Virtual Time: 2714
(62) 14:25:22,693  Initiated GVT calculation
(63) 14:25:22,708  Min time received from LP1: Infinite (unconfirmed end of simulation for time 6379)
(64) 14:25:22,709  Min time received from LP2: Infinite (unconfirmed end of simulation for time 2766)
(65) 14:25:22,747  Simulation finished
(66) 14:25:22,883  **************  Simulation report  **************
(67) 14:25:22,888  The simulation was completed at the simulation time: 2766
```

```
(68) 14:25:22,888 by the transaction xact(Id: 5532, move time: 2766, current block: (2,2), next block: (2,2))
(69) 14:25:22,890 The average simulation perfomance in simulation time per second real time was: 101,188950
(time units/s)
(70)
(71) 14:25:22,891 Block report section:
(72) Block                          total
(73)             xacts            xacts
(74) Partition: Partition1
(75) (1,1)            1            767    Block: GENERATE 1,0,2000,(no limit),0
(76) (1,2)            0            767    Block: TRANSFER 0.3,(Label: Label1, partition: 2, block: 2)
(77) (1,3)            0            533    Block: TERMINATE 1
(78) Partition: Partition2
(79) (2,1)            1            2766   Block: GENERATE 1,0,(no offset),(no limit),0
(80) (2,2)            0            3000   Block: TERMINATE 1
(81)
(82) 14:25:22,893 Summary entity report section:
(83)
(84) 14:25:23,176 SimulationController.terminateLPs() called
```

**Validation 6.2, output of LP1:**

(1) ------ StartNode --------------------------------------
(2) --- StartNode --------------------------------------
(3) 14:44:15,530 LogicalProcess.runActivity() started
(4) 14:44:16,183 LP1 with partition 'Partition1' initialized
(5) 14:44:16,494 LP simulation state changed from INITIALIZED to SIMULATING
(6) 14:44:25,977 Unconfirmed End Of Simulation reached by xact: xact(Id: 8475, move time: 6237, current block: (1,3), next block: (1,3))
(7) 14:44:26,011 Sent GVT calculation request to SimulationController
(8) 14:44:26,086 GVT parameter requested by SimulationController, min. time: Infinite (unconfirmed end of simulation for time 6237)
(9) 14:44:26,130 Received new GVT of 1802 from SimulationController, LPCC processing needed = false
(10) 14:44:32,860 Received a request from other LP for local LP to initiate GVT calculation when it passed the simulation time 2765
(11) 14:44:32,888 GVT parameter requested by SimulationController, min. time: Infinite (unconfirmed end of simulation for time 6237)
(12) 14:44:32,919 SimulationController reported confirmed End of Simulation by xact: xact(Id: 5528, move time: 2764, current block: (2,2), next block: (2,2))
(13) 14:44:32,943 Rollback for time 2764, state restored for time 2764, time before rollback 6238
(14) 14:44:32,945 Simulation stopped and simulation state changed to TERMINATED
(15) 14:44:33,790 LogicalProcess.runActivity() left (ActiveObject stopped)

**Validation 6.2, output of LP2:**

(1)
(2) ---- StartNode ---------------------------------------
(3) 14:44:15,765 LogicalProcess.runActivity() started
(4) 14:44:16,574 LP2 with partition 'Partition2' initialized
(5) 14:44:16,577 LP simulation state changed from INITIALIZED to SIMULATING
(6) 14:44:26,005 Received a request from other LP for local LP to initiate GVT calculation when it passed the simulation time 6238
(7) 14:44:26,082 GVT parameter requested by SimulationController, min. time: 1802
(8) 14:44:26,125 Received new GVT of 1802 from SimulationController, LPCC processing needed = false
(9) 14:44:26,156 Simulation state committed for time 1802
(10) 14:44:32,821 Unconfirmed End of Simulation reached by xact: xact(Id: 5528, move time: 2764, current block: (2,2), next block: (2,2))
(11) 14:44:32,868 Sent GVT calculation request to SimulationController
(12) 14:44:32,875 GVT parameter requested by SimulationController, min. time: Infinite (unconfirmed end of simulation for time 2764)
(13) 14:44:32,919 SimulationController reported confirmed End of Simulation by xact: xact(Id: 5528, move time: 2764, current block: (2,2), next block: (2,2))
(14) 14:44:32,919 Simulation stopped and simulation state changed to TERMINATED
(15) 14:44:33,789 LogicalProcess.runActivity() left (ActiveObject stopped)

**Validation 6.2, output of simulate:**

(1) 14:44:12,815 Simulation model file read and parsed successfully.
(2) 14:44:12,817 2 partition(s) found in simulation model file.
(3)
(4) 14:44:12,827 Creating SimulationController instance
(5) 14:44:12,827 parallelJavaGpssSimulator.SimulationController
(6) 14:44:14,346 SimulationController.runActivity() started
(7) 14:44:14,404 Creating LPs and starting simulation
(8) 14:44:16,527 Simulation started
(9)
(10) 14:44:16,528 Please press:
(11) G + <Enter> to force GVT calculation
(12) X + <Enter> to terminate the simulation
(13)
(14) 14:44:26,010 Initiated GVT calculation
(15) 14:44:26,108 Min time received from LP1: Infinite (unconfirmed end of simulation for time 6237)
(16) 14:44:26,109 Min time received from LP2: 1802
(17) 14:44:26,132 Simulation reached Global Virtual Time: 1802
(18) 14:44:32,867 Initiated GVT calculation
(19) 14:44:32,894 Min time received from LP1: Infinite (unconfirmed end of simulation for time 6237)
(20) 14:44:32,894 Min time received from LP2: Infinite (unconfirmed end of simulation for time 2764)
(21) 14:44:32,947 Simulation finished
(22) 14:44:33,071 ***************** Simulation report *****************
(23) 14:44:33,077 The simulation was completed at the simulation time: 2764
(24) 14:44:33,078 by the transaction xact(Id: 5528, move time: 2764, current block: (2,2), next block: (2,2))
(25) 14:44:33,082 The average simulation perfomance in simulation time per second real time was: 166,898132 (time units/s)
(26)
(27) 14:44:33,082 Block report section:
(28) Block            current      total
(29)                  xacts        xacts
(30) Partition: Partition1
(31) (1,1)            1            765      Block: GENERATE 1,0,2000,(no limit),0

```
(32) (1,2)                  0       765   Block: TRANSFER 0.3,(Label: Label1, partition: 2, block: 2)
(33) (1,3)                  0       529   Block: TERMINATE 1
(34) Partition: Partition2
(35) (2,1)                  1      2764   Block: GENERATE 1,0,(no offset),(no limit),0
(36) (2,2)                  0      3000   Block: TERMINATE 1
(37)
(38) 14:44:33,083 Summary entity report section:
(39)
(40) 14:44:33,774 SimulationController.terminateLPs() called
```